\documentclass[12pt]{report}

\usepackage{styles/thesis_packages}

\newif\ifcolorfigure
\colorfiguretrue

\usepackage{graphicx}

\usepackage{hyperref}

\usepackage[acronym,shortcuts]{glossaries}
\makeglossaries

\usepackage{amssymb}

\usepackage{amsmath}

\usepackage{xspace}

\usepackage{url}

\usepackage{xstring}

\newcommand{\absval}[1]{\ensuremath{|#1|}}

\newcommand{\intspace}{\;}

\newcommand{\derivative}{\ensuremath{\text{d}}\xspace}

\newcommand{\solidangle}{\ensuremath{\vec\Omega}\xspace}
\newcommand{\dsolidangle}{\ensuremath{d\Omega}\xspace}

\newcommand{\degree}{\ensuremath{^\circ}\xspace}

\newcommand\arcsec{\mbox{$^{\prime\prime}$}}%
\newcommand\fdg{\mbox{$.\!\!^\circ$}}%
\newcommand\arcmin{\mbox{$^\prime$}}%

\newcommand{\myvector}[1]{\ensuremath{\boldsymbol{#1}}\xspace}

\newcommand{\cross}[2]{\ensuremath{#1\!\boldsymbol{\times}\!#2}}

\newcommand{\minuit}{\ensuremath{\mathtt{minuit}}\xspace}

\newcommand{\hypothesis}{\ensuremath{\text{H}}\xspace}

\newcommand{\energy}{\ensuremath{E}\xspace}
\newcommand{\energydot}{\ensuremath{\dot{\energy}}\xspace}
\newcommand{\denergy}{\ensuremath{\derivative\energy}\xspace}

\newcommand{\EnergyDensity}{\ensuremath{U}\xspace}

\newcommand{\momentum}{\ensuremath{p}\xspace}

\renewcommand{\time}{\ensuremath{t}\xspace}
\newcommand{\dtime}{\ensuremath{\derivative\time}\xspace}
\newcommand{\dbydt}{\ensuremath{\frac{\derivative}{\derivative\time}}\xspace}

\renewcommand{\time}{\ensuremath{t}\xspace}

\newcommand{\CrossSection}{\ensuremath{\sigma}\xspace}

\newcommand{\power}{\ensuremath{P}\xspace}
\newcommand{\TotalPower}{\ensuremath{\power_\text{tot}}\xspace}

\newcommand{\speedoflight}{\ensuremath{c}\xspace}
\newcommand{\gravitationalconstant}{\ensuremath{G}\xspace}

\newcommand{\mass}{\ensuremath{m}\xspace}
\newcommand{\charge}{\ensuremath{q}\xspace}

\newcommand{\ParticleDistribution}{\ensuremath{N}\xspace}

\newcommand{\MassHydrogen}{\ensuremath{m_\text{H}}\xspace}

\newcommand{\proton}{\ensuremath{p}\xspace}
\newcommand{\hydrogen}{\ensuremath{H}\xspace}
\newcommand{\electron}{\ensuremath{e}\xspace}
\newcommand{\neutron}{\ensuremath{n}\xspace}
\newcommand{\neutrino}{\ensuremath{\nu}\xspace}
\newcommand{\electronneutrino}{\ensuremath{\neutrino_e}\xspace}

\newcommand{\NumberProtons}{\ensuremath{Z}\xspace}
\newcommand{\NumberNucleons}{\ensuremath{A}\xspace}

\newcommand{\AtomicNumber}{\ensuremath{Z}\xspace}

\newcommand{\ElectronDensity}{\ensuremath{n_\electron}\xspace}
\newcommand{\IonDensity}{\ensuremath{n_\AtomicNumber}\xspace}

\newcommand{\HydrogenDensity}{\ensuremath{n_\hydrogen}\xspace}

\newcommand{\positive}{\ensuremath{+}\xspace}
\newcommand{\negative}{\ensuremath{-}\xspace}
\newcommand{\processarrow}{\ensuremath{\rightarrow}\xspace}

\newcommand{\velocity}{\ensuremath{v}}
\newcommand{\VelocityVector}{\ensuremath{\myvector{\velocity}}}

\newcommand{\MagneticField}{\ensuremath{B}\xspace}
\newcommand{\MagneticFieldVector}{\ensuremath{\myvector{\MagneticField}}\xspace}

\newcommand{\Mass}{\ensuremath{M}\xspace}

\newcommand{\glon}{\text{GLON}\xspace}
\newcommand{\glat}{\text{GLAT}\xspace}

\newcommand{\MassChandrasekhar}{\ensuremath{\Mass_\text{Ch}}\xspace}
\newcommand{\MassNeutronStar}{\ensuremath{\Mass_\text{ns}}\xspace}
\newcommand{\RadiusNeutronStar}{\ensuremath{R_\text{ns}}\xspace}

\newcommand{\energyrotational}{\ensuremath{\energy_\text{rot}}\xspace}

\newcommand{\PulsarRotationAngle}{\ensuremath{\theta}\xspace}
\newcommand{\PulsarRadius}{\ensuremath{R_\text{NS}}\xspace}

\newcommand{\period}{\ensuremath{P}\xspace}
\newcommand{\perioddot}{\ensuremath{\dot{\period}}\xspace}

\newcommand{\PulsarAngularFrequency}{\ensuremath{\Omega}\xspace}
\newcommand{\PulsarAngularFrequencyDot}{\ensuremath{\dot{\PulsarAngularFrequency}}\xspace}
\newcommand{\PulsarAngularFrequencyDotDot}{\ensuremath{\ddot{\PulsarAngularFrequency}}\xspace}

\newcommand{\frequency}{\ensuremath{\nu}\xspace}
\newcommand{\angularfrequency}{\ensuremath{\omega}\xspace}

\newcommand{\breakingindex}{\ensuremath{n}\xspace}
\newcommand{\SpinDownTimescale}{\ensuremath{\tau_\text{dec}}\xspace}

\newcommand{\momentofinertia}{\ensuremath{I}\xspace}
\newcommand{\lifetime}{\ensuremath{\tau}\xspace}
\newcommand{\PulsarAge}{\ensuremath{\tau_c}\xspace}

\newcommand{\PulsarPotential}{\ensuremath{\Delta \Phi}\xspace}

\newcommand{\solarmass}{\ensuremath{\Mass_{\odot}}\xspace}

\newcommand{\radiusterminationshock}{\ensuremath{r_\text{ts}}\xspace}
\newcommand{\pointingflux}{\ensuremath{F_{E\times B}}\xspace}
\newcommand{\particleflux}{\ensuremath{F_\text{particle}}\xspace}
\newcommand{\magnetization}{\ensuremath{\sigma}\xspace}

\newcommand{\pressureISM}{\ensuremath{P_\text{ISM}}\xspace}

\newcommand{\KleinNishinaCrossSection}{\ensuremath{\CrossSection_\text{KN}}\xspace}

\newcommand{\pion}{\ensuremath{{\pi}^0}\xspace}

\newcommand{\luminosity}{\ensuremath{L}\xspace}

\newcommand{\prefactor}{\ensuremath{N_0}\xspace}
\newcommand{\spectralindex}{\ensuremath{\gamma}\xspace}
\newcommand{\Escale}{\ensuremath{E_0}\xspace}

\newcommand{\Ecutoff}{\ensuremath{E_c}\xspace}
\newcommand{\Ebreak}{\ensuremath{E_b}\xspace}

\newcommand{\explorerxi}{Explorer XI\xspace}
\newcommand{\fermi}{\textit{Fermi}\xspace}
\newcommand{\cosb}{COS-B\xspace}

\newcommand{\chandra}{\text{{\em Chandra}}\xspace}
\newcommand{\swiftxrt}{\text{{\em Swift}/XRT}\xspace}
\newcommand{\rosat}{\text{{\em ROSAT}}\xspace}
\newcommand{\suzaku}{\text{{\em Suzaku}}\xspace}
\newcommand{\asca}{\text{{\em ASCA}}\xspace}

\newcommand{\xmmnewton}{\text{{\em XMM-Newton}}\xspace}

\newcommand{\tevcat}{\text{TeVCat}\xspace}

\newcommand{\pointlike}{\ensuremath{\mathtt{pointlike}}\xspace}
\newcommand{\gtlike}{\ensuremath{\mathtt{gtlike}}\xspace}
\newcommand{\gtobssim}{\ensuremath{\mathtt{gtobssim}}\xspace}

\newcommand{\gtbin}{\ensuremath{\mathtt{gtbin}}\xspace}
\newcommand{\gtltcube}{\ensuremath{\mathtt{gtltcube}}\xspace}
\newcommand{\gtexpcubetwo}{\ensuremath{\mathtt{gtexpcube2}}\xspace}
\newcommand{\gtsrcmaps}{\ensuremath{\mathtt{gtsrcmaps}}\xspace}

\newcommand{\tsextpointlike}{\ensuremath{{\tsext}_{,\pointlike}}\xspace}
\newcommand{\tsextgtlike}{\ensuremath{{\tsext}_{,\gtlike}}\xspace}
\newcommand{\tsextaltdiff}{\ensuremath{{\tsext}_{\altdiff}}\xspace}

\newcommand{\tstev}{\ensuremath{\ts_\tev}\xspace}
\newcommand{\tspointlike}{\ensuremath{\ts_{\pointlike}}\xspace}
\newcommand{\tsgtlike}{\ensuremath{\ts_{\gtlike}}\xspace}
\newcommand{\tsaltdiff}{\ensuremath{\ts_{\altdiff}}\xspace}

\newcommand{\altdiff}{\text{alt,diff}\xspace}
\newcommand{\altpsf}{\text{alt,psf}\xspace}

\newcommand{\modelparams}{\ensuremath{\myvector{\lambda}}\xspace}
\newcommand{\fluxdensity}{\mathcal{F}\xspace}
\newcommand{\effectivearea}{\ensuremath{\epsilon}\xspace}
\newcommand{\dispersion}{\ensuremath{P}\xspace}
\newcommand{\response}{\ensuremath{R}\xspace}
\newcommand{\eventrate}{\ensuremath{\tau}\xspace}
\newcommand{\psf}{\ensuremath{\text{PSF}}\xspace}
\newcommand{\edisp}{\ensuremath{\text{E}_\text{disp}}\xspace}
\newcommand{\tdisp}{\ensuremath{\text{T}_\text{disp}}\xspace}

\newcommand{\dnde}{\ensuremath{\frac{dN}{d\energy}}\xspace}
\newcommand{\dndeinline}{\ensuremath{dN/d\energy}\xspace}

\newcommand{\psevensourcevsix}{\texttt{P7SOURCE\_V6}\xspace}
\newcommand{\galprop}{\texttt{GALPROP}\xspace}
\newcommand{\healpix}{\texttt{HEALPIX}\xspace}

\newcommand{\FluxPWNKeV}{\ensuremath{F_{2\unitspace\kev}^{10\unitspace\kev}}}
\newcommand{\FluxPWNTeV}{\ensuremath{F_{1\unitspace\tev}^{30\unitspace\tev}}}
\newcommand{\MeanLuminosityRatio}{\ensuremath{\bar{R}}}

\newcommand{\paperref}[1]{\begin{quote}\em{#1}\end{quote}}

\newcommand{\chapref}[1]{Chapter~\ref{chap:#1}}
\newcommand{\secref}[1]{Section~\ref{sec:#1}}
\newcommand{\subsecref}[1]{Section~\ref{subsec:#1}}
\newcommand{\tabref}[1]{Table~\ref{tab:#1}}
\newcommand{\figref}[1]{Figure~\ref{fig:#1}}
\newcommand{\eqnref}[1]{Equation~\ref{eqn:#1}}

\newcommand{\chaplabel}[1]{\label{chap:#1}}
\newcommand{\seclabel}[1]{\label{sec:#1}}
\newcommand{\subseclabel}[1]{\label{subsec:#1}}
\newcommand{\tablabel}[1]{\label{tab:#1}}
\newcommand{\figlabel}[1]{\label{fig:#1}}
\newcommand{\eqnlabel}[1]{\label{eqn:#1}}

\newcommand{\basehessj}[3]{HESS\allowbreak$\,$J#1$#2$\allowbreak#3\xspace}
\DeclareRobustCommand{\hessj}[1]{\IfEqCase{#1}{
{1825}{\basehessj{1825}{-}{137}}
{1640}{\basehessj{1640}{-}{465}}
{1837}{\basehessj{1837}{-}{069}}
{1632}{\basehessj{1632}{-}{478}}
{1614}{\basehessj{1614}{-}{518}}
{1616}{\basehessj{1616}{-}{508}}
{1119}{\basehessj{1119}{-}{614}}
{1303}{\basehessj{1303}{-}{631}}
{1420}{\basehessj{1420}{-}{607}}
{1841}{\basehessj{1841}{-}{055}}
{1356}{\basehessj{1356}{-}{645}}
{1023}{\basehessj{1023}{-}{575}}
{1848}{\basehessj{1848}{-}{018}}
{1514}{\basehessj{1514}{-}{591}}
{1857}{\basehessj{1857}{+}{026}}
{1708}{\basehessj{1708}{-}{443}}
{1804}{\basehessj{1804}{-}{216}}
{1634}{\basehessj{1634}{-}{472}}
{1018}{\basehessj{1018}{-}{589}}
{1507}{\basehessj{1507}{-}{622}}
{1834}{\basehessj{1834}{-}{087}}
}[\PackageError{hess}{Undefined option to \\hessj: #1}{}]}

\newcommand{\baseverj}[3]{VER\allowbreak$\,$J#1$#2$\allowbreak#3\xspace}
\DeclareRobustCommand{\verj}[1]{\IfEqCase{#1}{
{2016}{\baseverj{2016}{+}{372}}
}[\PackageError{psrj}{Undefined option to \\verj: #1}{}]}

\newcommand{\basepsrb}[3]{PSR$\,$B#1$#2$#3\xspace}
\DeclareRobustCommand{\psrb}[1]{\IfEqCase{#1}{
{1823}{\basepsrb{1823}{-}{13}}
{1509}{\basepsrb{1509}{-}{58}}
}[\PackageError{psrb}{Undefined option to \\psrb: #1}{}]}

\newcommand{\basepsrj}[3]{PSR\allowbreak$\,$J#1$#2$\allowbreak#3\xspace}
\DeclareRobustCommand{\psrj}[1]{\IfEqCase{#1}{
{1119}{\basepsrj{1119}{-}{6127}}
{1301}{\basepsrj{1301}{-}{6305}}
{1023}{\basepsrj{1023}{-}{5746}}
{1420}{\basepsrj{1420}{-}{6048}}
{1838}{\basepsrj{1838}{-}{0537}}
{1826}{\basepsrj{1826}{-}{1334}}
{1856}{\basepsrj{1856}{+}{0245}}
}[\PackageError{psrj}{Undefined option to \\psrj: #1}{}]}

\DeclareRobustCommand{\twofglj}[1]{
\IfEqCase{#1}{
}[\PackageError{hess}{Undefined option to \\twofglj: #1}{}]
}

\newcommand{\baseonefglj}[3]{1FGL\allowbreak$\,$J#1$#2$\allowbreak#3\xspace}
\DeclareRobustCommand{\onefglj}[1]{
\IfEqCase{#1}{
{1018.6}{\baseonefglj{1018.6}{-}{5856}}
}[\PackageError{hess}{Undefined option to \\onefglj: #1}{}]
}

\newcommand{\basesnrg}[3]{SNR\allowbreak$\,$G#1$#2$\allowbreak#3\xspace}
\DeclareRobustCommand{\snrg}[1]{
\IfEqCase{#1}{
{284.3}{\basesnrg{284.3}{-}{1.8}}
{320.4}{\basesnrg{320.4}{-}{1.2}}
{338.3}{\basesnrg{338.3}{-}{0.0}}
{8.7}{\basesnrg{8.7}{-}{0.1}}
}[\PackageError{hess}{Undefined option to \\snrg: #1}{}]
}

\newcommand{\mshfifteenfiftytwo}{MSH\,15$-$52\xspace}
\newcommand{\velax}{Vela$-$X\xspace}
\newcommand{\threecfiftyeight}{3C\,58\xspace}

\newcommand{\PWNClass}{``PWN''\xspace}
\newcommand{\PWNcClass}{``PWNc''\xspace}
\newcommand{\OtherClass}{``O''\xspace}
\newcommand{\PSRClass}{``PSR''\xspace}

\newcommand{\unitspace}{\,}

\newcommand{\ph}{\text{ph}\xspace}
\newcommand{\erg}{\text{erg}\xspace}

\newcommand{\cm}{\text{cm}\xspace}
\newcommand{\km}{\text{km}\xspace}
\newcommand{\parsec}{\text{pc}\xspace}

\newcommand{\kilo}{\ensuremath{\text{k}}\xspace}
\newcommand{\micro}{\ensuremath{\mu}\xspace}

\newcommand{\kiloparsec}{\kilo\parsec}

\newcommand{\gauss}{\text{G}\xspace}

\newcommand{\gram}{\text{g}\xspace}
\newcommand{\kg}{\text{kg}\xspace}

\newcommand{\kelvin}{\text{K}\xspace}

\newcommand{\hour}{\text{hr}\xspace}
\newcommand{\second}{\text{s}\xspace}
\newcommand{\millisecond}{\text{ms}\xspace}
\newcommand{\nanosecond}{\text{ns}\xspace}

\newcommand{\microsecond}{\micro\second}

\newcommand{\dayunit}{\ensuremath{\text{day}}\xspace}
\newcommand{\yearunit}{\ensuremath{\text{yr}}\xspace}

\newcommand{\kyr}{\kilo\yearunit\xspace}

\newcommand{\steradian}{\ensuremath{\text{sr}}\xspace}

\newcommand{\electronvolt}{\ensuremath{\text{eV}}\xspace}
\newcommand{\kev}{\text{keV}\xspace}
\newcommand{\mev}{\ensuremath{\text{MeV}}\xspace}
\newcommand{\gev}{\ensuremath{\text{GeV}}\xspace}
\newcommand{\tev}{\text{TeV}\xspace}
\newcommand{\fluxunits}{\ensuremath{\ph\;\cm^{-2}\second^{-1}}\xspace}
\newcommand{\efluxunits}{\ensuremath{\erg\;\cm^{-2}\second^{-1}}\xspace}
\newcommand{\prefunits}{\ensuremath{\fluxunits\erg^{-1}}\xspace}

\newcommand{\fluxdensityunits}{\ensuremath{\prefunits\steradian^{-1}}\xspace}

\newcommand{\pdfunits}{\ensuremath{\steradian^{-1}}\xspace}

\newcommand{\likelihood}{\mathcal{L}\xspace}
\newcommand{\data}{\ensuremath{\text{data}}\xspace}
\newcommand{\model}{\ensuremath{\text{model}}\xspace}
\newcommand{\pdf}{\ensuremath{\text{PDF}}\xspace}

\newcommand{\ts}{\ensuremath{\mathrm{TS}}\xspace}
\newcommand{\tspoint}{\ensuremath{\ts_\mathrm{point}}\xspace}
\newcommand{\tsext}{\ensuremath{\ts_\mathrm{ext}}\xspace}
\newcommand{\tsinc}{\ensuremath{\ts_{\mathrm{2pts}}}\xspace}

\newcommand{\tscutoff}{\ensuremath{\ts_\mathrm{cutoff}}\xspace}

\newcommand{\rsixeight}{{\ensuremath{\mathrm{r}_{68}}}\xspace}

\newcommand{\aic}{\text{AIC}\xspace}

\newcommand{\sys}{\text{sys}\xspace}
\newcommand{\stat}{\text{stat}\xspace}

\newcommand{\mynewacronym}[4][]{%
   \newacronym[user1=a,user2=a,user3=#4,user4=#4s, #1]{#2}{#3}{#4}}

\newcommand{\aac}[1]{%
   \ifglsused{#1}{\glsentryuserii{#1}}{\glsentryuseri{#1}}
   \gls{#1}%
}

\newcommand{\Actitle}[1]{\glsentryuseriii{#1}}

\newcommand{\Acptitle}[1]{\glsentryuseriv{#1}}

\newcommand{\Acstitle}[1]{\glsentryshort{#1}}
\newcommand{\Acsptitle}[1]{\glsentryshortpl{#1}}

\mynewacronym{SA}{SA}{solid angle}
\mynewacronym{FWHM}{FWHM}{full width at half maximum}

\mynewacronym{LAT}{LAT}{Large Area Telescope}

\mynewacronym{PSF}{PSF}{point spread function}

\mynewacronym{ACD}{ACD}{Anti-Coincidence Detector}
\mynewacronym{GBM}{GBM}{Gamma-ray Burst Monitor}

\mynewacronym[longplural=pulsar wind nebulae,
            shortplural=PWNe,
            plural=PWNe, 
            user3=Pulsar Wind Nebula,
            user4=Pulsar Wind Nebulae]{PWN}{PWN}{pulsar wind nebula}
\mynewacronym{SNR}{SNR}{supernova remnant}
\mynewacronym[longplural=active galactic nuclei,
             ]{AGN}{AGN}{active galactic nucleus}
\mynewacronym{SN}{SN}{supernova}
\mynewacronym[longplural=high-mass binaries,
             ]{HMB}{HMB}{high-mass binary}
\mynewacronym{NS}{NS}{neutron star}
\mynewacronym{MSP}{MSP}{millisecond pulsar}
\mynewacronym{MSC}{MSC}{massive star cluster}
\mynewacronym{UNID}{UNID}{unidentified source}

\mynewacronym{VHE}{VHE}{very high energy}

\mynewacronym[user3=Inverse Compton]{IC}{IC}{inverse Compoton}

\mynewacronym{2CG}{2CG}{the second \cosb catalog}
\mynewacronym[user3=The Second Fermi Catalog]{2FGL}{2FGL}{the second \fermi catalog}
\mynewacronym[user3=The Second Fermi Pulsar Catalog]{2PC}{2PC}{the second \fermi pulsar catalog}
\mynewacronym{1FHL}{1FHL}{the first \fermi hard-source list}
\mynewacronym{3EG}{3EG}{the third EGRET catalog}

\mynewacronym{CGS}{CGS}{the Centimetre-Gram-Second System of Units}

\mynewacronym{PL}{PL}{power law}
\mynewacronym{ECPL}{ECPL}{exponentially-cutoff power law}
\mynewacronym{BPL}{BPL}{broken-power law}

\mynewacronym{MIT}{MIT}{the Massachusetts Institute of Technology}

\mynewacronym{OSO-3}{OSO-3}{the Third Orbiting Solar Observatory}
\mynewacronym{SAS-2}{SAS-2}{the second Small Astronomy Satellite}
\mynewacronym{EGRET}{EGRET}{the Energetic Gamma Ray Experiment Telescope}
\mynewacronym{HESS}{H.E.S.S.}{the High Energy Stereoscopic System}
\mynewacronym{CGRO}{CGRO}{the Compton Gamma Ray Observatory}
\mynewacronym{AGILE}{AGILE}{Astro-rivelatore Gamma a Immagini LEggero}
\mynewacronym{CTA}{CTA}{the Cherenkov Telescope Array}

\mynewacronym{ESA}{ESA}{the European Space Agency}
\mynewacronym{NASA}{NASA}{the National Aeronautics and Space Administration}
\mynewacronym{NRL}{NRL}{the Naval Research Laboratory}
\mynewacronym{ASI}{ASI}{Italian Space Agency}
\mynewacronym{DOE}{DOE}{United States Department of Energy}

\mynewacronym{ATNF}{ATNF}{Australia Telescope National Facility}

\mynewacronym[longplural=seconds of arc]{arcsec}{arcsec}{second of arc}

\mynewacronym{IACT}{IACT}{Imaging air Cherenkov detector}

\mynewacronym{LMC}{LMC}{Large Magellanic Cloud}
\mynewacronym{CsI}{CsI}{cesium iodide}
\mynewacronym{GRB}{GRB}{gamma-ray burst}

\mynewacronym{CR}{CR}{cosmic ray}
\mynewacronym{SED}{SED}{spectral energy distribution}
\mynewacronym{LRT}{LRT}{likelihood-ratio test}

\mynewacronym{OG}{OG}{outer gap}
\mynewacronym{TPC}{TPC}{two pole caustic}
\mynewacronym{PC}{PC}{polar cap}
\mynewacronym{CMB}{CMB}{cosmic microwave background}
\mynewacronym{IRF}{IRF}{instrument response function}
\mynewacronym{2LAC}{2LAC}{the second LAT AGN catalog}

\usepackage{styles/deluxetable}

\title{Neutron Star Powered Nebulae: A New View on Pulsar Wind Nebulae with the Fermi Gamma-ray Space Telescope}
\author{Joshua Jeremy Lande}
\dept{Physics}
\principaladviser{Stefan Funk}
\firstreader{Elliott Bloom}
\secondreader{Roger Romani}
\submitdate{August 2013}

\begin{document}

\beforepreface

\vspace*{\fill}
\begingroup
\begin{quote}
\centering
\em{``Two things fill the mind with ever-increasing wonder and awe,
the more often and the more intensely the mind of thought is drawn to
them: the starry heavens above me and the moral law within me.'' 

-- Immanuel Kant}
\end{quote}
\endgroup
\vspace*{\fill}

\prefacesection{Abstract}

Pulsars are rapidly-rotating neutron stars born out of the death of
stars.  A diffuse nebula is formed when particles stream from these
neutron stars and interact with the ambient medium. These \acp{PWN}
are visible across the electromagnetic spectrum,  producing some of the
most brilliant objects ever observed.  The launch of the \fermi Gamma-ray
Space Telescope in 2008 has offered us an unprecedented view of the cosmic
$\gamma$-ray sky.  Using data from the \acrlong{LAT} on board \fermi, 
we search for new $\gamma$-ray-emitting \ac{PWN}.  With these
new observations, we vastly expand the number of \ac{PWN} observed
at these energies. We interpret the observed $\gamma$-ray emission
from these \ac{PWN} in terms of a model where accelerated electrons
produce $\gamma$-rays through inverse Compton upscattering when they
interact with interstellar photon fields.  We conclude by studying how
the observed \ac{PWN} evolve with the age and spin-down power
of the host pulsar.

\prefacesection{Acknowledgement}

This thesis was made possibly only by the incredible support and
mentorship of a large number of teachers, advisers, colleagues, and
friends.

I would first like to acknowledge the educational institutes I have
attended: my high school HB Woodlawn, my undergraduate institution
Marlboro College, and my graduate university Stanford University.  At HB
Woodlawn, I acknowledge my high school physics teacher Mark Dodge who
sparked my initial interest in physics. At Marlboro College, I would
like to thank the professors Travis Norsen, Matt Ollis, and Jim Mahoney
who fuled my interests in math and science.

I would next like to acknowledge the science advisers who brought the
science I was learning in my textbooks to life.  These people are Ron
Turner at ANSER, Tony Tyson at UC Davis, and Apurva Mehta and Sam Webb
at SLAC.

During my PhD I was helped by an almost overwhelmingly large number
of people in the \gls{LAT} collaboration.  These include Damien Parent,
David Smith, Heather Kelly, James Chiang Jean Ballet, Joanne Bogart, Johann
Cohen-Tanugi Junichiro Katsuta, Marianne Lemoine-Goumard, Marie-H\'el\`ene
Grondin, Markus Ackermann, Matthew Kerr, Ozlem Celik, Peter den Hartog,
Richard Dubois, Seth Digel, Tobias Jogler, Toby Burnett, Tyrel Johnson,
and Yasunobu Uchiyama.

I would like to acknowledge the Stanford and SLAC administrators and
technical support, including Glenn Morris, Stuart Marshall, Ken Zhou,
Martha Siegel, Chris Hall, Ziba Mahdavi, Maria Frank, Elva Carbajal,
and Violet Catindig.  They are awesome and really kept the place running!

I would next like to mention the large number of graduate students I
worked along side.  First, I acknowledge the Fermi Grad Students Adam Van
Etten, Alex Drlica-Wagner, Alice Allafort, Bijan Berenji, Eric Wallace,
Herman Lee, Keith Bechtol, Kyle Watters, Marshall Roth, Michael Shaw, Ping
Wang, Romain Rousseau, Warit Mitthumsiri, and Yvonne Edmonds.  Second,
I acknowledge my graduate student peers at Stanford Ahmed Ismail, Chris
Davis, Dan Riley, Joel Frederico, Joshua Cogan, Kristi Schneck, Kunal
Sahasrabuddhe, Kurt Barry, Mason Jiang, Matthew Lewandowski Paul Simeon,
Sarah Stokes Kernasovskiy, Steven Ehlert, Tony Li, and Yajie Yuan.

I would like to acknowledge my parents Jim Lande and Joyce Mason as well
as my brother Nathan Lande. They put up with my moving three time zones
away from home to follow my interests. I would also like to acknowledge
my girlfriend Helen Craig. She kept me sane over the long period of time
it took me to put this thesis together.

Finally, I acknowledge the great help of my thesis committee:
Elliott Bloom, Roger Romani, Stefan Funk, Persis Drell, Brad Efron.
In paritcular, I will always be indebted to my nurturing thesis adviser
Stefan Funk.  On many occasions, my PhD research felt insurmountable and
I doubted my abilities.  But even when I felt lost, Stefan never gave up
on me. He always encouraged me to keep pushing forward, to keep learning,
and to keep asking questions.  Stefan's faith in me never wavered, and
he always made me want to succeed. I hope that this thesis stands as a
testament to his mentoring.

\afterpreface

\glsresetall

\chapter{Overview}

In \chapref{gamma_ray_astro}, we discuss the history of $\gamma$-ray
astrophysics. First we present broadly the history of astronomy in
\secref{astronomy_and_the_atmosphere} and the history of $\gamma$-ray
astrophysics in \secref{history_gamma_ray_detectors}.  Then, we discuss
the \fermi Gamma-ray Space Telescope in \secref{fermi_telescope}.  Next,
we discuss historical developments in our understanding of pulsars and
\ac{PWN} in \secref{pulsars_and_pwn}.  We conclude by discussing the major
source classes detected by the \ac{LAT} in \secref{sources_detected_fermi}
adn the major radiation processes that occur in high-energy astrophysics
in \secref{radiation_processes}.

In \chapref{pulsar_pwn_system}, we discuss our current understanding
of the physics of pulsars and \ac{PWN}.  We discuss the formation of a
pulsar in \secref{neutron_star_formation} and the time evolution of a
pulsar in \secref{pulsar_evolution}.  Then, we describe the magnetosphere
of the pulsar in \secref{pulsar_magnetosphere} and the structure of a
typical \ac{PWN} in \secref{pwn_structure}.  Finally, we describe the
energy spectrum emitted from a typical \ac{PWN} in \secref{pwn_emission}.

In \chapref{maximum_likelihood_analysis}, we discuss maximum-likelihood
analysis and how it can be used to analyze \ac{LAT} data.
We describe the motivation for using maximum-likelihood analysis
in \secref{motivations_maximum_likelihood}
and the maximum-likelihood formulation in
\secref{description_maximum_likelihood}.  Then, we describe how to build
a model of the $\gamma$-ray sky in \secref{defining_model} and describe
the \ac{LAT} \acp{IRF} in \secref{lat_irfs}.
Finally, 
we describe the standard
package \gtlike for performing maximum-likelihood analysis of \ac{LAT}
data in \secref{binned_science_tools} and we describe \pointlike,
an alternate package for performing maximum-likelihood analysis of
\ac{LAT} data, in \secref{pointlike_package}.

In \chapref{extended_analysis}, we discuss a new method to study
spatially-extended sources.  We discuss the formulation of this
method in \secref{analysis_methods_section}.  We validate the
extension-significance calculation in \secref{validate_ts} and then
we compute the sensitivity of the \ac{LAT} to spatially-extended
sources in \secref{extension_sensitivity}.  We develop a new method
to compare the hypothesis of multiple point-like sources to one
spatially-extended source in \secref{dual_localization_method} and
finally in \secref{test_2lac_sources} we validate our method by testing
point-like sources from \ac{2LAC} for extension.

In \chapref{extended_search}, we apply the extension test developed in
\chapref{extended_analysis} to search for new spatially-extended sources.
First, we validate the method by studying known spatially-extended
\ac{LAT} sources in \secref{validate_known}.  

In \secref{systematic_errors_on_extension}, we develop a method
to estimate systematic errors associated with studying extended
sources and in \secref{extended_source_search_method} we develop
a method to search for new spatially-extended sources.  Finally,
we discuss the newly-discovered spatially-extended sources in
\secref{new_ext_srcs_section} and the population of spatially-extended
\ac{LAT} sources in \secref{extended_source_discussion}.

In \chapref{offpeak}, we perform a search for new \ac{PWN} in the
off-peak regions of \ac{LAT}-detected pulsars.  First, we develop
a new method to define the off-peak regions used for the search in
\secref{peak_definition}.  Then, we describe the analysis method we
used to search these regions in \secref{off_peak_analysis} and the
results of this search in \secref{off_peak_results}.  Finally,
we discuss some of the sources detected with this method in
\secref{off_peak_individual_source_discussion}.

In \chapref{tevcat}, we perform a search for $\gamma$-ray emission
from \ac{VHE} \ac{PWN}.  We discuss our list of \ac{VHE} candidates
in \secref{tevcat_list_vhe_pwn_candidates} and our analysis method
in \secref{tevcat_analysis_method}.  Finally, we several new \ac{PWN}
detected using this method in \secref{tevcat_sources_detected}.

In \chapref{population_study}, we describe the population of $\gamma$-ray
emitting \ac{PWN} and study how they evolve with the spin-down energy
and age of their associated pulsars.  Finally, in \chapref{outlook} we
remark on potential future opportunities to expand our understanding of
\acp{PWN} and their $\gamma$-ray emission.

\chapter{Gamma-ray Astrophysics}
\chaplabel{gamma_ray_astro}

\section{Astronomy and the Atmosphere}
\seclabel{astronomy_and_the_atmosphere}

From the very beginning, humans have surely stared into space and
contemplated its brilliance.  Stone circles in the Nabta Playa in
Egypt are likely the first observed astronomical observatory and
are believed to have acted as a prehistoric calendar.  Dating back
to the 5th century BC, they are 1,000 years older than Stonehenge
\citep{mck-mahille_2007_astronomy-nabta}.

Historically, the field of astronomy concerned the study of visible light
because it is not significantly absorbed in the atmosphere.  But slowly,
over time, astronomers expanded their view across the electromagnetic
spectrum.  Infrared radiation from the sun was first observed by William
Herschel in 1800 \citep{herschel_1800_experiments-refrangibility}.
The first extraterrestrial source of radio waves was detected
by Jansky in 1933 \citep{jansky_1933_electrical-disturbances}.
The expansion of astronomy to other wavelengths required the
development of rockets and satellites in the 20th century.
The first ultraviolet observation of the sun was performed in 1946
from a captured V-2 rocket \citep{baum_1946_ultraviolet-spectrum}.
Observations of x-rays from the sun were first performed in 1949
\citep{burnight_1949_x-radiation-atmosphere}.

\section{The History of Gamma-ray Astrophysics}
\seclabel{history_gamma_ray_detectors}

It was only natural to wonder about photons with even
higher energies. 
As is common in the field of physics, the prediction of
the detection of cosmic $\gamma$-rays proceeded their discovery.
\cite{feenberg_1948_interaction-cosmic-ray} theorized that the interaction
of starlight with cosmic rays could produce $\gamma$-rays through
\ac{IC} upscattering.  Following the discovery of the neutral
pion in 1949, \cite{hayakawa_1952_propagation-cosmic}
predicted that $\gamma$-ray emission could be observed from the
decay of neutral pions when cosmic rays interacted with interstellar
matter.  And in the same year, \cite{hutchinson_1952_possible-relation}
discussed the bremsstrahlung radiation of cosmic-ray electrons.
\cite{morrison_1958_gamma-ray-astronomy} predicted the detectability
of $\gamma$-ray emission from solar flares, \acp{PWN}, and active galaxies.

The first space-based $\gamma$-ray detector was \explorerxi
\citep{kraushaar_1965_explorer-experiment}.  \explorerxi operated in
the energy energy range above $100\unitspace\mev$.  It had an area of
$\sim45\cm^2$ but an effective area of only $\sim 7\cm^2$, corresponding
to a detector efficiency of $\sim 15\%$.  It was launched on April
27, 1961 and was in operation for 7 months.  \explorerxi observed 31
$\gamma$-rays but, because the distribution a distribution of these
$\gamma$-rays was consistent isotropy, the experiment could not firmly
identify the $\gamma$-rays as being cosmic.

The first definitive detection of $\gamma$-ray came in
1962 by an experiment on the Ranger 3 moon
probe \citep{arnold_1962_gamma-space}.  It detected an isotropic flux
of $\gamma$-rays in the 0.5 \mev to 2.1 \mev energy range.

\Ac{OSO-3} was the first experiment to firmly identify
$\gamma$-ray emission from the Galaxy
\citep{kraushaar_1972_high-energy-cosmic}.  
\Ac{OSO-3} was launched on March 8, 1967 and operated for 16 months, measuring
621 cosmic $\gamma$-rays.  
\figref{oso3_skymap} shows a sky map of these $\gamma$-rays.  This
experiment confirmed both a Galactic component to the $\gamma$-ray
sky as well as an additional isotropic component, hypothesised to be
extragalactic in origin.

\begin{figure}[htbp]
  \centering
  \includegraphics{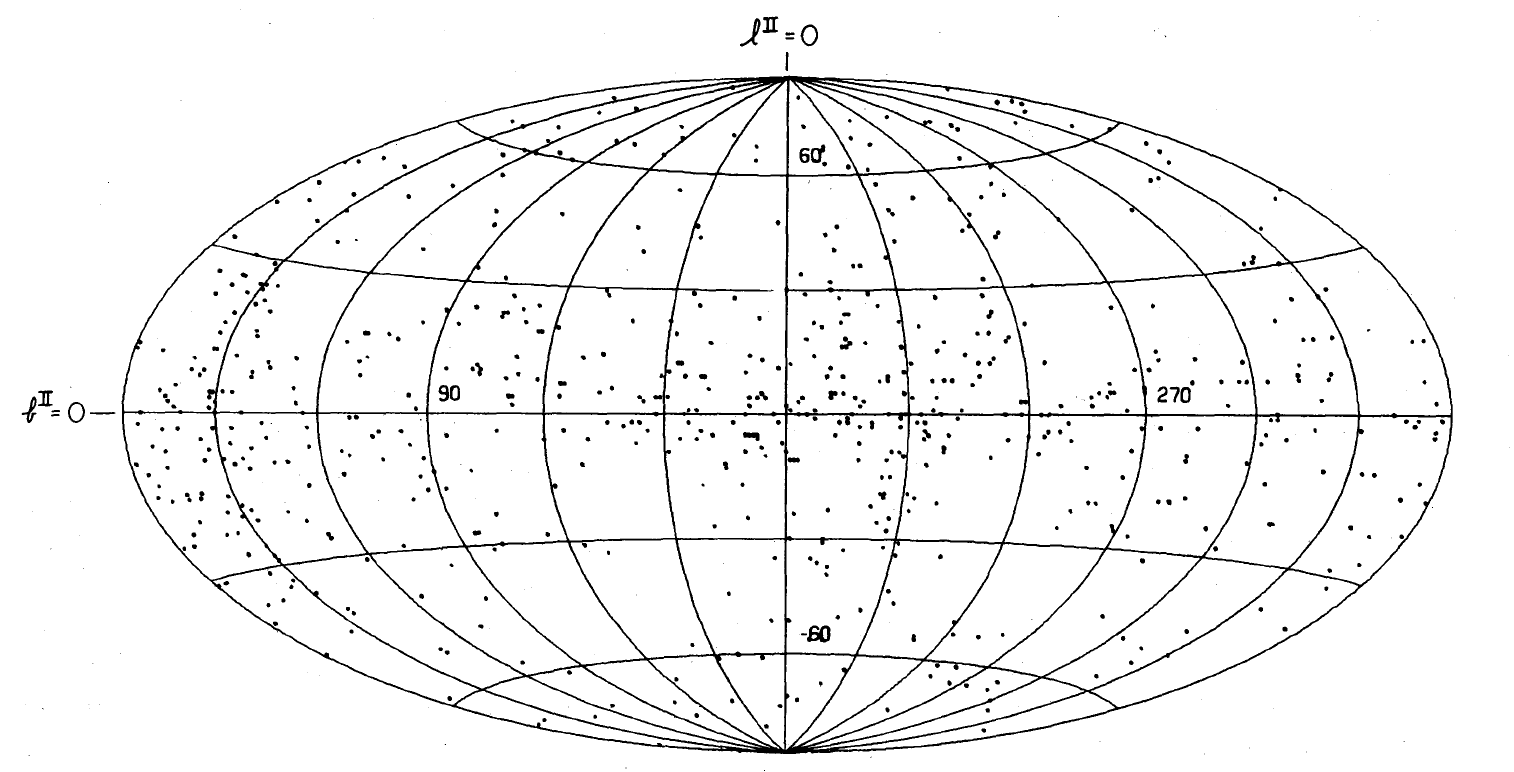}
  \figlabel{oso3_skymap}
  \caption{The position of all 621 cosmic $\gamma$-rays
  detected by \ac{OSO-3}. This figure is from
  \cite{kraushaar_1972_high-energy-cosmic}. }
\end{figure}

This anisotropic $gamma$-ray distribution
was confirmed by a balloon-based
$\gamma$-ray detector in 1970  \citep{kniffen_1970_study-gamma}.  In the
following year, the first $\gamma$-ray pulsar (the Crab pulsar) was detected
by another balloon-based detector \cite{browning_1971_detection-pulsed}.

The next major advancements in $\gamma$-ray astronomy
came from \ac{SAS-2} and \cosb.  \Ac{SAS-2} was a dedicated
$\gamma$-ray detector launched by \ac{NASA} in November 15, 1972
\cite{fichtel_1975_high-energy-gamma-ray}. It improved upon \ac{OSO-3}
by incorporating a spark chamber and having an overall larger size.
The size of the active area of the detector was 640 $\cm^2$ and the
experiment had a much improved effective area of $\sim 115\unitspace
cm^2$. The spark chamber allowed for a separate measurement of the
electron and positron tracks, which allowed for improved directional
reconstruction of the incident $\gamma$-rays. \Ac{SAS-2} had a PSF
$\sim5\degree$ at 30 \mev and $\sim1\degree$ at 1 \gev.

In 6 months, \ac{SAS-2} observed over 8,000 $\gamma$-ray
photons and covered $\sim55\%$ of the sky including most of
the Galactic plane.  It discovered pulsations from the Crab
\citep{fichtel_1975_high-energy-gamma-ray} and Vela pulsar
\citep{thompson_1977_sas-2-high-energy}.  In addition, \ac{SAS-2}
discovered Geminga, the first $\gamma$-ray source with no compelling
multiwavelength counterpart \citep{thompson_1977_final-sas-2}. Geminga
was eventually discovered to be a pulsar by \ac{EGRET}
\citep{bertsch_1992_pulsed-high-energy} and retroactively by \ac{SAS-2}
\citep{mattox_1992_observation-pulsed}.

\ac{ESA} launched \cosb on August 9, 1975.
\cosb improved upon
the design of \ac{SAS-2} by including a calorimeter below the spark
chamber which improved the energy resolution to $<100\%$ for energies
$\sim 3\unitspace\gev$ \citep{bignami_1975_cos-b-experiment}.  
\cosb operated successfully for over 6 years and produced the first
detailed catalog of the $\gamma$-ray sky.  In total, \cosb observed $\sim
80,000$ photons \citep{mayer-hasselwander_1982_large-scale-distribution}.
\Ac{2CG} detailed the detection 25 $\gamma$-ray sources for
$E>100\unitspace\mev$ \citep{swanenburg_1981_second-catalog}.
\figref{cos_b_skymap} shows a map of these sources.  Of these sources,
the vast majority lay along the galactic plane and could not be positively
identified with sources observed at other wavelengths.  In addition,
\cosb observed the first ever extragalactic $\gamma$-ray source,
\citep[3C273,][]{swanenburg_1978_observation-high-energy}.

\begin{figure}[htbp]
  \centering
  \includegraphics[width=\textwidth]{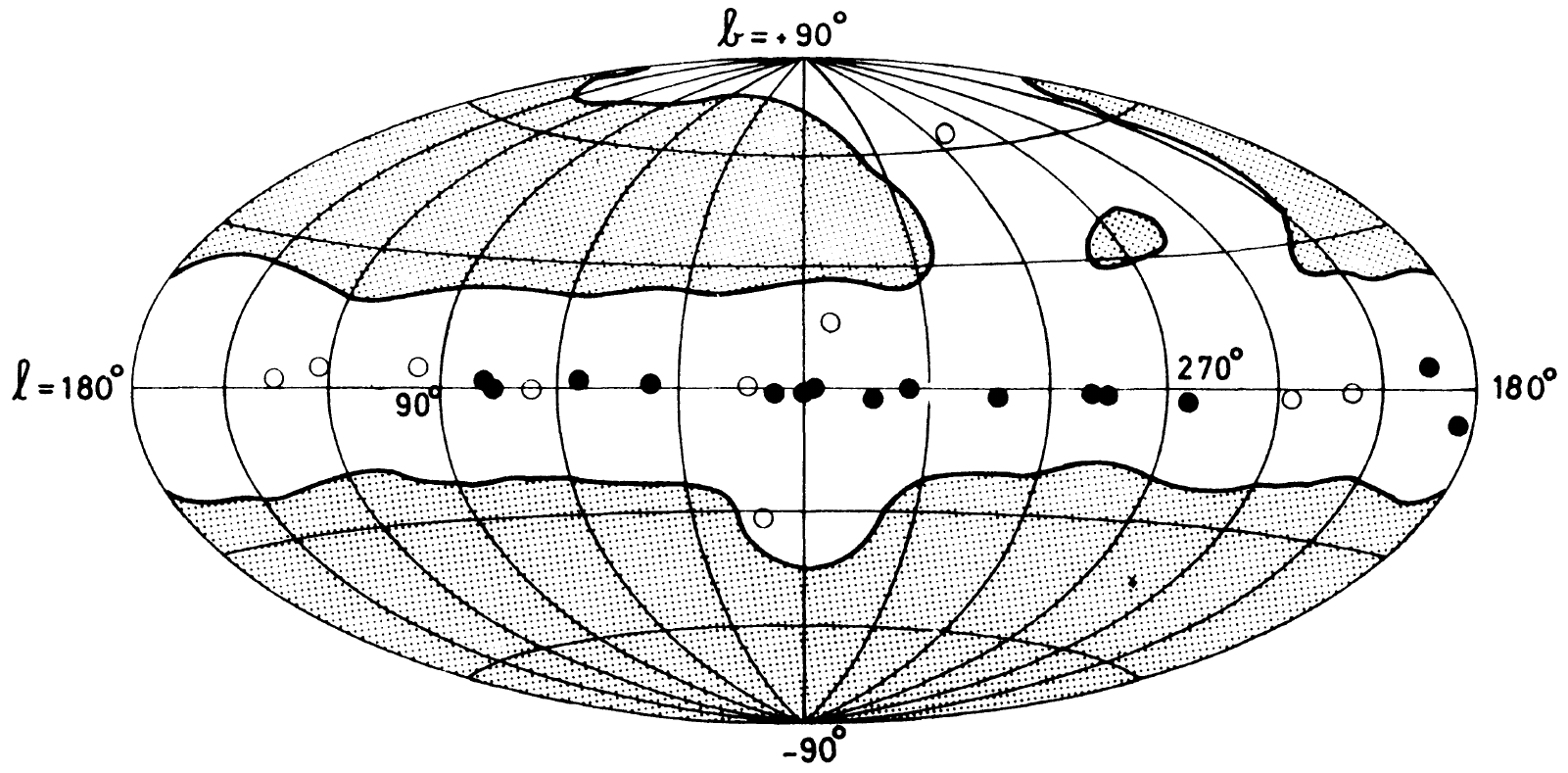}
  \figlabel{cos_b_skymap}
  \caption{
  A map of the sources observed by \cosb. The filled circles
  represent brighter sources. The unshaded region corresponds to
  the parts of the sky observed by \cosb.  This figure is from
  \cite{swanenburg_1981_second-catalog}.
  }
\end{figure}

The next major $\gamma$-ray experiment was \ac{EGRET},
launched
on board \ac{CGRO} in 1991. \ac{EGRET} had a design
similar to \ac{SAS-2}, but had an expanded energy range, operating from
$20\unitspace\mev$ to $30\unitspace\gev$, an improved effective area of
$\sim1500\unitspace\cm^2$ from $\sim500\mev$ to $\sim1\unitspace\gev$,
and an improved angular resolution, decreasing to $\sim0.5\degree$
at its highest energies \citep{thompson_1993a_calibration-energetic}.
At the time, \ac{CGRO} was the heaviest astrophysical experiment
launched into orbit, weighting $\sim17,000\unitspace\kg$. \ac{EGRET}
contributed $\sim6,000\unitspace\kg$ to the mass of \ac{CGRO}.
\figref{egret_detector} shows a schematic diagram of \ac{EGRET}.

\begin{figure}[htbp]
\centering
\includegraphics{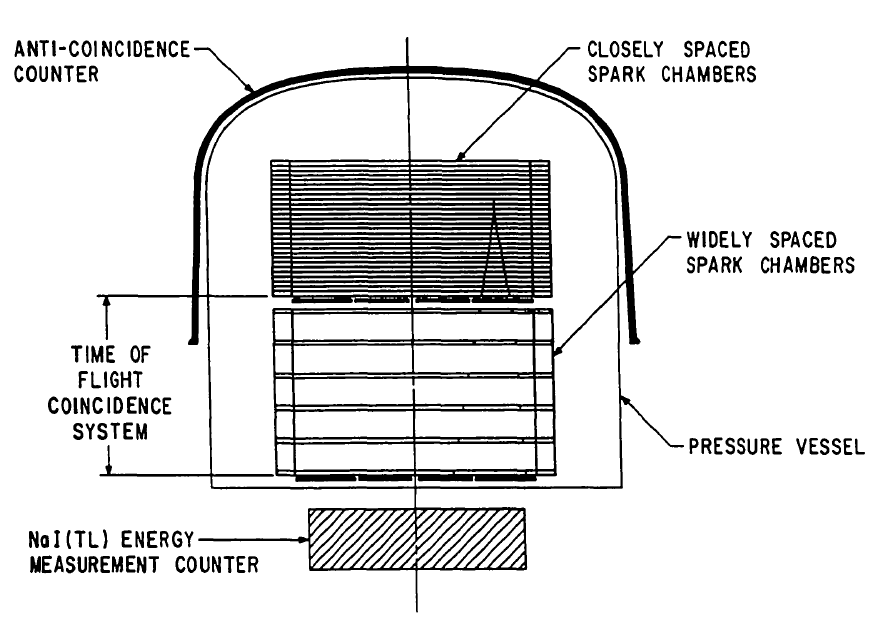}
\figlabel{egret_detector}
\caption{A diagram of the \ac{EGRET} detector.  This figure is from
\citep{thompson_1993a_calibration-energetic}.}
\end{figure}

\ac{EGRET} vastly expanded the field of $\gamma$-ray astronomy.
\ac{EGRET} detected six pulsars \citep{nolan_1996a_egret-observations} and
also the Crab Nebula \citep{nolan_1993a_observations-pulsar}.  \ac{EGRET}
also detected the LMC, the first normal galaxy outside of our galaxy to be
detected at $\gamma$-rays \citep{sreekumar_1992a_observations-large}.
\ac{EGRET} also detected Centarus A, the first radio galaxy
detected at $\gamma$-rays \citep{sreekumar_1999a_emission-nearby}.
In total, EGRET detected 271 $\gamma$-ray sources in \ac{3EG}
\citep{hartman_1999a_third-egret}. This catalog included 66
high confidence blazar identifications and 27 low-confidence AGN
identifications. \figref{third_egret_catalog_sources} plots the sources
observed by \ac{EGRET}.  In total, \ac{EGRET} detected over 1,500,000
celestial gamma rays \citep{thompson_2008a_gamma-astrophysics:}.

\begin{figure}[htbp]
\centering
\includegraphics{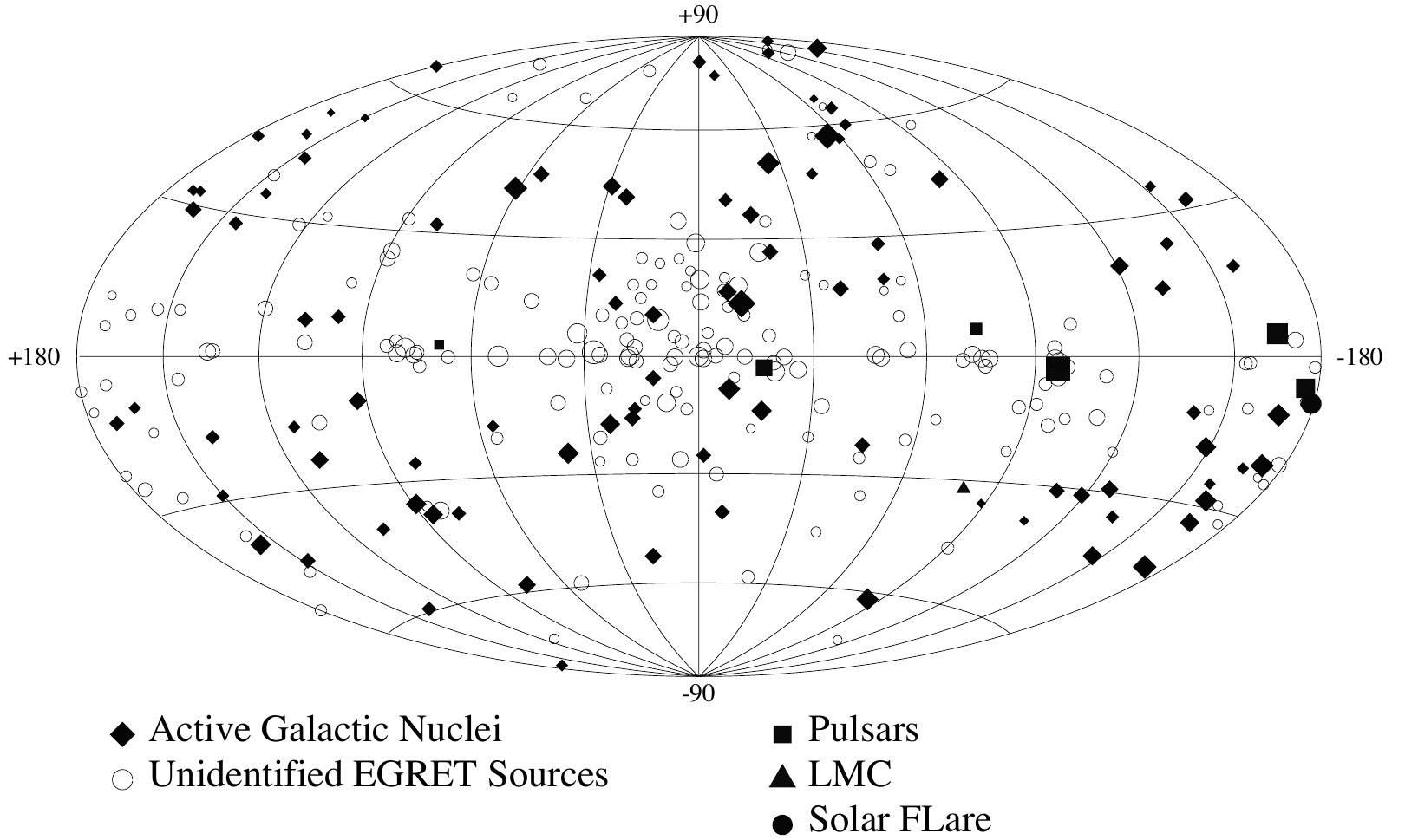}
\figlabel{third_egret_catalog_sources}
\caption{The position of \ac{EGRET} sources in the sky in galactic
coordinates.  The size of the source markers corresponds to the overall
source intensity.  This figure is from \citep{hartman_1999a_third-egret}.}
\end{figure}

Following \ac{EGRET}, the next major $\gamma$-ray observatories
were \ac{AGILE} \citep{pittori_2003a_gamma-ray-imaging}
and the \fermi Gamma-ray Space Telescope \citep{atwood_2009a_large-telescope}.  \ac{AGILE}
was an \ac{ASI} experiment launched in 2007 and \fermi was
a joint \ac{NASA} and \ac{DOE} experiment which was launched
in 2008.  The major difference between \ac{AGILE} and \fermi
was that \fermi has a significantly-improved effective area
\citep[$9,500\unitspace\cm^2$,][]{atwood_2009a_large-telescope}
compared to \ac{AGILE}
\citep[$\sim500\unitspace\cm^2$,][]{pittori_2003a_gamma-ray-imaging}.
We will discuss the \fermi detector in \secref{fermi_telescope}.

\section{The \fermi Gamma-ray Space Telescope}
\seclabel{fermi_telescope}

\begin{figure}[htbp]
  \centering
  \includegraphics{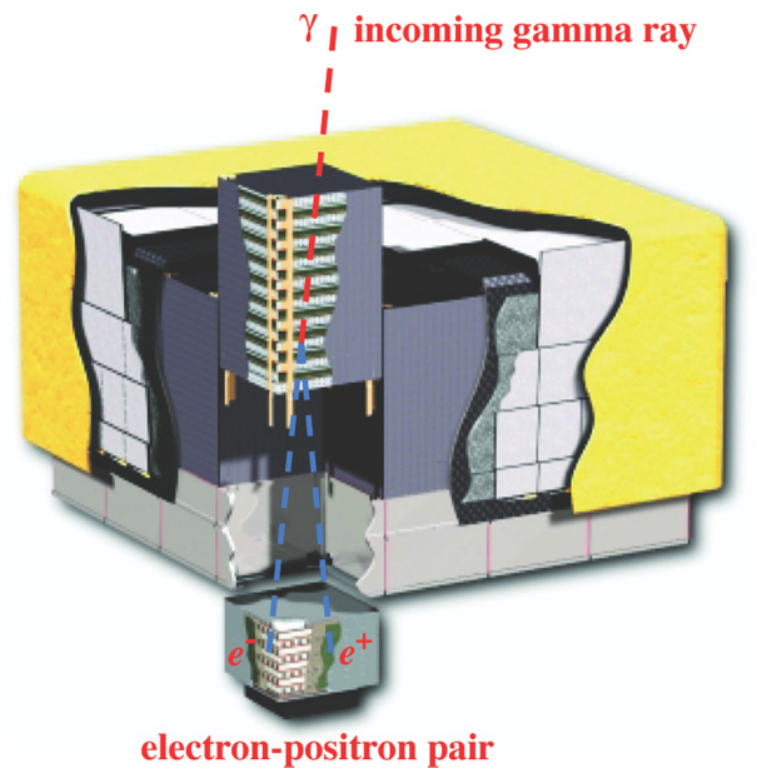}
  \caption{A schematic diagram of the \ac{LAT} with an incident
  $\gamma$-ray (red line) pair-converting into an electron
  and positron pair (blue lines).  This figure is taken from
  \citep{atwood_2009a_large-telescope}.}
  \figlabel{lat_detector_cutout}
\end{figure} 

The \fermi Gamma-ray Space telescope was launched on June 11, 2008 on
a Delta II heavy launch vehicle \citep{atwood_2009a_large-telescope}.
The primary since instrument on board \fermi is the \ac{LAT},
a pair-conversion telescope which detects $\gamma$-rays
in the energy range from $20\unitspace\mev$ to $>300\unitspace\gev$
(see \figref{lat_detector_cutout}).
In addition, \fermi contains the \Ac{GBM}, which is used to observe
\acp{GRB} in the energy range from $\sim8\unitspace\kev$ to $\sim40\unitspace\mev$.
See \cite{meegan_2009a_fermi-gamma-ray} for a description of the \ac{GBM}.

\subsection{The \acs{LAT} Detector}

The \ac{LAT} is composed of three major subsystems: the tracker, the
calorimeter, and the \ac{ACD}. Fundamentally, the detector operates by
inducing an incident $\gamma$-ray to pair convert in the tracker into
an electron and positron pair. The electron and position travel through
the tracker and into the \ac{CsI} calorimeter.  The tracks and energy
deposit can be used to infer the direction and energy of the incident
$\gamma$-ray.  Both the tracker and calorimeter are $4\times4$ arrays,
each composed of 16 modules.  Each tracker tower is divided into 18
tungsten converter layers and 16 dual-silicon tracker planes. Each
calorimeter module is composed of eight layers of 12 \ac{CsI} crystals.

The \ac{ACD} provides provides
background rejection of charged particles incident
on the \ac{LAT}.  The \ac{ACD} surrounds the tracker and is composed
of 89 plastic scintillator tiles ($5\times5$ on the top and 16
on each of the sides). The \ac{ACD} has a 0.9997 efficiency for
detecting singly-charged particles entering the \ac{LAT}.  A detailed
discussion of the various subsystems of the LAT can be found in
\citep{atwood_2009a_large-telescope}.

\subsection{Performance of the \acs{LAT}}
\subseclabel{performance_lat}

The \ac{LAT} 
has an unprecedented effective area ($\sim9,500\unitspace\cm^2$),
single-photon energy resolution ($\sim10\%$), and single-photon
angular resolution ($\sim3\fdg5$ at $\energy=100\unitspace\mev$
and decreasing to $\lesssim0\fdg15$ for $\energy>10\unitspace\gev$)
\citep{atwood_2009a_large-telescope}.

\begin{figure}[htbp]
  \centering
  \includegraphics{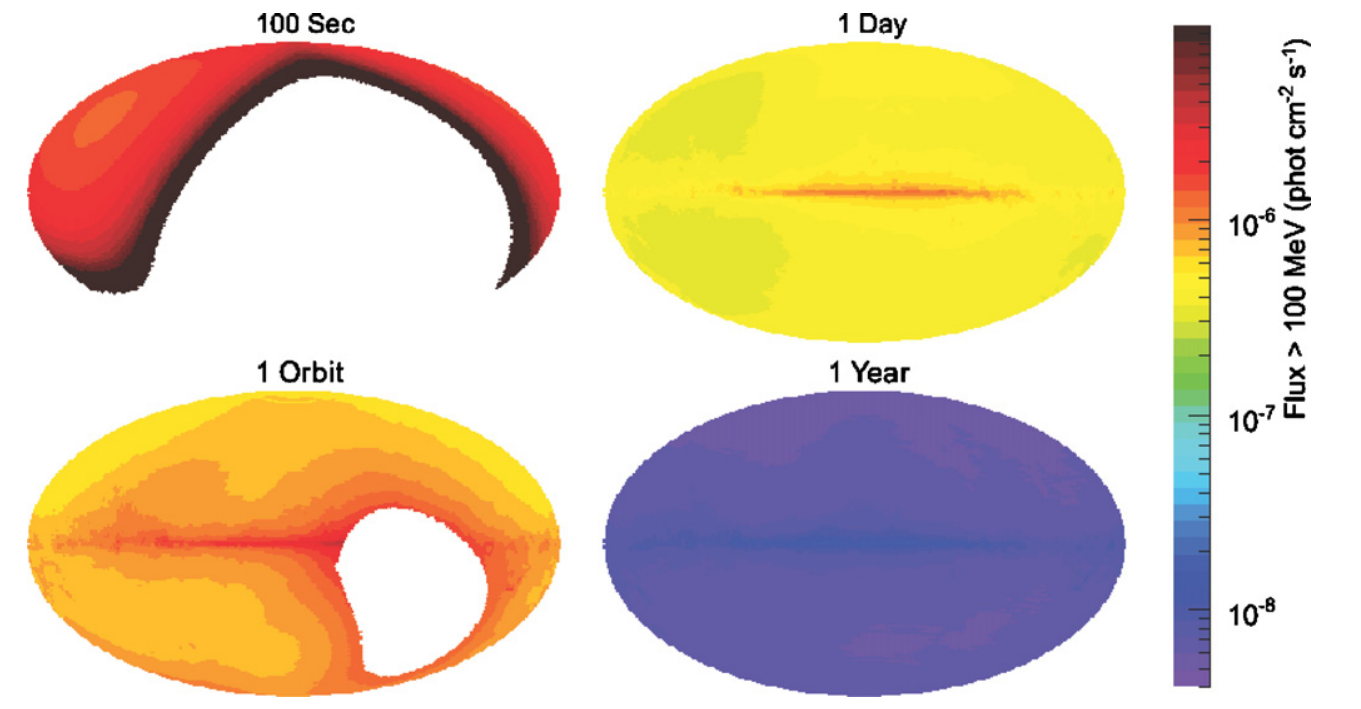}
  \caption{
  The \ac{LAT} point-source sensitivity for exposures of
  $100\unitspace\second$, 1 orbit, $1\unitspace\dayunit$,
  and $1\unitspace\yearunit$.  This figure is from
  \cite{atwood_2009a_large-telescope}.
  }
  \figlabel{lat_point_source_sensitivity}
\end{figure} 

With its $2.4\unitspace\steradian$ field of view, \fermi can observe
the entire sky almost uniformly every $\sim3\unitspace\hour$.
With one year of observations, the \ac{LAT} has a point-source
flux sensitivity $3 \times 10^{-9} ($E>100\unitspace\mev$)
\ph\unitspace\cm^{-2}\second{-1}$ assuming a high-latitude diffuse
flux of $1.5\times10^{-5}\cm^{-2}\second^{-1}\steradian^{-1}$
($E>100\unitspace\mev$) \figref{lat_point_source_sensitivity} plots the
sensitivity for exposures of varying timescales.

\begin{figure}[htbp]
  \centering
  \includegraphics{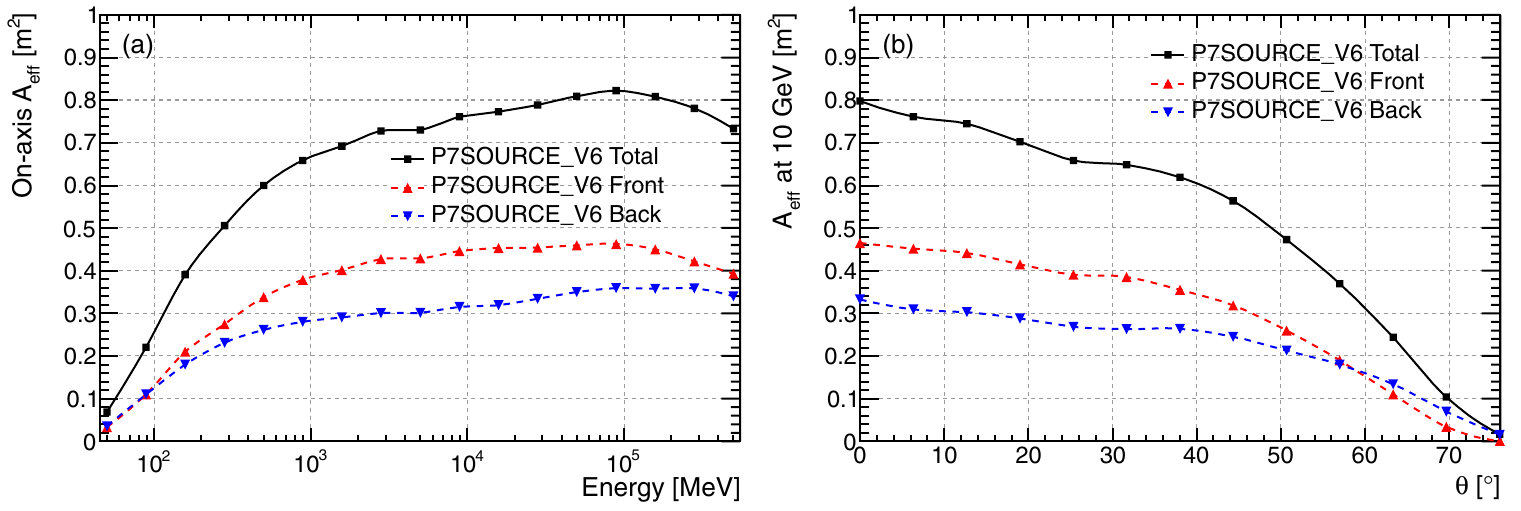}
  \caption{
  The \ac{LAT} effective area (a) as a function of energy for
  $\gamma$-rays that are incident on the \ac{LAT} perpendicularly from
  above and (b) as a function of incident angle for photons with an
  energy of $10\unitspace\gev$.  The \ac{LAT} performance is computed
  for the \psevensourcevsix event classification.  This figure is from
  \cite{ackermann_2012a_fermi-large}.
  }
  \figlabel{lat_effective_area}
\end{figure} 

\begin{figure}[htbp]
  \centering
  \includegraphics{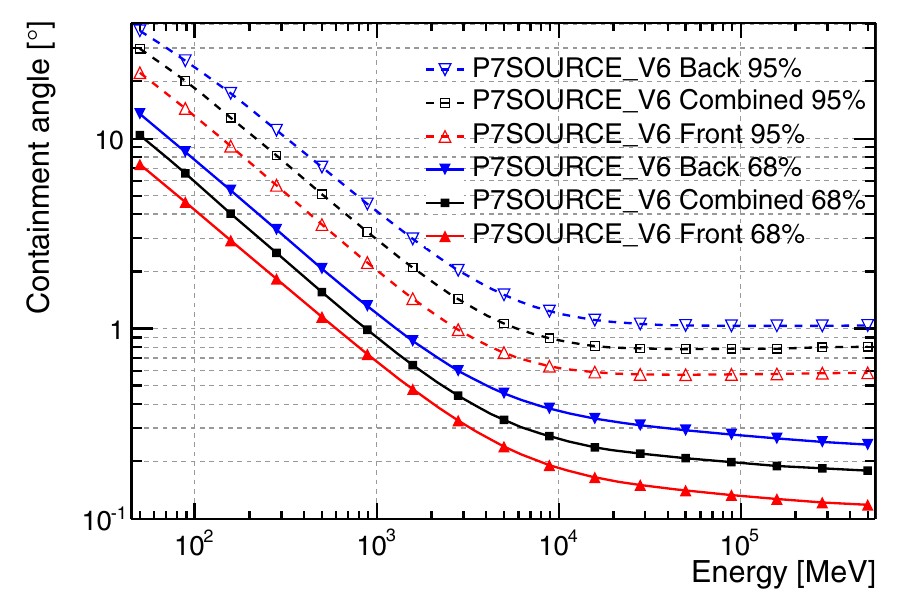}
  \caption{
  The angular resolution (68\% and 95\% containment radius)
  as a function of energy.
  The \ac{LAT} performance is computed for the \psevensourcevsix
  event classification.
  This figure is from \cite{ackermann_2012a_fermi-large}.
  }
  \figlabel{lat_psf}
\end{figure}

\begin{figure}[htbp]
  \centering
  \includegraphics{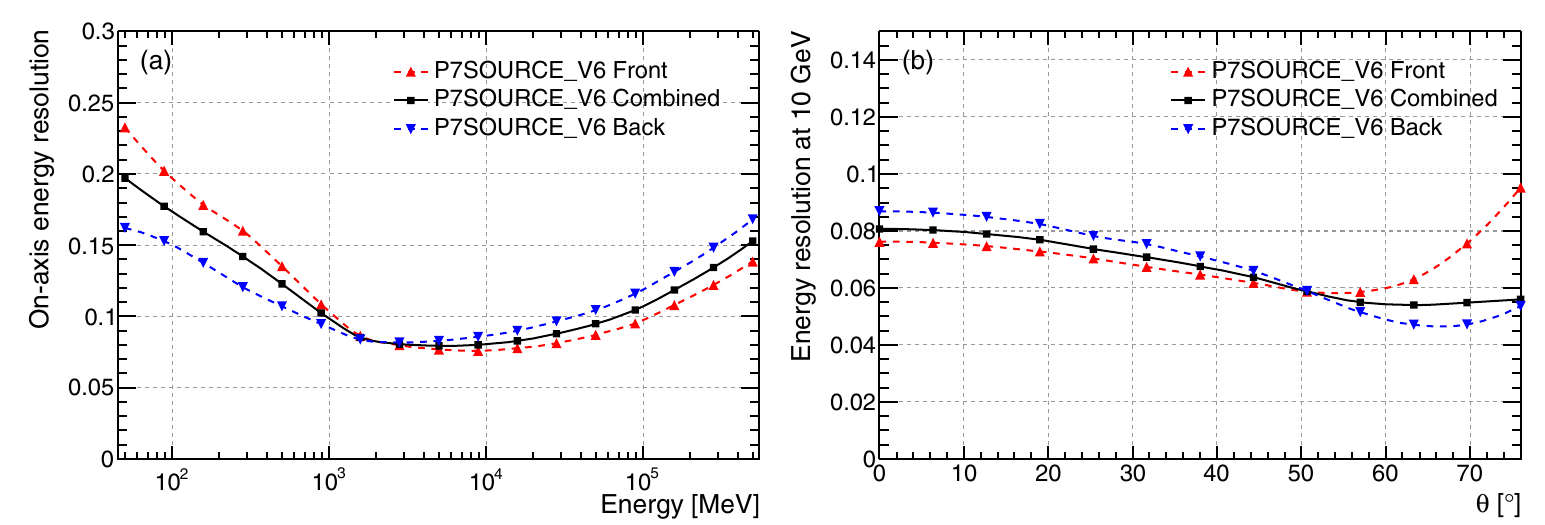}
  \caption{
  The energy dispersion (a) as a function of energy for $\gamma$-rays that
  are incident on the \ac{LAT} perpendicularly from above and (b) as a function
  of the incident angle for photons with an energy of $10\unitspace\gev$.
  The \ac{LAT} performance is computed for the \psevensourcevsix
  event classification.
  This figure is from \cite{ackermann_2012a_fermi-large}.
  }
  \figlabel{lat_energy_dispersion}
\end{figure} 

The effective area, \ac{PSF}, and energy dispersion are both a function
of energy and of incident angle.  \figref{lat_effective_area} plots
the effective area as a function of energy and incident angle.
\figref{lat_psf} plots the \ac{PSF} as a function of energy.
Finally, \figref{lat_energy_dispersion} plots the energy dispersion
as a function of energy and incident angle.  We will describe in
\chapref{maximum_likelihood_analysis} the analysis methods used to
analyze \ac{LAT} data.

\section{Pulsars and \Acptitle{PWN}}
\seclabel{pulsars_and_pwn}

\subsection{Pulsars}

It is widely accepted that in the collapse of a massive star, a large
amount of ejecta is released as a supernova powering a \ac{SNR}
and that much of the remaining mass collapses into a neutron star
\citep{baade_1934a_remarks-super-novae}.

Pulsars were first discovered observationally in 1967 by Jocelyn Bell
Burnell and Antony Hewish \citep{hewish_1968_observation-rapidly}. 
We note in that pulsars had been previously observed by the air force
\citep{brumfiel_2007_force-early}.
Even before the discovery, \cite{pacini_1967_energy-emission}
had predicted the existence of \acp{NS}.  Shortly following
the 1967 discovery, \cite{gold_1968_rotating-neutron} and
\cite{pacini_1968_rotating-neutron} argued the connection
between pulsars and rotating \acp{NS}.

The discovery of many more pulsars came quickly.  In 1968, and the
Vela pulsar \citep{large_1968_pulsar-supernova} and the Crab pulsar
\citep{staelin_1968_pulsating-radio} were discovered.
The first pulsar observed at optical frequencies was the
Crab \citep{cocke_1969_discovery-optical}.  In the same year, the first X-ray
pulsations were discovered from the same source
from an X-ray detector on a rocket.  The discovery was carried out almost concurrently by
a group at \gls{NRL} \citep{fritz_1969_x-ray-pulsar} and at \gls{MIT}
\citep{bradt_1969_x-ray-optical}.  Using proportional counters, these
experiments showed that the pulsed emission from the Crab extended to
X-ray energies and that, for this source, the X-rays emission was a
factor $>100$ more energetic than the observed visible emission.

From these early sources, pulsar physics has blossomed into a vast
field. In the on-line \ac{ATNF} catalog, there are currently over 2,200
pulsars \citep{manchester_2005a_australia-telescope}.

As was discussed in \secref{history_gamma_ray_detectors},
the first pulsar was observed in $\gamma$-ray in 1970
\citep{kniffen_1970_study-gamma}.  Observations by \ac{EGRET}
brought the total number of $\gamma$-ray-detected pulsars to six
\citep{nolan_1996a_egret-observations}.  \fermi has vastly expanded the
number of pulsars detected in $\gamma$-rays and we will discuss these
observations in \subsecref{2pc}

\subsection{\Acptitle{PWN}}
\subseclabel{pwn}

A \gls{PWN} is a diffuse nebula of shocked relativistic particles
that surrounds and is powered by an accompanying pulsar. 
\glspl{PWN} have been observed long before the discovery of pulsars, but
the pulsar/\gls{PWN} connection was not made until
after the detection of pulsars.

The most famous \glspl{PWN} is the Crab nebula, associated with the Crab
pulsar.  The Crab \ac{SN} (SN 1054) was observed by Chinese astrologers 
in 1054 AD \cite{hester_2008_nebula:-astrophysical}.
It was also likely observed in
Japan, Europe, by Native Americans,
and in the Arab world 
\citep[see][and references therein]{collins_1999a_reinterpretation-historical}.

\begin{figure}[htbp]
  \centering
  \includegraphics[width=\textwidth]{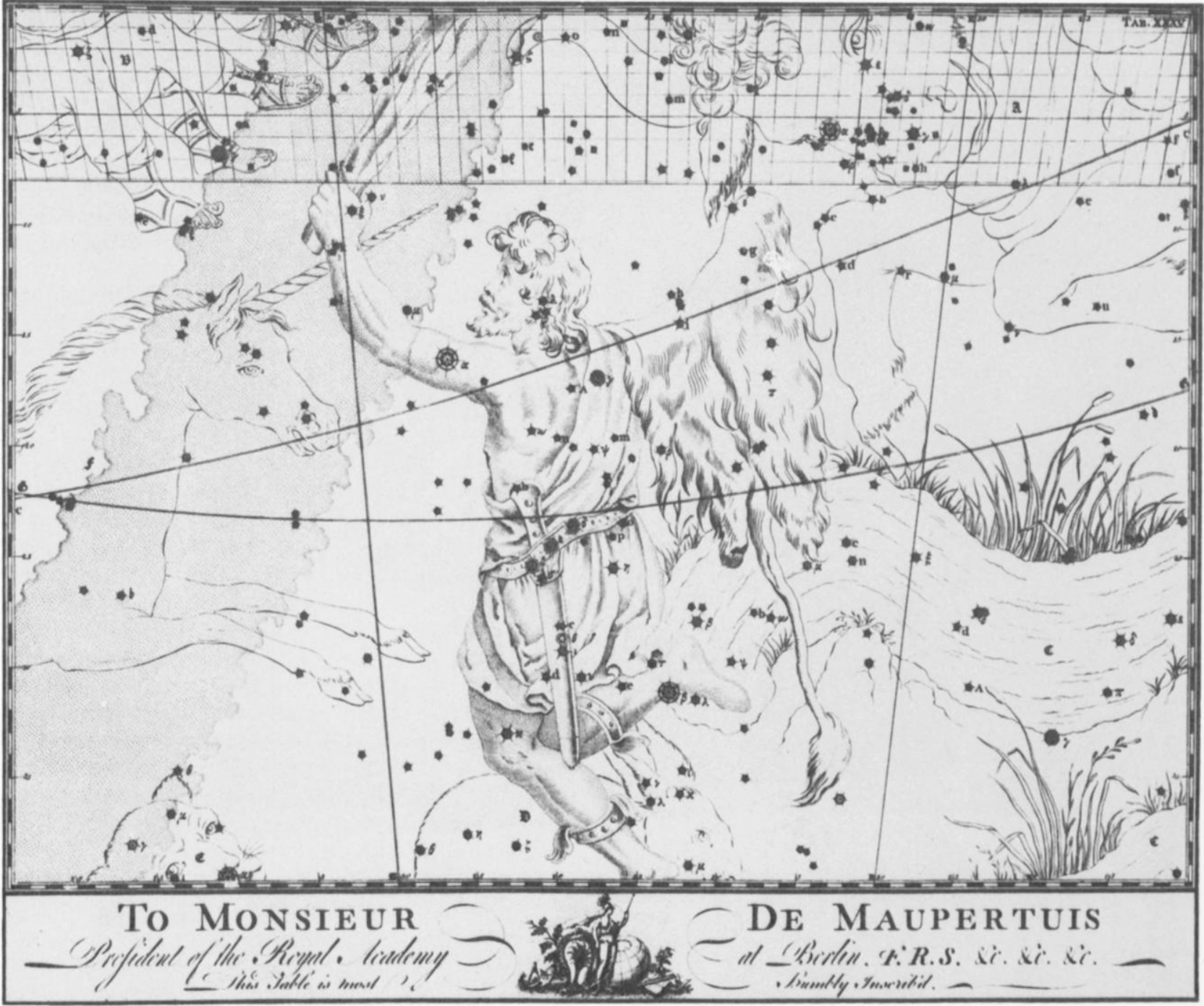}
  \figlabel{bevis_crab}
  \caption{The Orion plate from Bevis' book {\em Uranographia Britannica}.
  The Crab nebula can be found on the horn of Taurus the Bull 
  on the top of the figure and the source is marked by a 
  cloudy symbol.
  This figure was reproduced from \cite{ashworth_1981_bevis-uranographia}.}
\end{figure}

The Crab nebula, in the remains of SN 1054,
was first discovered in 1731 by physician and amateur astronomer
John Bevis.  This source was going to be published in his sky atlas
{\em Uranographia Britannica}, but the work was never published because
his publisher filed for bankruptcy in 1750.  \figref{bevis_crab}
shows Beavis' plate containing the Crab nebula.  A detailed history of
John Bevis' work can be found in \cite{ashworth_1981_bevis-uranographia}.
The Crab Nebulae was famously included in Charles Messier's catalog as
M1 in 1758 \cite{hester_2008_nebula:-astrophysical}.

In 1921, \cite{lampland_1921a_observed-changes} 
observed motions and changes in brightness of parts of the nebula.
In the same year, \cite{duncan_1921a_changes-observed} observed
that the entire nebula was expanding. Also in the same year
Knut Lundmark proposed a connection between the Crab Nebula
and the 1054 supernova \citep{lundmark_1921a_suspected-stars}.
In 1942, \cite{mayall_1942a_further-bearing} connected improved
historical observations with a detailed study of the historical record
to unmistakably connect the Crab nebula to SN 1054.

Radio emission from the Crab nebula was first detected
in 1949 \citep{bolton_1949a_positions-three}.  The
synchrotron hypothesis for the observed emission was
first proposed by \cite{shklovskii_1953a_nature-nebulas},
and quickly confirmed by optical polarization observations
\citep{dombrovsky_1954a_nature-radiation}.  X-rays from the object were
first detected by \cite{bowyer_1964a_lunar-occultation}.  
As was discussed in \secref{history_gamma_ray_detectors}, the Crab pulsar
was discovered in 1968.  In the discovery paper, the \ac{SN}, \ac{PWN},
\ac{NS} connection was proposed \citep{staelin_1968_pulsating-radio}.

The synchrotron and \ac{IC} model of the Crab nebula predicting observable
\ac{VHE} emission was first proposed by \cite{gould_1965a_energy-cosmic},
and improved in \cite{rieke_1969a_production-cosmic} and
\cite{grindlay_1971a_compton-synchrotron-spectrum}.  As was discussed
in \secref{history_gamma_ray_detectors}, $\gamma$-rays from the Crab
nebula were first observed by \cite{nolan_1993a_observations-pulsar}.
\ac{VHE} emission from the Crab nebula by an \ac{IACT} was first observed
by \cite{weekes_1989a_observation-gamma}.

\acp{PWN} are commonly observed to surround pulsars.
Some of the famous \acp{PWN} include \velax surrounding the Vela
pulsar \citep[first observed by][]{rishbeth_1958a_radio-emission},
\threecfiftyeight \citep{slane_2004a_constraints-structure},
and \mshfifteenfiftytwo \citep{seward_1982a_x-ray-pulsar}.
There are now over 50 sources identified as being \acp{PWN}
both inside our galaxy and in the \ac{LMC}
\cite{kaspi_2006_isolated-neutron}.
In addition, many \ac{PWN} have been detected at \ac{VHE} energies.
As of April 2013, the \tevcat\footnote{\tevcat 
is a catalog of \ac{VHE} sources compiled by the University of
Chicago. It can be found at \url{http://tevcat.uchicago.edu}.} 
includes 31 \ac{VHE} sources classified as \acp{PWN}.
We will discuss these \ac{VHE} \ac{PWN} in \chapref{tevcat}.

\section{Sources Detected by \Actitle{LAT}}
\seclabel{sources_detected_fermi} 

\begin{figure}[htbp]
  \centering
    \includegraphics[width=\textwidth]{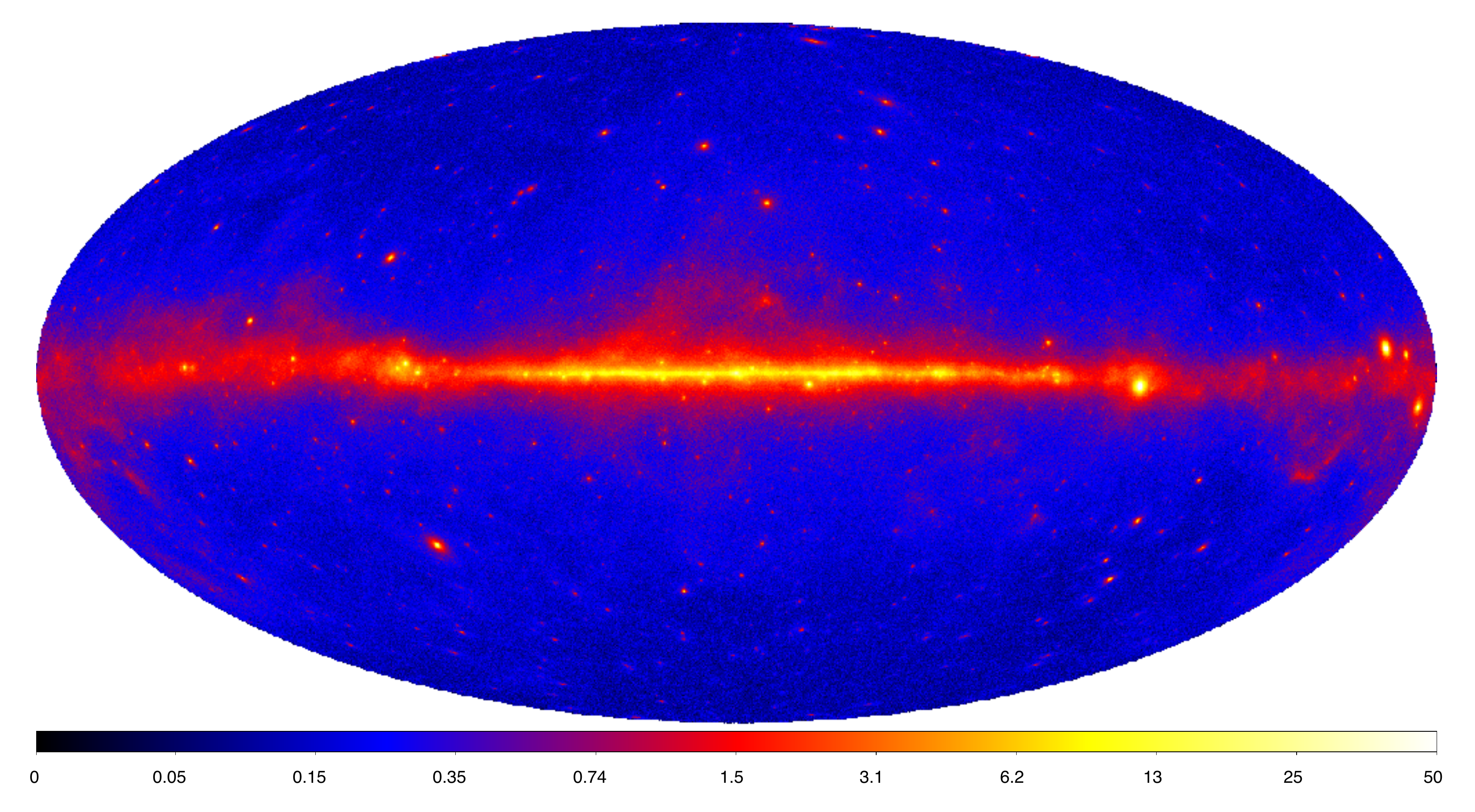}
  \caption{
  An Aitoff projection map of the $\gamma$-ray sky observed by the
  \ac{LAT} with a 2 year exposure.  This map is integrated in the
  energy range from $100\unitspace\mev$ to $10\unitspace\gev$ in units
  of $10^{-7}\erg\unitspace\cm^{-2}\second^{-1}\steradian^{-1}$.  This figure is
  from \cite{nolan_2012_fermi-large}.
  }
  \figlabel{lat_skymap_2fgl}
\end{figure}

\figref{lat_skymap_2fgl} shows a map of the $\gamma$-ray sky observed
by the \ac{LAT} with two years of data. One can clearly observed a
strongly-structured anisotropic component of the $\gamma$-ray emission
coming from the galaxy. In addition, many individual sources of $\gamma$-rays
can be viewed. In \subsecref{galactic_diffuse_and_isotropic}, we
discuss the Galactic diffuse and isotropic $\gamma$-ray background. In
\subsecref{2fgl}, we discuss \ac{2FGL}, a catalog of point-like
sources detected by the \ac{LAT}. In \subsecref{2pc}, we discuss
\ac{2PC}, a catalog of pulsars detected by the \ac{LAT}.  Finally,
in \subsecref{pwn_detected_lat} we discuss \acp{PWN} detected by the
\ac{LAT}.

\subsection{The Galactic Diffuse and Isotropic Gamma-ray Background}
\subseclabel{galactic_diffuse_and_isotropic}

The structured Galactic diffuse $\gamma$-ray emission in our galaxy is
caused by The interaction of cosmic-ray electrons and protons with the
gas in our Milky Way (through the \pion and bremsstrahlung process)
and with Galactic radiation fields (through the \ac{IC} process).

Much work has gone into theoretically modeling this
diffuse $\gamma$-ray emission. The most advanced
theoretical model of the Galactic emission is \galprop
\citep{strong_1998a_propagation-cosmic-ray,moskalenko_2000a_anisotropic-inverse}.
In addition, significant work has gone into comparing these
models to the observed $\gamma$-ray intensity
distribution observed by the \ac{LAT} \citep{abdo_2009a_fermi-large,
ackermann_2012a_fermi-lat-observations}.

In addition to the Galactic diffuse background, the \ac{LAT}
observes an isotropic component to the $\gamma$-ray distribution.
This emission is believed to be a composite of unresolved extragalactic
point-like sources as well as a residual charged-particle background.
\cite{abdo_2010a_spectrum-isotropic} presents detailed measurements of
the isotropic background observed by the \ac{LAT}.

The \galprop predictions for the $\gamma$-ray background are not
accurate enough for the analysis of point-like and $\sim1\degree$
large extended sources.  Therefore, an improved data-driven model
of the Galactic diffuse background has been devised where components
of the \galprop model are fit to the observed $\gamma$-ray emission.
This data-driven model is described in \cite{nolan_2012_fermi-large}.

\subsection{\Actitle{2FGL}}
\subseclabel{2fgl}

Using 2 years of observations, the \ac{LAT} collaboration produced a list
of 1873 $\gamma$-ray-emitting sources detected in the $100\unitspace\mev$
to $100\unitspace\gev$ energy range \citep{nolan_2012_fermi-large}.
Primarily, the catalog assumed sources to be point like. But twelve
previously-published sources were included as being spatially extended
with the spatial model taken from prior publications.

Of these 1873 sources, 127 were firmly identified with a multiwavelength
counterpart.  A source is only firmly identified if it meets one of
three criteria. First, it can have periodic variability (pulsars and
high-mass binaries).  Second, it could have a matching spatial morphology
(\acp{SNR} and \acp{PWN}). Finally, it could have correlated variability
(\ac{AGN}).  In total, \ac{2FGL} firmly identified 83 pulsars, 28
\acp{AGN}, 6 \acp{SNR}, 4 \acp{HMB}, 3 \acp{PWN}, 2 normal galaxies,
and one nova \cite{nolan_2012_fermi-large}.

In addition, 1171 sources are included in the looser criteria that
they were potentially associated with a multiwavelength counterpart.
Using this criteria, 86 sources are associated with pulsars, 25 with
\acp{PWN},  98 with \acp{SNR}, and 162 were flagged as being potentially
spurious due to residuals included by incorrectly modeling the galactic
diffuse emission.

\subsection{\Actitle{2PC}}
\subseclabel{2pc}

Using 3 years of data, the \ac{LAT} collaboration produced \acf{2PC},
a list of 117 pulsars significantly detected by the \ac{LAT}
\citep{abdo_2013a_second-fermi}.  Typically, a \ac{LAT}-detected pulsar is
first detected at either radio or X-ray energies.  This method was used
to discover 61 of the $\gamma$-ray emitting pulsars.  But some pulsars
are known to emit only $\gamma$-rays.  These sources can be searched
for blindly using $\gamma$-ray data.  This method was used to detect
36 pulsars.  Finally, in the third method, the positions of unidentified
\ac{LAT} sources which could potentially be associated with pulsars.
These regions are often searched for in radio to look for pulsar
emission. This method has lead to the detection of 20 new \acp{MSP}.
In total \ac{2PC} detected 42 radio-loud pulsars, 35 radio-quiet pulsars,
and 40 $\gamma$-ray \acp{MSP}.

\subsection{\Acptitle{PWN} Detected by \Actitle{LAT}}
\subseclabel{pwn_detected_lat}

In addition to detecting over 100 pulsars, the \ac{LAT} has detected
several \acp{PWN}.  In situations where the \acp{PWN} has an associated
\ac{LAT}-detected pulsar, typically the spectral analysis of the
\ac{PWN} is performed during times in the pulsar phase where the
pulsar emission is at a minimum. For some pulsars, such as \hessj{1825},
there is no associated \ac{LAT}-detected pulsar and the spectral analysis
can be performed without cutting on pulsar phase.

\subsubsection{Crab}

\begin{figure}[htbp]
  \centering
    \includegraphics{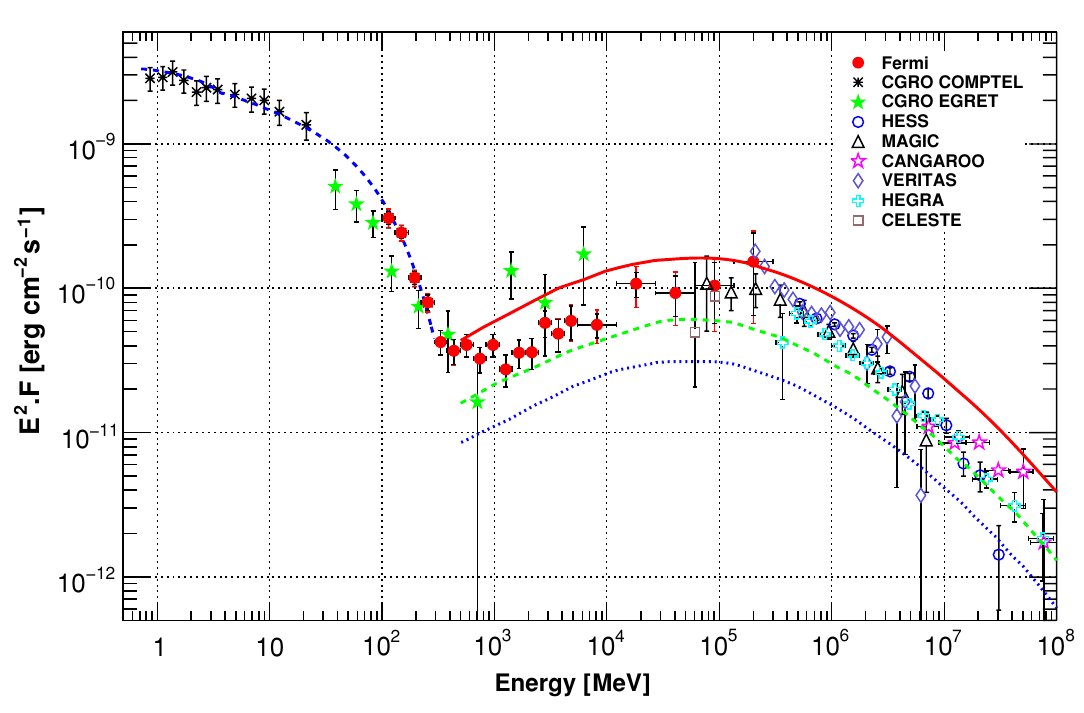}
  \caption{
    The \ac{SED} of the Crab nebula observed by the \ac{LAT} as
    well as several other instruments.
    This figure is from \cite{abdo_2010a_fermi-large}.
  }
  \figlabel{crab_spectrum}
\end{figure}

Observations of the Crab nebula by the
\ac{LAT} provided detailed spectral resolution of Crab's spectrum
\cite{abdo_2010a_fermi-large}.  The Crab nebula shows a very strong
spectral break in the \ac{LAT} energy band, and the $\gamma$-ray emission
is interpreted as being the combination of a synchrotron component at low
energy and an \ac{IC} component at high energy.

In addition, $\gamma$-ray emission from the Crab nebula has
been observed to be variability in time and have flaring periods
\citep{abdo_2011a_gamma-ray-flares}.  The Crab was observed to have
an extreme flare in 2011 \citep{buehler_2012a_gamma-ray-activity}.
This variability is challenging to understand given conventional models
of \ac{PWN} emission.

\subsubsection{\velax}

\begin{figure}[htbp]
  \centering
    \includegraphics{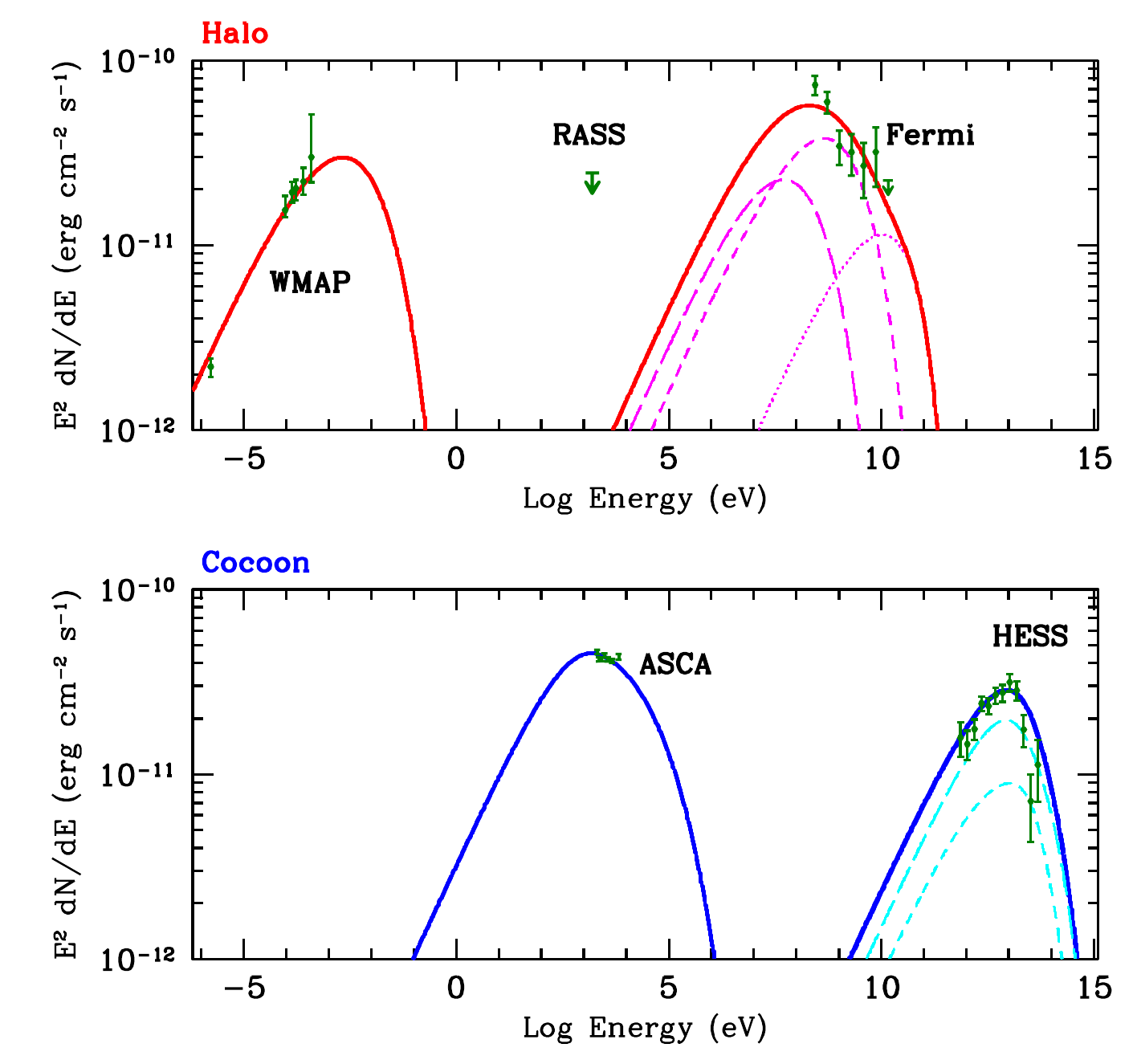}
    \caption{The \ac{SED} of \velax observed at radio, x-ray,
      $\gamma$-ray, and \ac{VHE} energies. The emission was suggested
      by \citep{abdo_2010c_fermi-large} to be driven by two pollutions
      of electrons.  In this model, the lower-energy electron population
      powers the radio and $\gamma$-ray emission adn the higher-energy
      electron population powers the x-ray and \ac{VHE} emission.
      This figure is from \cite{abdo_2010c_fermi-large}.}
  \figlabel{vela_x_sed_two_populations}
\end{figure}

\velax is a \ac{PWN} powered by the Vela pulsar.  It was first observed
by \cite{rishbeth_1958a_radio-emission}.  It was observed at \ac{VHE}
energies by \cite{aharonian_2006a_first-detection} and at \gev
energies by \ac{AGILE} \citep{pellizzoni_2010a_detection-gamma-ray}.
The detailed multiwavelength spectra of \velax is plotted in
\figref{vela_x_sed_two_populations}.  Based upon the morphological
and spectral disconnect between the \gev and \tev emission,
\citep{abdo_2010c_fermi-large} argued that emission was not consistent
with a single population of accelerated electrons.  They suggested
instead that the emission comes instead from two populations of electrons.

\subsubsection{\mshfifteenfiftytwo}

\ac{SNR}
\citep[\mshfifteenfiftytwo][]{caswell_1981a_high-resolution-radio} is
commonly associated with \psrb{1509} \citep{seward_1982a_x-ray-pulsar}.
A diffuse nebula was observed surrounding the pulsar
\citep{seward_1982a_x-ray-pulsar}, adn interpreted as an \ac{PWN}
\cite{trussoni_1996a_rosat-observations}.  The \ac{PWN} was detected at
\ac{VHE} energies by \cite{aharonian_2005a_discovery-extended} and at
\gev energies by \cite{abdo_2010a_detection-energetic}

\subsubsection{\hessj{1825}}

\hessj{1825} is an extended ($\sim0\fdg5$) \ac{VHE} sources
first detected during the \ac{HESS} survey of the inner
galaxy \citep{aharonian_2006a_h.e.s.s.-survey}.  It was
interpreted by \cite{aharonian_2005a_possible-association}
as being a \ac{PWN} powered by \psrj{1826} \citep[also known
as \psrb{1823},][]{clifton_1992a_high-frequency-survey}.
Surrounding the pulsar is a diffuse $\sim 5'$ nebula
\citep{finley_1996a_morphology-young}.  The large size difference
can be understood in terms of the different lifetimes for
the synchrotron-emitting and \ac{IC}-emitting electrons
\citep{aharonian_2006a_h.e.s.s.-survey}.

This source was subsequently detected by
\cite{grondin_2011a_detection-pulsar} at \gev energies.  Interestingly,
the \ac{VHE} emission from \hessj{1825} was observed to have an
energy-dependent morphology, with the size decreasing with increasing
energy \citep{aharonian_2006a_energy-dependent}.  This can be explained
by the \ac{IC} emission model if the electron injection decreases
with time.

\subsubsection{\hessj{1640}}

The \ac{VHE} source \hessj{1640} 
\citep{aharonian_2006a_h.e.s.s.-survey}
is spatially-coincident with \snrg{338.3}
\citep{shaver_1970a_galactic-radio}.  X-ray observations by
\xmmnewton uncovered a spatially-coincident X-ray nebula and within it
a point-like source \cite{funk_2007a_xmm-newton-observations}.
This point-like source is believed to be a neutron star powering the
\ac{PWN}, but pulsations have not yet been detected from it.
\cite{slane_2010_fermi-detection} discovered an associated \gev source.

\subsubsection{\hessj{1857}}

\hessj{1857} was also discovered by \ac{HESS}
\citep{aharonian_2008a_very-high-energy-gamma-ray}
\cite{hessels_2008a_j18560245:-arecibo} suggested that \hessj{1857}
is a \ac{PWN} powered by \psrj{1856}.  \hessj{1857} was also detected
by the \ac{LAT}.

\subsubsection{\hessj{1023}}

The \ac{VHE} source \hessj{1023} was discovered in the region of the young
stellar cluster Westerlund 2 \cite{aharonian_2007a_detection-extended}.
This same source was subsequently detected by the \ac{LAT} in the off-peak
region surrounding \psrj{1023} \citep{ackermann_2011a_fermi-lat-search}.
\cite{h.e.s.s.collaboration_2011a_revisiting-westerlund} proposed that
the emission could either be due to an \acp{PWN} or due to hadronic
interactions of cosmic rays accelerated in the open stellar cluster
interacting with molecular clouds.

\section{Radiation Processes in Gamma-ray Astrophysics}
\seclabel{radiation_processes}

Nonthermal radiation observed from astrophysical sources is
typically believed to originate in through synchrotron radiation, \ac{IC}
upscattering, and the
decay of neutral \pion particles.  We will discuss these processes
in \subsecref{synchrotron}, \subsecref{inverse_compton}, and
\subsecref{bremsstrahlung} respectively.

\subsection{Synchrotron}
\subseclabel{synchrotron}

\begin{figure}[htbp]
  \centering 
    \includegraphics{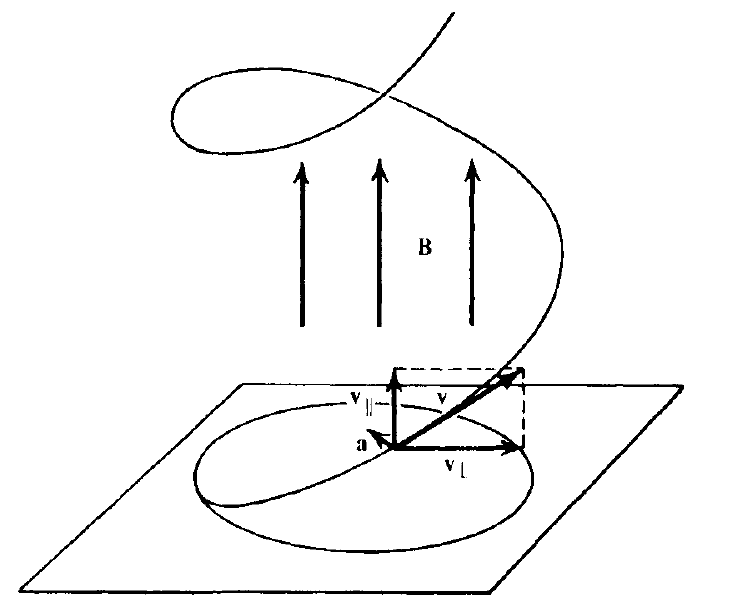}
    \caption{In synchrotron radiation, charged particles spiral along
      magnetic filed lines, radiating photons as they accelerate.
	This figure is from \cite{rybicki_1979a_radiative-processes}.
    }
  \figlabel{syncrotron_radiation_spiral}
\end{figure}

The synchrotron radiation processes is observed when charged particles
spiral around magnetic field lines.  This process is illustrated in
\figref{syncrotron_radiation_spiral}.  This emission is discussed
thoroughly in \cite{blumenthal_1970a_bremsstrahlung-synchrotron} and
\cite{rybicki_1979a_radiative-processes}.  In what follows, we adopt
the notation from \cite{houck_2006a_models-nonthermal}.

A charged particle of mass $\mass$ and charge $\charge$ in a magnetic
field of strength \MagneticFieldVector will experience an electromagnetic
force:
\begin{align}
  \dbydt (\gamma m \VelocityVector) = & \frac{q}{c} \cross{\VelocityVector}{\MagneticFieldVector}.
\end{align}
This force will cause a particle to accelerate around the magnetic field
lines, radiating due to maxwell's equations.  The power emitted at a
frequency $\frequency$ by one of these particles is
\begin{equation}
  \eqnlabel{power_emitted_particle_sync}
  \power_\text{emitted}(\frequency) = 
  \frac{\sqrt{3} \charge^3 B \sin\alpha}{\mass \speedoflight^2} F(\frequency/\frequency_c),
\end{equation}
where $\alpha$ is the angle between the particle's velocity vector and
the magnetic field vector. Here,
\begin{equation}
  F(x) \equiv x \int_x^\infty K_{\tfrac{5}{3}} (\xi) d\xi,
\end{equation}
and
\begin{equation}
  \eqnlabel{characteristic_freqnecy_synctotron}
\frequency_c = \frac{3q \MagneticField \gamma^2}{4\pi \mass \speedoflight} 
\sin\alpha \equiv \nu_0 \gamma^2 \sin\alpha
\end{equation}

Because power is inversely-proportional to mass, synchrotron radiation
is predominatly from electrons.

Now, we assume a population of particles and compute the total observed
emission. We say that $\ParticleDistribution(\momentum,\alpha)$ is the
number of particles per unit momentum and solid angle with a momentum
$\momentum$ and pitch angle $\alpha$.  We find the total power emitted
by integrating over particle momentum and distribution
\begin{equation}
  \frac{\derivative W}{\derivative\time}=
  \int \derivative \momentum 
  \int \derivative \solidangle
  \power_\text{emitted}(\frequency)
  \ParticleDistribution(\momentum,\alpha)
\end{equation}
If we assume the pitch angles of the particles to be isotropically
distributed and, including \eqnref{power_emitted_particle_sync}, we
find that the number of photons emitted per unit energy and time is
\begin{equation}
  \frac{\derivative \ParticleDistribution}{\derivative \omega \derivative \time} =
  \frac{\sqrt{3}q^3 B}{h m_e c^2 \angularfrequency}
  \int \derivative\momentum
  \ParticleDistribution(\momentum)
  R \left(\frac{\omega}{\omega_0 \gamma^2}\right)
\end{equation}
where
\begin{equation}
  R(x) \equiv \frac{1}{2} \int_0^\pi
  \derivative \alpha \sin^2 \alpha
  F\left(\frac{x}{\sin\alpha}\right)
\end{equation}

It is typical in astrophysics to assume a 
a power-law distribution of electrons:
\begin{equation}
\eqnlabel{ElectronPowerLawEnergyDistribution}
  \ParticleDistribution(\momentum) \derivative\momentum = 
  \kappa \momentum^{-\spectralindex} \derivative\momentum.
\end{equation}
For a power-law distribution of photons integrated over
pitch angle, we find the total power emitted to me 
\begin{equation}
\TotalPower(\angularfrequency) \propto \kappa \MagneticField^{(p+1)/2} 
\angularfrequency^{-(p-1)/2}.
\end{equation}
See, \cite{rybicki_1979a_radiative-processes} 
or \cite{longair_2013a_energy-astrophysics} for a full derivation.
This shows that, assuming a power-law electron distribution,
the electron spectral index can be related to the photon spectral
index.

\subsection{\Actitle{IC}}
\subseclabel{inverse_compton}

Normal Compton scattering involves a photon colliding with a free electron
and transferring energy to it. In \ac{IC} scattering, a high-energy 
electron interacts with a low-energy photon imparting energy to it.
This process occurs when highly-energetic electrons interact with
a dense photon field.

The derivation of \Ac{IC} emission requires a quantum
electrodynamical treatment. It was first derived by
\cite{klein_1929a_streuung-strahlung}. In what follows, we follow the
notational convention of \cite{houck_2006a_models-nonthermal}.
We assume a population of relativistic ($\gamma\gg1$) electrons
written as $\ParticleDistribution(\momentum)$ which is contained inside isotropic 
photon distribution with number density $n(\omega_i)$.

The distribution of photons 
emitted by \ac{IC} scatter is
written as
\begin{equation}
  \frac{\derivative\ParticleDistribution}{\derivative\omega\derivative\time} = 
  c \int d \omega_i n(\omega_i)
  \int_{p_\text{min}}^\infty  dp
  \ParticleDistribution(p) 
 \KleinNishinaCrossSection(\gamma,\omega_i,\omega)
\end{equation}
where $\omega$ is the outgoing photon energy written
in units of the electron rest mass energy, $\omega\equiv h\nu/(m_e c^2)$,
and $\KleinNishinaCrossSection$ is the Klein-Nishina cross section:
\begin{equation}
\KleinNishinaCrossSection(\gamma,\omega_i,\omega) = \frac{2\pi r_0^2}{\omega_i \gamma^2}
  \left[
  1 + q - 2q^2 + 2q\ln q + \frac{\tau^2 q^2 (1-q)}{2(1+\tau q)}.
  \right]
\end{equation}
Here,
\begin{equation}
  q \equiv \frac{\omega}{4 \omega_i \gamma (\gamma-\omega)},
\end{equation}
$\tau \equiv 4\omega_i \gamma$, and $r_0 = e^2/(m_e c^2)$ is the classical
electron radius.
The threshold electron Lorentz factor is
\begin{equation}
  \gamma_\text{min} =
  \frac{1}{2} 
  \left(
  \omega + \sqrt{\omega^2 + \frac{\omega}{\omega_i}}
  \right)
\end{equation}

Often, \ac{IC} emission happens when an accelerated power-law distribution
of electrons interacts with a thermal photon distribution
\begin{equation}
  n(\omega_i) = 
  \frac{1}{\pi^2\lambda^3} 
  \frac{\omega_i^2}{e^{\omega_i/\Theta} -1}
\end{equation}
where $\lambda=\hbar/(m_e c)$ and $\Theta=kT/(m_e c^2)$.
Often, the target photon distribution is the \ac{CMB}, with
$T=2.725\unitspace\kelvin$.

\subsection{Bremsstrahlung}
\subseclabel{bremsstrahlung}

Bremsstrahlung radiation is composed of electron-electron and electron-ion
interactions.  In either case, we assume a differential spectrum of
accelerated electrons $\ParticleDistribution_\electron(\energy)$ that
interacts with a target density of electrons (\ElectronDensity) or ions
(\IonDensity).

\begin{equation}
  \frac{\derivative \ParticleDistribution}{\derivative\energy\derivative\time} =
  \ElectronDensity \int \derivative\energy
  \ParticleDistribution_\electron(\energy) \velocity_\electron
  \frac{\derivative\CrossSection_{\electron\electron}}{\derivative\energy} +
  \IonDensity \int \denergy
  \ParticleDistribution_\electron(\energy) \velocity_\electron
  \frac{\derivative\CrossSection_{\electron\AtomicNumber}}{\derivative\energy}
\end{equation}

Here, $\velocity_\electron$ is the velocity of the
electron, and $\CrossSection_{\electron\electron}$ and
$\CrossSection_{\electron\AtomicNumber}$ are the electron-electron
and electron-ion cross sections.  The actual formulas for
$\derivative\CrossSection_{\electron\electron}/\derivative\energy$ and
$\derivative\CrossSection_{\electron\AtomicNumber}/\derivative\energy$
are quite involved.  The electron-electron cross section was worked out
in \cite{haug_1975a_bremsstrahlung-production}.  The electron-ion cross
section is called the Bethe-Heitler cross-section and is worked
out in the Born approximation in \cite{heitler_1954a_quantum-theory}
and \cite{koch_1959a_bremsstrahlung-cross-section}.  A more
accurate relativistic correction to this formula is given in
\cite{haug_1997a_nonrelativistic-bremsstrahlung}.  We refer to
\cite{houck_2006a_models-nonthermal} for a detailed numerical
implementation of these formulas.

\subsection{Pion Decay}

Neutral \pion decay occurs when highly-energetic protons interact with
thermal protons. This emission happens when protons decay into neutral
pions through $pp \processarrow \pion + X$ and the $\pion$ subsequently
decay through $\pion \processarrow 2\gamma$.  The gamma-ray emission
from neutral pion decay can be computed as
\begin{equation}
  \frac{\derivative\ParticleDistribution}{\derivative\energy\derivative\time} = 
  \HydrogenDensity \int \denergy \velocity_\proton \ParticleDistribution_\proton(\energy) 
  \frac{\derivative\CrossSection_{\proton\proton}}{\derivative\energy}
\end{equation}
Here, $\ParticleDistribution_\proton(\energy)$
is the differential proton distribution,
$\derivative\CrossSection_{\proton\proton}/\derivative\energy$ is
$\gamma$-ray cross section from proton-proton interactions, and
$\HydrogenDensity$ is the target hydrogen density.  The computation
of $\derivative\CrossSection_{\proton\proton}/\derivative\energy$
is rather involved. Typically, people employ a parameterization
of the calculations performed by
\cite{kamae_2006a_parameterization-gamma}.

\chapter{The Pulsar/\Actitle{PWN} System}
\chaplabel{pulsar_pwn_system}

\section{Neutron Star Formation}
\seclabel{neutron_star_formation}

As was discussed in \secref{pulsars_and_pwn}, pulsars, \acp{PWN},
and supernova remnants are all connected through the death of a star.
When a star undergoes a supernova, the ejecta forms a supernova remnant.
If the remaining stellar core has a mass above the Chandrasekhar
limit, then the core's electron degeneracy pressure cannot counteract
the core's gravitational force and the core will collapse into
a \ac{NS}.  The Chandrasekhar mass limit can be approximated as
\citep{chandrasekhar_1931_maximum-ideal}
\begin{equation}
  \MassChandrasekhar \approx 
  \frac{3\sqrt{2\pi}}{8}
  \left(\frac{\hbar\speedoflight}{\gravitationalconstant}\right)^{3/2}
  \left[
  \left(\frac{\NumberProtons}{\NumberNucleons}\right)
  \frac{1}{\MassHydrogen^2}
  \right],
\end{equation}
where $\hbar$ is the reduced Planck constant, \speedoflight is the
speed of light, \gravitationalconstant is the gravitational constant,
\MassHydrogen is the mass of hydrogen, \NumberProtons is the number of
protons, \NumberNucleons is the number of nucleons, and \solarmass is the
mass of the sun \citep{carroll_2006_introduction-modern}.  When this
formula is computed more exactly, one finds $\MassChandrasekhar =
1.44 \solarmass$.

Because \acp{NS} are supported by
neutron degeneracy pressure,
the radius of a neutron star can be approximated as
\citep{carroll_2006_introduction-modern}
\begin{equation}
  \RadiusNeutronStar \approx \frac{(18 \pi)^{2/3}}{10}
  \frac{\hbar^2}{\gravitationalconstant \MassNeutronStar^{1/3}}
  \left(\frac{1}{\MassHydrogen}\right)^{8/3}.
\end{equation}
The canonical radius for \acp{NS} is $\sim 10\unitspace\km$.

In these very dense environments, the protons and electrons in the \ac{NS}
form into neutrons through inverse $\beta$ decay:
\begin{equation}
  \proton^\positive + \electron^\negative 
  \processarrow \neutron + \electronneutrino.
\end{equation}

If a \ac{NS} had a sufficiency large mass, the gravitational
force would overpower the neutron degeneracy pressure and the
object would collapse into a black hole. The maximum mass of a
\ac{NS} is unknown because it depends on the equation of state
inside the star, but is commonly predicted to be $\sim 2.5\solarmass$
Recently, a pulsar with a mass of $\sim 2\solarmass$ was discovered
\citep{demorest_2010_two-solar-mass-neutron}, constraining
theories of the equation of state.

In addition to rotationally-powered pulsars, the primary class of observed
pulsars, there are two additional classes of pulsars with a different
emission mechanism.  For accretion-powered pulsars, also called X-ray
pulsars, the emission energy comes from the accretion of matter from a
donor star \citep{caballero_2012a_x-ray-pulsars:}.  They are bright and
populous at X-ray energies.  Magnetars have the strong magnetic field
which power their emission \cite{rea_2011a_magnetar-outbursts:}.

\section{Pulsar Evolution}
\seclabel{pulsar_evolution}

\begin{figure}[htbp]
  \centering
    \includegraphics[width=0.5\textwidth]{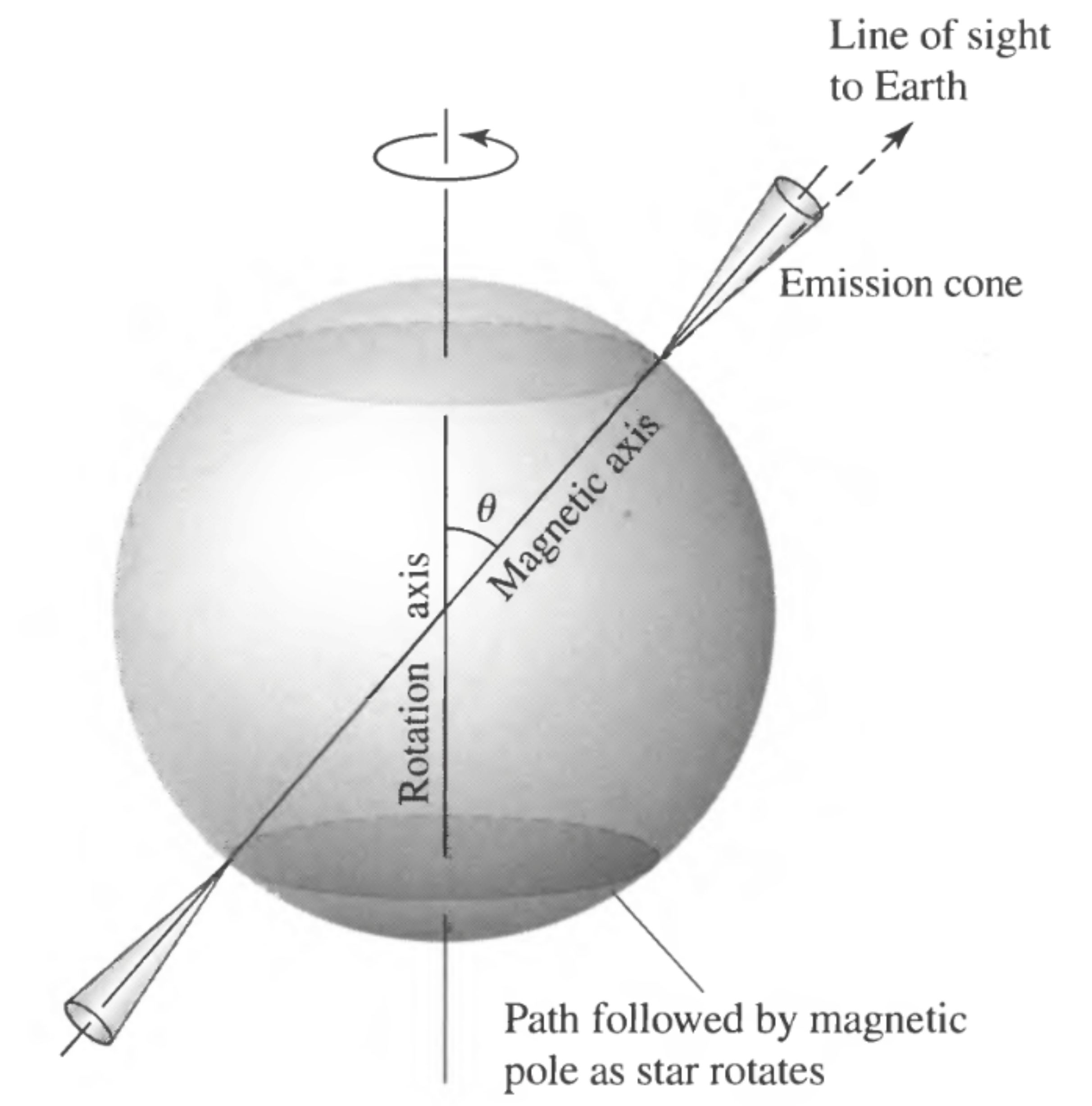}
    \caption{The rotating dipole model of a pulsar. This figure is taken
    from \citep{carroll_2006_introduction-modern}.}
  \figlabel{pulsar_model}
\end{figure}

The simplest model of a pulsar is that it is a rotating dipole magnet
with the rotation axis and the magnetic axis offset by an angle
$\PulsarRotationAngle$ (see \figref{pulsar_model}).  The energy output
from the pulsar is assumed to come from rotational kinetic energy
stored in the neutron star which is released as the pulsar spins down.

For a pulsar, both the period \period and the period derivative
$\perioddot=d\period/d\time$ can be directly observed.  Except in a
few \acp{MSP} which are being sped up through accretion \cite[see for
example][]{falanga_2005_integral-observations}), pulsars are slowing down
($\perioddot<0$).  We write the rotational kinetic energy as
\begin{equation}\eqnlabel{rotational_energy}
  \energyrotational = \tfrac{1}{2} I \PulsarAngularFrequency^2
\end{equation}
where $\PulsarAngularFrequency = 2\pi/\period$ is the angular frequency
of the pulsar and \momentofinertia is the moment of inertia.
For a uniform sphere,
\begin{equation}
  \momentofinertia = \frac{2}{5} M R^2.
\end{equation}
Assuming a canonical pulsar (see \secref{neutron_star_formation}),
we find a canonical moment of inertia of
$\momentofinertia=10^{45}\unitspace\gram\unitspace\cm^{-2}$.

We make the connection between the pulsar's spin-down
energy and the rotational kinetic energy as $\energydot
= - \derivative\energyrotational/\dtime$.
\eqnref{rotational_energy} can be rewritten as
\begin{equation}\eqnlabel{edot_from_rotation}
  \energydot = I \PulsarAngularFrequency \PulsarAngularFrequencyDot.
\end{equation}
It is believed that as the pulsar spins down, the this rotational energy
is released as pulsed electromagnetic radiation and also as a wind of
electrons and positrons accelerated in the magnetic field of the pulsar.

If the pulsar were a pure dipole magnet, its radiation would be described
as \citep{gunn_1969_magnetic-dipole}
\begin{equation}\eqnlabel{edot_pure_dipole}
  \energydot = \frac{2\MagneticField^2 \PulsarRadius^6 
  \PulsarAngularFrequency^4 \sin^2\PulsarRotationAngle}{
  3\speedoflight^3}.
\end{equation}
Combining equations \eqnref{edot_from_rotation} and
\eqnref{edot_pure_dipole}, we find that for a pure dipole magnet,
\begin{equation}\eqnlabel{breaking_index_dipole}
  \PulsarAngularFrequencyDot \propto \PulsarAngularFrequency^3.
\end{equation}

In the few situations where this relationship has been
conclusively measured, this relationship does not hold
\citep[See][ and references therein]{espinoza_2011_braking-index}.
We generalize \eqnref{breaking_index_dipole} as:
\begin{equation}\eqnlabel{angular_frequency_derivative_relation}
  \PulsarAngularFrequencyDot \propto \PulsarAngularFrequency^\breakingindex
\end{equation}
where $\breakingindex$ is what we call the breaking index.
We solve \eqnref{angular_frequency_derivative_relation}
for \breakingindex by taking the derivative:
\begin{equation}
  \breakingindex = \frac{\PulsarAngularFrequency \PulsarAngularFrequencyDotDot}{\PulsarAngularFrequencyDot^2}
\end{equation}

The breaking index is hard to measure due to timing noise and glitches
in the pulsar's phase. To this date, it has been measured in eight
pulsars \cite{espinoza_2011_braking-index}, and in all situations
$\breakingindex<3$. This suggests that there are additional processes
besides magnetic dipole radiation that contribute to the energy release
\citep{blandford_1988_interpretation-pulsar}.

\eqnref{angular_frequency_derivative_relation} is a Bernoulli differential
equation which can be integrated to solve for time:
\begin{equation}\eqnlabel{pulsar_age}
  T = \frac{\period}{(\breakingindex-1) \absval{\perioddot}}
  \left(
  1-\left(\frac{\period_0}{\period}\right)^{(\breakingindex-1)}
  \right)
\end{equation}
For a canonical $\breakingindex=3$ pulsars which is relatively old
$\period_0 \ll \period$, we obtain what is called the characteristic
age of the pulsar:
\begin{equation}
  \PulsarAge = \period/2\perioddot.
\end{equation}

Using \eqnref{edot_from_rotation} and \eqnref{breaking_index_dipole},
we can solve for the spin-down evolution of the pulsar as a function of
time \citep{pacini_1973_evolution-supernova}:
\begin{equation}\eqnlabel{energy_dot_vs_time}
    \energydot(t) = \energydot_0
    \left(
    1 + \frac{\time}{\SpinDownTimescale}
    \right)^{-\frac{(\breakingindex+1)}{(\breakingindex-1)}}.
\end{equation}
Here,
\begin{equation}
  \SpinDownTimescale \equiv \frac{\period_0}{(\breakingindex-1)\absval{\perioddot_0}}.
\end{equation}

\eqnref{edot_from_rotation}, \eqnref{pulsar_age}, and
\eqnref{energy_dot_vs_time} show us that given the current period,
period derivative, and breaking index, we can calculate the pulsar's
age and energy-emission history.

In a few situations, the pulsar's age is well known and the
breaking index can be measured, so $\period_0$ can be inferred. See
\cite{kaspi_2002_constraining-birth} for a review of the topic. For
other sources, attempts have been made to infer the initial spin-down
age based on the dynamics of an associated \acs{SNR}/\ac{PWN}
\citep{van-der-swaluw_2001_inferring-initial}.

Finally, if we assume dipole radiation is the only source of energy
release, we can combine equation \eqnref{edot_from_rotation} and
\eqnref{edot_pure_dipole} to solve for the magnetic field:
\begin{equation}
  \MagneticField = \sqrt{\frac{3\momentofinertia\speedoflight^3}{
  8\pi^2\PulsarRadius^6\sin^2\PulsarRotationAngle}\period\perioddot}
  = 3.2\times 10^{19} \sqrt{\period\perioddot} \unitspace\gauss
\end{equation}
where in the last step we assumed the canonical values of
$\momentofinertia=10^{45}\unitspace\gram\unitspace\cm^{-2}$,
$\PulsarRadius=10\unitspace\km$, $\PulsarRotationAngle=90\degree$, and we
assume that $\period$ is measured in units of seconds.  For example,
for the Crab nebula, $\period\approx33\unitspace\millisecond$
\citep{staelin_1968_pulsating-radio} and
$\perioddot\approx36\unitspace\nanosecond$
per day \citep{richards_1969a_period-pulsar} so
$\MagneticField\approx10^{12}\unitspace\gauss$.

\section{Pulsar Magnetosphere}
\seclabel{pulsar_magnetosphere}

The basic picture of a pulsar magnetosphere was first presented in
\cite{goldreich_1969_pulsar-electrodynamics}.  The magnetic dipole of
the rotating \ac{NS} creates a quadrupole electric field.

The potential generated by this field is given as
\citep{goldreich_1969_pulsar-electrodynamics}:
\begin{equation}
  \PulsarPotential = \frac{\MagneticField \PulsarAngularFrequency^2 \PulsarRadius^2}{2c^2}
  \approx 6\times 10^{12} 
  \left(\frac{\MagneticField}{10^{12}\unitspace\gauss}\right)
  \left(\frac{\PulsarRadius}{10\unitspace\km}\right)^3
  \left(\frac{\period}{1\unitspace\second}\right).
\end{equation}
For \acp{NS}, this potential produces a magnetic field that is much larger
than the gravitational force and acts as a powerful particle accelerators.

Pulsars typically release only a small percent of their overall
energy budget as pulsed emission. The efficiency of converting
spin-down energy into pulsed $\gamma-$rays is typically $\sim$
0.1\% to 10\% \citep{abdo_2010a_first-fermi}.  For example, the Crab
nebulae is estimated to release 0.1\% of it's spin-down energy as
pulsed $\gamma$-rays \citep{abdo_2010a_fermi-large}.  Typically,
the energy released as radio and optical photons is much less.
The optical flux of the Crab is a factor of $\sim100$ smaller
\citep{cocke_1969_discovery-optical} and the radio flux is a factor of
$\sim 10^4$ smaller.  Therefore, the vast majority of the energy output
of the pulsar is carried away as a pulsar wind, which will be described
in the next section.

\begin{figure}[htbp]
  \centering
    \includegraphics{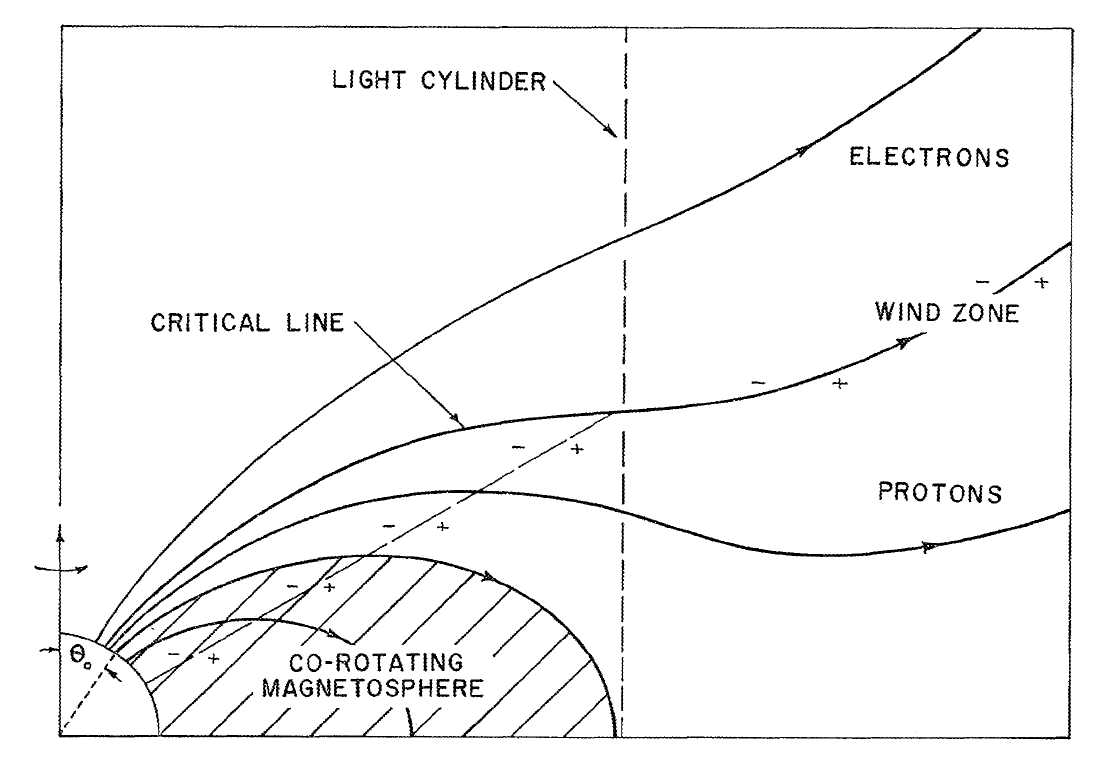}
  \caption{The magnetosphere for a rotating pulsar.
  The pulsar is on the bottom left of the plot. This figure is
  from \cite{goldreich_1969_pulsar-electrodynamics}.}
  \figlabel{pulsar_magnetosphere}
\end{figure}

\figref{pulsar_magnetosphere} shows a schematic diagram of this
magnetosphere. It is commonly believed that the radio emission
from pulsars originates within 10\% of the light cylinder radius
\citep[see][and references therein]{kijak_2003a_radio-emission}

On the other hand, there is still much debate about the location
of the $\gamma$-ray emission.  Three locations have been proposed.
In the \ac{PC} model, the $\gamma$-ray emission arises from within one
stellar radius \citep{daugherty_1996a_gamma-ray-pulsars:}.  This model
was disfavored based upon the predicted $\gamma$-ray spectrum
\citep{abdo_2009b_fermi-large}.  In the \ac{OG} model,
$\gamma$-ray emission is predicted near the pulsar's light cylinder
\citep{cheng_1986a_energetic-radiation,romani_1996a_gamma-ray-pulsars:}.
Finally, in the \ac{TPC} model, the $\gamma$-ray emission
comes from an intermediate region in the pulsar magnetosphere
\citep{dyks_2003a_two-pole-caustic,muslimov_2004a_high-altitude-particle}
Much work has gone into comparing the \ac{TPC} and
\ac{OG} models in the context of detailed \ac{LAT}
observations of $\gamma$-ray pulsars \citep[See for
example][]{watters_2011a_galactic-population,romani_2011a_sub-luminous-gamma-ray}.

\section{Pulsar Wind Nebulae Structure}
\seclabel{pwn_structure}

The basic picture of \acp{PWN}
comes from \cite{rees_1974_origin-magnetic} and
\cite{kennel_1984_magnetohydrodynamic-model}.  More 
sophisticated models have emerged over the years.  See, for example,
\cite{gelfand_2009_dynamical-model} and references therein.

The wind ejected from the pulsar's magnetosphere is initially
cold which means that it flows radially out from the pulsar.  This
unshocked pulsar wind only emits radiation through \ac{IC} scattering
\citep{bogovalov_2000_very-high-energy-gamma}.  This pulsar wind forms
a bubble as it presses into the \ac{SNR} and forms a termination shock
where the particle wind is further accelerated.

As the wind leaves the magnetosphere, it is believed to be dominated
by the energy carried off in electromagnetic fields (the pointing flux
\pointingflux).  The rest of the energy is released as a particle flux
(\particleflux).  We define the magnetization of the pulsar wind as
\begin{equation}
  \magnetization = \frac{\pointingflux}{\particleflux}
\end{equation}
Outside the pulsar light curve, typically $\magnetization>10^4$, but
at the termination shock typical values for $\sigma$ are $\lesssim0.01$
\citep{kennel_1984a_confinement-pulsars}.  The cause of this transition
is uncertain \citep{gaensler_2006_evolution-structure}.

\begin{figure}[htbp]
  \centering
    \includegraphics{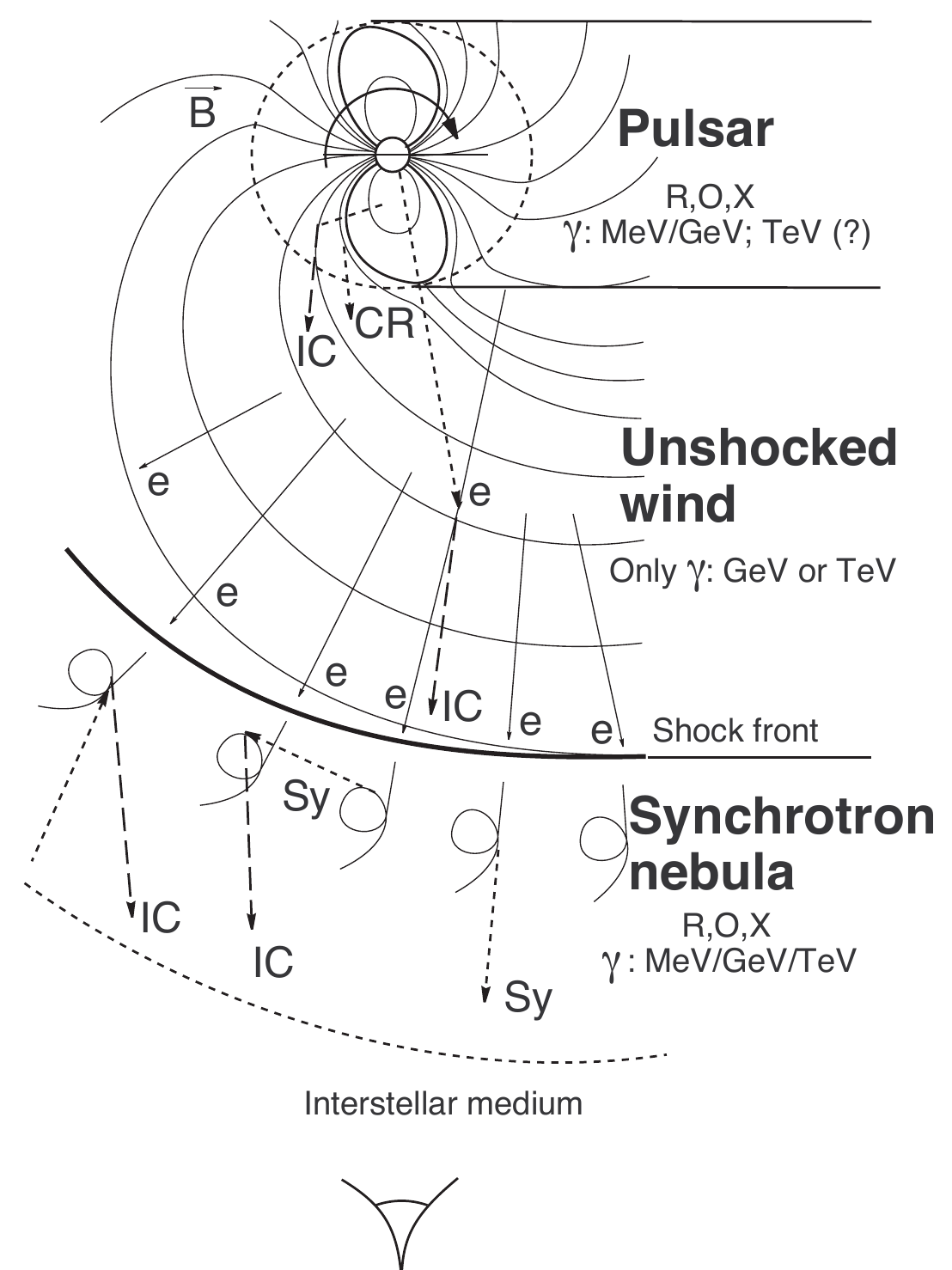}
  \caption{The regions of emission in a pulsar/\ac{PWN} system. 
  This figure shows (top) the pulsar's magnetosphere, (middle), the
  unshocked pulsar wind and (bottom) the shocked pulsar wind which can
  be observed as the \ac{PWN}.
  ``R'', ``O'', ``X'', and ``$\gamma$'' describe sites of radio, optical, X-ray, and
  $\gamma$-ray emission respectively.
  ``CR'', ``Sy'', and ``IC'' refer to regions of curvature, inverse Compton, and
  synchrotron emission.
  Figure is taken from \cite{aharonian_2003_exploring-physics}.
  }
  \figlabel{termination_shock}
\end{figure}

The radius of the bubble (\radiusterminationshock) can be computed as the
radius where the ram pressure from the wind equals the pressure of the
gas in the \ac{SNR}.  The ram pressure is computed as the energy in the
bubble $\energydot \radiusterminationshock/c$ (assuming the particles
travel with a velocity $\approx \speedoflight$) divided by the volume
$4\pi\radiusterminationshock^3/3$: \begin{equation}
  \radiusterminationshock = \sqrt{\frac{\energydot}{\tfrac{4}{3}\pi \pressureISM \speedoflight}}.
\end{equation}
Here, \pressureISM is the pressure in the SNR.  Typical values
for the termination shock are 0.1\unitspace\parsec which is an
angular size $\sim$ \ac{arcsec} for distances $\sim\kiloparsec$
\citep{gaensler_2006_evolution-structure}.

At the termination shock, the particles are thermalized
(given a random pitch angle), and accelerated to energies of
$10^{15}\unitspace\electronvolt$ \citep{arons_1996_pulsars-gamma-rays}.
Downstream of the shock, the particles emit synchrotron and \ac{IC}
radiation as the thermalized electron population interacts with the
magnetic filed and seed photons \citep{gaensler_2006_evolution-structure}.
\figref{termination_shock} shows a diagram describing the pulsar
magnetosphere, the unshocked wind, and the synchrotron nebula which make
up the Pulsar/\ac{PWN} system.

\section{\Actitle{PWN} Emission}
\seclabel{pwn_emission}

In a \acp{PWN}, accelerated electrons emit radiation across the
electromagnetic spectrum through synchrotron and \ac{IC} emission.
Typical photon energies for synchrotron and \ac{IC} emission from
\acp{PWN} are $\sim1\kev$ and $\sim1\tev$ respectively.  A
typical magnetic field strength is $\sim10\micro\gauss$.

Using \eqnref{characteristic_freqnecy_synctotron}, we can show that
photons with an energy $\energy_\kev$ in a magnetic field of strength
$\MagneticField$ radiate electrons with a typical energy of
$E_\electron$ given by
\begin{equation}
  \energy_\electron \approx  70\unitspace\tev \MagneticField_{-5}^{-1/2} \energy_\kev^{1/2},
\end{equation}
where the magnetic field is $\MagneticField=10^{-5}
B_{-5}\gauss$ and $\energy_\kev$ is written in units of \kev
\citep{de-jager_2009a_implications-observations}.

Similarly, if we assume the \ac{PWN} \ac{IC} emission is due to scattering
off the \ac{CMB}, the electron energy which will produce characteristic
\tev $\gamma$-rays is:
\begin{equation}
  \energy_\electron \approx 20\unitspace\tev \energy_\tev^{1/2}
\end{equation}
where $\energy_\tev$ is the scattered photon energy in units of $\tev$
\citep{de-jager_2009a_implications-observations}.  This shows that for a
typical \ac{PWN}, $\sim70\unitspace\tev$ electrons power the synchrotron
emission and $\sim20\unitspace\tev$ electrons power the \ac{IC} emission.

Similarly, we can write down the lifetime of electrons due to
synchrotron and \ac{IC} emission. We define the lifetime as $\lifetime =
\energy/\energydot$ and, using \cite{rybicki_1979a_radiative-processes}:
\begin{equation}
  \lifetime(\energy_\electron) = 
  \left(
  \tfrac{4}{3} \CrossSection_T \speedoflight (\EnergyDensity_\MagneticField + \EnergyDensity_\text{ph}) 
  \energy_\electron/ \mass_\electron^2\speedoflight^4
  \right)^{-1}
\end{equation}
where $\EnergyDensity_\MagneticField = \MagneticField^2/8\pi$
is the magnetic field energy density and
$\EnergyDensity_\text{ph}$ is the energy density of the photon field
($\EnergyDensity_\text{ph}=0.25\unitspace\electronvolt\unitspace\cm^{-3}$
for the \ac{CMB} radiation field).  If $B>3\micro\gauss$, the synchrotron
radiation dominates the cooling ($\EnergyDensity_\MagneticField >
\EnergyDensity_\text{ph}$).

For synchrotron-emitting electrons, the cooling time is
\begin{equation}
  \lifetime_\text{sync} = (1.2\unitspace\kyr) B_{-5}^{-3/2} \energy_\kev^{-1/2}.
\end{equation}
For \ac{IC}-scattering electrons, the cooling time (in the Thomson
limit) is
\begin{equation}
  \eqnlabel{ic_timescale}
  \lifetime_\text{IC} = (4.8\kyr) B_{-5}^{-2} \energy_\tev^{-1/2}.
\end{equation}
From this, we see that the typical timescale for cooling of
synchrotron-emitting electrons ($1\unitspace\kyr$) is much shorter
than the timescale for cooling of \ac{IC}-emitting electrons
($5\kyr$). Because of this, the \ac{IC}-emitting electrons have a
longer time to diffuse away from the pulsar.  For older \ac{PWN}, we
therefore expected the observed \ac{VHE} emission to be larger than
the observed X-ray emission.  This has been observed in many \acp{PWN}
such as \hessj{1825} \citep{aharonian_2006a_h.e.s.s.-survey} and in can
also make identification of \ac{VHE} sources as \ac{PWN} difficult.

We also note that \eqnref{ic_timescale} predicts that the timescale
of \ac{IC}-emitting electrons scales with inverse square root of
the emitted photon energy. This leads to the prediction that
the size of the \ac{VHE} $\gamma$-ray emission should decrease
with increasing energy. This has been observed for \hessj{1825}
\citep{aharonian_2006a_energy-dependent}.

Finally, we mention that \cite{mattana_2009_evolution-gamma-} discussed
the relationship between the X-ray and $\gamma$-ray luminosity as a
function of the pulsar spin-down energy \energydot and age \PulsarAge.
The time integral of \eqnref{energy_dot_vs_time} can be used to compute
the total number of particles that emit synchrotron and \ac{IC} photons.

Most $\gamma$-ray-emitting \ac{PWN} are connected to pulsars with a
characteristic age $\PulsarAge\sim1-20\unitspace\kyr$. So for most
\ac{PWN}:
\begin{equation}
  \lifetime_\text{sync} < \PulsarAge < \lifetime_\text{IC}
\end{equation}
Therefore, for synchrotron emission the number of synchrotron-emitting
particles $n_\text{sync}$ goes as
\begin{equation}
  n_\text{sync} \approx \energydot(\PulsarAge) \lifetime_\text{sync} \sim \energydot_0 \PulsarAge^{-2}
\end{equation}
where in the last step we have assumed a pure dipole magnetic field
($\breakingindex=3$) and used \eqnref{energy_dot_vs_time}.

On the other hand, the number of \ac{IC}-emitting particles $n_\text{IC}$
is approximately independent of time because $\PulsarAge \gg
\SpinDownTimescale$.  Furthermore, \cite{mattana_2009_evolution-gamma-}
argues that $n_\text{IC}$ should be independent of \energydot because it
more-strongly depends on other environmental factors.  Combining these
relations, \cite{mattana_2009_evolution-gamma-} proposes
\begin{equation}
  \luminosity_\text{IC}/\luminosity_\text{sync} \approx n_\text{IC}/n_\text{sync}
  \propto \PulsarAge^2 \propto \energydot^{-1}
\end{equation}
\cite{mattana_2009_evolution-gamma-} 
observed empirically for \ac{VHE} sources
that $\luminosity_\text{IC}/\luminosity_\text{sync} \propto
\PulsarAge^{2.2}$ and $\luminosity_\text{IC}/\luminosity_\text{sync}
\propto \energydot^{-1.9}$. This is qualitatively consistent with the
simple picture described above.  We will compare these simple scaling
relations with \ac{LAT} observations in \chapref{population_study}

\chapter{Maximum-likelihood Analysis of \Acstitle{LAT} Data}
\chaplabel{maximum_likelihood_analysis}

In this chapter, we discuss maximum-likelihood analysis, the primary
analysis method used to perform spectral and spatial analysis of
\ac{LAT} data.  In \secref{motivations_maximum_likelihood}, we discuss the
reasons necessary for employing this analysis procedure compared to 
simpler analysis methods.  In \secref{description_maximum_likelihood},
we describe the benefits of a maximum-likelihood analysis.
In \secref{defining_model}, we discuss the steps invovled in defining
a complete model of the sky, a necessary part of any likelihood analysis.

In \secref{binned_science_tools}, we discuss the standard implementation
of binned maximum likelihood in the \ac{LAT} Science Tools and
in particular the tool \gtlike.  In \secref{pointlike_package},
we then discuss the \pointlike pacakge, an alterate package for
maximum-likelihood analysis of \ac{LAT} data.  In the next chapter
(\chapref{extended_analysis}), we functionality written into \pointlike
for studying spatially-extended sources.  That much of the notation
and formulation of likelihood analysis in this chapter follows
\cite{kerr_2010a_likelihood-methods}.

\section{Motivations for Maximum-Likelihood Analysis of Gamma-ray Data}
\seclabel{motivations_maximum_likelihood}

Traditionally, spectral and spatial analysis of astrophysical data
relies on a process known as aperture photometry.  This process is
done by measuring the counts within a given radius of the source and
subtracting from it a background level estimated from a nearby region.
Often, the source's flux is calibrated by measurements of nearby objects
with known fluxes.  Otherwise, the flux can be obtained by dividing the
number of source counts by the telescope's size, the observation time,
and the telescope's conversion efficiency.  The application of this
method to \ac{VHE} data is described in \cite{li_1983_analysis-methods}.

Unfortunately, this simpler analysis method is inadequate for dealing with
the complexities introduced in analyzing LAT data.  Most importantly,
aperture photometry assumed that the background is isotropic so that
the background level can be estimated from nearby regions.  As was
discussed in \subsecref{galactic_diffuse_and_isotropic}, the Galactic
diffuse emission is highly anisotropic, rendering this assumption invalid.

In addition, this method is not optimal due to the high density of sources
detected in the $\gamma$-ray sky.  \Ac{2FGL} reported on the detection
of 1873 sources, which corresponds to an average source spacing of
$\sim5\degree$.  But within the inner $45\degree$ of the galactic plane
in longitude and $0.5\degree$ of the galactic plane in latitude, there
are 73 sources, corresponding to a source density of $\sim 1$ source per
square degree.  The aperture photometry method is unable to effectively
fit multiple sources when the tails of their \acp{PSF} overlap.

Finally, this method is suboptimal due to the large energy range of
\ac{LAT} observations.  A typical spectral analysis studies a source in an
energy from 100 \mev to above 100 \gev.  As was shown
in \secref{fermi_telescope}, the \ac{PSF} of the \ac{LAT} is rather broad ($\gtrsim
1\degree$) at low energy and much narrower ($\sim 0.1\degree$) at higher
energies. Therefore, higher 
energy photons coming from a source are much more sensitive, which
is discarded by simple aperture photometry methods.

\section{Description of Maximum-Likelihood Analysis}
\seclabel{description_maximum_likelihood}

The field of $\gamma$-ray astrophysics has generally adopted
maximum-likelihood analysis to avoid the issues discussed in
\secref{motivations_maximum_likelihood}.  The term likelihood
was first introduced by \cite{fisher_1925_statistical-methods}.
Maximum-likelihood was applied to astrophysical photon-counting
experiments by \cite{cash_1979_parameter-estimation}.
\cite{mattox_1996a_likelihood-analysis} described the maximum-likelihood
analysis framework developed to analyze \ac{EGRET} data.

In the formulation, one defines the likelihood, denoted $\likelihood$,
as the probability of obtaining the observed data given an assumed model:
\begin{equation}
  \likelihood = P(\data|\model).
\end{equation}
Generally, a model of the sky is a function of a list of parameters
that we denote as $\modelparams$.  The likelihood function can
be written as:
\begin{equation}
  \likelihood = \likelihood(\modelparams).
\end{equation}
In a maximum-likelihood analysis, one typically fits parameters of a model
by maximizing the likelihood as a function of the parameters of the model.
\begin{equation}
\modelparams_\text{max} = \underset{}{\text{arg }}\underset{\modelparams}{\text{max}} \likelihood(\modelparams)
\end{equation}

Assuming that you have a good model for your data and that
you understand the distribution of the data, maximum-likelihood
analysis can be used to very sensitively test for new features in
your model.  This is because the likelihood function naturally
incorporates data with different significance levels.

Typically, a \ac{LRT} is used to determine the significance of a new
feature in a model. A common use case is searching for a new source
or testing for a spectral break. In a \ac{LRT}, the likelihood under two
hypothesis are compared. We define $\hypothesis_0$ to be a background
model and $\hypothesis_1$ to be a model including the background and in
addition a feature that is being tested for.
Under the assumption that $\hypothesis_0$ is nested
within $\hypothesis_1$, we use Wilks' theorem to
compute the significance of the detection of this feature
\citep{wilks_1938a_large-sample-distribution}.  We define the test
statistic as
\begin{equation}
  \ts = 2\log(\likelihood_{\hypothesis_1}/\likelihood_{\hypothesis_0})
\end{equation}
Here, $\likelihood_{\hypothesis_0}$ and $\likelihood_{\hypothesis_1}$ are the
likelihoods maximized by varying all the parameters of $\hypothesis_0$
and $\hypothesis_1$ respectively.  According to Wilks' theorem,
if $\hypothesis_1$ has $n$ additional degrees of freedom compared to
$\hypothesis_0$, if none of the additional parameters lie on the edge of
parameter space, and if the true data is distributed as $\hypothesis_0$,
then the distribution of \ts should be
\begin{equation}
  \pdf(\ts) = \chi^2_n(\ts)
\end{equation}
Therefore, if one obtains a particular value of \ts, they can use
this this chi-squared distribution to determine the significance of
the detection.

\section{Defining a Model of the Sources in the Sky}
\seclabel{defining_model}

In order to perform a maximum-likelihood analysis, one requires a
parameterized model of the sky. A model of the sky is composed of a set
of $\gamma$-ray sources, each characterized by its photon flux density
$\fluxdensity(\energy,\time,\solidangle|\modelparams)$.  This represents
the number of photons emitted per unit energy, per unit time, per
units solid angle at a given energy, time, and position in the sky.
In \ac{CGS}, it has units of \fluxdensityunits.

Often, the spatial and spectral part of the source model are separable
and independent of time. When that is the case, we like to write the
source model as
\begin{equation}
  \fluxdensity(\energy,\time,\solidangle|\modelparams) = \dnde \times \pdf(\solidangle).
\end{equation}
Here, \dndeinline is a function of energy and \pdf(\solidangle) is 
a function of position (\solidangle).  In this formulation, some of
the model parameters \modelparams are taken by the $\dndeinline$ function and
some by the $\pdf(\solidangle)$ function.  In \ac{CGS}, \dndeinline has units
of \prefunits.

The spectrum \dndeinline is typically modeled by simple geometric functions.
The most popular spectral model is a \ac{PL}:
\begin{equation}
  \dnde = \prefactor \left(\frac{E}{\Escale}\right)^{-\spectralindex}
\end{equation}
Here, \dndeinline is a function of energy and also of the two model parameters
(the prefactor $\prefactor$ and the spectral index $\spectralindex$). The
parameter \Escale is often called the energy scale or the pivot energy
and is not fit to the data (sinde it is degenerate with \prefactor).

Another common spectral model is the \ac{BPL}:
\begin{equation}
  \dnde = \prefactor \times
    \begin{cases}
      (E/\Ebreak)^{-\spectralindex_1} &\text{ if }E<\Ebreak \\
      (E/\Ebreak)^{-\spectralindex_2} &\text{ if }E\ge\Ebreak.
    \end{cases}
\end{equation}
This model represents a \ac{PL} with an index of $\spectralindex_1$ which
breaks at energy \Ebreak to having an index of $\spectralindex_2$.

Finally, the \ac{ECPL} spectral model is often used to model the
$\gamma$-ray emission from pulsars:
\begin{equation}
  \dnde = \prefactor \left(\frac{E}{\Escale}\right)^{-\spectralindex}
  \exp\left(-\frac{E}{\Ecutoff}\right).
\end{equation}
For energies much below \Ecutoff, the \ac{ECPL} is a \ac{PL} with
spectral index \spectralindex.  For energies much larger than \Ecutoff,
the \ac{ECPL} spectrum exponentially decreases.

\pdf represents the spatial distribution of the emission.  It is
traditionally normalized as though it was a probability:
\begin{equation}
  \int \dsolidangle \intspace \pdf(\solidangle).
\end{equation}
Therefore, \pdf has units of \pdfunits For a point-like source at a
position $\solidangle'$, the spatial model is:
\begin{equation}
  \pdf(\solidangle) = \delta(\solidangle - \solidangle')
\end{equation}
and is a function of the position of the source ($\solidangle'$).
Example spatial models for spatially-extended sources will be presented
in \subsecref{extension_fitting}.

In some situations, the spatial and spectral part of a source do not
nicely decouple.  An example of this could be a spatially-extended
\acs{SNR} or \acp{PWN} which show a spectral variation
across the source, or equivalently show an energy-dependent
morphology.  \cite{katsuta_2012_fermi-lat-observation} and
\cite{hewitt_2012_fermi-lat-observations} have avoided this issue by
dividing the extended source into multiple non-overlapping extended
source templates which are each allowed to have a different spectra.

\section{The \Acstitle{LAT} Instrument Response Functions}
\seclabel{lat_irfs}

The performance of the \ac{LAT} is quantified by its effective area and
its dispersion. The effective area represents the collection area of the
\ac{LAT} and the dispersion represents the probability of misreconstructing
the true parameters of the incident $\gamma$-ray.
The effective area $\effectivearea(\energy,\time,\solidangle)$ is a
function of energy, time, and \ac{SA} and is measured in units of $\cm^2$.

The dispersion is the probability of a photon with true energy
\energy and incoming direction $\solidangle$ at time \time being
reconstructed to have an energy $\energy'$, an incoming direction
$\solidangle'$ at a time $\time'$.  The dispersion is written as
$\dispersion(\energy',\time',\solidangle'|\energy,\time,\solidangle)$.
It represents a probability and is therefore normalized such that
\begin{equation}
  \int \int \int \denergy \dsolidangle \dtime 
  \dispersion(\energy',\time',\solidangle'|\energy,\time,\solidangle) = 1
\end{equation}
Therefore,
$\dispersion(\energy',\time',\solidangle'|\energy,\time,\solidangle)$
has units of 1/energy/\acs{SA}/time

The convolution of the model a source with the \acp{IRF} produces the
expected differential counts (counts per unit energy/time/\acs{SA})
that are reconstructed to have an energy $\energy'$ at a position
$\solidangle'$ and at a time $\time'$:
\begin{equation}
  \eqnlabel{eventrate}
  \eventrate(\energy',\solidangle',\time'|\modelparams)
  = \int \int \int \denergy \, \dsolidangle \, \dtime \,
  \fluxdensity(\energy,\time,\vec\Omega|\modelparams) 
  \effectivearea(\energy,\time,\solidangle) \dispersion(\energy',\time',\solidangle'|\energy,\time,\solidangle)
\end{equation}
Here, this integral is performed over all energies, \acp{SA}, and times.

For \ac{LAT} analysis, we conventionally make the simplifying assumption that
the energy, spatial, and temporal dispersion decouple:
\begin{equation}
  \dispersion(\energy',\time',\solidangle'|\energy,\time,\solidangle) = 
  \psf(\solidangle'|E,\solidangle) \edisp(\energy'|\energy) \tdisp(\time'|\time)
\end{equation}
\edisp represents the energy dispersion of the \ac{LAT}. The energy
dispersion is a function of both the incident energy and angle of
the photon. It varies from $\sim$ 5\% to 20\%, degrading at lower
energies due to energy losses in the tracker and at higher energy due
to electromagnetic shower losses outside the calorimeter. Similarly,
it improves for photons with higher incident angles which are allowed a
longer path through the calorimeter \citep{ackermann_2012a_fermi-large}.
\subsecref{performance_lat} includes a plot of the \edisp of the \ac{LAT}.

$\psf(\solidangle'|E,\solidangle)$ is the probability of reconstructing
a $\gamma$-ray to have a position $\solidangle'$ if the true position of
the $\gamma$-ray has a position $\solidangle$.  For the \ac{LAT}, the
\ac{PSF} is a strong function of energy.  \subsecref{performance_lat}
plots the \ac{PSF} of the \ac{LAT}.

Finally, we note that in principle, there is a finite timing resolution
of $\gamma$-rays measured by the \ac{LAT}. But the timing accuracy is
$<10\unitspace\microsecond$ \citep{atwood_2009a_large-telescope}. Since
this is much less than the smallest timing signal which is expected to
be observed by the \ac{LAT} (millisecond pulsars), issues with timing
accuracy are typically ignored.

For a typical analysis of \ac{LAT} data, we also ignore the inherent
energy dispersion of the \ac{LAT}.  \cite{ackermann_2012a_fermi-large}
performed a monte carlo simulation to show that for power-law point-like
sources, the bias introduced by ignoring energy dispersion was on the
level of a few percent.  Therefore, the instrument response is typically
approximated as
\begin{equation}
  \response(\energy',\solidangle',\time'|\energy,\solidangle,) = 
  \effectivearea(\energy,\time',\solidangle) \psf(\solidangle'|E,\solidangle)
\end{equation}
We caution that for analysis of sources extended to energies below
$100\unitspace\mev$, the effects of energy dispersion are be more
severe.

The differential count rate is typically integrated over time 
assuming that the source model is time independent:
\begin{equation}\eqnlabel{differential_model_counts}
  \eventrate(\energy',\solidangle'|\modelparams)
  = \int \dsolidangle \,
  \fluxdensity(\energy',\vec\Omega|\modelparams) 
\left(
\int \dtime \intspace \effectivearea(\energy',\time,\solidangle) 
\right)
\psf(\solidangle'|E,\solidangle)
\end{equation}
This equation says that the counts expected by the \ac{LAT} from a
given model is the product of the source's flux with the effective
area and then convolved with the \ac{PSF}.  Finally, we note that the
\psf and effective area are also functions of the conversion type of the
$\gamma$-ray (front-entering or back-entering photons), and the azimuthal
angle of the $\gamma$-ray.  \eqnref{differential_model_counts} can be
generalized to include these effects.

\section{Binned Maximum-Likelihood of \Acstitle{LAT} Data with the Science Tools}
\seclabel{binned_science_tools}

We typically use a binned maximum-likelihood analysis to analyze \ac{LAT}
data.  In this analysis, $\gamma$-rays are binned in position and energy (and
sometimes also separately into front-entering and back-entering events).
The likelihood function comes from the Poisson nature of the observed
emission:
\begin{equation}\eqnlabel{poisson_likelihood_counts_model}
  \likelihood=\prod_j \frac{\theta_j^{k_j} e^{-\theta_j}}{k_j!}.
\end{equation}
Here, $j$ refers to a sum over position and energy bins, $k_j$ are the counts
observed in bin $j$, and $\theta_j$ are the model counts predicted in
the same bin.

The model counts in bin $j$ are computed by integrating the differential
model counts over the bin:
\begin{equation}
  \theta_{ij} = \int_j \intspace \denergy \intspace 
  \dsolidangle \intspace \dtime \intspace 
  \eventrate(\energy,\solidangle,\time|\modelparams_i).
\end{equation}
Here, $j$ represents the integral over the $j$th position/energy
bin, $i$ represents the $i$th source, $\modelparams_i$
refers to the parameters defining the $i$th source, and
$\eventrate(\energy,\solidangle,\time|\modelparams_i)$ is defined in
\eqnref{eventrate}. The total model counts is computed by summing over
all sources:
\begin{equation}
  \theta_j = \sum_i \theta_{ij}
\end{equation}

In most situations, it is more convenient to work with the log of
the likelihood because the log of the likelihood varies more slowly.
In addition, typically a statistical analysis requires either maximizing the
likelihood or looking at a change in the likelihood, which is
arbitrary except for an overall additive constant. So we typically
write the log of the likelihood as
\begin{equation}\eqnlabel{log_likelihood_sum_theta}
  \log\likelihood = -\sum_j\theta_j + \sum_j k_j\log\theta_j 
\end{equation}
where we have dropped the arbitrary additive constant $-\log k_j!$.

In the standard \fermi science tools, \gtbin can be used to perform
basic cuts on the $\gamma$-ray photon list.  The binning of photons
over position in the sky and energy is performed with \gtbin.  The tools
required to compute exposure are \gtltcube and \gtexpcubetwo. Finally,
the likelihood itself is computed with a combination of \gtsrcmaps and
\gtlike.  Essentially, \gtsrcmaps is used to perform the two-dimensional
convolution integral in equation \eqnref{differential_model_counts} and
\gtlike is used to compute the likelihood function defined in equation
\eqnref{log_likelihood_sum_theta}.

As we discussed in \secref{description_maximum_likelihood},
we typically use \acp{LRT} to test for significant features in
the $\gamma$-ray data.  For example, we compare a model with and
without a source of interest to test if that source is significant.
\cite{mattox_1996a_likelihood-analysis} shows that for \ac{EGRET} data,
assuming the position of the source was known and that the spectral shape was
fixed, the distribution of \ts in the null hypothesis was
\begin{equation}
  \pdf(\ts) = \tfrac{1}{2} (\delta(\ts) + \chi^2_n(\ts))
\end{equation}
From this, one finds that $\ts^{1/2}$ can be used as a measure of the
statistical significance of the detection of a source.

We finally mention that this formulation assumed that the source models
are time independent.  In principle, these formulas could be generalized
so that the data was binned also in time. But this would almost never be
useful because it is rarely possible to have a simple parameterized model
for the time dependence of a source. Instead, the analysis of a variable
sources is typically performed by dividing the analysis into multiple time
intervals and performing the likelihood fits independently in each time
range. See \cite{nolan_2012_fermi-large} for an example implementation.

\section{The Alternate Maximum-Likelihood Package \pointlike}
\seclabel{pointlike_package}

\pointlike is an alternative maximum-likelihood framework developed
for analyzing \ac{LAT} data. In principle, both \pointlike and \gtlike
perform the same binned maximum-likelihood analysis described in
\secref{binned_science_tools}. \pointlike's major design difference
is that it was written with efficiency in mind. The primary use case
for \pointlike is fitting procedures which require multiple iterations
such as source finding, position and extension fitting, computing large
residual \ts maps.

What makes maximum-likelihood analysis of \ac{LAT} data difficult
is the strongly non-linear performance of the \ac{LAT} (see
\subsecref{performance_lat}). At lower energies, one typically finds
lots of photons but each photon is not very significant due to the
poor angular resolution of the instrument. At these energies, a binned
analysis with coarse bins is perfectly adequate to study the sky.
But at higher energies, there are limited numbers of photons due to
the limited source fluxes. But because the angular resolution is much
improved, these photons become much more important. At these energies,
an unbinned analysis which loops over each photon is more appropriate.

The primary efficiency gain of \pointlike comes from scaling the bin
size with energy so that the bin size is always comparable to the
\ac{PSF}.  To do this, \pointlike bins the sky into \healpix pixels
\citep{gorski_2005_healpix:-framework}, but only keeps bins with counts
in them.  At low energy, the bins are large and essentially every 
bin has many counts in it.  But at high energy, bins are very small and rarely
have more than one count in them.  So \pointlike essentially does a binned
analysis at low energy, approximates an unbinned analysis at high energy,
and naturally interpolates between the two extremes.

There is one obvious trade-off for keeping only bins with counts in them.
Using \eqnref{log_likelihood_sum_theta}, we note that the evaluation
of the $\sum_j k_j\log\theta_j$ term can easily be evaluation if only
the counts and model counts are computed in bins with counts in them.
But the $\sum_j \theta_j$ term (the overall model predicted counts in
all bins). To avoid this, \pointlike has to independently compute the
integral of the model counts.

More details about the implementation of \pointlike can be found in
\cite{kerr_2010a_likelihood-methods}. We will discuss the implementation
of extended sources in \pointlike in \chapref{extended_analysis}.

\chapter{Analysis of Spatially Extended \Acstitle{LAT} Sources}
\chaplabel{extended_analysis}

\paperref{This chapter is based the first part of the paper
  ``Search for Spatially Extended Fermi-LAT Sources Using Two Years of Data''
  \citep{lande_2012_search-spatially}.}

As we discussed in \subsecref{2fgl}, spatial extension is an important
characteristic for correctly associating $\gamma$-ray-emitting
sources with their counterparts at other wavelengths.  It is also
important for obtaining an unbiased model of their spectra.  And this
is particularly important for studying $\gamma$-ray-emitting \ac{PWN}
which are expected to be spatially-extended at $\gamma$-ray energies.
We present a new method for quantifying the spatial extension of
sources detected by the Large Area Telescope (LAT).  We perform a
series of Monte Carlo simulations to validate this tool and calculate
the LAT threshold for detecting the spatial extension of sources. In
\chapref{extended_search}, we apply the tools developed in this section
to search for new spatially-extended sources.

\section{Introduction}

A number of astrophysical source classes including supernova remnants
(SNRs), pulsar wind nebulae (PWNe), molecular clouds, normal galaxies,
and galaxy clusters are expected to be spatially resolvable by the
Large Area Telescope (LAT), the primary instrument on the {\em
\fermi Gamma-ray Space Telescope} (\fermi).  Additionally, dark
matter satellites are also hypothesized to be spatially extended. See
\cite{atwood_2009a_large-telescope} for pre-launch predictions.  The LAT
has detected seven SNRs which are significantly extended at \gev energies:
W51C, W30, IC~443, W28, W44, RX\,J1713.7$-$3946, and the Cygnus Loop
\citep{abdo_2009a_fermi-discovery,ajello_2012a_fermi-large,abdo_2010a_observation-supernova,abdo_2010d_fermi-large,abdo_2010a_gamma-ray-emission,abdo_2011a_observations-young,katagiri_2011a_fermi-large}.
In addition, three extended PWN have been detected
by the LAT: MSH\,15$-$52, Vela~X, and HESS\,J1825$-$137
\citep{abdo_2010a_detection-energetic,abdo_2010c_fermi-large,grondin_2011a_detection-pulsar}.
Two nearby galaxies, the Large and Small Magellanic Clouds, and the lobes
of one radio galaxy, Centaurus A, were spatially resolved at \gev energies
\citep{abdo_2010a_observations-large,abdo_2010a_detection-small,abdo_2010a_fermi-gamma-ray}.
A number of additional sources detected at \gev energies are positionally
coincident with sources that exhibit large enough extension at other
wavelengths to be spatially resolvable by the LAT at \gev energies.
In particular, there are 59 \gev sources in the second Fermi
Source Catalog (2FGL) that might be associated with extended SNRs
\citep[2FGL,][]{nolan_2012_fermi-large}.  Previous analyses of extended
LAT sources were performed as dedicated studies of individual sources
so we expect that a systematic scan of all LAT-detected sources could
uncover additional spatially extended sources.

The current generation of \acp{IACT} have made it apparent that many
sources can be spatially resolved at even higher energies.  Most prominent
was a survey of the Galactic plane using \ac{HESS} which reported 14
spatially extended sources with extensions varying from $\sim0\fdg1$
to $\sim0\fdg25$ \citep{aharonian_2006a_h.e.s.s.-survey}.  Within our
Galaxy very few sources detected at \tev energies, most notably the
$\gamma$-ray binaries LS\,5039 \citep{aharonian_2006a_orbital-modulation},
LS I+61$-$303 \citep{albert_2006a_variable-very-high-energy,
acciari_2011a_veritas-observations}, HESS\,J0632+057
\citep{aharonian_2007a_discovery-point-like}, and the Crab nebula
\citep{weekes_1989a_observation-gamma}, have no detectable extension.
High-energy $\gamma$-rays from \tev sources are produced by the decay
of $\pi^0$s produced by hadronic interactions with interstellar matter
and by relativistic electrons due to Inverse Compton (IC) scattering and
bremsstrahlung radiation.  It is plausible that the \gev and \tev emission
from these sources originates from the same population of high-energy
particles and so at least some of these sources should be detectable at
\gev energies.  Studying these \tev sources at \gev energies would help
to determine the emission mechanisms producing these high energy photons.

The LAT is a pair conversion telescope that has been surveying
the $\gamma$-ray sky since 2008 August.  The LAT has broad energy
coverage (20 \mev to $>300$ \gev), wide field of view ($\sim 2.4$
sr), and large effective area ($\sim 8000\ \cm^2$ at $>1$ \gev)
Additional information about the performance of the LAT can be found
in \cite{atwood_2009a_large-telescope}.

Using 2 years of all-sky survey data, the LAT Collaboration published
2FGL \citep[2FGL,][]{nolan_2012_fermi-large}.  The possible counterparts
of many of these sources can be spatially resolved when observed at other
frequencies. But detecting the spatial extension of these sources at \gev
energies is difficult because the size of the point-spread function (PSF)
of the LAT is comparable to the typical size of many of these sources.

The capability to spatially resolve \gev $\gamma$-ray sources is
important for several reasons.  Finding a coherent source extension
across different energy bands can help to associate a LAT source to an
otherwise confused counterpart.  Furthermore, $\gamma$-ray emission from
dark matter annihilation has been predicted to be detectable by the LAT.
Some of the dark matter substructure in our Galaxy could be spatially
resolvable by the LAT \citep{baltz_2008a_pre-launch-estimates}.
Characterization of spatial extension could help to identify this
substructure.  Also, due to the strong energy dependence of the LAT
PSF, the spatial and spectral characterization of a source cannot be
decoupled. An inaccurate spatial model will bias the spectral model of the
source and vice versa. Specifically, modeling a spatially extended source
as point-like will systematically soften measured spectra. Furthermore,
correctly modeling source extension is important for understanding an
entire region of the sky. For example, an imperfect model of the spatially
extended LMC introduced significant residuals in the surrounding region
\citep{abdo_2010b_fermi-large,nolan_2012_fermi-large}.  Such residuals
can bias the significance and measured spectra of neighboring sources
in the densely populated Galactic plane.

 For these reasons, in \secref{analysis_methods_section}
we present a new systematic method for analyzing spatially extended LAT
sources.  In \secref{validate_ts}, we demonstrate that this method
can be used to test the statistical significance of the extension of a LAT
source and we assess the expected level of bias introduced by assuming
an incorrect spatial model.  In \secref{extension_sensitivity},
we calculate the LAT detection threshold to resolve the extension
of a source.  In \secref{dual_localization_method}, we study
the ability of the LAT to distinguish between a single extended
source and unresolved closely-spaced point-like sources In
\secref{test_2lac_sources}, we further demonstrate that our
detection method does not misidentify point-like sources as being
extended by testing the extension of active Galactic nuclei (AGN)
believed to be unresolvable.  
In \chapref{extended_search}, we take the analysis
method developed in this chapter and use it to 
search for new spatially-extended sources.

\section{Analysis Method}
\seclabel{analysis_methods_section}

Morphological studies of sources using the LAT are challenging
because of the strongly energy-dependent PSF that is comparable in
size to the extension of many sources expected to be detected at
\gev energies.  Additional complications arise for sources along
the Galactic plane due to systematic uncertainties in the model for
Galactic diffuse emission.  

For energies below $\sim$300~\mev, the angular resolution is limited by
multiple scattering in the silicon strip tracking section
of the detector and is several degrees at 100 \mev.  The PSF improves
with energy approaching a 68\% containment radius of $\sim0\fdg2$ at
the highest energies (when averaged over the acceptance of the LAT)
and is limited by the ratio of the strip pitch to the height of the tracker
\citep{atwood_2009a_large-telescope,abdo_2009a_on-orbit-calibration,ackermann_2012a_fermi-large}.\footnote{More
information about the performance of the LAT can be found at the \fermi
Science Support Center (FSSC, \url{http://fermi.gsfc.nasa.gov}).} However,
since most high energy astrophysical sources have spectra that decrease
rapidly with increasing energy, there are typically fewer higher
energy photons with improved angular resolution. Therefore sophisticated
analysis techniques are required to maximize the sensitivity of the LAT
to extended sources.

\subsection{Modeling Extended Sources in the \pointlike Package}

A new maximum-likelihood analysis tool has been developed to address the
unique requirements for studying spatially extended sources with the LAT.
It works by maximizing the Poisson 
likelihood to detect the observed distributions of $\gamma$-rays (referred to as counts)
given a parametrized spatial and spectral model of the sky.  
The data are binned spatially, using a \healpix pixelization and spectrally 
\citep{gorski_2005_healpix:-framework} and the likelihood is maximized over all bins in
a region.
The extension of a source can be modeled by a geometric shape
(e.g. a disk or a two-dimensional Gaussian) and the position, extension,
and spectrum of the source can be simultaneously fit.

This type of analysis is unwieldy using the standard
LAT likelihood analysis tool \gtlike\footnote{\gtlike is
distributed publicly by the FSSC.} because it can only fit the
spectral parameters of the model unless a more sophisticated
iterative procedure is used.  We note that \gtlike has been used in
the past in several studies of source extension in the LAT Collaboration
\citep{abdo_2010a_observations-large,abdo_2010a_detection-small,abdo_2010d_fermi-large,abdo_2009a_fermi-discovery}.
In these studies, a set of \gtlike maximum likelihood fits at fixed
extensions was used to build a profile of the likelihood as a function
of extension.  The \gtlike likelihood profile approach has been shown to
correctly reproduce the extension of simulated extended sources assuming
that the true position is known \citep{giordano_2011a_extension-studies}.
But it is not optimal because the position, extension, and spectrum of
the source must be simultaneously fit to find the best fit parameters and
to maximize the statistical significance of the detection.  Furthermore,
because the \gtlike approach is computationally intensive, no large-scale
Monte Carlo simulations have been run to calculate its false detection
rate.

The approach presented here is based on a second maximum likelihood
fitting package developed in the LAT Collaboration called \pointlike
\citep{abdo_2010b_fermi-large,kerr_2010a_likelihood-methods}.  The choice
to base the spatial extension fitting on \pointlike rather than \gtlike
was made due to considerations of computing time.  The \pointlike
algorithm was optimized for speed to handle larger numbers of sources
efficiently, which is important for our catalog scan and for being
able to perform large-scale Monte Carlo simulations to validate
the analysis.  Details on the \pointlike package can be found in
\cite{kerr_2010a_likelihood-methods}.  We extended the code to allow a
simultaneous fit of the source extension together with the position and
the spectral parameters.

\subsection{Extension Fitting}
\subseclabel{extension_fitting}

In \pointlike, one can fit the position and extension of a source
under the assumption that the source model can be factorized:
$M(x,y,E)=S(x,y)\times X(E)$, where $S(x,y)$ is the spatial distribution
and $X(E)$ is the spectral distribution.  To fit an extended source,
\pointlike convolves the extended source shape with the PSF (as a function
of energy) and uses the \minuit library \citep{james_1975a_minuit-system}
to maximize the likelihood by simultaneously varying the position,
extension, and spectrum of the source.  As will be described in
\subsecref{monte_carlo_validation}, simultaneously fitting the position,
extension, and spectrum is important to maximize the statistical
significance of the detection of the extension of a source.  To avoid
projection effects, the longitude and latitude of the source are not
directly fit but instead the displacement of the source in a reference
frame centered on the source.

The significance of the extension of a source can be calculated from the
likelihood-ratio test. The likelihood ratio defines the test statistic
(TS) by comparing the likelihood of a simpler hypothesis to a more
complicated one:
\begin{equation}
  \ts=2\log(\likelihood(H_1)/\likelihood(H_0)),
\end{equation}
where $H_1$ is the more complicated hypothesis and $H_0$ the simpler one.
For the case of the extension test, we compare the likelihood when
assuming the source has either a point-like or spatially extended
spatial model:
\begin{equation}
  \tsext=2\log(\likelihood_\text{ext}/\likelihood_\text{ps}).
\end{equation}
\pointlike calculates \tsext by fitting a source first with a spatially
extended model and then as a point-like source.  The interpretation
of \tsext in terms of a statistical significance is discussed in
\subsecref{monte_carlo_validation}.

For extended sources with an assumed radially-symmetric shape, we
optimized the calculation by performing one of the integrals analytically.
The expected photon distribution can be written as
\begin{equation}
  \text{PDF}(\vec r) = \int  \text{PSF}(|\vec r - \vec r'|)I_\text{src}(\vec r') r' dr' d\phi'
\end{equation}
where $\vec r$ represents the position in the sky and $I_\text{src}(\vec
r)$ is the spatial distribution of the source.  The PSF of the LAT is
currently parameterized in the Pass~7\_V6 (P7\_V6) Source Instrument
Response Function \citep[IRFs,][]{ackermann_2012a_fermi-large} by a King
function \citep{king_1962a_structure-clusters.}:
\begin{equation}
  \text{PSF}(r) = 
  \frac{1}{2\pi\sigma^2}
  \left(1-\frac{1}{\gamma}\right)
  \left(1+\frac{u}{\gamma}\right)^{-\gamma},
\end{equation}
where $u=(r/\sigma)^2/2$ and $\sigma$ and $\gamma$ are free parameters
\citep{kerr_2010a_likelihood-methods}.  For radially-symmetric extended
sources, the angular part of the integral can be evaluated analytically
\begin{align}
  \text{PDF}(u) & = \int_0^\infty r' dr'
  I_\text{src}(v) 
  \int_0^{2\pi} d\phi' 
  \text{PSF}(\sqrt{2\sigma^2(u+v-2\sqrt{uv}\cos(\phi-\phi'))})
  \\
  & = \int_0^\infty dv
  I_\text{src}(v) 
  \left(\frac{\gamma-1}{\gamma}\right)
  \left( \frac{\gamma}{\gamma + u + v}\right)^\gamma 
  \times {}_2F_1 \left(\gamma/2,\frac{1+\gamma}{2},1,\frac{4uv}{(\gamma+u+v)^2}\right),
\end{align}
where $v=(r'/\sigma)^2/2$ and ${}_2F_1$ is the Gaussian hypergeometric
function.  This convolution formula reduces the expected photon
distribution to a single numerical integral.

There will always be a small numerical discrepancy between the expected
photon distribution derived from a true point-like source and a very
small extended source due to numerical error in the convolution.  In most
situations, this error is insignificant.  But in particular for very
bright sources, this numerical error has the potential to bias the TS
for the extension test. Therefore, when calculating \tsext, we compare
the likelihood fitting the source with an extended spatial model to the
likelihood when the extension is fixed to a very small value ($10^{-10}$
degrees in radius for a uniform disk model).

\begin{figure}[htbp]
  \includegraphics{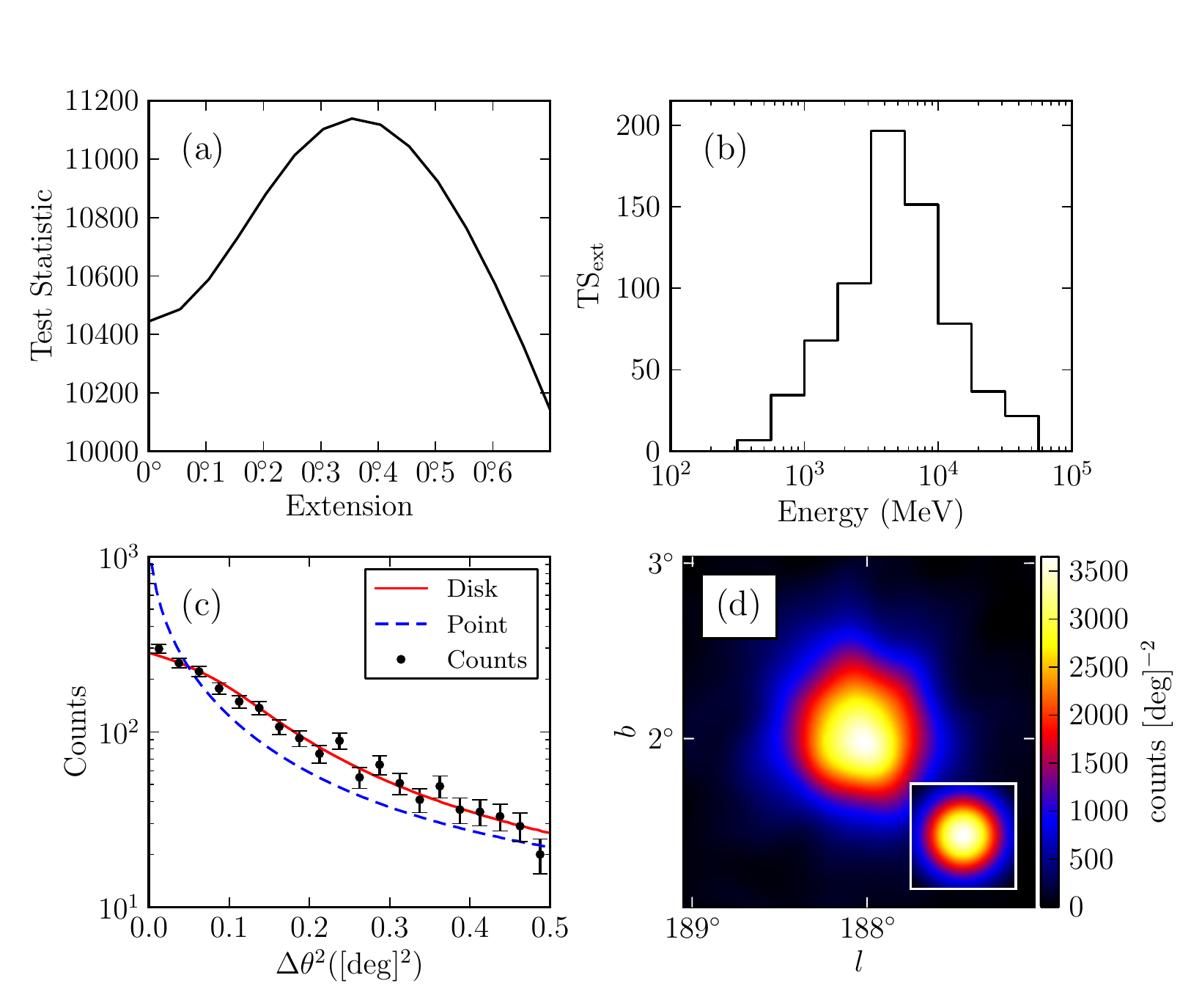}
  \caption{Counts maps and TS profiles for the SNR IC~443. (a)
  \ts vs. extension of the source. (b) \tsext for individual energy
  bands. (c) observed radial profile of counts in comparison to the
  expected profiles for a spatially extended source (solid and colored
  red in the online version) and for a point-like source (dashed and
  colored blue in the online version).  (d) smoothed counts map after
  subtraction of the diffuse emission compared to the smoothed LAT
  PSF (inset). Both were smoothed by a 0\fdg1 2D Gaussian kernel.
  Plots (a), (c), and (d) use only photons with energies between 1
  \gev and 100 \gev.  Plots (c) and (d) include only photons which
  converted in the front part of the tracker and have an improved
  angular resolution \citep{atwood_2009a_large-telescope}.}
  \figlabel{four_plots_ic443}
\end{figure}

We estimate the error on the extension of a source by fixing the
position of the source and varying the extension until the log of the
likelihood has decreased by 1/2, corresponding to a $1\sigma$ error
\citep{eadie_1971a_statistical-methods}.  \figref{four_plots_ic443}
demonstrates this method by showing the change in the log of the
likelihood when varying the modeled extension of the SNR IC~443.
The localization error is calculated by fixing the extension and spectrum
of the source to their best fit values and then fitting the log of the
likelihood to a 2D Gaussian as a function of position. This localization
error algorithm is further described in \cite{nolan_2012_fermi-large}.

\subsection{\gtlike Analysis Validation}
\subseclabel{gtlike_crosscheck}

\pointlike is important for analyses of LAT data that require many
iterations such as source localization and extension fitting.  On the
other hand, because \gtlike makes fewer approximations in calculating
the likelihood we expect the spectral parameters found with \gtlike to
be slightly more accurate.  Furthermore, because \gtlike is the standard
likelihood analysis package for LAT data, it has been more extensively
validated for spectral analysis.  For those reasons, in the following
analysis we used \pointlike to determine the position and extension of a
source and subsequently derived the spectrum using \gtlike. Both \gtlike
and \pointlike can be used to estimate the statistical significance of
the extension of a source and we required that both methods agree for a
source to be considered extended.  There was good agreement between the
two methods.  Unless explicitly mentioned, all \ts, \tsext, and spectral
parameters were calculated using \gtlike with the best-fit positions
and extension found by \pointlike.

\subsection{Comparing Source Sizes}
\subseclabel{compare_source_size}

\begin{figure}[htbp]
  \includegraphics{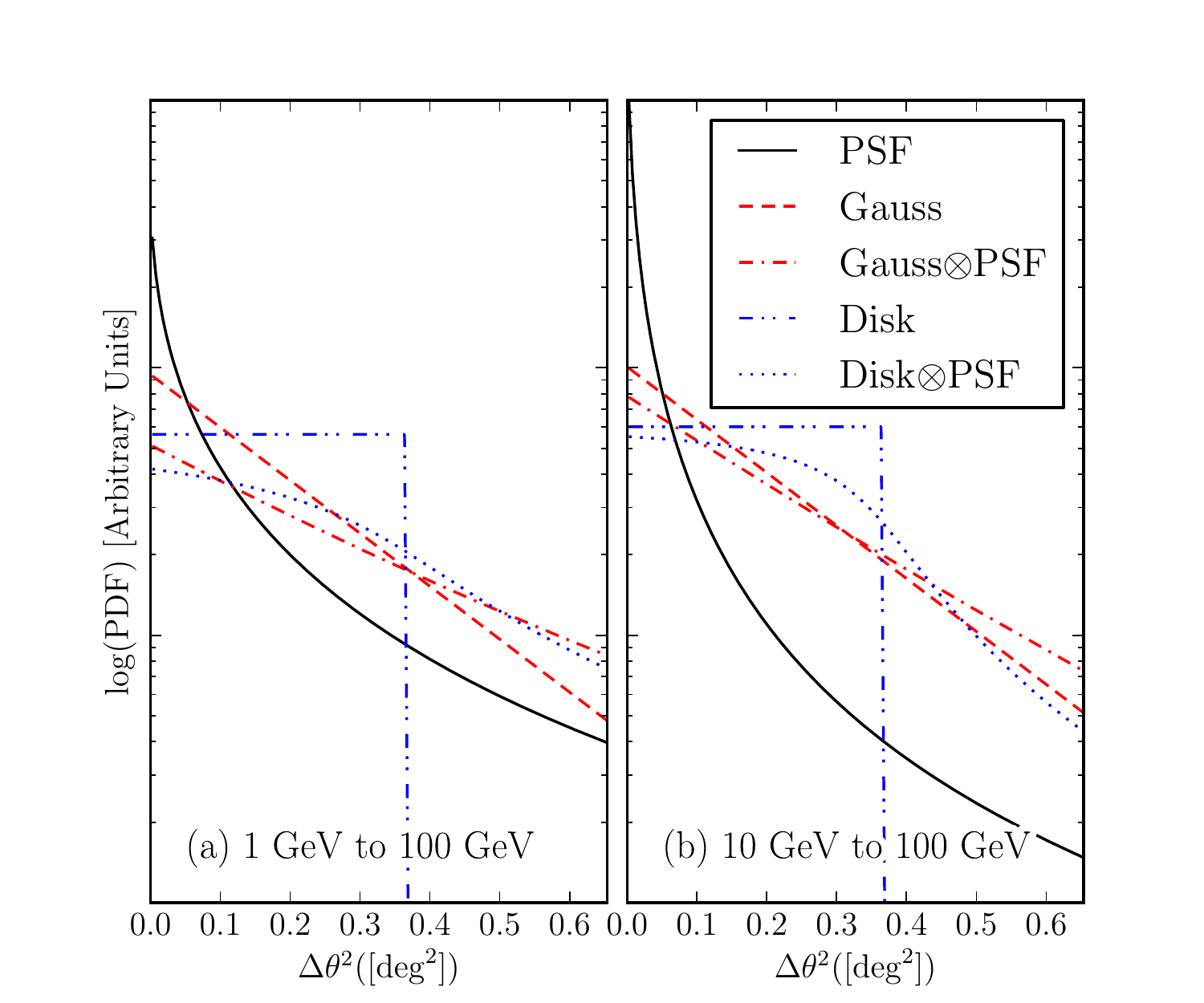}
  \caption{
  A comparison of a 2D Gaussian and uniform disk spatial model
  of extended sources before and after convolving with the PSF for two
  energy ranges.  The solid black line is the PSF that would be observed
  for a power-law source of spectral index 2. The dashed line
  and the dash-dotted lines are 
  the brightness profile of a Gaussian with $\rsixeight=0\fdg5$
  and the convolution of this profile with the LAT PSF respectively
  (colored red in the online version).
  The dash-dot-dotted and the dot-dotted lines are the brightness profile
  of a uniform disk with $\rsixeight=0\fdg5$ and the convolution
  of this profile with the LAT PSF respectively (colored blue in the online version).
  }\figlabel{compare_disk_gauss}
\end{figure}

We considered two models for the
surface brightness profile for extended sources: a 2D Gaussian model
\begin{equation}\eqnlabel{gauss_pdf}
  I_\text{Gaussian}(x,y)=\tfrac{1}{2\pi\sigma^2}\exp\left(-(x^2+y^2)/2\sigma^2\right)
\end{equation}
or a uniform disk model
\begin{equation}\eqnlabel{disk_pdf}
  I_\text{disk}(x,y)=
  \begin{cases}
    \frac{1}{\pi\sigma^2} & x^2+y^2\le\sigma^2 \\
    0                      & x^2+y^2>\sigma^2.
  \end{cases}
\end{equation}
Although these shapes are significantly different,
\figref{compare_disk_gauss} shows that, after convolution with the
LAT PSF, their PDFs are similar for a source that has a 0\fdg5 radius
typical of LAT-detected extended sources.  To allow a valid comparison
between the Gaussian and the uniform disk models, we define the source
size as the radius containing 68\% of the intensity ($\rsixeight$).
By direct integration, we find
\begin{align}
  \rsixeight_\text{,Gaussian}=&1.51\sigma, \\
  \rsixeight_\text{,disk}=&0.82\sigma, 
\end{align}
where $\sigma$ is defined
in \eqnref{gauss_pdf} and \eqnref{disk_pdf} respectively.
For the example above, $\rsixeight=0\fdg5$ so $\sigma_\text{disk}=0.61\degree$
and $\sigma_\text{Gaussian}=0.33\degree$.

For sources that are comparable in size to the PSF, the differences in
the PDF for different spatial models are lost in the noise and the LAT
is not sensitive to the detailed spatial structure of these sources.
In \subsecref{bias_wrong_spatial_model}, we perform a dedicated Monte
Carlo simulation that shows there is little bias due to incorrectly
modeling the spatial structure of an extended source.  Therefore, in our
search for extended sources we use only a radially-symmetric uniform disk
spatial model. Unless otherwise noted, we quote the radius to the edge
($\sigma$) as the size of the source.

\section{Validation of the TS Distribution}
\seclabel{validate_ts}

\subsection{Point-like Source Simulations Over a Uniform Background}
\subseclabel{monte_carlo_validation}

We tested the theoretical distribution for \tsext to evaluate
the false detection probability for measuring source extension.
To do so, we tested simulated point-like sources for extension.
\cite{mattox_1996a_likelihood-analysis} discuss that the \ts distribution
for a likelihood-ratio test on the existence of a source at a given
position is
\begin{equation}\eqnlabel{ts_ext_distribution}
  P(\ts)=\tfrac{1}{2}(\chi^2_1(\ts)+\delta(\ts)),
\end{equation}
where $P(\ts)$ is the probability density to get a particular value
of TS, $\chi^2_1$ is the chi-squared distribution with one degree of
freedom, and $\delta$ is the Dirac delta function.  The particular
form of \eqnref{ts_ext_distribution} is due to the null hypothesis
(source flux $\Phi=0$) residing on the edge of parameter space and the
model hypothesis adding a single degree of freedom (the source flux).
It leads to the often quoted result $\sqrt{TS}=\sigma$, where $\sigma$
here refers to the significance of the detection. It is plausible
to expect a similar distribution for the TS in the test for source
extension since the same conditions apply (with the source flux $\Phi$
replaced by the source radius $r$ and $r<0$ being unphysical).  To verify
\eqnref{ts_ext_distribution}, we evaluated the empirical distribution
function of \tsext computed from simulated sources.

We simulated point-like sources with various spectral forms using the
LAT on-orbit simulation tool \gtobssim\footnote{\gtobssim is distributed
publicly by the FSSC.} and fit the sources with \pointlike using both
point-like and extended source hypotheses.  These point-like sources
were simulated with a power-law spectral model with integrated fluxes
above 100 \mev ranging from $3\times10^{-9}$ to $1\times10^{-5}$
\fluxunits and spectral indices ranging from 1.5 to 3.  These values
were picked to represent typical parameters of LAT-detected sources. The
point-like sources were simulated on top of an isotropic background
with a power-law spectral model with integrated flux above 100 \mev
of $1.5\times10^{-5}$ \fluxunits sr$^{-1}$ and spectral index 2.1.
This was taken to be the same as the isotropic spectrum measured by EGRET
\citep{sreekumar_1998a_egret-observations}.  This spectrum is comparable
to the high-latitude background intensity seen by the LAT.  The Monte
Carlo simulation was performed over a one-year observation period using
a representative spacecraft orbit and livetime.  The reconstruction
was performed using the P7\_V6 Source class event selection and IRFs
\citep{ackermann_2012a_fermi-large}. For each significantly detected
point-like source ($\ts\ge25$), we used \pointlike to fit the source
as an extended source and calculate \tsext.  This entire procedure was
performed twice, once fitting in the 1 \gev to 100 \gev energy range
and once fitting in the 10 \gev to 100 \gev energy range.

\begin{figure}[htbp]
    \includegraphics{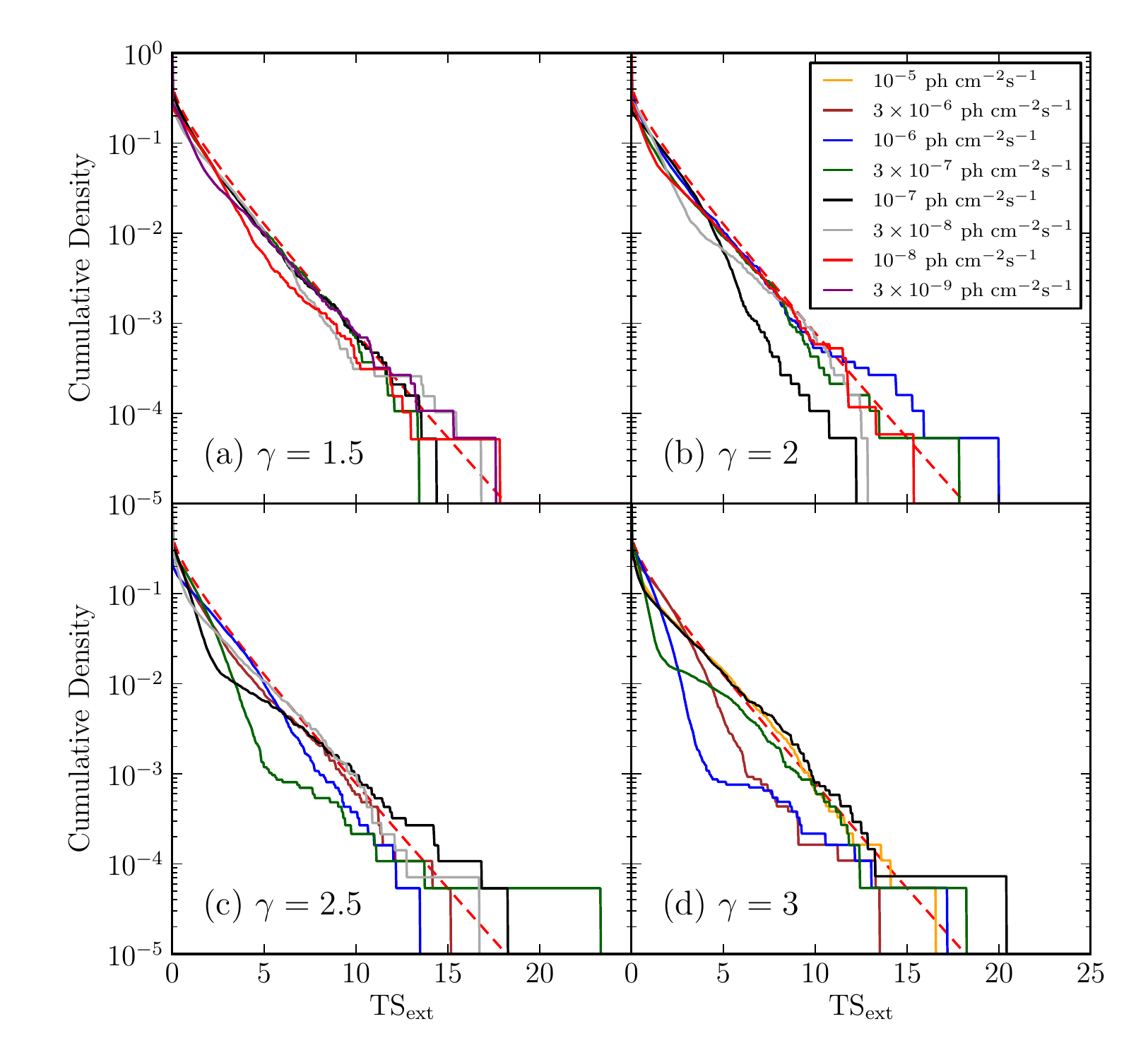}
    \caption{Cumulative distribution of the TS for the extension test
    when fitting simulated point-like sources in the 1 GeV to 100
    GeV energy range.  The four plots represent simulated sources of
    different spectral indices and the different lines (colored in the
    online version) represent point-like sources with different 100
    \mev to 100 \gev integral fluxes.  The dashed line (colored red)
    is the cumulative density function of \eqnref{ts_ext_distribution}.}
    \figlabel{ts_ext_mc_1000}
  \end{figure}

\begin{figure}[htbp]
    \includegraphics{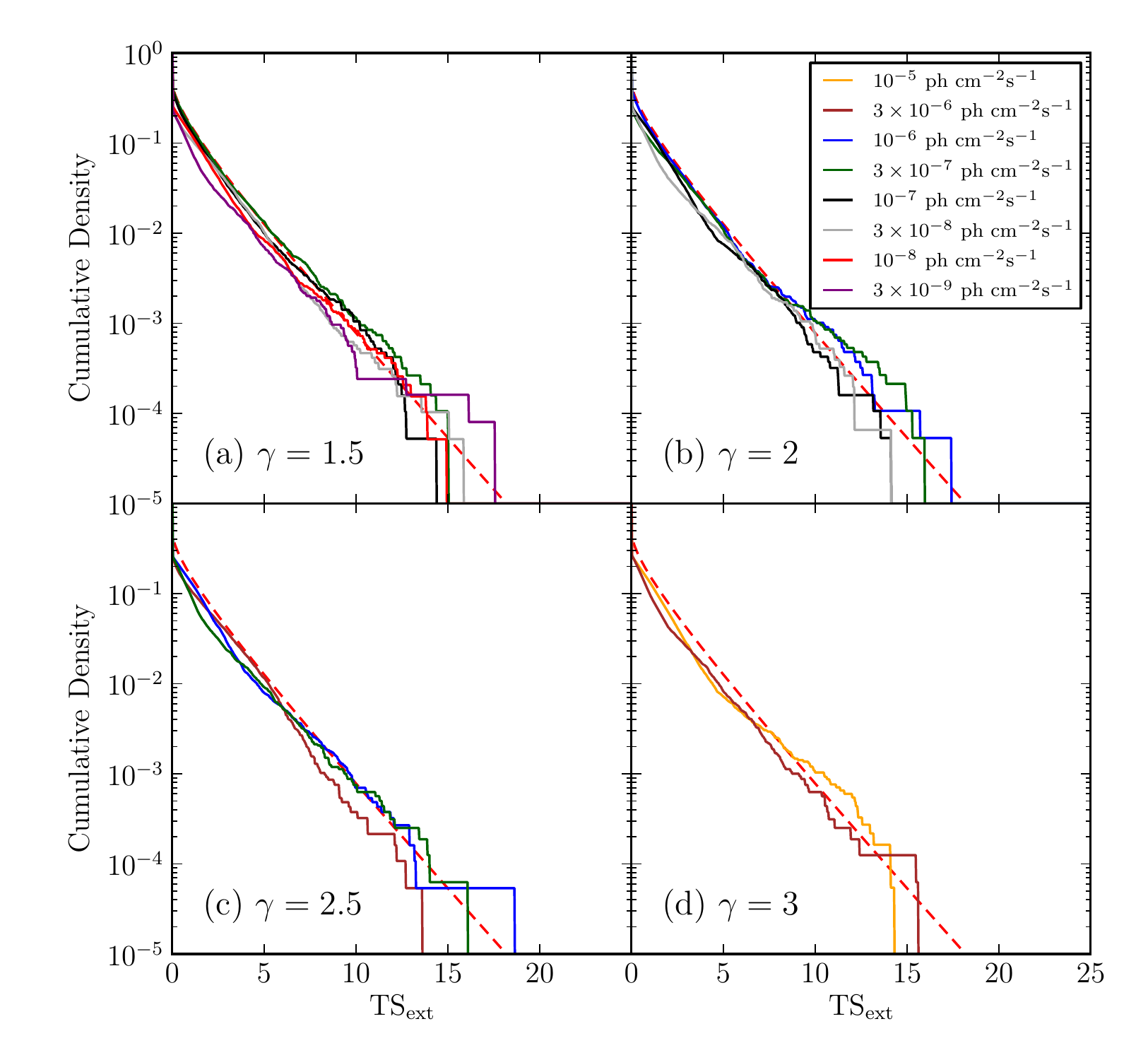}
    \caption{The same plot as \figref{ts_ext_mc_1000} but fitting in
    the 10 \gev to 100 \gev energy range.}
    \figlabel{ts_ext_mc_10000}
  \end{figure}

\begin{deluxetable}{rrrrrr}
\tabletypesize{\scriptsize}
\tablecaption{Monte Carlo Spectral Parameters
\tablabel{ts_ext_num_sims}
}
\tablecolumns{6}
\tablewidth{0pt}
\tablehead{
\colhead{Spectral Index}&
\colhead{Flux\tablenotemark{(a)}}&
\colhead{$N_{1-100\gev}$}&
\colhead{$\langle\ts\rangle_{1-100\gev}$}&
\colhead{$N_{10-100\gev}$}&
\colhead{$\langle\ts\rangle_{10-100\gev}$}\\
\colhead{}&
\colhead{(\fluxunits)}&
\colhead{}&
\colhead{}&
\colhead{}&
\colhead{}
}

\startdata
\multicolumn{6}{c}{Isotropic Background} \\
\hline
      1.5 &  $3\times 10^{-7}$ &           18938 &  22233    &     18938   &    8084     \\
          &          $10^{-7}$ &           19079 &   5827    &     19079   &    2258     \\
          &  $3\times 10^{-8}$ &           19303 &   1276    &     19303   &     541     \\
          &          $10^{-8}$ &           19385 &    303    &     19381   &     142     \\
          &  $3\times 10^{-9}$ &           18694 &     62    &     12442   &      43     \\
      \hline
        2 &          $10^{-6}$ &           18760 &  22101    &     18760   &    3033     \\
          &  $3\times 10^{-7}$ &           18775 &   4913    &     18775   &     730     \\
          &          $10^{-7}$ &           18804 &   1170    &     18803   &     192     \\
          &  $3\times 10^{-8}$ &           18836 &    224    &     15256   &      50     \\
          &          $10^{-8}$ &           17060 &     50    &   \nodata   & \nodata     \\
      \hline
      2.5 &  $3\times 10^{-6}$ &           18597 &  19036    &     18597   &     786    \\
          &          $10^{-6}$ &           18609 &   4738    &     18608   &     208    \\
          &  $3\times 10^{-7}$ &           18613 &    954    &     15958   &      53    \\
          &          $10^{-7}$ &           18658 &    203    &   \nodata   & \nodata    \\
          &  $3\times 10^{-8}$ &           14072 &     41    &   \nodata   & \nodata    \\
      \hline                                                
        3 &          $10^{-5}$ &           18354 &  19466    &     18354   &     215    \\
          &  $3\times 10^{-6}$ &           18381 &   4205    &     15973   &      54    \\
          &          $10^{-6}$ &           18449 &    966    &   \nodata   & \nodata    \\
          &  $3\times 10^{-7}$ &           18517 &    174    &   \nodata   & \nodata    \\
          &          $10^{-7}$ &           13714 &     41    &   \nodata   & \nodata    \\
\cutinhead{Galactic Diffuse and Isotropic Background\tablenotemark{(b)}}
      1.5 &  $2.3\times 10^{-8}$ &           90741 &     63    &   \nodata   & \nodata     \\
        2 &  $1.2\times 10^{-7}$ &           92161 &     60    &   \nodata   & \nodata     \\
      2.5 &  $4.5\times 10^{-7}$ &           86226 &     47    &   \nodata   & \nodata    \\
        3 &  $2.0\times 10^{-6}$ &           94412 &     61    &   \nodata   & \nodata    \\
\enddata

\tablenotetext{(a)}{Integral 100 \mev to 100 \gev flux.}
\tablenotetext{(b)}{
For the Galactic simulations, the quoted fluxes are
the fluxes for sources placed 
in the Galactic center. The actual fluxes are scaled by \eqnref{scale_flux_by_background}.
}

\tablecomments{
    A list of the spectral models of the simulated point-like sources
    which were tested for extension.  For each model, the number of
    statistically independent simulations and the average value of \ts is
    also tabulated.  
    The top rows are the simulations on top of an isotropic background and
    the bottom rows are the simulations on top of the Galactic diffuse and isotropic
    background.
}
\end{deluxetable}

For each set of spectral parameters, $\sim20,000$ statistically
independent simulations were performed. For lower-flux spectral models,
many of the simulations left the source insignificant ($\ts<25$) and
were discarded.  \tabref{ts_ext_num_sims} shows the different spectral
models used in our study as well as the number of simulations and the
average point-like source significance.  The cumulative density of \tsext
is plotted in \figref{ts_ext_mc_1000} and \figref{ts_ext_mc_10000} and
compared to the $\chi^2_1/2$ distribution of \eqnref{ts_ext_distribution}.

Our study shows broad agreement between simulations and
\eqnref{ts_ext_distribution}. To the extent that there is a discrepancy,
the simulations tended to produce smaller than expected values of \tsext
which would make the formal significance conservative.  Considering the
distribution in \figref{ts_ext_mc_1000} and \figref{ts_ext_mc_10000},
the choice of a threshold \tsext set to 16 (corresponding to a formal
$4\sigma$ significance) is reasonable.

\subsection{Point-like Source Simulations Over a Structured Background}
\subseclabel{validation_over_plane}

We performed a second set of simulations to show that the theoretical
distribution for \tsext is still preserved when the point-like sources
are present over a highly-structured diffuse background.  Our simulation
setup was the same as above except that the sources were simulated on top
of and analyzed assuming the presence of the standard Galactic diffuse
and isotropic background models used in 2FGL.  In our simulations, we
selected our sources to have random positions on the sky such that they
were within 5\degree of the Galactic plane. This probes the brightest
and most strongly contrasting areas of the Galactic background.

To limit the number of tests, we selected only one flux level for each of
the four spectral indices and we performed this test only in the 1 \gev
to 100 \gev energy range.  As described below, the fluxes were selected
so that $\ts\sim50$. We do not expect to be able to spatially resolve
sources at lower fluxes than these, and the results for much brighter
sources are less likely to be affected by the structured background.

Because the Galactic diffuse emission is highly structured with strong
gradients, the point-source sensitivity can vary significantly across
the Galactic plane.  To account for this, we scaled the flux (for a
given spectral index) so that the source always has approximately the
same signal-to-noise ratio:
\begin{equation}
  \eqnlabel{scale_flux_by_background}
  F(\vec{x}) = F(\text{GC}) \times \left(
  \frac{B(\vec{x})}{B(\text{GC})}\right)^{1/2}.
\end{equation}
Here, $\vec{x}$ is the position of the simulated source, $F$ is the
integral flux of the source from 100 \mev to 100 \gev, $F(\text{GC})$
is the same quantity if the source was at the Galactic center, $B$
is the integral of the Galactic diffuse and isotropic emission from 1
\gev to 100 \gev at the position of the source, and $B(\text{GC})$ is
the same quantity if the source was at the Galactic center.  For the
four spectral models, \tabref{ts_ext_num_sims} lists $F(\text{GC})$
and the average value of \ts.

\begin{figure}[htbp]
  \includegraphics{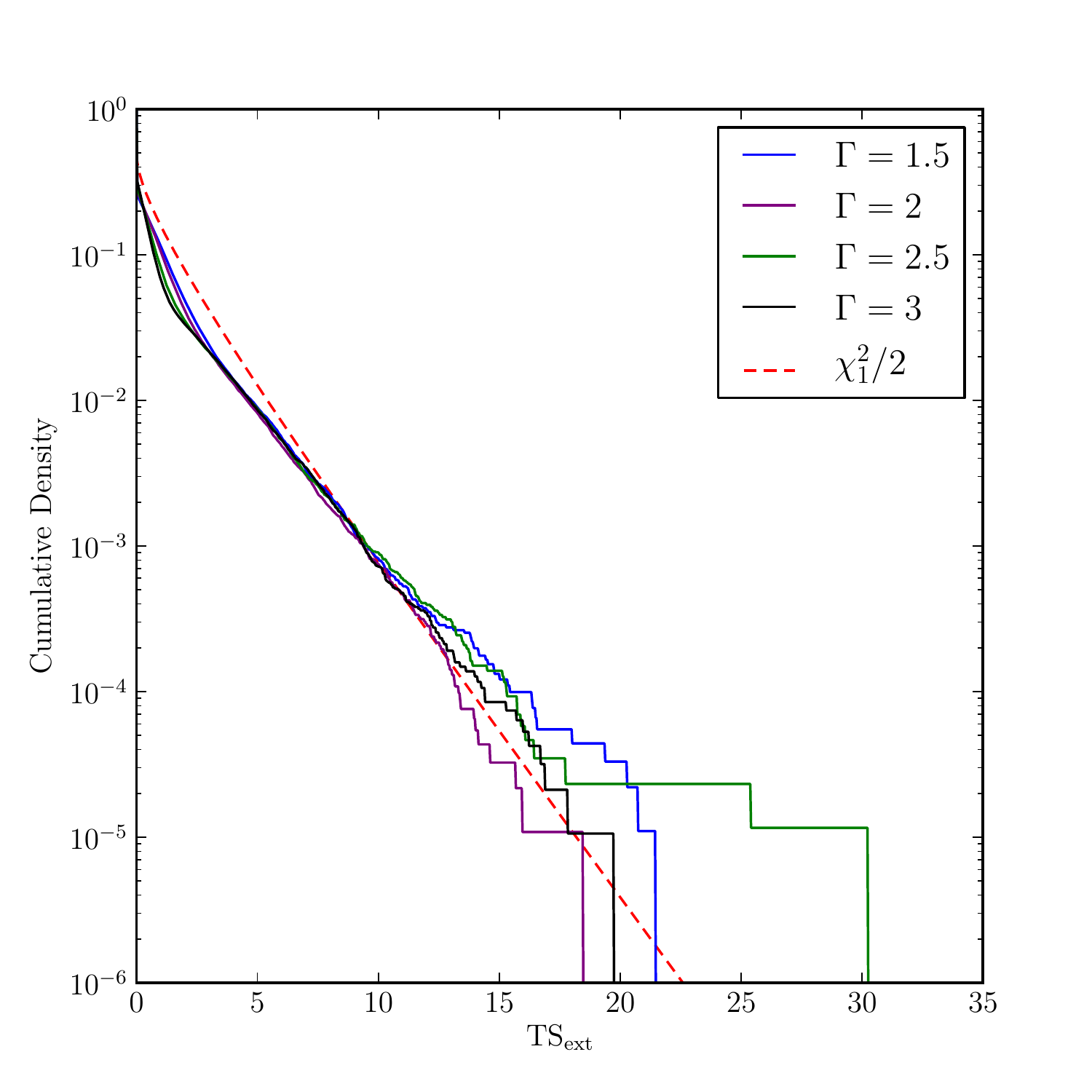}
  \caption{Cumulative distribution of \tsext for sources simulated on
  top of the Galactic diffuse and isotropic background.}
  \figlabel{tsext_plane_plot}
\end{figure}

For each spectrum, we performed $\sim90,000$ simulations.
\figref{tsext_plane_plot} shows the cumulative density of \tsext for each
spectrum. For small values of \tsext, there is good agreement between
the simulations and theory.  For the highest values of \tsext, there is
possibly a small discrepancy, but the discrepancy is not statistically
significant.  Therefore, we are confident we can use \tsext as a robust
measure of statistical significance when testing LAT-detected sources
for extension.

\subsection{Extended Source Simulations Over a Structured Background}
\subseclabel{bias_wrong_spatial_model}

We also performed a Monte Carlo study to show that incorrectly modeling
the spatial extension of an extended source does not substantially bias
the spectral fit of the source, although it does alter the value of
the \ts.  To assess this, we simulated the spatially extended ring-type
SNR W44.  We selected W44 because it is the most significant extended
source detected by the LAT that has a non-radially symmetric photon
distribution \citep{abdo_2010a_gamma-ray-emission}.

W44 was simulated with a power-law spectral model with an integral flux
of $7.12\times10^{-8}$ \fluxunits in the energy range from 1 \gev to
100 \gev and a spectral index of 2.66 (see \secref{validate_known}).

W44 was simulated with the elliptical ring spatial model described in
\cite{abdo_2010a_gamma-ray-emission}. For reference, the ellipse has
a semi-major axis of 0\fdg3, a semi-minor axis of 0\fdg19, a position
angle of $147\degree$ measured East of celestial North, and the ring's
inner radius is 75\% of the outer radius.

We used a simulation setup similar to that described in
\subsecref{validation_over_plane}, but the simulations were over the
2-year interval of the 2FGL catalog.  In the simulations, we did not
include the finite energy resolution of the LAT to isolate any effects
due to changing the assumed spatial model.  The fitting code we use also
ignores this energy dispersion and the potential bias introduced by
this will be discussed in an upcoming paper by the LAT collaboration
\citep{ackermann_2012a_fermi-large}.  In total, we performed 985
independent simulations.

The simulated sources were fit using a point-like spatial model,
a radially-symmetric Gaussian spatial model, a uniform disk spatial
model, an elliptical disk spatial model, and finally with an elliptical
ring spatial model.  We obtained the best fit spatial parameters using
\pointlike and, with these parameters, obtained the best fit spectral
parameters using \gtlike.

\begin{figure}[htbp]
  \includegraphics{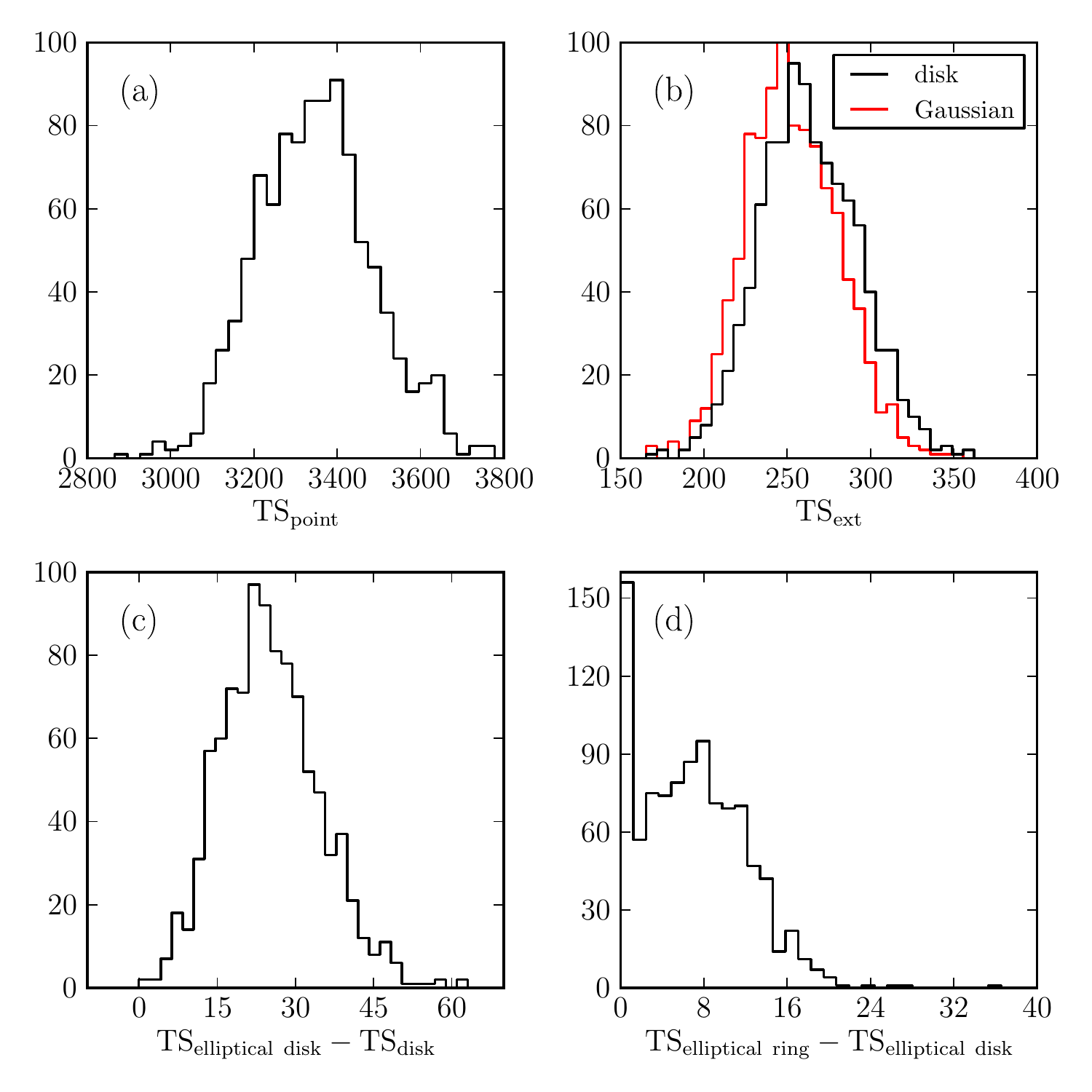}
  \caption{The distribution of \ts values when fitting 985 statistically
  independent simulations of W44. (a) is the distribution of \ts
  values when fitting W44 as a point-like source and (b) is the
  distribution of \tsext when fitting the source with a uniform disk or
  a radially-symmetric Gaussian spatial model. (c) is the distribution
  of the change in TS when fitting the source with an elliptical disk
  spatial model compared to fitting it with a radially-symmetric disk
  spatial model and (d) when fitting the source with an elliptical
  ring spatial model compared to an elliptical disk spatial model.}
  \figlabel{ts_comparison_w44sim}
\end{figure}

\figref{ts_comparison_w44sim}a shows that the significance of W44 in the
simulations is very large ($\ts\sim3500$) for a model with a point-like
source hypothesis.  \figref{ts_comparison_w44sim}b shows that the
significance of the spatial extension is also large ($\tsext\sim250$).
On average \tsext is somewhat larger when fitting the sources with
more accurate spatial models.  This shows that assuming an incorrect
spatial model will cause the source's significance to be underestimated.
\figref{ts_comparison_w44sim}c shows that the sources were fit better
when assuming an elliptical disk spatial model compared to a uniform disk
spatial model ($\ts_\text{elliptical\ disk}-\ts_\text{disk}\sim30$).
Finally, \figref{ts_comparison_w44sim}d shows that the sources
were fit somewhat better assuming an elliptical ring spatial model
compared to an elliptical disk spatial model ($\ts_\text{elliptical\
ring}-\ts_\text{elliptical\ disk}\sim9$). This shows that the LAT has
some additional power to resolve substructure in bright extended sources.

\begin{figure}[htbp]
  \includegraphics{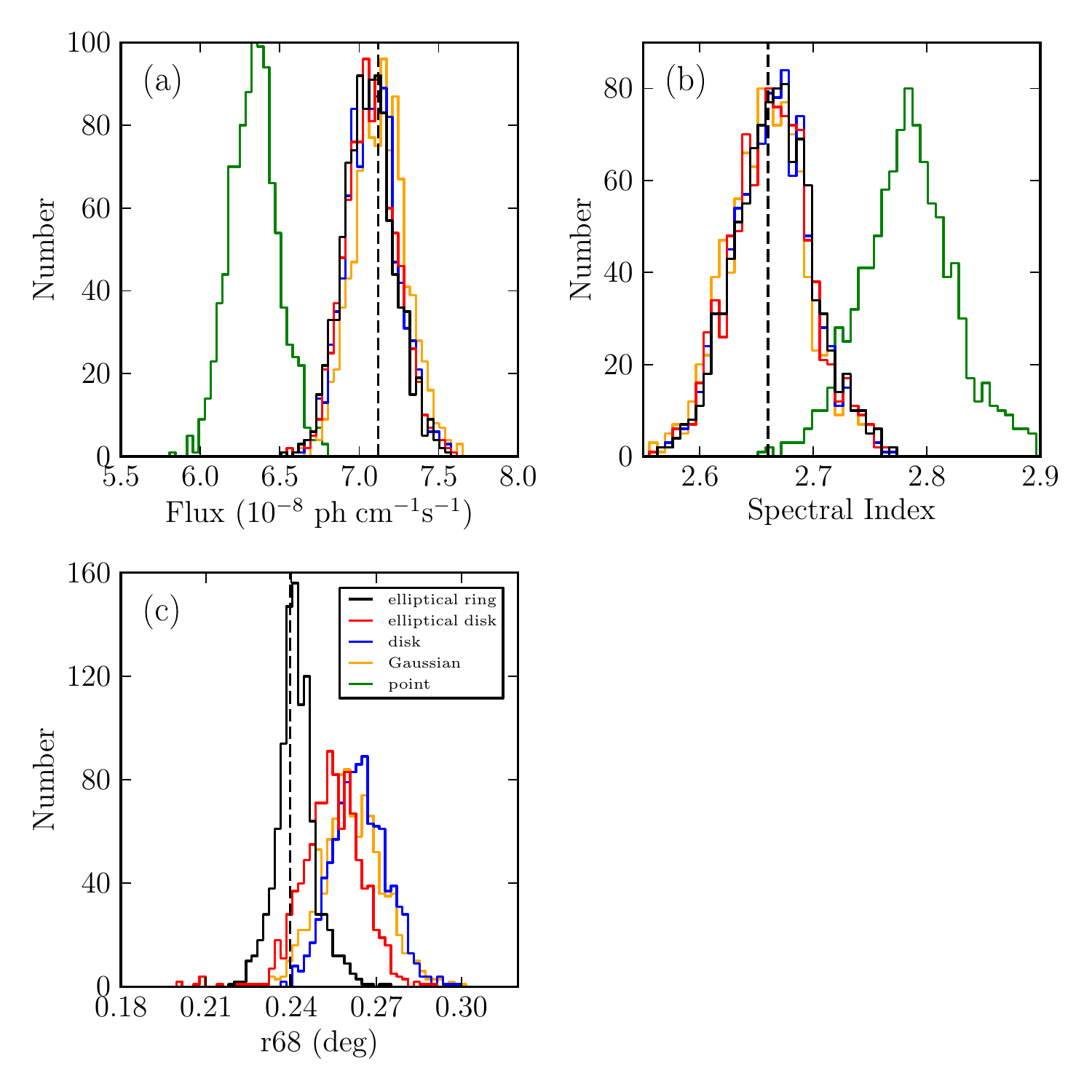}
  \caption{The distribution of fit parameters for the Monte Carlo
  simulations of W44.  The plots show the distribution of best fit (a)
  flux (b) spectral index and (c) 68\% containment radius. The dashed
  vertical lines represent the simulated values of the parameters.}
  \figlabel{bias_w44sim}
\end{figure}

\figref{bias_w44sim}a and \figref{bias_w44sim}b clearly show that no
significant bias is introduced by modeling the source as extended but with
an inaccurate spatial model, while a point-like source modeling results in
a $\sim10\%$ and $\sim0.125$ bias in the fit flux and index, respectively.
Furthermore, \figref{bias_w44sim}c shows that the \rsixeight estimate of
the extension size is very mildly biased ($\sim10\%$) toward higher values
when inaccurate spatial models are used, and thus represents a reasonable
measurement of the true 68\% containment radius for the source.  For the
elliptical spatial models, \rsixeight is computed by numeric integration.

\section{Extended Source Detection Threshold}
\seclabel{extension_sensitivity}

We calculated the LAT flux threshold to detect spatial extent. We define
the detection threshold as the flux at which the value of $\tsext$
averaged over many statistical realizations is $\langle\tsext\rangle=16$
(corresponding to a formal $4\sigma$ significance) for a source of a
given extension.

We used a simulation setup similar to that described in
\subsecref{monte_carlo_validation}, but instead of point-like sources we
simulated extended sources with radially-symmetric uniform disk spatial
models. Additionally, we simulated our sources over the two-year time
range included in the 2FGL catalog.  For each extension and spectral
index, we selected a flux range which bracketed $\tsext=16$ and performed
an extension test for $>100$ independent realizations of ten fluxes in
the range.  We calculated $\langle\tsext\rangle=16$ by fitting a line
to the flux and $\tsext$ values in the narrow range.

\begin{figure}[htbp]
    \includegraphics{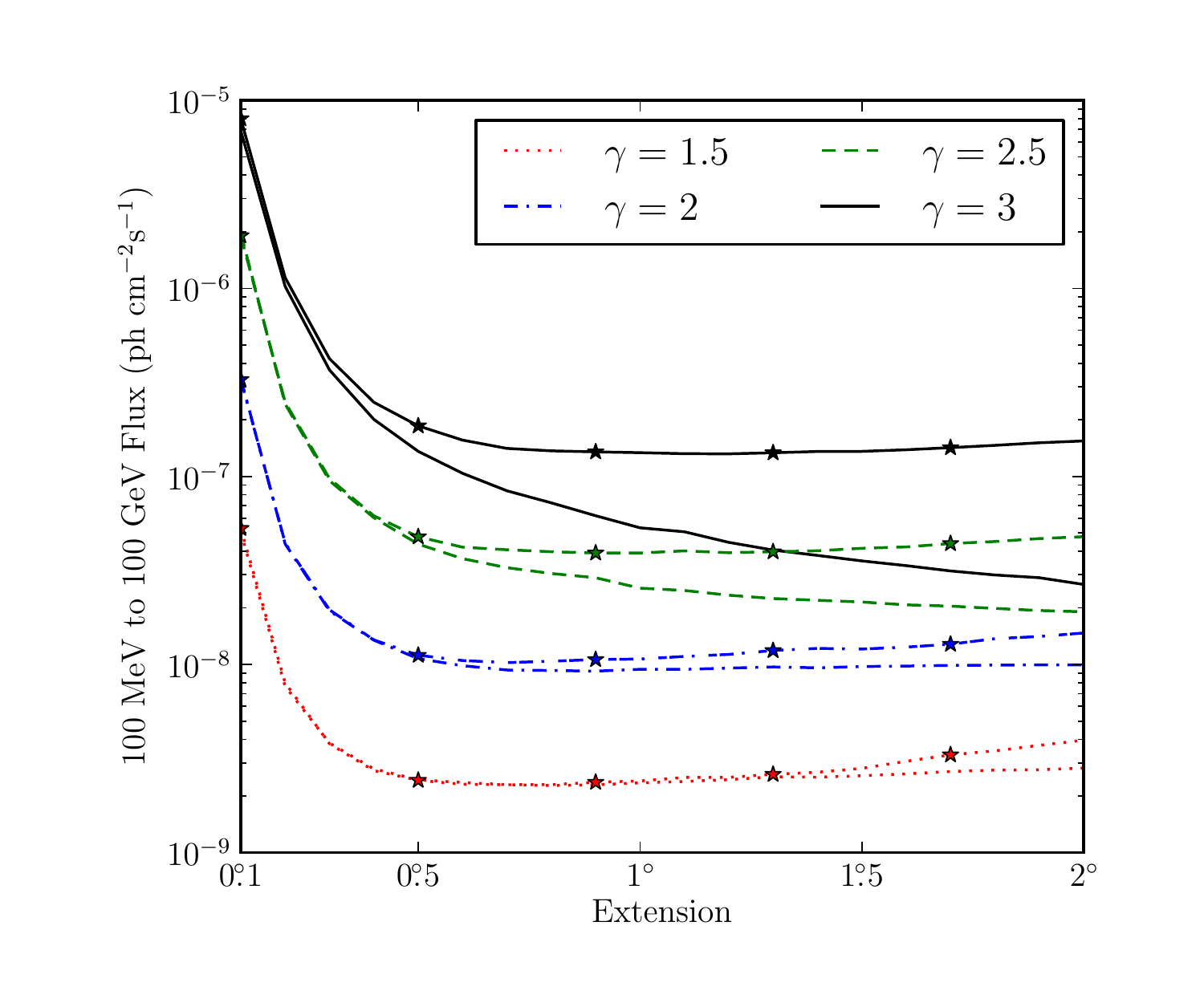}
    \caption{The detection threshold to resolve an extended source with
    a uniform disk model for a two-year exposure.  All sources have an
    assumed power-law spectrum and the different line styles (colors in
    the electronic version) correspond to different simulated spectral
    indices.  The lines with no markers correspond to the detection
    threshold using photons with energies between 100 \mev and 100 \gev,
    while the lines with star-shaped markers correspond to the threshold
    using photons with energies between 1 \gev and 100 \gev.}
    \figlabel{index_sensitivity}
  \end{figure}

\figref{index_sensitivity} shows the threshold for sources of four
spectral indices from 1.5 to 3 and extensions varying from $\sigma=0\fdg1$
to $2\fdg0$.  The threshold is high for small extensions when the
source is small compared to the size of the PSF.  It drops quickly
with increasing source size and reaches a minimum around 0\fdg5.  The
threshold increases for large extended sources because the source becomes
increasingly diluted by the background.  \figref{index_sensitivity}
shows the threshold using photons with energies between 100 \mev and
100 \gev and also using only photons with energies between 1 \gev and
100 \gev.  Except for very large or very soft sources, the threshold is
not substantially improved by including photons with energies between 100
\mev and 1 \gev.  This is also demonstrated in \figref{four_plots_ic443}
which shows \tsext for the SNR IC~443 computed independently in twelve
energy bins between 100 \mev and 100 \gev. For IC~443, which has a
spectral index $\sim2.4$ and an extension $\sim0\fdg35$, almost the
entire increase in likelihood from optimizing the source extent in the
model comes from energies above 1 \gev.  Furthermore, other systematic
errors become increasingly large at low energy. For our extension search
(\secref{extended_source_search_method}), we therefore used only photons
with energies above 1 \gev.

  \thispagestyle{empty}
\begin{deluxetable}{rr|rrrrrrrrrrrrrrrrrrrrr}
\tablecolumns{22}
\tabletypesize{\tiny}
\rotate
\tablewidth{0pt}
\tablecaption{Extension Detection Threshold
\tablabel{all_sensitivity_table}
}
\tablehead{
\colhead{$\gamma$}&       
\colhead{BG}&
\colhead{$0.1$}&
\colhead{$0.2$}&
\colhead{$0.3$}&
\colhead{$0.4$}&
\colhead{$0.5$}&
\colhead{$0.6$}&
\colhead{$0.7$}&
\colhead{$0.8$}&
\colhead{$0.9$}&
\colhead{$1.0$}&
\colhead{$1.1$}&
\colhead{$1.2$}&
\colhead{$1.3$}&
\colhead{$1.4$}&
\colhead{$1.5$}&
\colhead{$1.6$}&
\colhead{$1.7$}&
\colhead{$1.8$}&
\colhead{$1.9$}&
\colhead{$2.0$}
}
\startdata
\multicolumn{22}{c}{E$>$1 \gev} \\
\hline
     1.5 &      $1\times$ &      148.1 &       23.3 &       11.3 &        8.0 &        7.2 &        6.9 &        6.7 &        6.8 &        7.1 &        7.4 &        7.6 &        7.9 &        8.1 &        8.5 &        9.2 &        9.9 &        9.1 &        9.2 &        9.0 &       10.3 \\
         &     $10\times$ &      148.4 &       29.0 &       18.7 &       15.2 &       15.4 &       15.0 &       16.1 &       16.0 &       16.8 &       17.7 &       18.2 &       19.3 &       20.9 &       22.5 &       23.8 &       24.8 &       21.3 &       22.8 &       23.4 &       23.7 \\
         &    $100\times$ &      186.8 &       55.0 &       43.4 &       40.7 &       41.0 &       41.8 &       40.9 &       40.9 &       42.7 &       43.6 &       38.4 &       39.9 &       40.6 &       38.4 &       36.9 &       36.3 &       37.1 &       38.8 &       37.2 &       37.6 \\
       2 &      $1\times$ &      328.4 &       43.4 &       18.9 &       13.4 &       11.2 &       10.4 &       10.2 &       10.2 &       10.2 &       10.4 &       10.7 &       10.9 &       11.2 &       11.5 &       12.4 &       12.6 &       13.0 &       13.4 &       14.0 &       14.4 \\
         &     $10\times$ &      341.0 &       55.9 &       32.3 &       27.6 &       26.5 &       25.4 &       25.6 &       25.9 &       27.4 &       26.8 &       27.8 &       28.7 &       29.8 &       30.1 &       31.0 &       31.5 &       31.7 &       34.0 &       34.3 &       35.9 \\
         &    $100\times$ &      420.5 &      128.3 &       90.2 &       77.3 &       73.3 &       70.8 &       67.5 &       64.3 &       64.2 &       64.1 &       62.8 &       63.6 &       61.7 &       61.9 &       58.4 &       59.0 &       61.4 &       63.3 &       60.1 &       58.1 \\
     2.5 &      $1\times$ &      627.1 &       75.6 &       29.8 &       19.3 &       15.5 &       13.5 &       12.8 &       12.6 &       12.5 &       12.5 &       12.6 &       12.9 &       12.9 &       13.1 &       13.5 &       13.7 &       14.3 &       14.8 &       15.2 &       15.8 \\
         &     $10\times$ &      638.9 &       99.1 &       52.1 &       39.1 &       34.6 &       33.0 &       32.5 &       32.5 &       32.8 &       33.2 &       34.1 &       34.3 &       34.5 &       35.1 &       36.6 &       36.9 &       35.5 &       36.0 &       36.5 &       37.3 \\
         &    $100\times$ &      795.0 &      262.1 &      140.9 &      104.3 &       90.4 &       81.2 &       77.2 &       75.1 &       69.7 &       70.9 &       66.5 &       65.6 &       64.9 &       64.0 &       58.9 &       58.1 &       60.2 &       58.4 &       57.5 &       55.8 \\
       3 &      $1\times$ &      841.5 &      110.6 &       43.2 &       25.5 &       18.7 &       16.1 &       14.4 &       13.6 &       13.3 &       13.2 &       13.1 &       13.1 &       13.4 &       13.6 &       13.5 &       13.8 &       14.2 &       14.4 &       14.8 &       15.4 \\
         &     $10\times$ &      921.6 &      151.3 &       69.1 &       47.8 &       40.7 &       37.1 &       35.5 &       34.5 &       35.1 &       35.5 &       35.3 &       35.3 &       35.4 &       35.5 &       36.8 &       37.6 &       35.3 &       35.4 &       36.3 &       36.6 \\
         &    $100\times$ &     1124.1 &      282.9 &      181.1 &      119.8 &      100.7 &       91.1 &       84.3 &       77.9 &       73.3 &       71.8 &       67.6 &       66.4 &       65.5 &       63.9 &       59.0 &       58.6 &       58.8 &       57.5 &       55.4 &       54.4 \\
\cutinhead{E$>$10 \gev}
     1.5 &      $1\times$ &       44.6 &        8.0 &        4.3 &        3.2 &        2.7 &        2.6 &        2.5 &        2.5 &        2.4 &        2.5 &        2.5 &        2.6 &        2.7 &        2.8 &        2.9 &        2.9 &        3.1 &        3.2 &        3.3 &        3.4 \\
         &     $10\times$ &       45.2 &        9.2 &        5.8 &        5.0 &        4.9 &        4.9 &        5.0 &        5.2 &        5.3 &        5.7 &        5.9 &        6.3 &        6.6 &        6.5 &        6.8 &        7.6 &        7.8 &        8.2 &        8.5 &        8.7 \\
         &    $100\times$ &       47.3 &       13.4 &       11.6 &       10.6 &       10.8 &       10.8 &       12.0 &       12.7 &       13.2 &       13.7 &       15.3 &       16.1 &       17.2 &       18.2 &       18.9 &       19.5 &       20.4 &       21.0 &       21.7 &       22.9 \\
       2 &      $1\times$ &       49.7 &        8.4 &        4.4 &        3.3 &        2.8 &        2.6 &        2.6 &        2.6 &        2.6 &        2.6 &        2.7 &        2.7 &        2.8 &        2.9 &        3.0 &        3.2 &        3.2 &        3.4 &        3.5 &        3.5 \\
         &     $10\times$ &       48.6 &        9.5 &        6.0 &        5.2 &        5.0 &        5.2 &        5.2 &        5.3 &        5.4 &        5.8 &        6.4 &        6.6 &        7.0 &        7.1 &        7.5 &        8.0 &        8.3 &        8.6 &        9.0 &        9.2 \\
         &    $100\times$ &       51.8 &       14.7 &       11.8 &       11.5 &       11.5 &       11.9 &       13.2 &       14.0 &       14.3 &       15.3 &       16.2 &       16.9 &       18.4 &       19.2 &       19.8 &       21.0 &       22.0 &       22.8 &       23.2 &       24.3 \\
     2.5 &      $1\times$ &       53.1 &        9.1 &        4.5 &        3.3 &        2.8 &        2.7 &        2.6 &        2.5 &        2.5 &        2.6 &        2.7 &        2.7 &        2.8 &        2.8 &        2.9 &        3.1 &        3.2 &        3.3 &        3.5 &        3.6 \\
         &     $10\times$ &       53.7 &       10.5 &        6.3 &        5.4 &        5.1 &        5.1 &        5.3 &        5.4 &        5.7 &        6.0 &        6.3 &        6.6 &        6.8 &        6.9 &        7.5 &        8.1 &        8.3 &        8.6 &        8.9 &        9.2 \\
         &    $100\times$ &       57.0 &       15.6 &       12.7 &       11.9 &       11.8 &       12.2 &       13.1 &       14.3 &       14.6 &       15.2 &       16.3 &       17.0 &       18.8 &       19.2 &       19.9 &       21.0 &       21.9 &       22.3 &       23.3 &       23.7 \\
       3 &      $1\times$ &       55.5 &        9.4 &        4.8 &        3.4 &        2.9 &        2.7 &        2.6 &        2.5 &        2.5 &        2.5 &        2.6 &        2.7 &        2.7 &        2.8 &        2.9 &        3.0 &        3.1 &        3.2 &        3.4 &        3.4 \\
         &     $10\times$ &       56.0 &       10.5 &        6.2 &        5.3 &        5.1 &        5.1 &        5.1 &        5.3 &        5.5 &        5.7 &        5.9 &        6.4 &        6.4 &        6.6 &        7.0 &        7.8 &        8.0 &        8.3 &        8.6 &        8.9 \\
         &    $100\times$ &       60.3 &       16.2 &       12.7 &       11.7 &       11.8 &       12.2 &       12.6 &       13.8 &       14.2 &       14.6 &       15.8 &       16.5 &       17.6 &       18.5 &       19.4 &       19.8 &       20.7 &       21.0 &       21.8 &       22.5 \\
\enddata
\tablecomments{
      The detection threshold to resolve spatially extended
      sources with a uniform disk spatial model for a two-year exposure
      The threshold is calculated for
      sources of varying energy
      ranges, spectral indices, and background levels.  The sensitivity
      was calculated against
      a Sreekumar-like isotropic background
      and the second column is the factor that the simulated background
      was scaled by. The remaining columns are varying sizes of the source. 
      The table quotes
      integral fluxes in the analyzed energy range (1 \gev to 100 \gev or 10 \gev to 100
      \gev) in units of $10^{-10}$ \fluxunits.
      }
\end{deluxetable}

\begin{figure}[htbp]
  \includegraphics{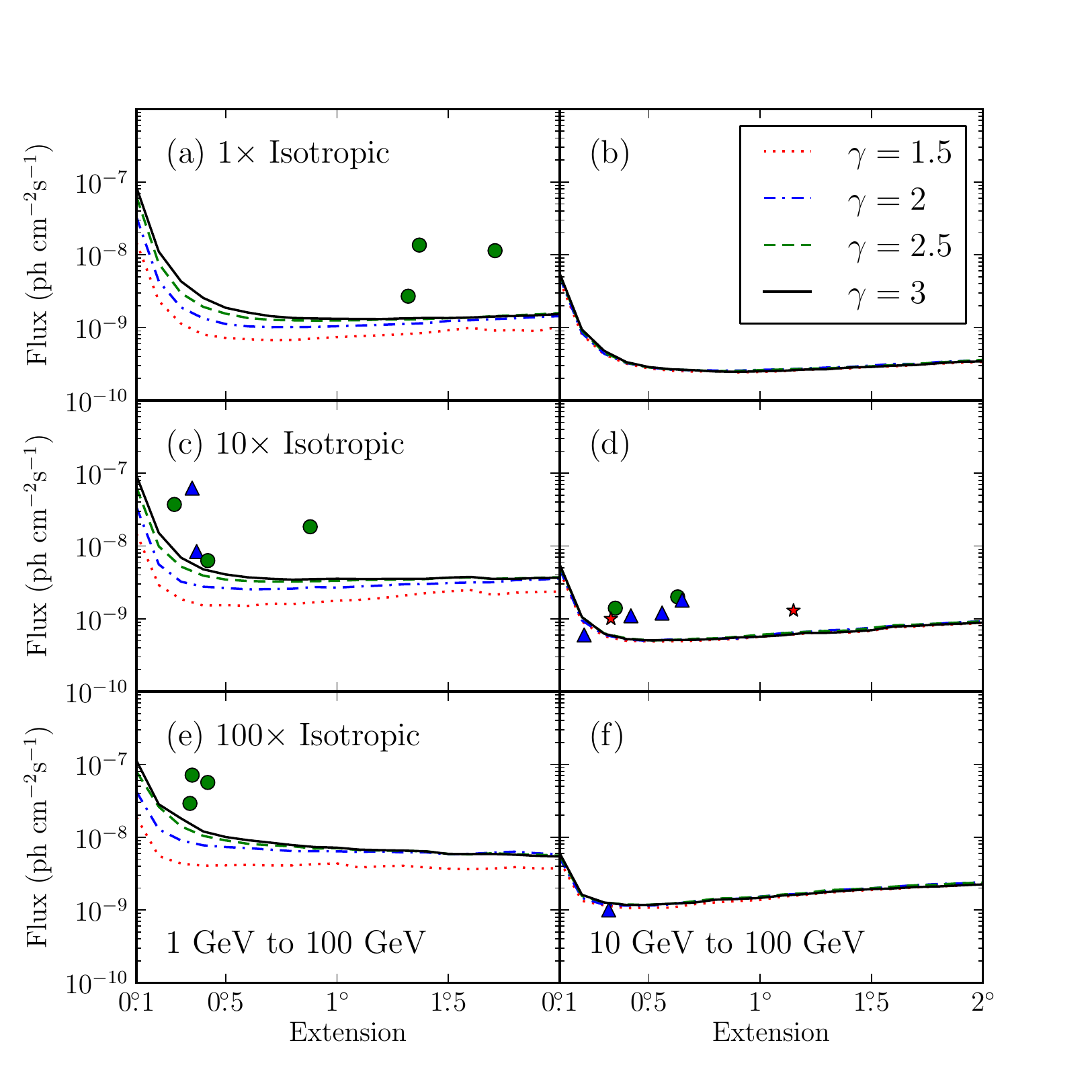}
  \caption{The LAT detection threshold for four spectral indices and three
  backgrounds ($1\times$, $10\times$, and $100\times$ the Sreekumar-like
  isotropic background) for a two-year exposure. The left-hand plots are
  the detection threshold when using photons with energies between 1 \gev
  and 100 \gev and the right-hand plots are the detection threshold when
  using photons with energies between 10 \gev and 100 \gev.  The flux
  is integrated only in the selected energy range.  Overlaid on this
  plot are the LAT-detected extended sources placed by the magnitude of
  the nearby Galactic diffuse emission and the energy range they were
  analyzed with.  The star-shaped markers (colored red in the electronic
  version) are sources with a spectral index closer to 1.5, the triangular
  markers (colored blue) an index closer to 2, and the circular markers
  (colored green) an index closer to 2.5.  The triangular marker in plot
  (d) below the sensitivity line is MSH\,15$-$52.}
  \figlabel{all_sensitivity}
\end{figure}

\figref{all_sensitivity} shows the flux threshold as a function of
source extension for different background levels ($1\times$, $10\times$,
and $100\times$ the nominal background), different spectral indices,
and two different energy ranges (1 \gev to 100 \gev and 10 \gev to
100 \gev).  The detection threshold is higher for sources in regions of
higher background.  When studying sources only at energies above 1 \gev,
the LAT detection threshold (defined as the 1 \gev to 100 \gev flux at
which $\langle\tsext\rangle=16$) depends less strongly on the spectral
index of the source.  The index dependence of the detection threshold is
even weaker when considering only photons with energies above 10 \gev
because the PSF changes little from 10 \gev to 100 \gev.  Overlaid on
\figref{all_sensitivity} are the LAT-detected extended sources that will
be discussed in \secref{validate_known} and \secref{new_ext_srcs_section}.
The extension thresholds are tabulated in \tabref{all_sensitivity_table}.

\begin{figure}[htbp]
  \includegraphics{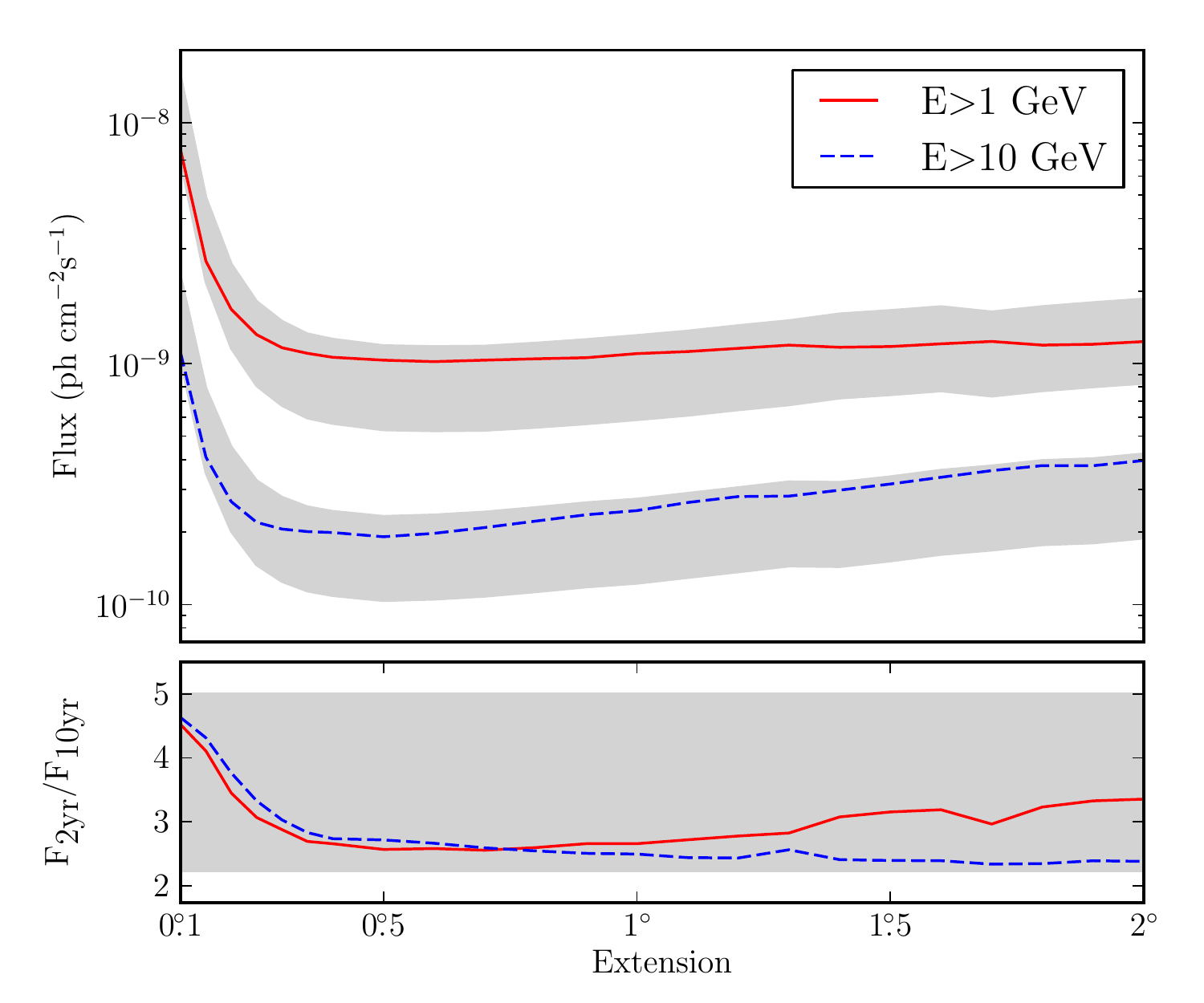}
  \caption{The projected detection threshold of the LAT to extension
  after 10 years for a power-law source of spectral index 2 against 10
  times the isotropic background in the energy range from 1 \gev to
  100 \gev (solid line colored red in the electronic version) and 10
  \gev to 100 \gev (dashed line colored blue). The shaded gray regions
  represent the detection threshold assuming the sensitivity improves
  from 2 to 10 years by the square root of the exposure (top edge)
  and linearly with exposure (bottom edge).  The lower plot shows
  the factor increase in sensitivity.  For small extended sources,
  the detection threshold of the LAT to the extension of a source will
  improve by a factor larger than the square root of the exposure.}
  \figlabel{time_sensitivity}
\end{figure}

Finally, \figref{time_sensitivity} shows the projected detection
threshold of the LAT to extension with a 10 year exposure against 10
times the isotropic background measured by EGRET. This background is
representative of the background near the Galactic plane.  For small
extended sources, the threshold improves by a factor larger than the
square root of the relative exposures because the LAT is signal-limited
at high energies where the present analysis is most sensitive. For large
extended sources, the relevant background is over a larger spatial range
and so the improvement is closer to a factor corresponding to the square
root of the relative exposures that is caused by Poisson fluctuations
in the background.

\section{Testing Against Source Confusion}
\seclabel{dual_localization_method}

It is impossible to discriminate using only LAT data between a spatially
extended source and multiple point-like sources separated by angular
distances comparable to or smaller than the size of the LAT PSF.
To assess the plausibility of source confusion for sources with
$\tsext\ge16$, we developed an algorithm to test if a region contains
two point-like sources.  The algorithm works by simultaneously fitting
in \pointlike the positions and spectra of the two point-like sources.
To help with convergence, it begins by dividing the source into two
spatially coincident point-like sources and then fitting the sum and
difference of the positions of the two sources without any limitations
on the fit parameters.

After simultaneously fitting the two positions and two spectra, we define
\tsinc as twice the increase in the log of the likelihood fitting the
region with two point-like sources compared to fitting the region with
one point-like source:
\begin{equation}
  \tsinc=2\log(\likelihood_\text{2pts}/\likelihood_\text{ps}).
\end{equation} 
For the following analysis of LAT data, \tsinc was computed by fitting
the spectra of the two point-like sources in \gtlike using the best fit
positions of the sources found by \pointlike.

\tsinc cannot be quantitatively compared to \tsext using
a simple likelihood-ratio test to evaluate which model
is significantly better because the models are not nested
\citep{protassov_2002a_statistics-handle}.  Even though the comparison of
\tsext with \tsinc is not a calibrated test, $\tsext>\tsinc$ indicates
that the likelihood for the extended source hypothesis is higher than
for two point-like sources and we only consider a source to be extended
if $\tsext>\tsinc$.

We considered using the Bayesian information criterion
\citep[BIC,][]{schwarz_1978a_estimating-dimension} as an alternative
Bayesian formulation for this test, but it is difficult to apply to LAT
data because it contains a term including the number of data points.
For studying $\gamma$-ray sources in LAT data, we analyze relatively
large regions of the sky to better define the contributions from diffuse
backgrounds and nearby point sources. This is important for accurately
evaluating source locations and fluxes but the fraction of data directly
relevant to the evaluation of the parameters for the source of interest
is relatively small.

As an alternative, we considered the Akaike information criterion test
\citep[\aic,][]{akaike_1974a_statistical-model}.  The \aic is defined
as $\aic=2k-2\log\likelihood$, where $k$ is the number of parameters
in the model.  In this formulation, the best hypothesis is considered
to be the one that minimizes the \aic.  The first term penalizes models
with additional parameters.

The two point-like sources hypothesis has three more parameters than
the single extended source hypothesis (two more spatial parameters and
two more spectral parameters compared to one extension parameter), so the
comparison $\aic_\text{ext} < \aic_\text{2pts}$  is formally equivalent to
$\tsext + 6 > \tsinc$.  Our criterion for accepting extension ($\tsext >
\tsinc$) is thus equivalent to requesting that the AIC-based empirical
support for the two point-like sources model be ``considerably less''
than for the extended source model, following the classification by
\cite{burnham_2002a_model-selection}.

We assessed the power of the $\tsext>\tsinc$ test with a Monte Carlo
study.  We simulated one spatially extended source and fit it as both
an extended source and as two point-like sources using \pointlike.
We then simulated two point-like sources and fit them with the same
two hypotheses. By comparing the distribution of \tsinc and \tsext
computed by \pointlike for the two cases, we evaluated how effective
the $\tsext>\tsinc$ test is at rejecting cases of source confusion as
well as how likely it is to incorrectly reject that an extended source is
spatially extended.  All sources were simulated using the same time range
as in \secref{extension_sensitivity} against a background 10 times the
isotropic background measured by EGRET, representative of the background
near the Galactic plane.

\begin{figure}[htbp]
  \includegraphics{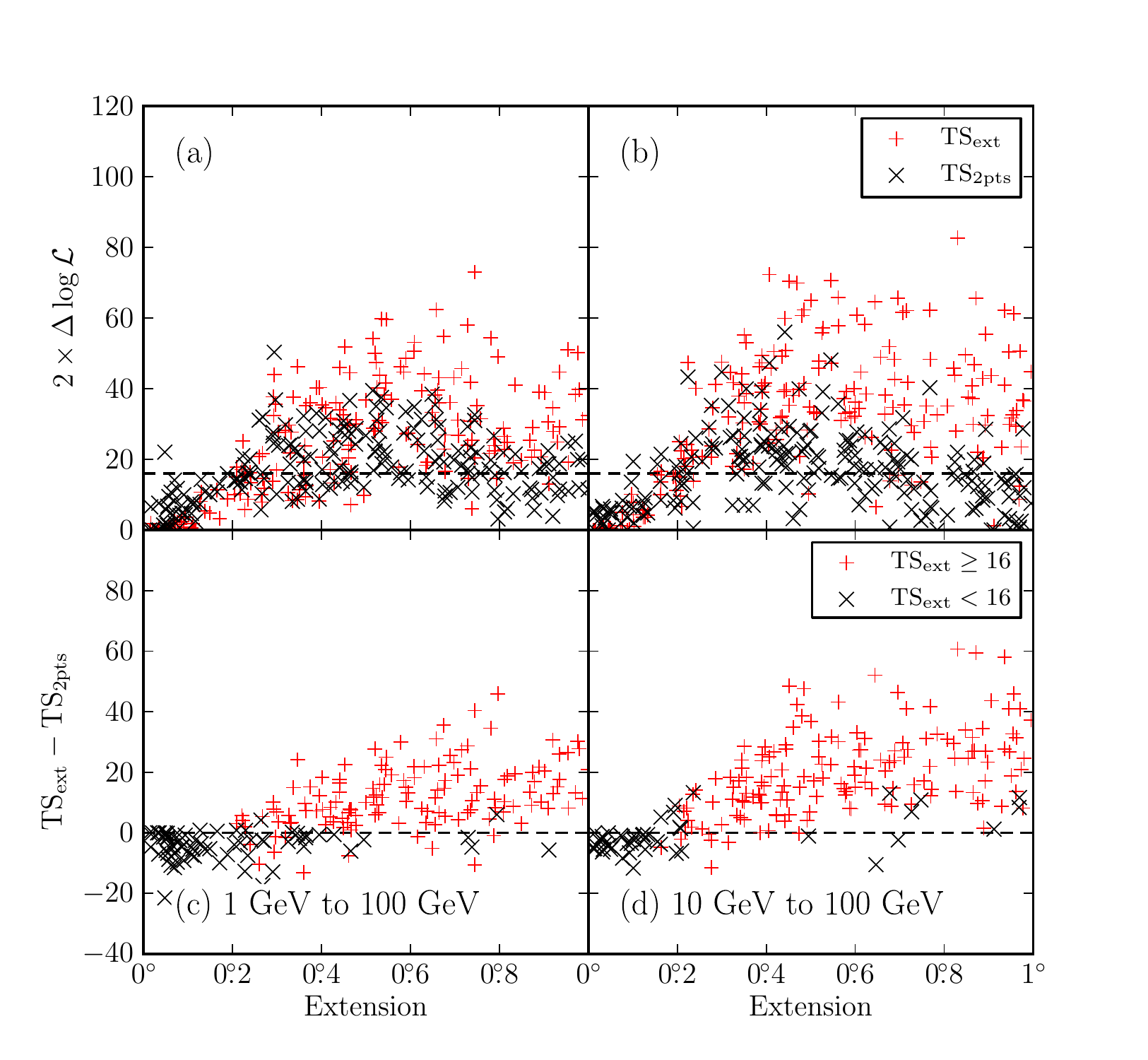}
  \caption{(a) and (b) are the distribution of \tsext and of \tsinc
  when fitting simulated spatially extended sources of varying sizes
  as both an extended source and as two point-like sources.  (c) and
  (d) are the distribution of $\tsext-\tsinc$ for the same simulated
  sources.  (a) and (c) represent sources fit in the 1 \gev to 100 \gev
  energy range and (b) and (d) represent sources fit in the 10 \gev
  to 100 \gev energy range.  In (c) and (d), the plus-shaped markers
  (colored red in the electronic version) are fits where $\tsext\ge16$.}
  \figlabel{confusion_extended_plot}
\end{figure}

We did this study first in the energy range from 1 \gev to 100 \gev
by simulating extended sources of flux $4\times10^{-9}$ \fluxunits
integrated from 1 \gev to 100 \gev and a power-law spectral model
with spectral index 2.  This spectrum was picked to be representative
of the new extended sources that were discovered in the following
analysis when looking in the 1 \gev to 100 \gev energy range (see
\secref{new_ext_srcs_section}).  We simulated these sources using
uniform disk spatial models with extensions varying up to $1\degree$.
\figref{confusion_extended_plot}a shows the distribution of \tsext and
\tsinc and \figref{confusion_extended_plot}c shows the distribution of
$\tsext-\tsinc$ as a function of the simulated extension of the source
for 200 statistically independent simulations.

\begin{figure}[htbp]
  \includegraphics{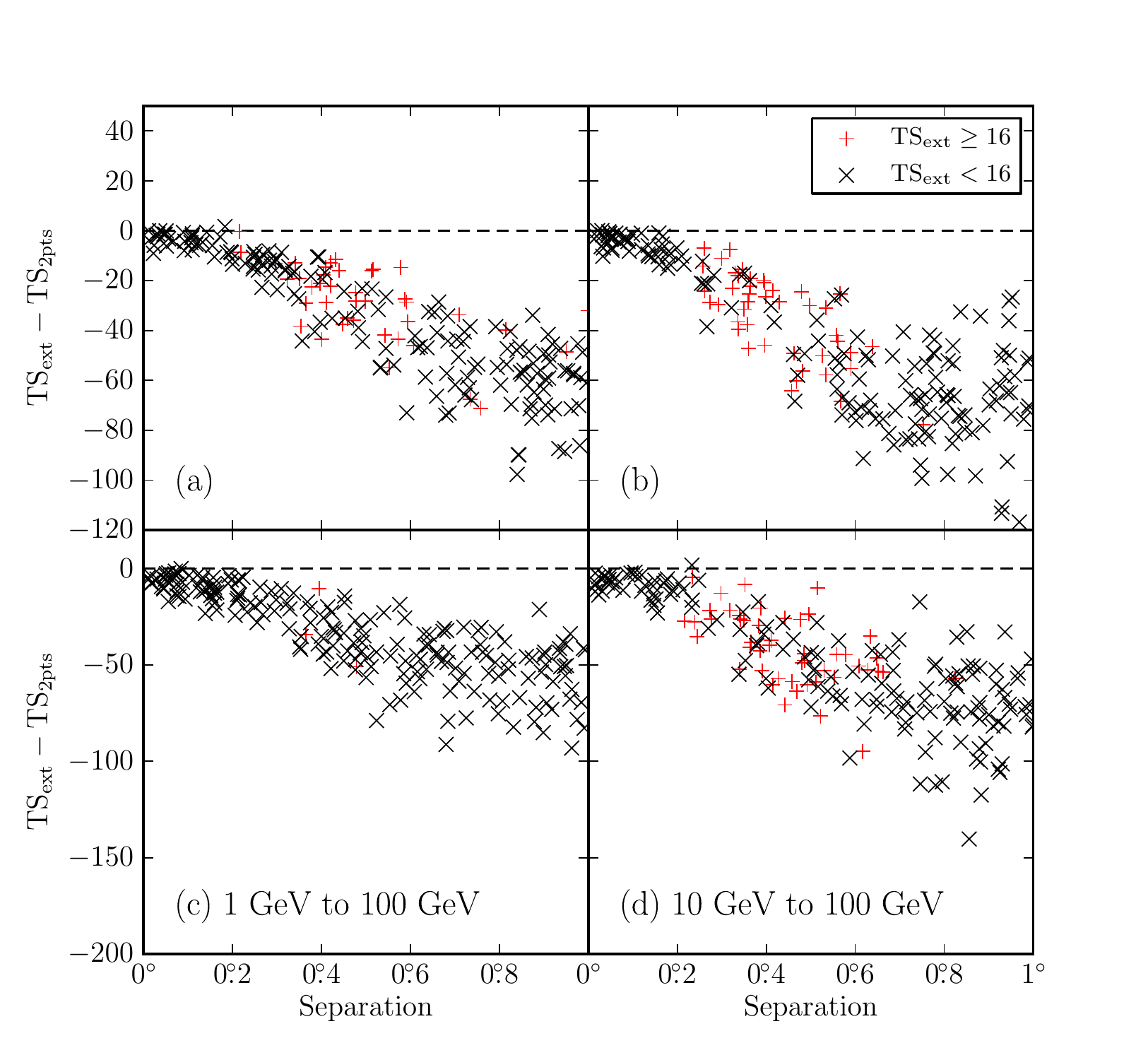}
  \caption{The distribution of $\tsext-\tsinc$ when fitting two simulated
  point-like sources of varying separations as both an extended source
  and as two point-like sources.  (a), and (b) represent simulations
  of two point-like sources with the same spectral index and (c) and
  (d) represent simulations of two point-like sources with different
  spectral indices.  (a) and (c) fit the simulated sources in the 1
  \gev to 100 \gev energy range and (b) and (d) fit in the 10 \gev to
  100 \gev energy range.  The plus-shaped markers (colored red in the
  electronic version) are fits where $\tsext\ge16$.}
  \figlabel{confusion_2pts_plot}
\end{figure}

\figref{confusion_2pts_plot}a shows the same plot but when fitting two
simulated point-like sources each with half of the flux of the spatially
extended source and with the same spectral index as the extended source.
Finally, \figref{confusion_2pts_plot}c shows the same plot with each
point-like source having the same flux but different spectral indices.
One point-like source had a spectral index of 1.5 and the other an index
of 2.5.  These indices are representative of the range of indices of
LAT-detected sources.

The same four plots are shown in \figref{confusion_extended_plot}b,
\figref{confusion_extended_plot}d, \figref{confusion_2pts_plot}b, and
\figref{confusion_2pts_plot}d but this time when analyzing a source of
flux $10^{-9}$ \fluxunits (integrated from 10 \gev to 100 \gev) only in
the 10 \gev to 100 \gev energy range.  This flux is typical of the new
extended sources discovered using only photons with energies between 10
\gev and 100 \gev (see \secref{new_ext_srcs_section}).

Several interesting conclusions can be made from this study.  As one would
expect, $\tsext-\tsinc$ is mostly positive when fitting the simulated
extended sources.  In the 1 \gev to 100 \gev analysis, only 11 of the
200 simulated extended sources had $\tsext>16$ but were incorrectly
rejected due to \tsinc being greater than \tsext.  In the 10 \gev to 100
\gev analysis, only 7 of the 200 sources were incorrectly rejected. From
this, we conclude that this test is unlikely to incorrectly reject truly
spatially extended sources.

On the other hand, it is often the case that $\tsext>16$ when testing
the two simulated point-like sources for extension.  This is especially
the case when the two sources had the same spectral index. Forty out
of 200 sources in the 1 \gev to 100 \gev energy range and 43 out of
200 sources in the 10 \gev to 100 \gev energy range had $\tsext>16$.
But in these cases, we always found the single extended source fit to
be worse than the two point-like source fit.  From this, we conclude
that the $\tsext>\tsinc$ test is powerful at discarding cases in which
the true emission comes from two point-like sources.

The other interesting feature in \figref{confusion_extended_plot}a
and \figref{confusion_extended_plot}b is that for simulated extended
sources with typical sizes ($\sigma\sim0\fdg5$), one can often obtain
almost as large an increase in likelihood fitting the source as two
point-like sources ($\tsinc\sim\tsext$).  This is because although
the two point-like sources represent an incorrect spatial model, the
second source has four additional degrees of freedom (two spatial and
two spectral parameters) and can therefore easily model much of the
extended source and statistical fluctuations in the data.  This effect
is most pronounced when using photons with energies between 1 \gev and
100 \gev where the PSF is broader.

From this Monte Carlo study, we can see the limits of an analysis with LAT
data of spatially extended sources.  \subsecref{monte_carlo_validation}
showed that we have a statistical test that finds when a LAT source is
not well described by the PSF.  But this test does not uniquely prove
that the emission originates from spatially extended emission instead
of from multiple unresolved sources.  Demanding that $\tsext>\tsinc$
is a powerful second test to avoid cases of simple confusion of two
point-like sources. But it could always be the case that an extended
source is actually the superposition of multiple point-like or extended
sources that could be resolved with deeper observations of the region.
There is nothing about this conclusion unique to analyzing LAT data,
but the broad PSF of the LAT and the density of sources expected to be
\gev emitters in the Galactic plane makes this issue more significant
for analyses of LAT data.  When possible, multiwavelength information
should be used to help select the best model of the sky.

\section{Test of 2LAC Sources}
\seclabel{test_2lac_sources}

For all following analyses of LAT data, we used the same two-year dataset
that was used in the 2FGL catalog spanning from 2008 August 4 to 2010
August 1. We applied the same acceptance cuts and we used the same P7\_V6
Source class event selection and IRFs \citep{ackermann_2012a_fermi-large}.
When analyzing sources in \pointlike, we used a circular $10\degree$
region of interest (ROI) centered on our source and eight energy bins
per logarithmic decade in energy.  When refitting the region in \gtlike
using the best fit spatial and spectral models from \pointlike, we used
the `binned likelihood' mode of \gtlike on a $14\degree\times14\degree$
ROI with a pixel size of 0\fdg03.

Unless explicitly mentioned, we used the same background model as 2FGL
to represent the Galactic diffuse, isotropic, and Earth limb emission.
To compensate for possible residuals in the diffuse emission model,
the Galactic emission was scaled by a power-law and the normalization
of the isotropic component was left free.  Unless explicitly mentioned,
we used all 2FGL sources within $15\degree$ of our source as our list of
background sources and we refit the spectral parameters of all sources
within $2\degree$ of the source.

To validate our method, we tested LAT sources associated with AGN
for extension.  \gev emission from AGN is believed to originate from
collimated jets.  Therefore AGN are not expected to be spatially
resolvable by the LAT and provide a good calibration source to
demonstrate that our extension detection method does not misidentify
point-like sources as being extended.  We note that megaparsec-scale
$\gamma$-ray halos around AGNs have been hypothesized to be
resolvable by the LAT \citep{aharonian_1994a_energy-gamma}. However,
no such halo has been discovered in the LAT data so far
\citep{neronov_2011a_evidence-gamma-ray}.

Following 2FGL, the LAT Collaboration published the Second LAT AGN Catalog
(2LAC), a list of high latitude ($|b|>10\degree$) sources that had a high
probability association with AGN \citep{ackermann_2011a_second-catalog}.
2LAC associated 1016 2FGL sources with AGN.  To avoid systematic problems
with AGN classification, we selected only the 885 AGN which made it into
the clean AGN sub-sample defined in the 2LAC paper.  An AGN association is
considered clean only if it has a high probability of association $P\ge
80\%$, if it is the only AGN associated with the 2FGL source, and if no
analysis flags have been set for the source in the 2FGL catalog. These
last two conditions are important for our analysis. Source confusion
may look like a spatially extended source and flagged 2FGL sources may
correlate with unmodeled structure in the diffuse emission.

\begin{figure}[htbp]
  \includegraphics{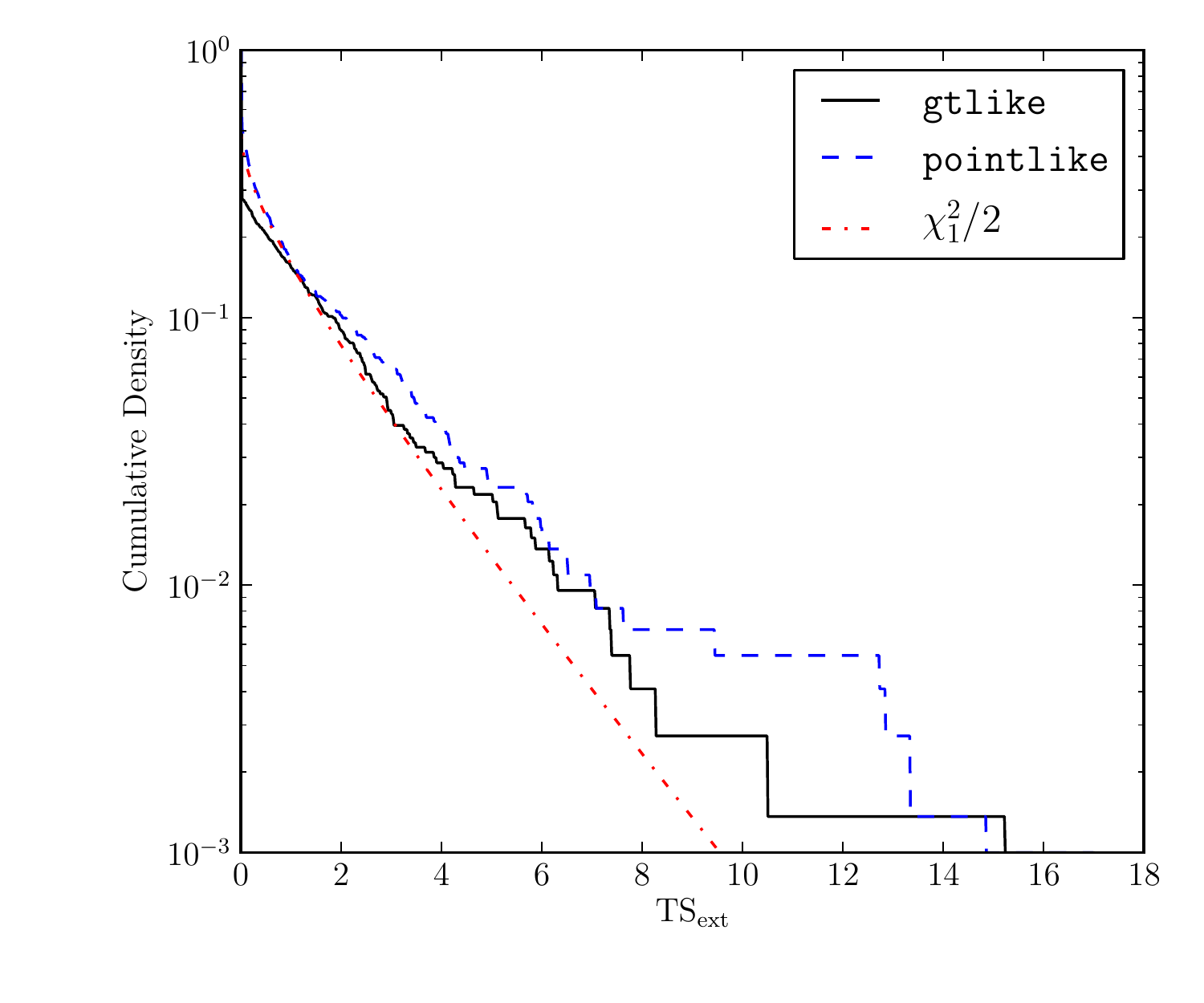}
  \caption{The cumulative density of \tsext for the 733 clean AGN in 2LAC
  that were significant above 1 \gev calculated with \pointlike (dashed
  line colored blue in the electronic version) and with \gtlike (solid
  line colored black).  AGN are too far and too small to be resolved
  by the LAT. Therefore, the cumulative density of $\tsext$ is expected
  to follow a $\chi^2_1/2$ distribution (\eqnref{ts_ext_distribution},
  the dash-dotted line colored red).}
  \figlabel{agn_ts_ext}
\end{figure}

Of the 885 clean AGN, we selected the 733 of these 2FGL sources which
were significantly detected above 1 \gev and fit each of them for
extension.  The cumulative density of \tsext for these AGN is compared
to the $\chi^2_1/2$ distribution of \eqnref{ts_ext_distribution}
in \figref{agn_ts_ext}.  The \tsext distribution for the AGN shows
reasonable agreement with the theoretical distribution and no AGN
was found to be significantly extended ($\tsext>16$).  The observed
discrepancy from the theoretical distribution is likely due to small
systematics in our model of the LAT PSF and the Galactic diffuse emission
(see \secref{systematic_errors_on_extension}).  The discrepancy could
also in a few cases be due to confusion with a nearby undetected source.
We note that the Monte Carlo study of \subsecref{monte_carlo_validation}
effectively used perfect IRFs and a perfect model of the sky.  The overall
agreement with the expected distribution demonstrates that we can use
\tsext as a measure of the statistical significance of the detection of
the extension of a source.

We note that the LAT PSF used in this study was determined empirically
by fitting the distributions of gamma rays around bright AGN (see
\secref{systematic_errors_on_extension}). Finding that the AGN we test
are not extended is not surprising.  This validation analysis is not
suitable to reject any hypotheses about the existence of megaparsec-scale
halos around AGN.

\chapter{Search for Spatially-extended \Acstitle{LAT} Sources}
\chaplabel{extended_search}

\paperref{
  This chapter is based the second part of the paper
  ``Search for Spatially Extended Fermi-LAT Sources Using Two Years of Data''
  \citep{lande_2012_search-spatially}.}

In \chapref{extended_analysis}, we developed a new method to study
spatially-extended sources.  In this chapter, we apply this method to
search for new spatially-extended sources.  In \secref{validate_known},
we systematically reanalyze the twelve extended sources included in
the \ac{2FGL} catalog and in \secref{systematic_errors_on_extension}
we describe a method to estimate a systematic error on the spatial
extension of a source.  In \secref{extended_source_search_method}, we
describe a search for new spatially extended \ac{LAT} sources. Finally,
in \secref{new_ext_srcs_section} we present the detection of the extension
of nine spatially extended sources that were reported in the \ac{2FGL}
catalog but were treated as being point-like in the previous analysis.
Two of these sources have been previously analyzed in dedicated
publications.

\section{Analysis of Extended Sources Identified in the 2FGL Catalog}
\seclabel{validate_known}

As further validation of our method for studying spatially extended
sources, we reanalyzed the twelve spatially extended sources which were
included in the 2FGL catalog \citep{nolan_2012_fermi-large}.  Even though
these sources had all been the subjects of dedicated analyses and separate
publications, and had been fit with a variety of spatial models, it is
valuable to show that these sources are significantly extended using our
systematic method assuming radially-symmetric uniform disk spatial models.
On the other hand, for some of these sources a uniform disk spatial
model does not well describe the observed extended emission and so the
dedicated publications by the LAT collaboration provide better models
of these sources.

Six extended SNRs were included in the
2FGL catalog: W51C, IC~443, W28, W44, the Cygnus Loop, and W30
\citep{abdo_2009a_fermi-discovery,abdo_2010a_observation-supernova,abdo_2010d_fermi-large,abdo_2010a_gamma-ray-emission,katagiri_2011a_fermi-large,ajello_2012a_fermi-large}.
Using photons with energies between 1 \gev and 100 \gev, our analysis
significantly detected that these six SNRs are spatially extended.

Two nearby satellite galaxies of the Milky Way the Large
Magellanic Cloud (LMC) and the Small Magellanic Cloud (SMC)
were included in the 2FGL catalog as spatially extended sources
\citep{abdo_2010a_observations-large,abdo_2010a_detection-small}.
Their extensions were significantly detected using photons with energies
between 1 \gev and 100 \gev. Our fit extensions are comparable to
the published result, but we note that the previous LAT Collaboration
publication on the LMC used a more complicated two 2D Gaussian surface
brightness profile when fitting it \citep{abdo_2010a_observations-large}.

Three PWNe, MSH\,15$-$52, Vela X, and HESS\,J1825$-$137,
were fit as extended sources in the 2FGL analysis
\citep{abdo_2010a_detection-energetic,abdo_2010c_fermi-large,grondin_2011a_detection-pulsar}.
In the present analysis, HESS\,J1825$-$137 was significantly detected
using photons with energies between 10 \gev and 100 \gev.  To avoid
confusion with the nearby bright pulsar PSR\,J1509$-$5850, MSH\,15$-$52
had to be analyzed at high energies.  Using photons with energies above
10 \gev, we fit the extension of MSH\,15$-$52 to be consistent with the
published size but with \tsext=6.6.

Our analysis was unable to resolve Vela X which would have required first
removing the pulsed photons from the Vela pulsar which was beyond the
scope of this paper.  Our analysis also failed to detect a significant
extension for the Centaurus A Lobes because the shape of the source
is significantly different from a uniform radially-symmetric disk
\citep{abdo_2010a_fermi-gamma-ray}.

\begin{deluxetable}{lrrrrrrrr}
\tablecolumns{9}
\rotate
\tabletypesize{\footnotesize}
\tablewidth{0pt}
\tablecaption{Analysis of the twelve extended sources included in the 2FGL catalog
\tablabel{known_extended_sources}
}
\tablehead{
\colhead{Name}&
\colhead{\glon}&
\colhead{\glat}&
\colhead{$\sigma$}&
\colhead{\ts}&
\colhead{\tsext}&
\colhead{Pos Err}&
\colhead{Flux\tablenotemark{(a)}}&
\colhead{Index}\\
\colhead{}&
\colhead{(deg.)}&
\colhead{(deg.)}&
\colhead{(deg.)}&
\colhead{}&
\colhead{}&
\colhead{(deg.)}&
\colhead{}&
\colhead{}
}

\startdata
\multicolumn{9}{c}{E$>$1 \gev} \\[3pt]
\hline
SMC                                  &     302.59 &   $-$44.42 & $  1.32 \pm   0.15 \pm   0.31 $ &       95.0 &       52.9 &   0.14 & $    2.7 \pm     0.3$ & $   2.48 \pm    0.19$  \\
LMC                                  &     279.26 &   $-$32.31 & $  1.37 \pm   0.04 \pm   0.11 $ &     1127.9 &      909.9 &   0.04 & $   13.6 \pm     0.6$ & $   2.43 \pm    0.06$  \\
IC~443                               &     189.05 &       3.04 & $  0.35 \pm   0.01 \pm   0.04 $ &    10692.9 &      554.4 &   0.01 & $   62.4 \pm     1.1$ & $   2.22 \pm    0.02$  \\
Vela X                               &     263.34 &    $-$3.11 & $                         0.88$ &            &            &        &                       &                        \\
Centaurus A                          &     309.52 &      19.42 &                        $\sim10$ &            &            &        &                       &                        \\
W28                                  &       6.50 &    $-$0.27 & $  0.42 \pm   0.02 \pm   0.05 $ &     1330.8 &      163.8 &   0.01 & $   56.5 \pm     1.8$ & $   2.60 \pm    0.03$  \\
W30                                  &       8.61 &    $-$0.20 & $  0.34 \pm   0.02 \pm   0.02 $ &      464.8 &       76.0 &   0.02 & $   29.1 \pm     1.5$ & $   2.56 \pm    0.05$  \\
W44                                  &      34.69 &    $-$0.39 & $  0.35 \pm   0.02 \pm   0.02 $ &     1917.0 &      224.8 &   0.01 & $   71.2 \pm     0.5$ & $   2.66 \pm    0.00$  \\
W51C                                 &      49.12 &    $-$0.45 & $  0.27 \pm   0.02 \pm   0.04 $ &     1823.4 &      118.9 &   0.01 & $   37.2 \pm     1.3$ & $   2.34 \pm    0.03$  \\
Cygnus Loop                          &      74.21 &    $-$8.48 & $  1.71 \pm   0.05 \pm   0.06 $ &      357.9 &      246.0 &   0.06 & $   11.4 \pm     0.7$ & $   2.50 \pm    0.10$  \\
\hline\\[-8pt]
\multicolumn{9}{c}{E$>$10 \gev} \\[3pt]
\hline
MSH\,15$-$52\tablenotemark{(b)}      &     320.39 &    $-$1.22 & $  0.21 \pm   0.04 \pm   0.04 $ &       76.3 &        6.6 &   0.03 & $    0.6 \pm     0.1$ & $   2.20 \pm    0.22$  \\
HESS\,J1825$-$137\tablenotemark{(b)} &      17.56 &    $-$0.47 & $  0.65 \pm   0.04 \pm   0.02 $ &       59.7 &       33.8 &   0.05 & $    1.6 \pm     0.2$ & $   1.63 \pm    0.22$  \\
\enddata

\tablenotetext{(a)}{
Integral Flux in units of $10^{-9}$ \fluxunits and integrated in the fit
energy range (either 1 \gev to 100 \gev or 10 \gev to 100 \gev).
}
\tablenotetext{(b)}{
The discrepancy in the best fit spectra of MSH\,15$-$52 and HESS\,J1825$-$137
compared to \cite{abdo_2010a_detection-energetic} and
\cite{grondin_2011a_detection-pulsar} is due to fitting over a different energy range.
}

\tablecomments{
All sources were fit using a radially-symmetric uniform disk spatial model.
\glon and \glat are Galactic longitude
and latitude of the best fit extended source respectively.  The first
error on $\sigma$ is statistical and the second is systematic (see
\secref{systematic_errors_on_extension}).  
The errors on the integral fluxes and the spectral indices
are statistical only.
Pos Err is the error on
the position of the source.  Vela X and the Centaurus A Lobes were
not fit in our analysis but are included for completeness.
}
\end{deluxetable}

Our analysis of these sources is summarized in
\tabref{known_extended_sources}.  This table includes the best
fit positions and extensions of these sources when fitting them
with a radially-symmetric uniform disk model. It also includes
the best fit spectral parameters for each source.  The positions
and extensions of Vela X and the Centaurus A Lobes were taken from
\cite{abdo_2010c_fermi-large,abdo_2010a_fermi-gamma-ray} and are included
in this table for completeness.

\section{Systematic Errors on Extension}
\seclabel{systematic_errors_on_extension}

We developed two criteria for estimating systematic errors on the
extensions of the sources.  First, we estimated a systematic error due
to uncertainty in our knowledge of the LAT PSF.  Before launch, the
LAT PSF was determined by detector simulations which were verified in
accelerator beam tests \citep{atwood_2009a_large-telescope}. However,
in-flight data revealed a discrepancy above 3 \gev in the PSF
compared to the angular distribution of photons from bright AGN
\citep{ackermann_2012a_fermi-large}.  Subsequently, the PSF was
fit empirically to bright AGN and this empirical parameterization
is used in the P7\_V6 IRFs.  To account for the uncertainty in our
knowledge of the PSF, we refit our extended source candidates using
the pre-flight Monte Carlo representation of the PSF and consider
the difference in extension found using the two PSFs as a systematic
error on the extension of a source.  The same approach was used
in \cite{abdo_2010a_observation-supernova}.  We believe that our
parameterization of the PSF from bright AGN is substantially better
than the Monte Carlo representation of the PSF so this systematic error
is conservative.

We estimated a second systematic error on the extension of a source
due to uncertainty in our model of the Galactic diffuse emission by
using an alternative approach to modeling the diffuse emission which
takes as input templates calculated by GALPROP\footnote{GALPROP is a
software package for calculating the Galactic $\gamma$-ray emission
based on a model of cosmic-ray propagation in the Galaxy and maps
of the distributions of the components of the interstellar medium
\citep{strong_1998a_propagation-cosmic-ray,vladimirov_2011a_galprop-webrun:}.
See also \url{http://galprop.stanford.edu/} for details.} but then fits
each template locally in the surrounding region.  The particular GALPROP
model that was used as input is described in the analysis of the isotropic
diffuse emission with LAT data \citep{abdo_2010a_spectrum-isotropic}.
The intensities of various components of the Galactic diffuse emission
were fitted individually using a spatial distribution predicted by the
model.  We considered separate contributions from cosmic-ray interactions
with the molecular hydrogen, the atomic and ionized hydrogen, residual
gas traced by dust \citep{grenier_2005a_unveiling-extensive}, and
the interstellar radiation field. We further split the contributions
from interactions with molecular and atomic hydrogen to the Galactic
diffuse emission according to the distance from the Galactic center
in which they are produced. Hence, we replaced the standard diffuse
emission model by 18 individually fitted templates to describe
individual components of the diffuse emission.  A similar crosscheck
was used in an analysis of RX\,J1713.7$-$3946 by the LAT Collaboration
\citep{abdo_2011a_observations-young}.

It is not expected that this diffuse model is superior to the standard
LAT model obtained through an all-sky fit.  However, adding degrees of
freedom to the background model can remove likely spurious sources that
correlate with features in the Galactic diffuse emission.  Therefore,
this tests systematics that may be due to imperfect modeling of the
diffuse emission in the region.  Nevertheless, this alternative approach
to modeling the diffuse emission does not test all systematics related
to the diffuse emission model. In particular, because the alternative
approach uses the same underlying gas maps, it is unable to be used to
assess systematics due to insufficient resolution of the underlying
maps. Structure in the diffuse emission that is not correlated with
these maps will also not be assessed by this test.

We do not expect the systematic error due to uncertainties in the PSF
to be correlated with the systematic error due to uncertainty in the
Galactic diffuse emission. Therefore, the total systematic error on the
extension of a source was obtained by adding the two errors in quadrature.

There is another systematic error on the size of a source due to issues
modeling nearby sources in crowded regions of the sky. It is beyond
the scope of this paper to address this systematic error. Therefore,
for sources in crowded regions the systematic errors quoted in this
paper may not represent the full set of systematic errors associated
with this analysis.

\section{Extended Source Search Method}
\seclabel{extended_source_search_method}

Having demonstrated that we understand the statistical
issues associated with analyzing spatially extended sources
(\subsecref{monte_carlo_validation} and \secref{test_2lac_sources}) and
that our method can correctly analyze the extended sources included in
2FGL (\secref{validate_known}), we applied this method to search for new
spatially extended \gev sources.  The data and general analysis setting
is as described in \secref{test_2lac_sources}.

Ideally, we would apply a completely blind and uniform search that
tests the extension of each 2FGL source in the presence of all other
2FGL sources to find a complete list of all spatially extended sources.
As our test of AGN in \secref{test_2lac_sources} showed, at high Galactic
latitude where the source density is not as large and the diffuse emission
is less structured, this method works well.

But this is infeasible in the Galactic plane where we are most likely to
discover new spatially extended sources.  In the Galactic plane, this
analysis is challenged by our imperfect model of the diffuse emission
and by an imperfect model of nearby sources.  The Monte Carlo study in
\secref{dual_localization_method} showed that the overall likelihood
would greatly increase by fitting a spatially extended source as two
point-like sources so we expect that spatially extended sources would be
modeled in the 2FGL catalog as multiple point-like sources. Furthermore,
the positions of other nearby sources in the region close to an extended
source could be biased by not correctly modeling the extension of
the source.  The 2FGL catalog contains a list of sources significant at
energies above 100 \mev whereas we are most sensitive to spatial extension
at higher energies.  We therefore expect that at higher energies our
analysis would be complicated by 2FGL sources no longer significant
and by 2FGL sources whose positions were biased by diffuse emission at
lower energies.

To account for these issues, we first produced a large list of possibly
extended sources employing very liberal search criteria and then refined
the analysis of the promising candidates on a case by case basis.  Our
strategy was to test all point-like 2FGL sources for extension assuming
they had a uniform radially-symmetric disk spatial model and a power-law
spectral model.  Although not all extended sources are expected to have
a shape very similar to a uniform disk, \subsecref{compare_source_size}
showed that for many spatially extended sources the wide PSF of the LAT
and limited statistics makes this a reasonable approximation.  On the
other hand, choosing this spatial model biases us against finding
extended sources that are not well described by a uniform disk model
such as shell-type SNRs.

Before testing for extension, we automatically removed from the
background model all other 2FGL sources within 0\fdg5 of the source.
This distance is somewhat arbitrary, but was picked in hopes of finding
extended sources with sizes on the order of $\sim1\degree$ or smaller. On
the other hand, by removing these nearby background sources we expect to
also incorrectly add to our list of extended source candidates point-like
sources that are confused with nearby sources.  To screen out obvious
cases of source confusion, we performed the dual localization procedure
described in \secref{dual_localization_method} to compare the extended
source hypothesis to the hypothesis of two independent point-like sources.

As was shown in \secref{extension_sensitivity}, little sensitivity is
gained by using photons with energies below 1 \gev. In addition, the
broad PSF at low energy makes the analysis more susceptible to systematic
errors arising from source confusion due to nearby soft point-like sources
and by uncertainties in our modeling of the Galactic diffuse emission.
For these reasons, we performed our search using only photons with
energies between 1 \gev and 100 \gev.

We also performed a second search for extended sources using only
photons with energies between 10 \gev and 100 \gev.  Although this
approach tests the same sources, it is complementary because the Galactic
diffuse emission is even less dominant above 10 \gev and because source
confusion is less of an issue.  A similar procedure was used to detect
the spatial extensions of MSH\,15$-$52 and HESS\,J1825$-$137 with the
LAT \citep{abdo_2010a_detection-energetic,grondin_2011a_detection-pulsar}.

When we applied this test to the 1861 point-like sources in the 2FGL
catalog, our search found 117 extended source candidates in the 1 \gev
to 100 \gev energy range and 11 extended source candidates in the 10
\gev to 100 \gev energy range. Most of the extended sources found above
10 \gev were also found above 1 \gev and in many cases multiple nearby
point-like sources were found to be extended even though they fit the
same emission region.  For example, the sources 2FGL\,J1630.2$-$4752,
2FGL\,J1632.4$-$4753c 2FGL\,J1634.4$-$4743c, and 2FGL\,J1636.3$-$4740c
were all found to be spatially extended in the 10 \gev to 100 \gev energy
range even though they all fit to similar positions and sizes.  For these
situations, we manually discarded all but one of the 2FGL sources.

\begin{figure}[htbp]
  \includegraphics{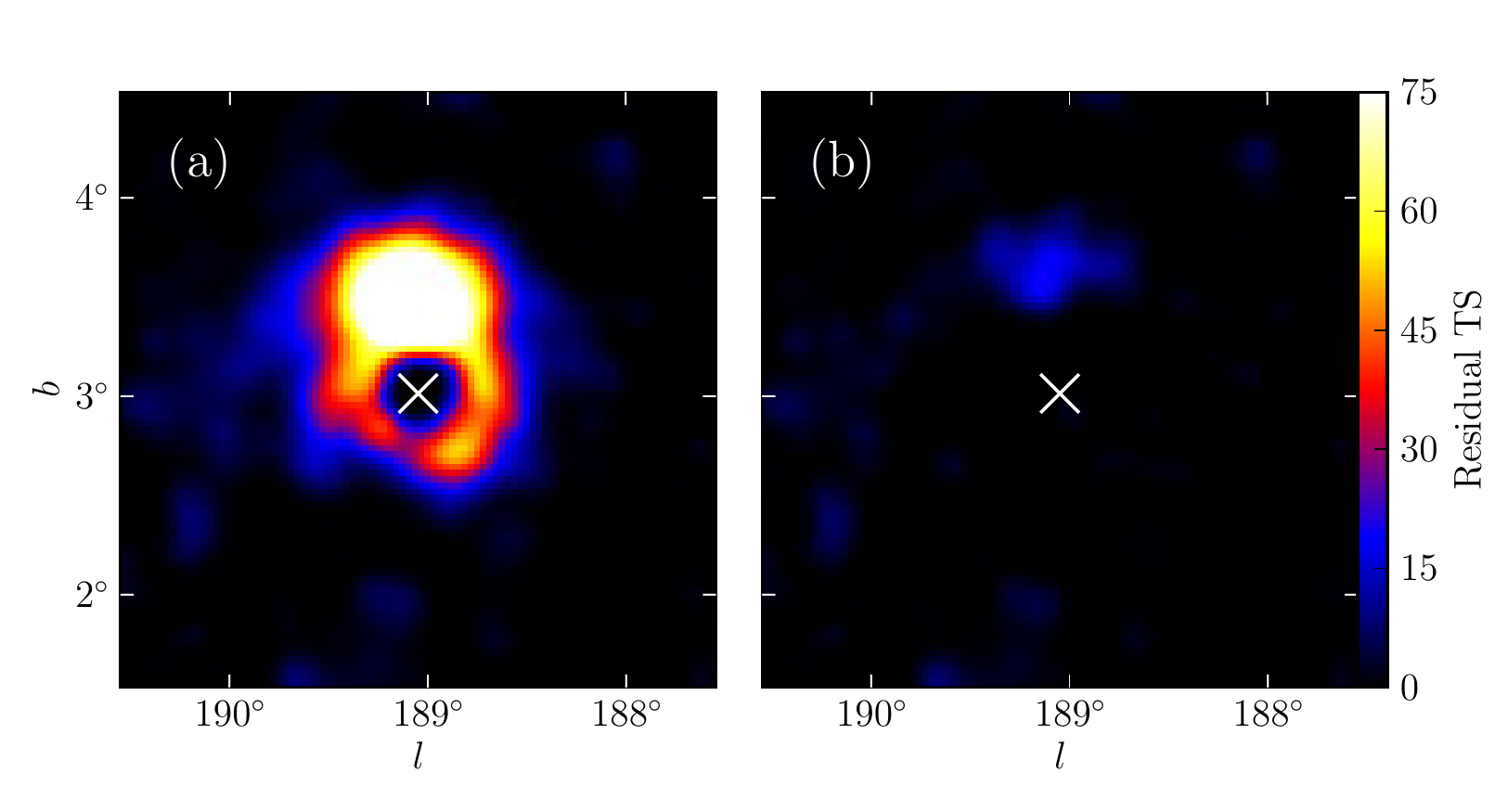}
  \caption{A TS map generated for the region around the SNR IC~443 using
  photons with energies between 1 \gev and 100 \gev.  (a) TS map after
  subtracting IC~443 modeled as a point-like source. (b) same as (a),
  but IC~443 modeled as an extended source. The cross represents the
  best fit position of IC~443.}
  \figlabel{res_tsmaps}
\end{figure}

Similarly, many of these sources were confused with nearby point-like
sources or influenced by large-scale residuals in the diffuse emission.
To help determine which of these fits found truly extended sources
and when the extension was influenced by source confusion and diffuse
emission, we generated a series of diagnostic plots.  For each candidate,
we generated a map of the residual TS by adding a new point-like source
of spectral index 2 into the region at each position and finding the
increase in likelihood when fitting its flux. \figref{res_tsmaps} shows
this map around the most significantly extended source IC~443 when
it is modeled both as a point-like source and as an extended source.
The residual TS map indicates that the spatially extended model for
IC~443 is a significantly better description of the observed photons
and that there is no $\ts>25$ residual in the region after modeling the
source as being spatially extended.  We also generated plots of the
sum of all counts within a given distance of the source and compared
them to the model predictions assuming the emission originated from a
point-like source.  An example radial integral plot is shown for the
extended source IC~443 in \figref{four_plots_ic443}.  For each source,
we also made diffuse-emission-subtracted smoothed counts maps (shown
for IC~443 in \figref{four_plots_ic443}).

\begin{figure}[htbp]
    \includegraphics{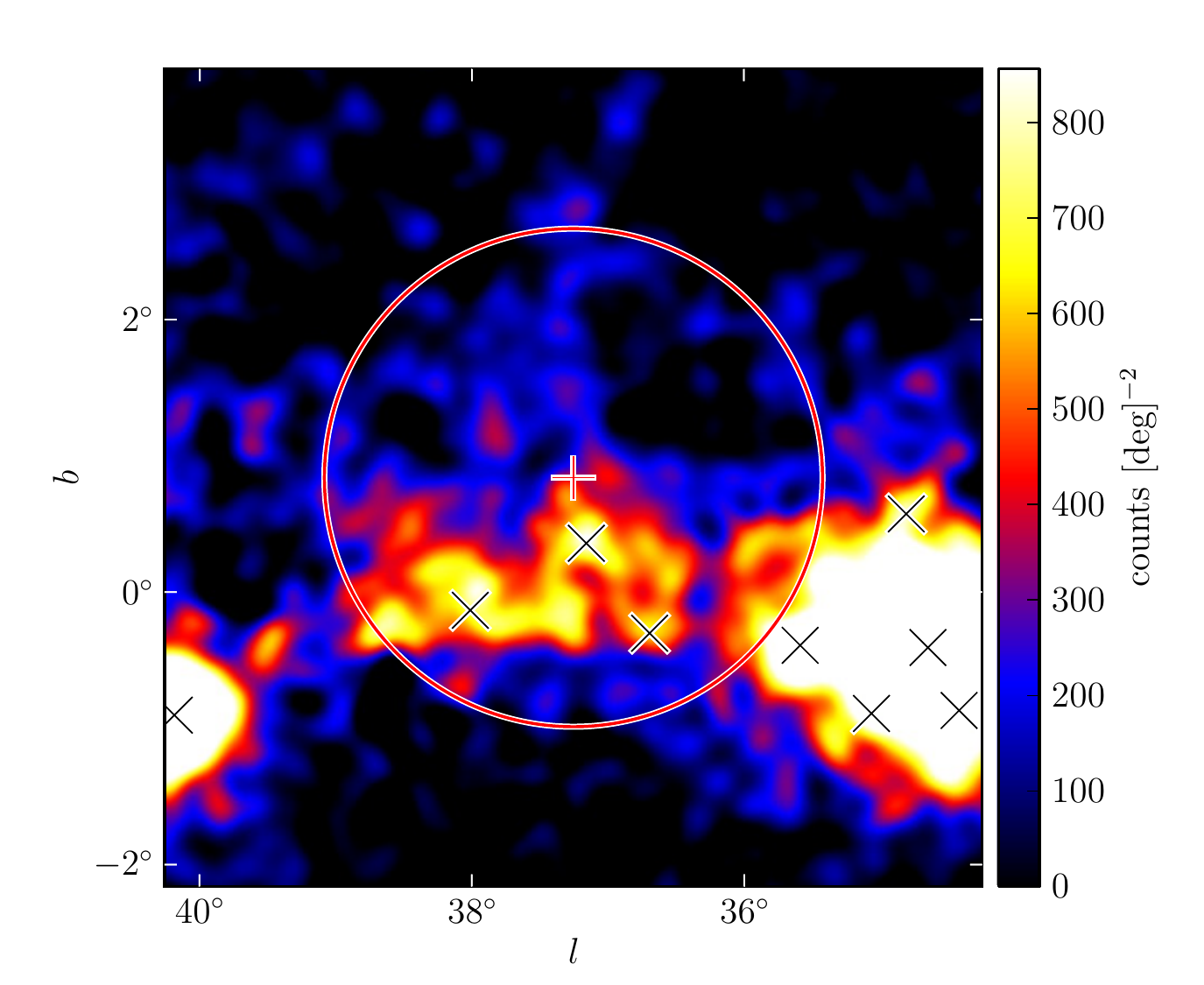}
    \caption{A diffuse-emission-subtracted 1 \gev to 100 \gev counts
    map of the region around 2FGL\,J1856.2+0450c smoothed by a 0\fdg1
    2D Gaussian kernel. The plus-shaped marker and circle (colored
    red in the online version) represent the center and size of the
    source fit with a radially-symmetric uniform disk spatial model.
    The black crosses represent the positions of other 2FGL sources.
    The extension is statistically significant, but the extension
    encompasses many 2FGL sources and the emission does not look to be
    uniform. Although the fit is statistically significant, it likely
    corresponds to residual features of inaccurately modeled diffuse
    emission picked up by the fit.}
    \figlabel{example_bad_fit}
\end{figure}

We found by visual inspection that in many cases our results were strongly
influenced by large-scale residuals in the diffuse emission and hence
the extension measure was unreliable.  This was especially true in our
analysis of sources in the 1 \gev to 100 \gev energy range.  An example
of such a case is 2FGL\,J1856.2+0450c analyzed in the 1 \gev to 100 \gev
energy range. \figref{example_bad_fit} shows a diffuse-emission-subtracted
smoothed counts map for this source with the best fit extension of
the source overlaid. There appear to be large-scale residuals in
the diffuse emission in this region along the Galactic plane.  As a
result, 2FGL\,J1856.2+0450c is fit to an extension of $\sim2\degree$
and the result is statistically significant with \tsext=45.4. However,
by looking at the residuals it is clear that this complicated region is
not well modeled. We manually discard sources like this.

We only selected extended source candidates in regions that did not
appear dominated by these issues and where there was a multiwavelength
counterpart. Because of these systematic issues, this search can not be
expected to be complete and it is likely that there are other spatially
extended sources that this method missed.

\begin{deluxetable}{lr}
\tablecolumns{2}
\tablewidth{0pt}
\tabletypesize{\small}
\tablecaption{Nearby Residual-induced Sources
\tablabel{fake_2fgl_sources}
}
\tablehead{
\colhead{Extended Source}&
\colhead{Residual-induced Sources}
}
\startdata
2FGL\,J0823.0$-$4246    & 2FGL\,J0821.0$-$4254, 2FGL\,J0823.4$-$4305 \\
2FGL\,J1627.0$-$2425c   & \nodata \\
2FGL\,J0851.7$-$4635    & 2FGL\,J0848.5$-$4535, 2FGL\,J0853.5$-$4711, 2FGL\,J0855.4$-$4625 \\
2FGL\,J1615.0$-$5051    & \nodata \\
2FGL\,J1615.2$-$5138    & 2FGL\,J1614.9$-$5212  \\
2FGL\,J1632.4$-$4753c   & 2FGL\,J1634.4$-$4743c \\
2FGL\,J1712.4$-$3941    & \nodata \\
2FGL\,J1837.3$-$0700c   & 2FGL\,J1835.5$-$0649 \\
2FGL\,J2021.5+4026      & 2FGL\,J2019.1+4040 \\
\enddata

\tablecomments{
For each new extended source, we list nearby 2FGL
soruces that we have concluded here correspond to residuals induced by not
modeling the extensions of nearby extended sources.
}
\end{deluxetable}

For each candidate that was not biased by neighboring point-like
sources or by large-scale residuals in the diffuse emission model, we
improved the model of the region by deciding on a case by case basis
which background point-like sources should be kept.  We kept in our
model the sources that we believed represented physically distinct
sources and we removed sources that we believed were included in the
2FGL catalog to compensate for residuals induced by not modeling the
extension of the source.  Soft nearby point-like 2FGL sources that
were not significant at higher energies were frozen to the spectras
predicted by 2FGL.  When deciding which background sources to keep and
which to remove, we used multiwavelength information about possibly
extended source counterparts to help guide our choice. For each extended
source presented in \secref{new_ext_srcs_section}, we describe any
modifications from 2FGL of the background model that were performed.
In \tabref{fake_2fgl_sources}, we summarize the sources in the 2FGL
catalog that we have concluded here correspond to residuals induced by
not modeling the extensions of nearby extended sources.

The best fit positions of nearby point-like sources can be influenced by
the extended source and vice versa.  Similarly, the best fit positions
of nearby point-like sources in the 2FGL catalog can be biased by
systematic issues at lower energies.  Therefore, after selecting the list
of background sources, we iteratively refit the positions and spectra
of nearby background sources as well as the positions and extensions of
the analyzed spatially extended sources until the overall fit converged
globally.  For each extended source, we will describe the positions of
any relocalized background sources.

After obtaining the overall best fit positions and extensions of all
of the sources in the region using \pointlike, we refit the spectral
parameters of the region using \gtlike.  With \gtlike, we obtained a
second measure of \tsext.  We only consider a source to be extended
when both \pointlike and \gtlike agree that $\tsext\ge16$.  We further
required that $\tsext\ge16$ using the alternative approach to modeling
the diffuse emission presented in \secref{systematic_errors_on_extension}.
We then replaced the spatially extended source with two point-like sources
and refit the positions and spectra of the two point-like sources to
calculate \tsinc.  We only consider a source to be spatially extended,
instead of being the result of confusion of two point-like sources,
if $\tsext>\tsinc$.  As was shown in \secref{dual_localization_method},
this test is fairly powerful at removing situations in which the emission
actually originates from two distinct point-like sources instead of one
spatially extended source.  On the other hand, it is still possible that
longer observations could resolve additional structure or new sources that
the analysis cannot currently detect.  Considering the very complicated
morphologies of extended sources observed at other wavelengths and
the high density of possible sources that are expected to emit at \gev
energies, it is likely that in some of these regions further observations
will reveal that the emission is significantly more complicated than
the simple radially-symmetric uniform disk model that we assume.

\section{New Extended Sources}
\seclabel{new_ext_srcs_section}

\thispagestyle{empty}
\begin{deluxetable}{lrrrrrrrrl}
\tablecolumns{10}
\tabletypesize{\footnotesize}
\rotate
\tablewidth{0pt}
\tablecaption{Extension fit for the nine additional extended sources
\tablabel{new_ext_srcs_table}
}
\tablehead{
\colhead{Name}&
\colhead{\glon}&
\colhead{\glat}&
\colhead{$\sigma$}&
\colhead{\ts}&
\colhead{\tsext}&
\colhead{Pos Err}&
\colhead{Flux\tablenotemark{(a)}}&
\colhead{Index}&
\colhead{Counterpart}\\
\colhead{}&
\colhead{(deg.)}&
\colhead{(deg.)}&
\colhead{(deg.)}&
\colhead{}&
\colhead{}&
\colhead{(deg.)}&
\colhead{}&
\colhead{}&
\colhead{}
}

\startdata
\multicolumn{10}{c}{E$>$1 \gev} \\[3pt]
\hline
2FGL\,J0823.0$-$4246                       &     260.32 &    $-$3.28 & $  0.37 \pm   0.03 \pm   0.02 $ &      322.2 &       48.0 &   0.02 & $    8.4 \pm     0.6$ & $   2.21 \pm    0.09$ &                  Puppis A \\
2FGL\,J1627.0$-$2425c                      &     353.07 &      16.80 & $  0.42 \pm   0.05 \pm   0.16 $ &      139.9 &       32.4 &   0.04 & $    6.3 \pm     0.6$ & $   2.50 \pm    0.14$ &                 Ophiuchus \\
\hline\\[-8pt]
\multicolumn{10}{c}{E$>$10 \gev} \\[3pt]
\hline
2FGL\,J0851.7$-$4635                       &     266.31 &    $-$1.43 & $  1.15 \pm   0.08 \pm   0.02 $ &      116.6 &       86.8 &   0.07 & $    1.3 \pm     0.2$ & $   1.74 \pm    0.21$ &                  Vela Jr. \\
2FGL\,J1615.0$-$5051                       &     332.37 &    $-$0.13 & $  0.32 \pm   0.04 \pm   0.01 $ &       50.4 &       16.7 &   0.04 & $    1.0 \pm     0.2$ & $   2.19 \pm    0.28$ &         HESS\,J1616$-$508 \\
2FGL\,J1615.2$-$5138                       &     331.66 &    $-$0.66 & $  0.42 \pm   0.04 \pm   0.02 $ &       76.1 &       46.5 &   0.04 & $    1.1 \pm     0.2$ & $   1.79 \pm    0.26$ &         HESS\,J1614$-$518 \\
2FGL\,J1632.4$-$4753c                      &     336.52 &       0.12 & $  0.35 \pm   0.04 \pm   0.02 $ &       64.4 &       26.9 &   0.04 & $    1.4 \pm     0.2$ & $   2.66 \pm    0.30$ &         HESS\,J1632$-$478 \\
2FGL\,J1712.4$-$3941\tablenotemark{(b)}    &     347.26 &    $-$0.53 & $  0.56 \pm   0.04 \pm   0.02 $ &       59.4 &       38.5 &   0.05 & $    1.2 \pm     0.2$ & $   1.87 \pm    0.22$ &        RX\,J1713.7$-$3946 \\
2FGL\,J1837.3$-$0700c                      &      25.08 &       0.13 & $  0.33 \pm   0.07 \pm   0.05 $ &       47.0 &       18.5 &   0.07 & $    1.0 \pm     0.2$ & $   1.65 \pm    0.29$ &         HESS\,J1837$-$069 \\
2FGL\,J2021.5+4026                         &      78.24 &       2.20 & $  0.63 \pm   0.05 \pm   0.04 $ &      237.2 &      128.9 &   0.05 & $    2.0 \pm     0.2$ & $   2.42 \pm    0.19$ &            $\gamma$-Cygni \\
\enddata

\tablenotetext{(a)}{
Integral Flux in units of $10^{-9}$ \fluxunits and integrated in the fit
energy range (either 1 \gev to 100 \gev or 10 \gev to 100 \gev).
}

\tablenotetext{(b)}{
The discrepancy in the best fit spectra of 2FGL\,J1712.4$-$3941 compared
to \cite{abdo_2011a_observations-young} is due to fitting over a different energy range.
}

\tablecomments{
    The columns in this table have the same meaning as those in
    \tabref{known_extended_sources}.
    RX\,J1713.7$-$3946 and Vela Jr. were previously studied in dedicated publications \citep{abdo_2011a_observations-young,tanaka_2011a_gamma-ray-observations}.
}
\end{deluxetable}

\thispagestyle{empty}
\begin{deluxetable}{lrrrrrrrrrr}
  \tabletypesize{\scriptsize}
  \tablecolumns{11}
  \rotate
  \tablewidth{0pt}
  \tablecaption{Dual localization, alternative PSF, and alternative approach to modeling the diffuse emission
  \tablabel{alt_diff_model_results}
  }
  \tablehead{
  \colhead{Name}&
  \colhead{$\tspointlike$}&
  \colhead{$\tsgtlike$}&
  \colhead{$\tsaltdiff$}&
  \colhead{$\tsextpointlike$}&
  \colhead{$\tsextgtlike$}&
  \colhead{$\tsextaltdiff$}&
  \colhead{$\sigma$}&
  \colhead{$\sigma_\altdiff$}&
  \colhead{$\sigma_\altpsf$}&
  \colhead{$\tsinc$}\\
  \colhead{}&
  \colhead{}&
  \colhead{}&
  \colhead{}&
  \colhead{}&
  \colhead{}&
  \colhead{}&
  \colhead{(deg.)}&
  \colhead{(deg.)}&
  \colhead{(deg.)}&
  \colhead{}
  }
  \startdata
\multicolumn{11}{c}{E$>$1 \gev} \\[3pt]
\hline
2FGL\,J0823.0$-$4246      &                331.9 &                322.2 &                356.0 &                 60.0 &                 48.0 &                 56.0 &                 0.37 &                 0.39 &                 0.39 &                 23.0 \\
2FGL\,J1627.0$-$2425c     &                154.8 &                139.9 &                105.7 &                 39.4 &                 32.4 &                 24.8 &                 0.42 &                
 0.40 &                 0.58 &                 24.5 \\
\hline\\[-4pt]
\multicolumn{11}{c}{E$>$10 \gev} \\[3pt]
\hline
2FGL\,J0851.7$-$4635      &                115.2 &                116.6 &                123.1 &                 83.9 &                 86.8 &                 89.8 &                 1.15 &                 1.16 &                 1.17 &                 15.5 \\
2FGL\,J1615.0$-$5051\tablenotemark{(a)} &                 48.2 &                 50.4 &                 56.6 &                 15.2 &                 16.7 &                 17.8 &                 0.32 &                 0.33 &                 0.32 &                 13.1 \\
2FGL\,J1615.2$-$5138      &                 75.0 &                 76.1 &                 83.8 &                 42.9 &                 46.5 &                 54.1 &                 0.42 &                 0.43 &                 0.43 &                 35.1 \\
2FGL\,J1632.4$-$4753c     &                 64.5 &                 64.4 &                 66.8 &                 23.0 &                 26.9 &                 25.5 &                 0.35 &                 0.36 &                 0.37 &                 10.9 \\
2FGL\,J1712.4$-$3941      &                 59.8 &                 59.4 &                 39.9 &                 38.4 &                 38.5 &                 30.7 &                 0.56 &                 0.55 &                 0.53 &                  2.7 \\
2FGL\,J1837.3$-$0700c     &                 44.5 &                 47.0 &                 39.2 &                 17.6 &                 18.5 &                 16.1 &                 0.33 &                 0.32 &                 0.38 &                 10.8 \\
2FGL\,J2021.5+4026        &                239.1 &                237.2 &                255.8 &                139.1 &                128.9 &                138.0 &                 0.63 &                 0.65 &                 0.59 &                 37.3 \\
  \enddata

\tablenotetext{(a)}{
Using \pointlike, \tsext for 2FGL\,J1615.0$-$5051 was sligthly below
16 when the source was fit in the 10 \gev to 100 \gev energy range. To
confirm the extension measure, the extension was refit in \pointlike
using a slightly lower energy.  In the 5.6 \gev to 100 \gev energy range,
we obtained a consistent extension and \tsext=28.0.  In the rest of
this paper, we quote the $E>10 \gev$ results for consistency with the
other sources.
}

\tablecomments{
$\tspointlike$, $\tsgtlike$, and $\tsaltdiff$ are the test
statistic values from \pointlike, \gtlike, and \gtlike with the alternative
approach to modeling the diffuse emission respectively.  $\tsextpointlike$, $\tsextgtlike$,
and $\tsextaltdiff$ are the TS values from \pointlike, \gtlike, and \gtlike with the alternative 
approach to modeling the diffuse emission respectively.  $\sigma$, $\sigma_\altdiff$, and $\sigma_\altpsf$
are the fit sizes assuming
a radially-symmetric uniform disk model
with the standard analysis, the alternative 
approach to modeling the diffuse emission, and the alternative PSF respectively.  
}
\end{deluxetable}

Nine extended sources not included in the 2FGL catalog were found
by our extended source search. Two of these have been previously
studied in dedicated publications: RX\,J1713.7$-$3946 and Vela Jr.
\citep{abdo_2011a_observations-young,tanaka_2011a_gamma-ray-observations}.
Two of these sources were found when using photons with energies
between 1 \gev and 100 \gev and seven were found when using photons
with energies between 10 \gev and 100 \gev.  For the sources found at
energies above 10 \gev, we restrict our analysis to higher energies
because of the issues of source confusion and diffuse emission modeling
described in \secref{extended_source_search_method}.  The spectral
and spatial properties of these nine sources are summarized in
\tabref{new_ext_srcs_table} and the results of our investigation of
systematic errors are presented in \tabref{alt_diff_model_results}.
\tabref{alt_diff_model_results} also compares the likelihood assuming
the source is spatially extended to the likelihood assuming that the
emission originates from two independent point-like sources. For
these new extended sources, $\tsext>\tsinc$ so we conclude that
the \gev emission does not originate from two physically distinct
point-like sources (see \secref{dual_localization_method}).
\tabref{alt_diff_model_results} also includes the results of the
extension fits using variations of the PSF and the Galactic diffuse model
described in \secref{systematic_errors_on_extension}.  There is good
agreement between \tsext and the fit size using the standard analysis,
the alternative approach to modeling the diffuse emission, and the
alternative PSF.  This suggests that the sources are robust against
mis-modeled features in the diffuse emission model and uncertainties in
the PSF.

\subsection{2FGL\,J0823.0$-$4246}
\subseclabel{section_2FGL_J0823.0-4246}

\begin{figure}[htbp]
  \includegraphics{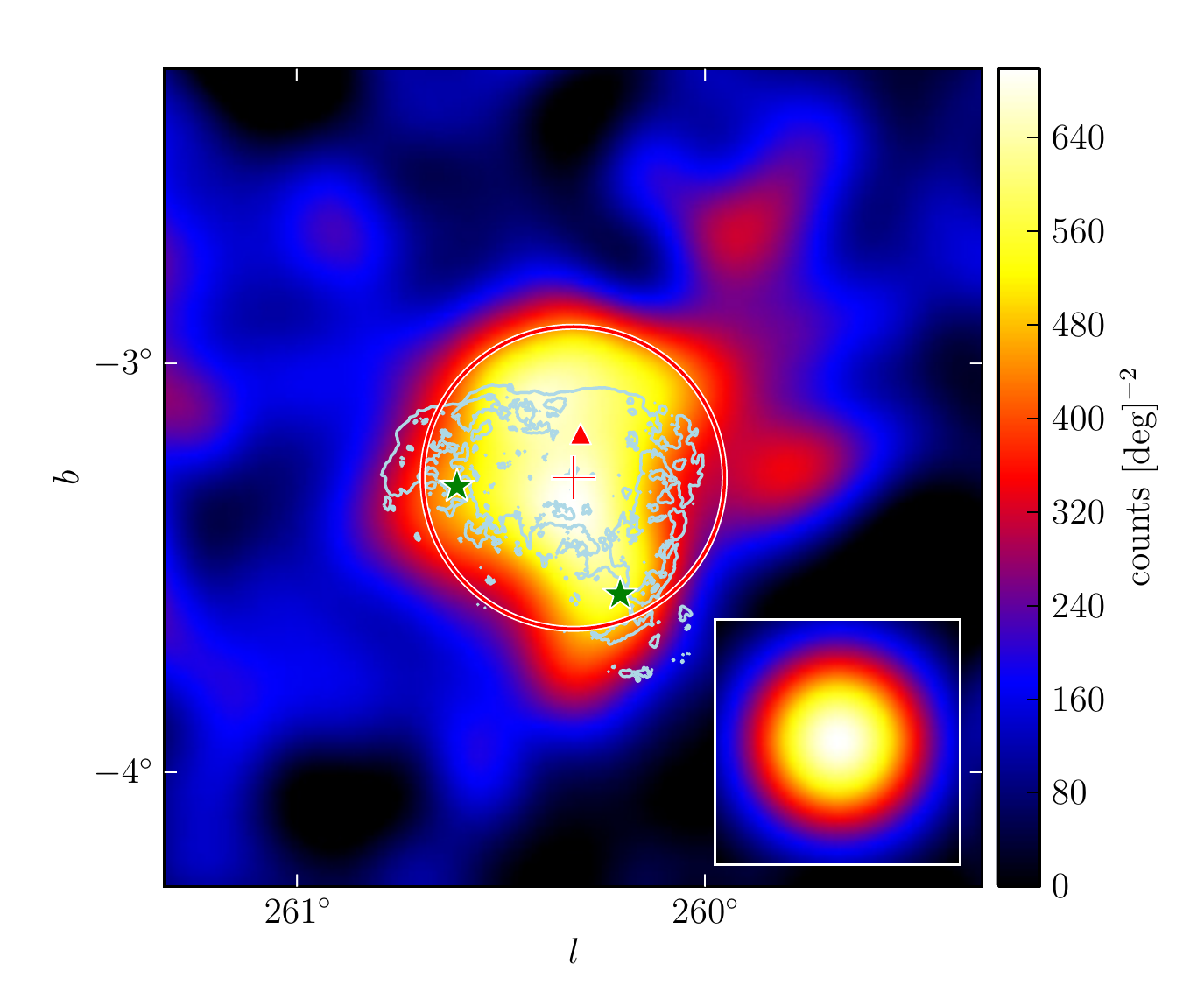}
  \caption{A diffuse-emission-subtracted 1 \gev to 100 \gev counts
  map of 2FGL\,J0823.0$-$4246 smoothed by a 0\fdg1 2D Gaussian kernel.
  The triangular marker (colored red in the online version) represents the
  2FGL position of this source.  The plus-shaped marker and the circle
  (colored red) represent the best fit position and extension of this
  source assuming a radially-symmetric uniform disk model.  The two
  star-shaped markers (colored green) represent 2FGL sources that were
  removed from the background model.  From left to right, these sources
  are 2FGL\,J0823.4$-$4305 and 2FGL\,J0821.0$-$4254.  The lower right
  inset is the model predicted emission from a point-like source with
  the same spectrum as 2FGL\,J0823.4$-$4305 smoothed by the same kernel.
  This source is spatially coincident with the Puppis A SNR. The light
  blue contours correspond to the X-ray image of Puppis A observed by
  \rosat \citep{petre_1996a_central-stellar}.}
  \figlabel{1FGL_J0823.3-4248}
\end{figure}

2FGL\,J0823.0$-$4246 was found by our search to be an extended source
candidate in the 1 \gev to 100 \gev energy range and is spatially
coincident with the SNR Puppis A.  \figref{1FGL_J0823.3-4248}
shows a counts map of this source. There are two nearby 2FGL sources
2FGL\,J0823.4$-$4305 and 2FGL\,J0821.0$-$4254 that are also coincident
with the SNR but that do not appear to represent physically distinct
sources.  We conclude that these nearby point-like sources were included
in the 2FGL catalog to compensate for residuals induced by not modeling
the extension of this source and removed them from our model of the sky.
After removing these sources, 2FGL\,J0823.0$-$4246 was found to have
an extension $\sigma=0\fdg37\pm0\fdg03_\stat\pm0\fdg02_\sys$ with
$\tsext=48.0$.  \figref{snr_seds} shows the spectrum of this source.

Puppis A has been studied in detail in radio
\citep{castelletti_2006a_observations-puppis}, and  X-ray
\citep{petre_1996a_central-stellar,hwang_2008a_x-ray-emitting-ejecta}.
The fit extension of 2FGL\,J0823.0$-$4246 matches well the size of
Puppis A in X-ray.  The distance of Puppis A was estimated at 2.2 kpc
\citep{reynoso_1995a_observations-neutral,reynoso_2003a_observations-neutral}
and leads to a 1 \gev to 100 \gev luminosity of $\sim 3\times 10^{34}$
ergs$\,\second^{-1}$.  No molecular clouds have been observed directly
adjacent to Puppis A \citep{paron_2008a_high-resolution-observations},
similar to the LAT-detected Cygnus Loop SNR
\citep{katagiri_2011a_fermi-large}.  The luminosity of Puppis A is also
smaller than that of other SNRs believed to interact with molecular clouds
\citep{abdo_2009a_fermi-discovery,abdo_2010a_observation-supernova,abdo_2010a_gamma-ray-emission,abdo_2010d_fermi-large,abdo_2010a_fermi-lat-study}.

\subsection{2FGL\,J0851.7$-$4635}
\subseclabel{section_2FGL_J0851.7-4635}

\begin{figure}[htbp]
  \includegraphics{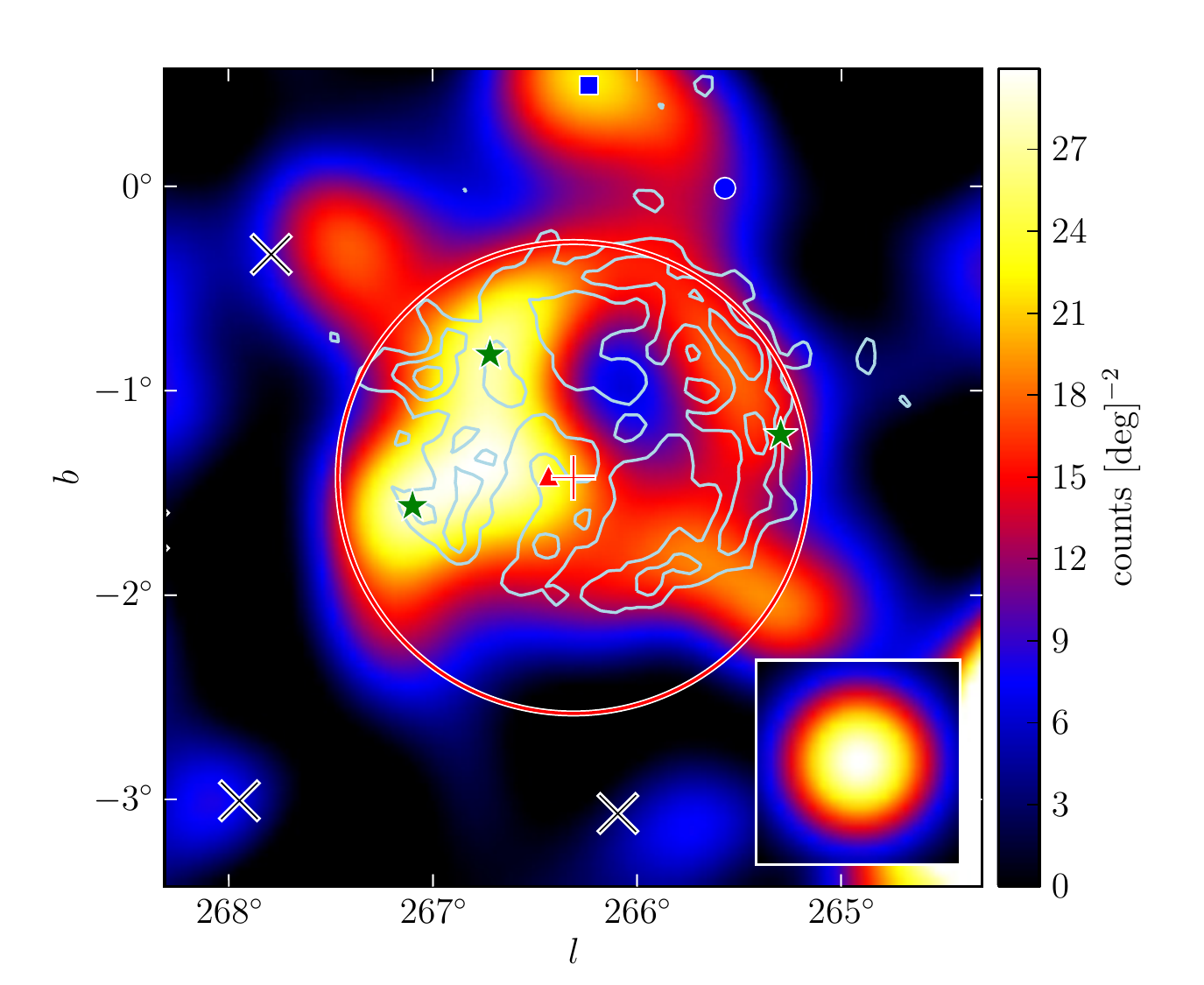}
  \caption{A diffuse-emission-subtracted 10 \gev to 100 \gev counts map
  of 2FGL\,J0851.7$-$4635 smoothed by a 0\fdg25 2D Gaussian kernel. The
  triangular marker (colored red in the electronic version) represents the
  2FGL position of this source.  The plus-shaped marker and the circle
  (colored red) are the best fit position and extension of this source
  assuming a radially-symmetric uniform disk model.  The three black
  crosses represent background 2FGL sources.  The three star-shaped
  markers (colored green) represent other 2FGL sources that were
  removed from the background model.  They are (from left to right)
  2FGL\,J0853.5$-$4711, 2FGL\,J0855.4$-$4625, and 2FGL\,J0848.5$-$4535.
  The circular and square-shaped marker (colored blue) represents the
  2FGL and relocalized position of another 2FGL source.  This extended
  source is spatially coincident with the Vela Jr. SNR.  The contours
  (colored light blue) correspond to the \tev image of Vela Jr.
  \citep{aharonian_2007a_h.e.s.s.-observations}.}
  \figlabel{Vela_Jr}
\end{figure}

2FGL\,J0851.7$-$4635 was found by our search to be an extended source
candidate in the 10 \gev to 100 \gev energy range and is spatially
coincident with the SNR Vela Jr. This source was recently studied by
the LAT Collaboration in \cite{tanaka_2011a_gamma-ray-observations}.
\figref{Vela_Jr} shows a counts map of the source.
Overlaid on \figref{Vela_Jr} are \tev contours of Vela
Jr. \citep{aharonian_2007a_h.e.s.s.-observations}.  There are three
point-like 2FGL sources 2FGL\,J0848.5$-$4535, 2FGL\,J0853.5$-$4711, and
2FGL\,J0855.4$-$4625 which correlate with the multiwavelength emission of
this SNR but do not appear to be physically distinct sources.  They were
most likely included in the 2FGL catalog to compensate for residuals
induced by not modeling the extension of Vela Jr. and were removed from
our model of the sky.

With this model of the background, 2FGL\,J0851.7$-$4635 was found to
have an extension of $\sigma=1\fdg15\pm0\fdg08_\stat\pm0\fdg02_\sys$
with $\tsext=86.8$.  The LAT size matches well the \tev morphology
of Vela Jr.  While fitting the extension of 2FGL\,J0851.7$-$4635,
we iteratively relocalized the position of the nearby point-like 2FGL
source 2FGL\,J0854.7$-$4501 to $(l,b)=(266\fdg24,0\fdg49)$ to better
fit its position at high energies.
    
\subsection{2FGL\,J1615.0$-$5051}
\subseclabel{section_2FGL_J1615.0-5051}

\begin{figure}[htbp]
  \includegraphics{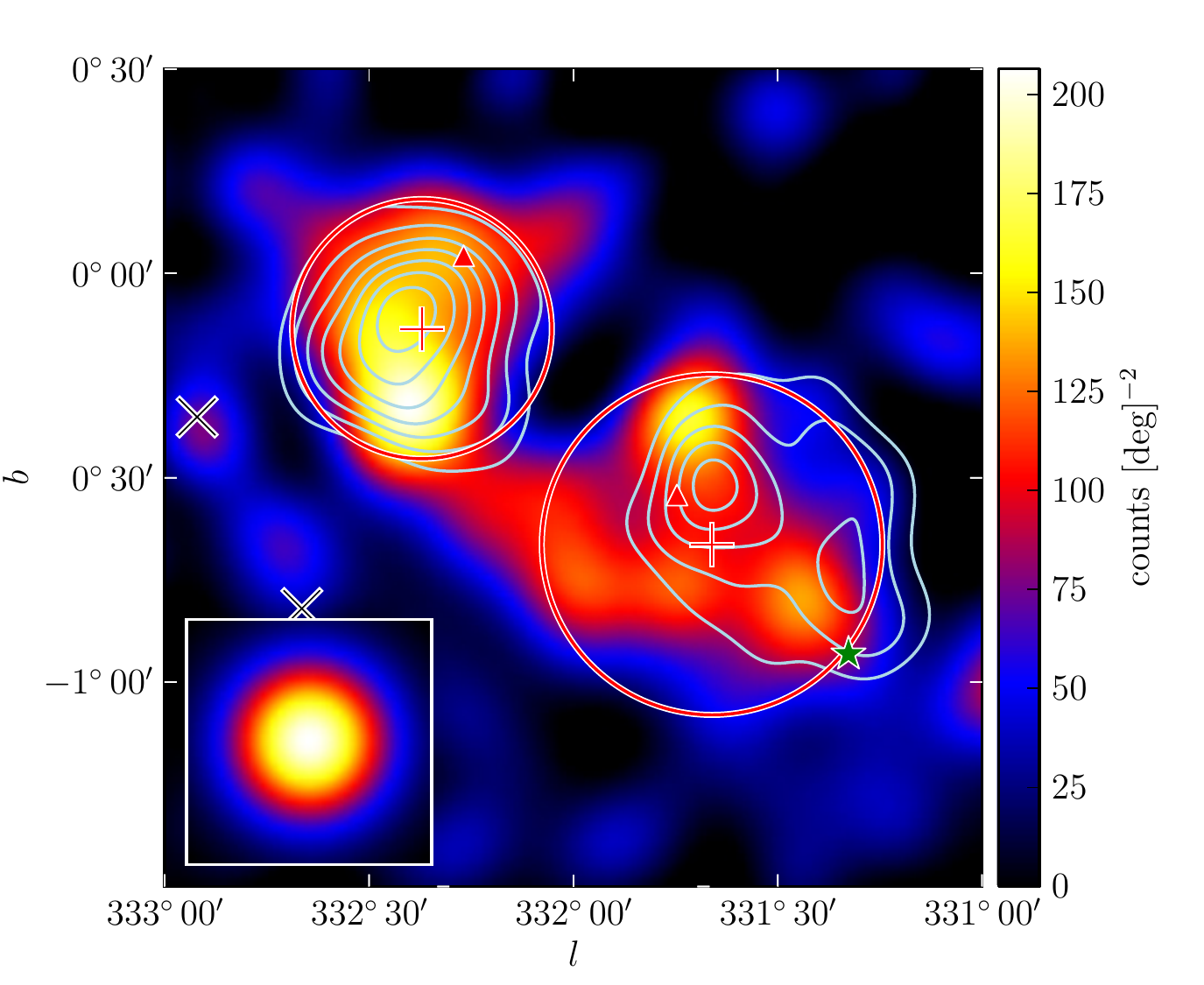}
  \caption{A diffuse-emission-subtracted 10 \gev to 100 \gev counts map
  of 2FGL\,J1615.0$-$5051 (upper left) and 2FGL\,J1615.2$-$5138 (lower
  right) smoothed by a 0\fdg1 2D Gaussian kernel.  The triangular markers
  (colored red in the electronic version) represent the 2FGL positions
  of these sources.  The cross-shaped markers and the circles (colored
  red) represent the best fit positions and extensions of these sources
  assuming a radially symmetric uniform disk model.  The two black
  crosses represent  background 2FGL sources and the star-shaped marker
  (colored green) represents 2FGL J1614.9-5212, another 2FGL source
  that was removed from the background model. The contours (colored
  light blue) correspond to the \tev image of HESS\,J1616$-$508 (left)
  and HESS\,J1614$-$518 (right) \citep{aharonian_2006a_h.e.s.s.-survey}.}
  \figlabel{1FGL_J1613.6-5100c}
\end{figure}

2FGL\,J1615.0$-$5051 and 2FGL\,J1615.2$-$5138 were both found to
be extended source candidates in the 10 \gev to 100 \gev energy
range. Because they are less than $1\degree$ away from each other,
they needed to be analyzed simultaneously.  2FGL\,J1615.0$-$5051 is
spatially coincident with the extended \tev source HESS\,J1616$-$508
and 2FGL\,J1615.2$-$5138 is coincident with the extended \tev source
HESS\,J1614$-$518.  \figref{1FGL_J1613.6-5100c} shows a counts map of
these sources and overlays the \tev contours of HESS\,J1616$-$508 and
HESS\,J1614$-$518 \citep{aharonian_2006a_h.e.s.s.-survey}.  The figure
shows that the 2FGL source 2FGL\,J1614.9$-$5212 is very close to
2FGL\,J1615.2$-$5138 and correlates with the same extended \tev source
as 2FGL\,J1615.2$-$5138.  We concluded that this source was included
in the 2FGL catalog to compensate for residuals induced by not modeling
the extension of 2FGL\,J1615.2$-$5138 and removed it from our model of
the sky.

With this model of the sky, we iteratively fit the extensions of
2FGL\,J1615.0$-$5051 and 2FGL\,J1615.2$-$5138.  2FGL\,J1615.0$-$5051 was
found to have an extension $\sigma=0\fdg32\pm0\fdg04_\stat\pm0\fdg01_\sys$
and \tsext=16.7.

The \tev counterpart of 2FGL\,J1615.0$-$5051 was fit with
a radially-symmetric Gaussian surface brightness profile with
$\sigma=0\fdg136\pm0\fdg008$ \citep{aharonian_2006a_h.e.s.s.-survey}. This
\tev size corresponds to a 68\% containment radius of
$\rsixeight=0\fdg21\pm0\fdg01$, comparable to the LAT size
$\rsixeight=0\fdg26\pm0\fdg03$.  \figref{hess_seds} shows that the
spectrum of 2FGL\,J1615.0$-$5051 at \gev energies connects to the spectrum
of HESS\,J1616$-$508 at \tev energies.

HESS\,J1616$-$508 is located in the region of two SNRs RCW103 (G332.4-04)
and Kes~32 (G332.4+0.1) but is not spatially coincident with either of
them \citep{aharonian_2006a_h.e.s.s.-survey}.  HESS\,J1616$-$508 is near
three pulsars PSR\,J1614$-$5048, PSR\,J1616$-$5109, and PSR\,J1617$-$5055.
\citep{torii_1998a_discovery-millisecond,landi_2007a_j1616-508:-likely}.
Only PSR\,J1617$-$5055 is energetically capable of powering the
\tev emission and \cite{aharonian_2006a_h.e.s.s.-survey} speculated
that HESS\,J1616$-$508 could be a PWN powered by this young pulsar.
Because HESS\,J1616$-$508 is $9\arcmin$ away from PSR\,J1617$-$5055,
this would require an asymmetric X-ray PWNe to power the \tev
emission.  \chandra ACIS observations revealed an underluminous
PWN of size $\sim1\arcmin$ around the pulsar that was not oriented
towards the \tev emission, rendering this association uncertain
\citep{kargaltsev_2008a_young-energetic}.  No other promising counterparts
were observed at X-ray and soft $\gamma$-ray energies by \suzaku
\citep{matsumoto_2007a_suzaku-observations}, \swiftxrt, IBIS/ISGRBI,
BeppoSAX and \xmmnewton \citep{landi_2007a_j1616-508:-likely}.
\cite{kargaltsev_2008a_young-energetic} discovered additional diffuse
emission towards the center of HESS\,J1616$-$508 using archival radio
and infared observations. Deeper observations will likely be necessary
to understand this $\gamma$-ray source.

\subsection{2FGL\,J1615.2$-$5138}
\subseclabel{section_2FGL_J1615.2-5138}

2FGL\,J1615.2$-$5138 was found
 to have an extension $\sigma=0\fdg42\pm0\fdg04_\stat\pm0.02_\sys$
with $\tsext=46.5$.  To test for the possibility that 2FGL\,J1615.2$-$5138
is not spatially extended but instead composed of two point-like sources
(one of them represented in the 2FGL catalog by 2FGL\,J1614.9$-$5212),
we refit 2FGL\,J1615.2$-$5138 as two point-like sources.  Because
$\tsinc=35.1$ is less than $\tsext=46.5$, we conclude that this emission
does not originate from two closely-spaced point-like sources.

2FGL\,J1615.2$-$5138 is spatially coincident with the extended \tev
source HESS\,J1614$-$518.  \ac{HESS} measured a 2D Gaussian extension
of $\sigma=0\fdg23\pm0\fdg02$ and $\sigma=0\fdg15\pm0\fdg02$ in the
semi-major and semi-minor axis. This corresponds to a 68\% containment
size of $\rsixeight=0\fdg35\pm0\fdg03$ and $0\fdg23\pm0\fdg03$,
consistent with the LAT size $\rsixeight=0\fdg34\pm0\fdg03$.
\figref{hess_seds} shows that the spectrum of 2FGL\,J1615.2$-$5138
at \gev energies connects to the spectrum of HESS\,J1614$-$518 at
\tev energies.  Further data collected by \ac{HESS} in 2007 resolve
a double peaked structure at \tev energies but no spectral variation
across this source, suggesting that the emission is not the confusion of
physically separate sources \citep{rowell_2008a_closer-unidentified}.
This double peaked structure is also hinted at in the LAT
counts map in \figref{1FGL_J1613.6-5100c} but is not very
significant.  The \tev source was also detected by CANGAROO-III
\citep{mizukami_2011a_cangaroo-iii-observation}.

There are five nearby pulsars, but none are luminous enough to provide
the energy output required to power the $\gamma$-ray emission
\citep{rowell_2008a_closer-unidentified}.  HESS\,J1614$-$518
is spatially coincident with a young open cluster Pismis 22
\citep{landi_2007a_j1614-518:-detection,rowell_2008a_closer-unidentified}.
\suzaku detected two promising X-ray candidates. Source A is an extended
source consistent with the peak of HESS\,J1614$-$518 and source B
coincident with Pismis 22 and towards the center but in a relatively dim
region of HESS\,J1614$-$518 \citep{matsumoto_2008a_discovery-extended}.
Three hypotheses have been presented to explain this emission: either
source A is an SNR powering the $\gamma$-ray emission; source A is a PWN
powered by an undiscovered pulsar in either source A or B; and finally
that the emission may arise from hadronic acceleration in the stellar
winds of Pismis 22 \citep{mizukami_2011a_cangaroo-iii-observation}.

\subsection{2FGL\,J1627.0$-$2425c}
\subseclabel{section_2FGL_J1627.0-2425c}

\begin{figure}[htbp]
  \includegraphics{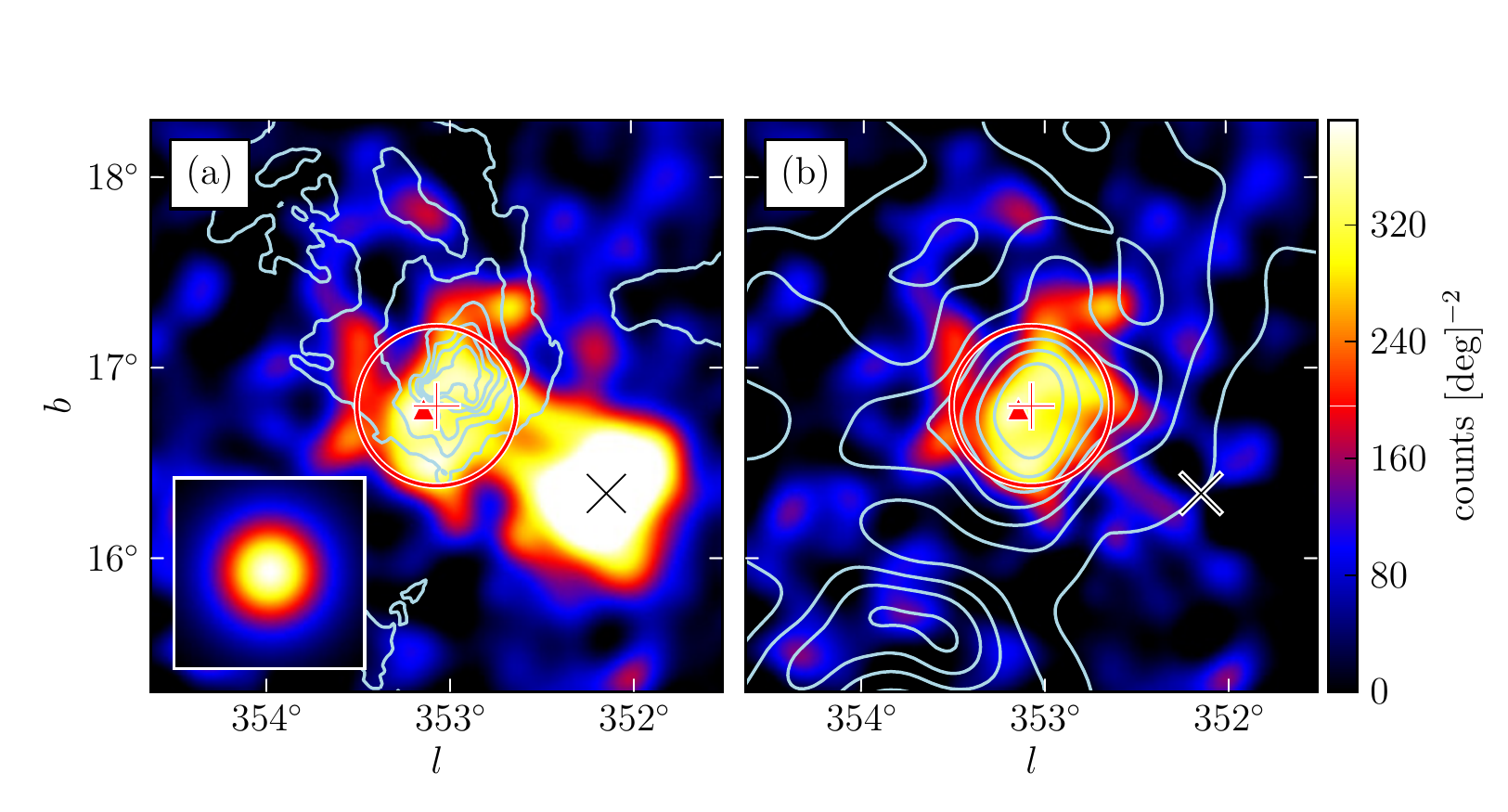}
  \caption{A diffuse-emission-subtracted 1 \gev to 100 \gev counts map
  of (a) the region around 2FGL\,J1627.0$-$2425 smoothed by a 0\fdg1 2D
  Gaussian kernel and (b) with the emission from 2FGL\,J1625.7$-$2526
  subtracted.  The triangular marker (colored red in the online version)
  represents the 2FGL position of this source.  The plus-shaped
  marker and the circle (colored red) represent the best fit position
  and extension of this source assuming a radially-symmetric uniform
  disk model and the black cross represents a background 2FGL source.
  The contours in (a) correspond to the 100 $\mu$m image observed by
  IRAS \citep{young_1986a_high-resolution-observations}.  The contours
  in (b) correspond to CO ($J=1\rightarrow 0$) emission integrated
  from $-$8 $\km\,\second^{-1}$ to 20 $\km\,\second^{-1}$.  They are
  from \cite{de-geus_1990a_survey-clouds}, were cleaned using the
  moment-masking technique \citep{dame_2011a_optimization-moment},
  and have been smoothed by a 0\fdg25 2D Gaussian kernel.}
  \figlabel{1FGL_J1628.6-2419c}
\end{figure}

2FGL\,J1627.0$-$2425c was found by our search to have an extension
$\sigma=0\fdg42\pm0\fdg05_\stat\pm0\fdg16_\sys$ with $\tsext=32.4$
using photons with energies between 1 \gev and 100 \gev.
\figref{1FGL_J1628.6-2419c} shows a counts map of this source.

This source is in a region of remarkably complicated diffuse emission.
Even though it is $16\degree$ from the Galactic plane, this source
is on top of the core of the Ophiuchus molecular cloud which
contains massive star-forming regions that are bright in infrared.
The region also has abundant molecular and atomic gas traced by CO
and H~I and significant dark gas found only by its association with
dust emission \citep{grenier_2005a_unveiling-extensive}. Embedded
star-forming regions make it even more challenging to measure
the column density of dust.  Infared and CO ($J=1\rightarrow
0$) contours are overlaid on \figref{1FGL_J1628.6-2419c}
and show good spatial correlation with the \gev emission
\citep{young_1986a_high-resolution-observations,de-geus_1990a_survey-clouds}.
This source might represent $\gamma$-ray emission from the interactions
of cosmic rays with interstellar gas which has not been accounted for
in the LAT diffuse emission model.

\subsection{2FGL\,J1632.4$-$4753c}
\subseclabel{section_2FGL_J1632.4-4753c}

\begin{figure}[htbp]
  \includegraphics{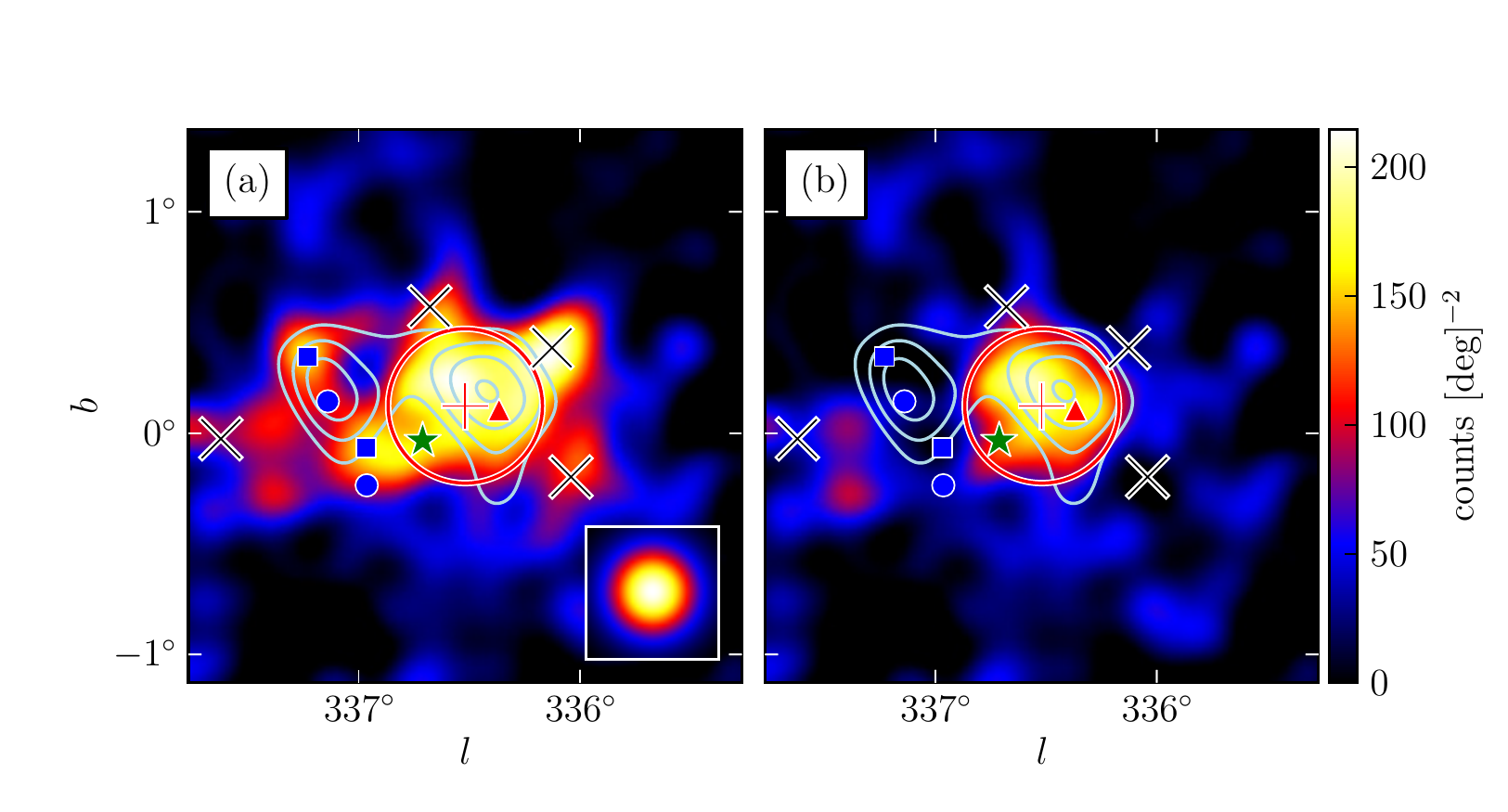}
  \caption{A diffuse-emission-subtracted 10 \gev to 100 \gev counts map
  of 2FGL\,J1632.4$-$4753c (a) smoothed by a 0\fdg1 2D Gaussian kernel
  and (b) with the emission from the background sources subtracted.
  The triangular marker (colored red in the electronic version)
  represents the 2FGL position of this source.  The plus-shaped marker
  and the circle (colored red) are the best fit position and extension of
  2FGL\,J1632.4$-$4753c assuming a radially-symmetric uniform disk model.
  The four black crosses represent background 2FGL sources subtracted in
  (b).  The circular and square-shaped markers (colored blue) represent
  the 2FGL and relocalized positions respectively of two additional
  background 2FGL sources subtracted in (b).  The star-shaped marker
  (colored green) represents 2FGL\,J1634.4$-$4743c, another 2FGL
  source that was removed from the background model.  The contours
  (colored light blue) correspond to the \tev image of HESS\,J1632$-$478
  \citep{aharonian_2006a_h.e.s.s.-survey}.}
  \figlabel{1FGL_J1632.9-4802c}
\end{figure}

2FGL\,J1632.4$-$4753c was found by our search to be an extended
source candidate in the 10 \gev to 100 \gev energy range but is in a
crowded region of the sky.  It is spatially coincident with the \tev
source HESS\,J1632$-$478.  \figref{1FGL_J1632.9-4802c}a shows a counts
map of this source and overlays \tev contours of HESS\,J1632$-$478
\citep{aharonian_2006a_h.e.s.s.-survey}.  There are six nearby
point-like 2FGL sources that appear to represent physically distinct
sources and were included in our background model: 2FGL\,J1630.2$-$4752,
2FGL\,J1631.7$-$4720c, 2FGL\,J1632.4$-$4820c, 2FGL\,J1635.4$-$4717c,
2FGL\,J1636.3$-$4740c, and 2FGL\,J1638.0$-$4703c. On the other hand,
one point-like 2FGL source 2FGL\,J1634.4$-$4743c  correlates with the
extended \tev source and at \gev energies does not appear physically
separate.  It is very close to the position of 2FGL\,J1632.4$-$4753c
and does not show spatially separated emission in the observed
photon distribution.  We therefore removed this source from our
model of the background.  \figref{1FGL_J1632.9-4802c}b shows the
same region with the background sources subtracted.  With this
model, 2FGL\,J1632.4$-$4753c was found to have an extension
$\sigma=0\fdg35\pm0\fdg04_\stat\pm0\fdg02_\sys$ with $\tsext=26.9$.
While fitting the extension of 2FGL\,J1632.4$-$4753c, we iteratively
relocalized 2FGL\,J1635.4$-$4717c to $(l,b)=(337\fdg23,0\fdg35)$ and
2FGL\,J1636.3$-$4740c to $(l,b)=(336\fdg97,-0\fdg07)$.

\begin{figure}[htbp]
  \includegraphics{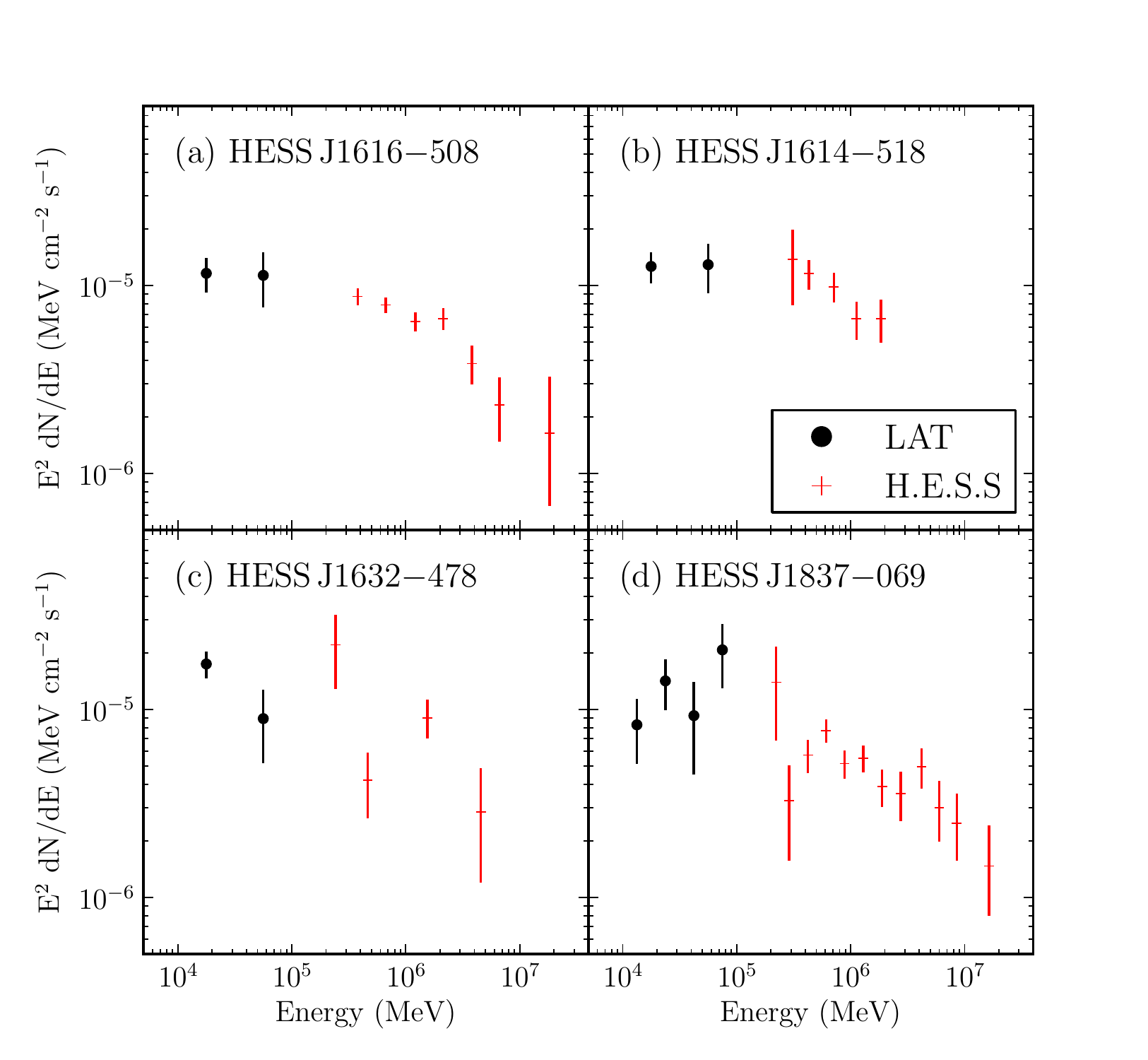}
  \caption{The spectral energy distribution of four extended
  sources associated with unidentified extended \tev sources.
  The black points with circular markers are obtained by the
  LAT. The points with plus-shaped markers (colored red in the
  electronic version) are for the associated \ac{HESS} sources.
  (a) the LAT SED of 2FGL\,J1615.0$-$5051 together with the
  \ac{HESS} SED of HESS\,J1616$-$508. (b) 2FGL\,J1615.2$-$5138 and
  HESS\,J1614$-$518. (c) 2FGL\,J1632.4$-$4753c and HESS\,J1632$-$478. (d)
  2FGL\,J1837.3$-$0700c and HESS\,J1837$-$069. The \ac{HESS} data points
  are from \citep{aharonian_2006a_h.e.s.s.-survey}. Both LAT and \ac{HESS}
  spectral errors are statistical only.}
    \figlabel{hess_seds}
  \end{figure}

\ac{HESS} measured an extension of $\sigma=0\fdg21\pm0\fdg05$ and
$0\fdg06\pm0\fdg04$ along the semi-major and semi-minor axes when
fitting HESS\,J1632$-$478 with an elliptical 2D Gaussian surface
brightness profile.  This corresponds to a 68\% containment size
$\rsixeight=0\fdg31\pm0\fdg08$ and $0\fdg09\pm0\fdg06$ along
the semi-major and semi-minor axis, consistent with the LAT size
$\rsixeight=0\fdg29\pm0\fdg04$.  \figref{hess_seds} shows that the
spectrum of 2FGL\,J1632.4$-$4753c at \gev energies connects to the
spectrum of HESS\,J1632$-$478 at \tev energies.

\cite{aharonian_2006a_h.e.s.s.-survey} argued that
HESS\,J1632$-$478 is positionally coincident with the hard X-ray
source IGR\,J1632$-$4751 observed by \asca, INTEGRAL, and \xmmnewton
\citep{sugizaki_2001a_faint-x-ray,tomsick_2003a_j16320-4751,rodriguez_2003a_xmm-newton-observation},
but this source is suspected to be a Galactic X-Ray Binary so the
$\gamma$-ray extension disfavors the association.  Further observations
by \xmmnewton discovered point-like emission coincident with the
peak of the \ac{HESS} source surrounded by extended emission of size
$\sim32\arcsec\times15\arcsec$ \citep{balbo_2010a_j1632-478:-energetic}.
They found in archival MGPS-2 data a spatially coincident extended
radio source \citep{murphy_2007a_second-epoch} and argued for a
single synchrotron and inverse Compton process producing the radio,
X-ray, and \tev emission, likely due to a PWN.  The increased
size at \tev energies compared to X-ray energies has previously
been observed in several aging PWNe including HESS\,J1825$-$137
\citep{gaensler_2003a_xmm-newton-observations,aharonian_2006a_energy-dependent},
HESS\,J1640$-$465
\citep{aharonian_2006a_h.e.s.s.-survey,funk_2007a_xmm-newton-observations},
and Vela X
\citep{markwardt_1995a_x-ray-pulsar,aharonian_2006a_first-detection}
and can be explained by different synchrotron cooling times for the
electrons that produce X-rays and $\gamma$-rays.

\subsection{2FGL\,J1712.4$-$3941}
\subseclabel{section_2FGL_J1712.4-3941}

\begin{figure}[htbp]
  \includegraphics{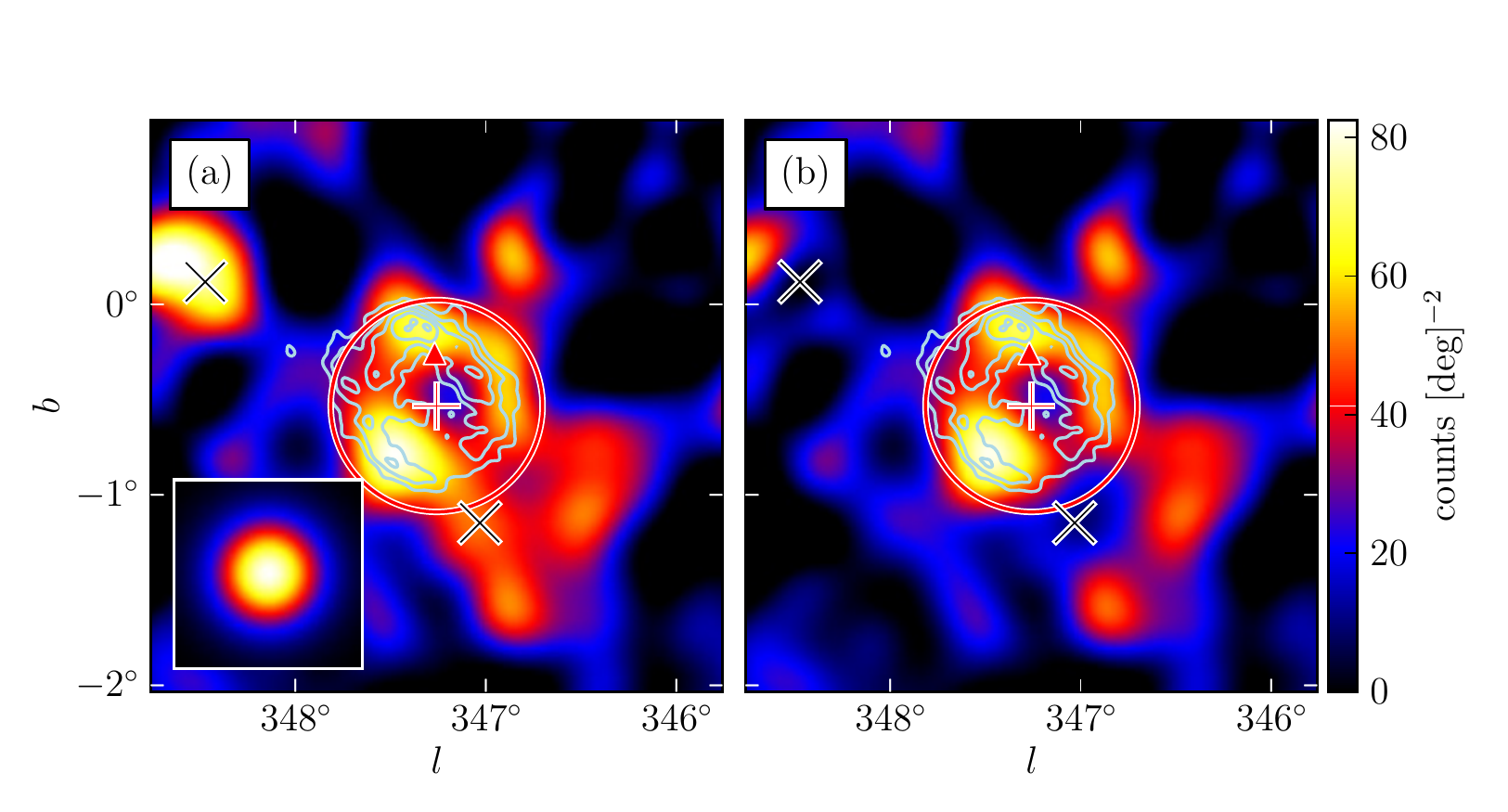}
  \caption{A diffuse-emission-subtracted 10 \gev to 100 \gev counts map
  of 2FGL\,J1712.4$-$3941 (a) smoothed by a 0\fdg15 2D Gaussian kernel
  and (b) with the emission from the background sources subtracted.
  This source is spatially coincident with RX\,J1713.7$-$3946 and
  was recently studied in \cite{abdo_2011a_observations-young}.
  The triangular marker (colored red in the online version) represents
  the 2FGL position of this source.  The plus-shaped marker and the
  circle (colored red) are the best fit position and extension of this
  source assuming a radially symmetric uniform disk model.  The two
  black crosses represent background 2FGL sources subtracted in (b).
  The contours (colored light blue) correspond to the \tev image
  \citep{aharonian_2007a_primary-particle}.}
  \figlabel{2FGL_J1712.4-3941}
\end{figure}

2FGL\,J1712.4$-$3941 was found by our search to be spatially extended
using photons with energies between 1 \gev and 100 \gev.  This source
is spatially coincident with the SNR RX\,J1713.7$-$3946 and was recently
studied by the LAT Collaboration in \cite{abdo_2011a_observations-young}.
To avoid issues related to uncertainties in the nearby Galactic
diffuse emission at lower energy, we restricted our analysis only
to energies above 10 \gev.  \figref{2FGL_J1712.4-3941} shows a
smoothed counts map of the source. Above 10 \gev, the \gev emission
nicely correlates with the \tev contours of RX\,J1713.7$-$3946
\citep{aharonian_2007a_primary-particle} and 2FGL\,J1712.4$-$3941 fit
to an extension $\sigma=0\fdg56\pm0\fdg04_\stat\pm0\fdg02_\sys$ with
$\tsext=38.5$.

\subsection{2FGL\,J1837.3$-$0700c}
\subseclabel{section_2FGL_J1837.3-0700c}

\begin{figure}[htbp]
  \includegraphics{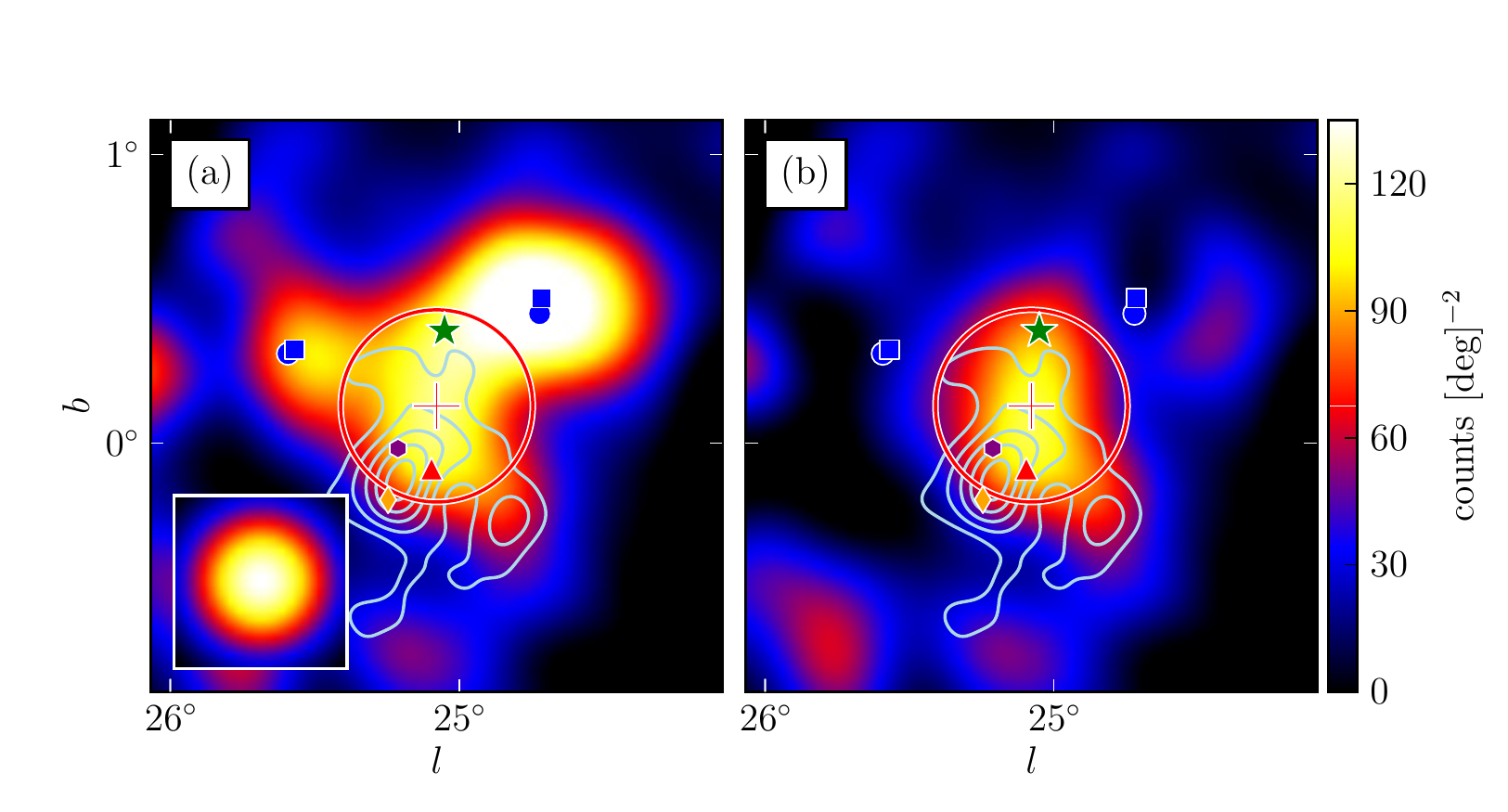}
  \caption{A diffuse-emission-subtracted 10 \gev to 100 \gev counts map
  of the region around 2FGL\,J1837.3$-$0700c (a) smoothed by a 0\fdg15
  2D Gaussian kernel and (b) with the emission from the background
  sources subtracted.  The triangular marker (colored red in the online
  version) represents the 2FGL position of this source.  The plus-shaped
  marker and the circle (colored red) represent the best fit position
  and extension of 2FGL\,J1837.3$-$0700c assuming a radially-symmetric
  uniform disk model.  The circular and square-shaped markers (colored
  blue) represent the 2FGL and the relocalized positions respectively
  of two background 2FGL sources subtracted in (b).  The star-shaped
  marker (colored green) represents 2FGL\,J1835.5$-$0649, another 2FGL
  source that was removed from the background model.  The contours
  (colored light blue) correspond to the \tev image of HESS\,J1837$-$069
  \citep{aharonian_2006a_h.e.s.s.-survey}.  The diamond-shaped marker
  (colored orange) represents the position of PSR\,J1838$-$0655 and
  the hexagonal-shaped marker (colored purple) represents the position
  AX\,J1837.3$-$0652 \citep{gotthelf_2008a_discovery-young}.}
  \figlabel{1FGL_J1837.5-0659c}
\end{figure}

2FGL\,J1837.3$-$0700c was found by our search to be an extended source
candidate in the 10 \gev to 100 \gev energy range and is spatially
coincident with the \tev source HESS\,J1837$-$069.  This source
is in a complicated region.  \figref{1FGL_J1837.5-0659c}a shows a
smoothed counts map of the region and overlays the \tev contours of
HESS\,J1837$-$069 \citep{aharonian_2006a_h.e.s.s.-survey}.  There
are two very nearby point-like 2FGL sources, 2FGL\,J1836.8$-$0623c
and 2FGL\,J1839.3$-$0558c, that clearly represent distinct sources.
On the other hand, there is another source 2FGL\,J1835.5$-$0649 located
between the three sources that appears to correlate with the \tev
morphology of HESS\,J1837$-$069 but at \gev energies does not appear
to represent a physically distinct source.  We concluded that this
source was included in the 2FGL catalog to compensate for residuals
induced by not modeling the extension of this source and removed
it from our model of the sky.  \figref{1FGL_J1837.5-0659c}b shows a
counts map of this region after subtracting these background sources.
After removing 2FGL\,J1835.5$-$0649, we tested for source confusion
by fitting 2FGL\,J1837.3$-$0700c instead as two point-like sources.
Because $\tsinc=10.8$ is less than $\tsext=18.5$, we conclude that this
emission does not originate from two nearby point-like sources.

With this model, 2FGL\,J1837.3$-$0700c was found to have an extension
$\sigma=0\fdg33\pm0\fdg07_\stat\pm0\fdg05_\sys$.  While fitting
the extension of 2FGL\,J1837.3$-$0700c, we iteratively relocalized
the two closest background sources along with the extension of
2FGL\,J1837.3$-$0700c but their positions did not significantly
change.  2FGL\,J1834.7$-$0705c moved to $(l,b)=(24\fdg77,0\fdg50)$,
2FGL\,J1836.8$-$0623c moved to $(l,b)=(25\fdg57,0\fdg32)$.

\ac{HESS} measured an extension of $\sigma=0\fdg12\pm0\fdg02$ and
$0\fdg05\pm0\fdg02$ of the coincident \tev source HESS\,J1837$-$069
along the semi-major and semi-minor axis when fitting this source with
an elliptical 2D Gaussian surface brightness profile.  This corresponds
to a 68\% containment radius of $\rsixeight=0\fdg18\pm0\fdg03$ and
$0\fdg08\pm0\fdg03$ along the semi-major and semi-minor axis. The size
is not significantly different from the LAT 68\% containment radius of
$\rsixeight=0\fdg27\pm0\fdg07$ (less than $2\sigma$).  \figref{hess_seds}
shows that the spectrum of 2FGL\,J1837.3$-$0700c at \gev energies connects
to the spectrum of HESS\,J1837$-$069 at \tev energies.

HESS\,J1837$-$069 is coincident with the hard and steady X-ray source
AX\,J1838.0$-$0655 \citep{bamba_2003a_diffuse-x-ray}.  This source
was discovered by RXTE to be a pulsar (PSR J1838-0655) sufficiently
luminous to power the \tev emission and was resolved by \chandra
to be a bright point-like source surrounded by a $\sim2\arcmin$
nebula \citep{gotthelf_2008a_discovery-young}. The $\gamma$-ray
emission may be powered by this pulsar.  The hard spectral index
and spatial extension of 2FGL\,J1837.3$-$0700c disfavor a pulsar
origin of the LAT emission and suggest instead that the \gev and \tev
emission both originate from the pulsar's wind.  There is another
X-ray point-like source AX\,J1837.3$-$0652 near HESS\,J1837$-$069
\citep{bamba_2003a_diffuse-x-ray} that was also resolved into a point-like
and diffuse component \citep{gotthelf_2008a_discovery-young}.  Although no
pulsations have been detected from it, it could also be a pulsar powering
some of the $\gamma$-ray emission.

\subsection{2FGL\,J2021.5+4026}
\subseclabel{section_2FGL_J2021.5+4026}

The source 2FGL\,J2021.5+4026 is associated with the $\gamma$-Cygni
SNR and has been speculated to originate from the interaction
of accelerated particles in the SNR with dense molecular clouds
\citep{pollock_1985a_probable-identification,gaisser_1998a_gamma-ray-production}.
This association was disfavored when the \gev
emission from this source was detected to be pulsed
\citep[PSR\,J2021+4026,][]{abdo_2010a_first-fermi}.  This pulsar was
also observed by AGILE \citep{chen_2011a_study-gamma-ray}.

\begin{figure}[htbp]
  \includegraphics{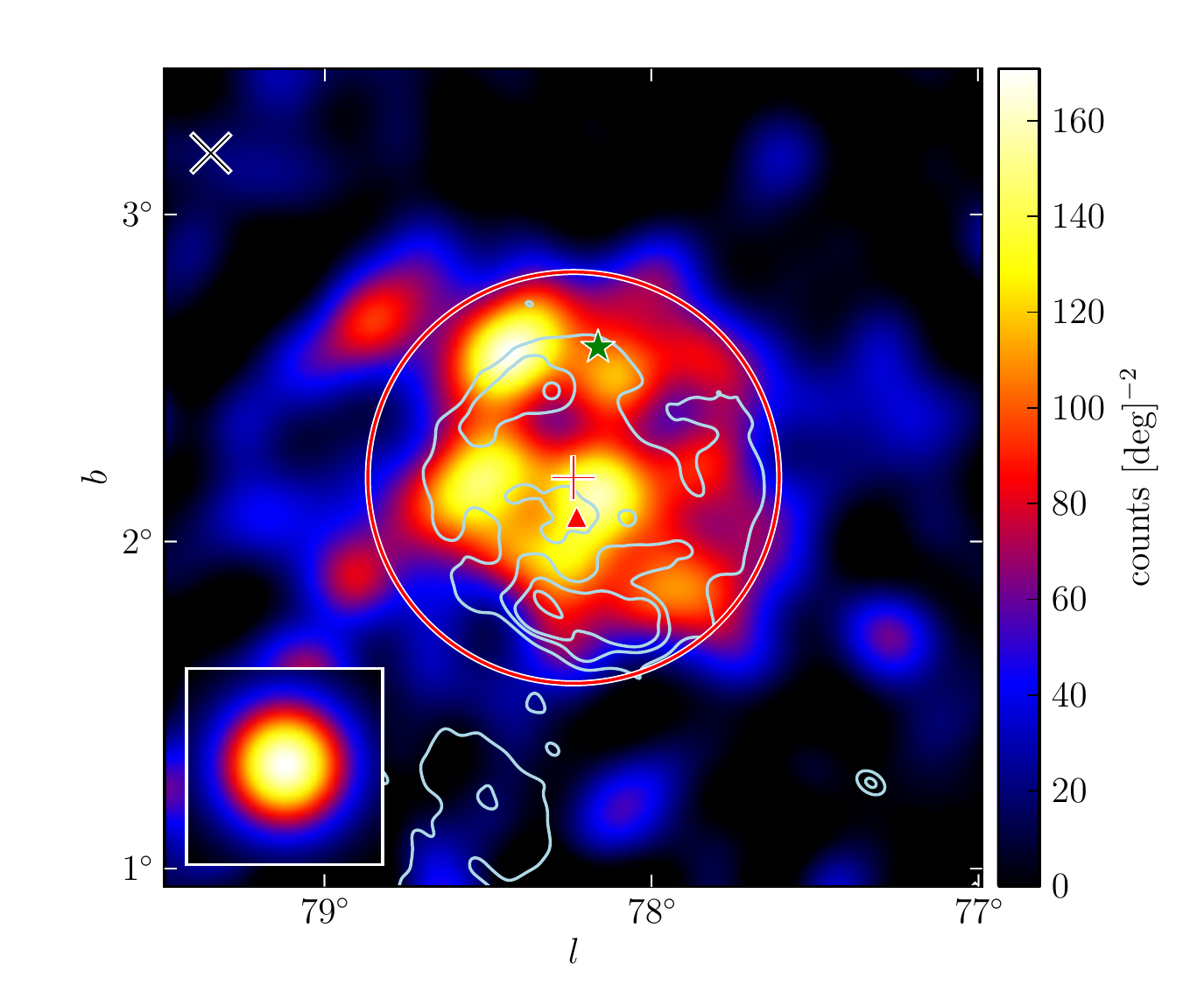}
  \caption{A diffuse-emission-subtracted 10 \gev to 100 \gev counts
  map of the region around 2FGL\,J2021.5+4026 smoothed by a 0\fdg1
  2D Gaussian kernel. The triangular marker (colored red in the
  online version) represents the 2FGL position of this source.
  The plus-shaped marker and the circle (colored red) represent the
  best fit position and extension of 2FGL\,J2021.5+4026 assuming
  a radially-symmetric uniform disk model.  The star-shaped marker
  (colored green) represents 2FGL\,J2019.1+4040, a 2FGL source that was
  removed from the background model.  2FGL\,J2021.5+4026 is spatially
  coincident with the $\gamma$-Cygni SNR.  The contours (colored light
  blue) correspond to the 408MHz image of $\gamma$-Cygni observed by the
  Canadian Galactic Plane Survey \citep{taylor_2003a_canadian-galactic}.}
  \figlabel{1FGL_J2020.0+4049}
\end{figure}

Looking at the same region at energies above 10 \gev, the pulsar is
no longer significant but we instead found in our search an extended
source candidate.  \figref{1FGL_J2020.0+4049} shows a counts map of this
source and overlays radio contours of $\gamma$-Cygni from the Canadian
Galactic Plane Survey \citep{taylor_2003a_canadian-galactic}.  There is
good spatial overlap between the SNR and the \gev emission.

\begin{figure}[htbp]
  \includegraphics{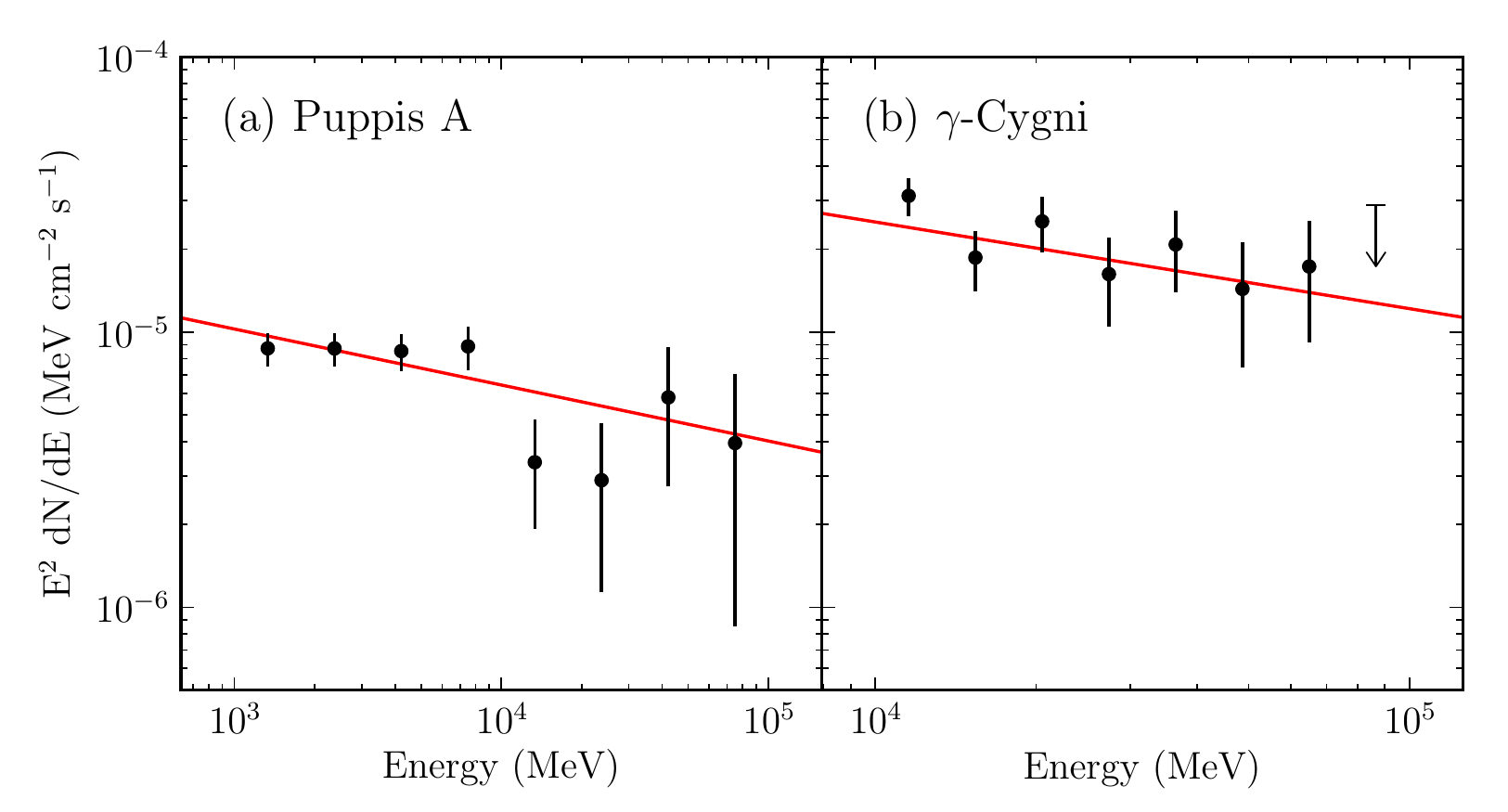}
  \caption{The spectral energy distribution of the extended sources Puppis
  A (2FGL\,J0823.0$-$4246) and $\gamma$-Cygni (2FGL\,J2021.5+4026).
  The lines (colored red in the online version) are the best fit
  power-law spectral models of these sources. Puppis A has a spectral
  index of $2.21\pm0.09$ and $\gamma$-Cygni has an index of $2.42\pm0.19$.
  The spectral errors are statistical only.  The upper limit is at the
  95\% confidence level.}
  \figlabel{snr_seds}
\end{figure}

There is a nearby source 2FGL\,J2019.1+4040 that correlates with the
radio emission of $\gamma$-Cygni and at \gev energies does not appear
to represent a physically distinct source.  We concluded that it was
included in the 2FGL catalog to compensate for residuals induced by
not modeling the extension of $\gamma$-Cygni and removed it from our
model of the sky.  With this model, 2FGL\,J2021.5+4026 was found to
have an extension $\sigma=0\fdg63\pm0\fdg05_\stat\pm0\fdg04_\sys$
with $\tsext=128.9$.  \figref{snr_seds} shows its spectrum.
The inferred size of this source at \gev energies well
matches the radio size of $\gamma$-Cygni.  Milagro detected a
$4.2\sigma$ excess at energies $\sim 30$ \tev from this location
\citep{abdo_2009a_fermi/large-telescope,abdo_2009a_milagro-observations}.
VERITAS also detected an extended source VER\,J2019+407
coincident with the SNR above 200 \gev and suggested that the
\tev emission could be a shock-cloud interaction in $\gamma$-Cygni
\citep{weinstein_2009a_veritas-survey}.

\section{Discussion}
\seclabel{extended_source_discussion}

\begin{figure}[htbp]
  \includegraphics{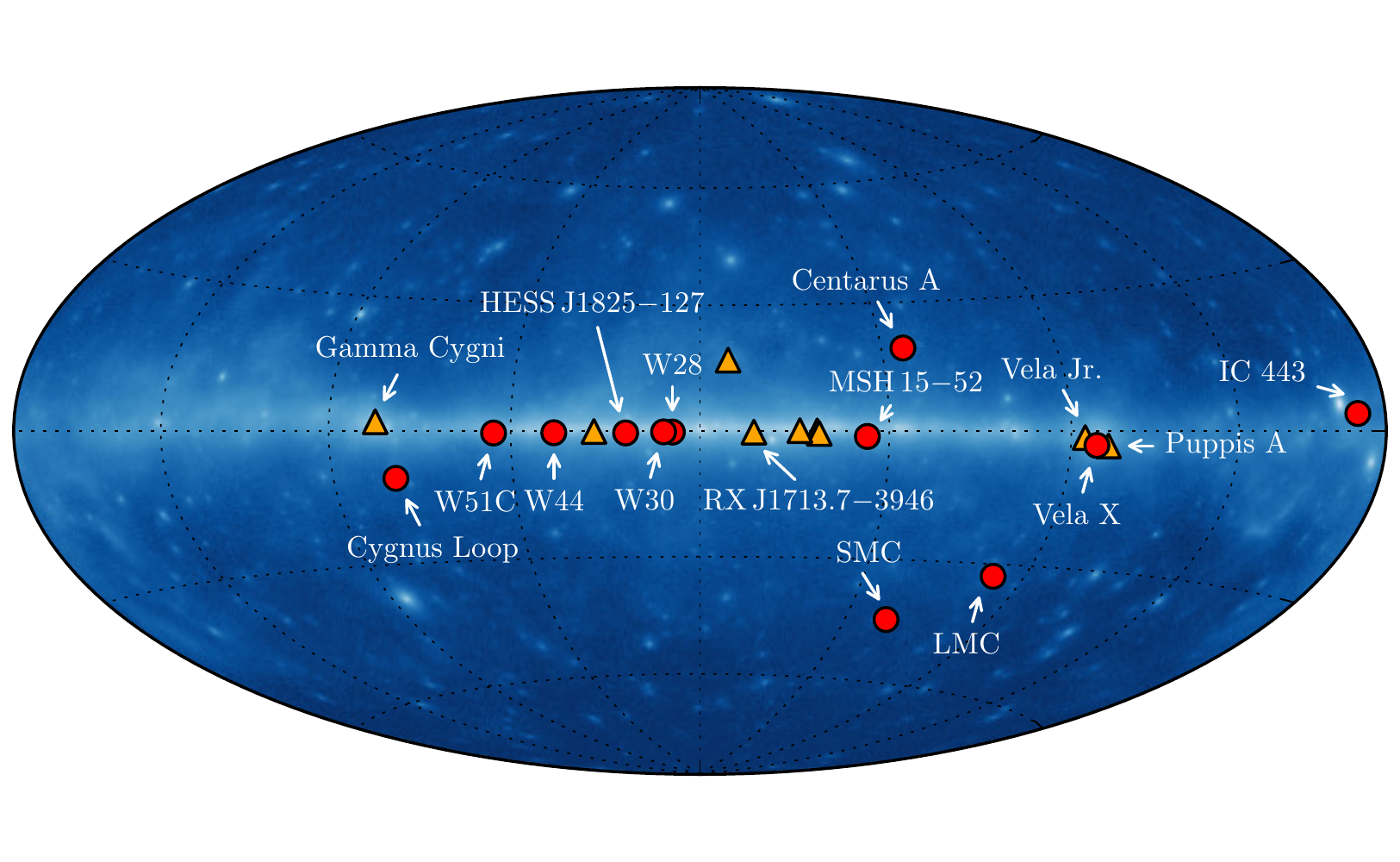}
  \caption{The 21 spatially extended sources detected by the LAT at \gev
  energies with 2 years of data.  The twelve extended sources included
  in 2FGL are represented by the circular markers (colored red in the
  online version).  The nine new extended sources are represented by
  the triangular markers (colored orange).  The source positions are
  overlaid on a 100 \mev to 100 \gev Aitoff projection sky map of the
  LAT data in Galactic coordinates.}
  \figlabel{allsky_extended_sources}
\end{figure}

Twelve extended sources were included in the 2FGL catalog and two
additional extended sources were studied in dedicated publications.
Using 2 years of LAT data and a new analysis method, we presented the
detection of seven additional extended sources.  We also reanalyzed the
spatial extents of the twelve extended sources in the 2FGL catalog and
the two additional sources.  The 21 extended LAT sources are located
primarily along the Galactic plane and their locations are shown in
\figref{allsky_extended_sources}.  Most of the LAT-detected extended
sources are expected to be of Galactic origin as the distances of
extragalactic sources (with the exception of the local group Galaxies) are
typically too large to be able to resolve them at $\gamma$-ray energies.

\begin{figure}[htbp]
  \includegraphics{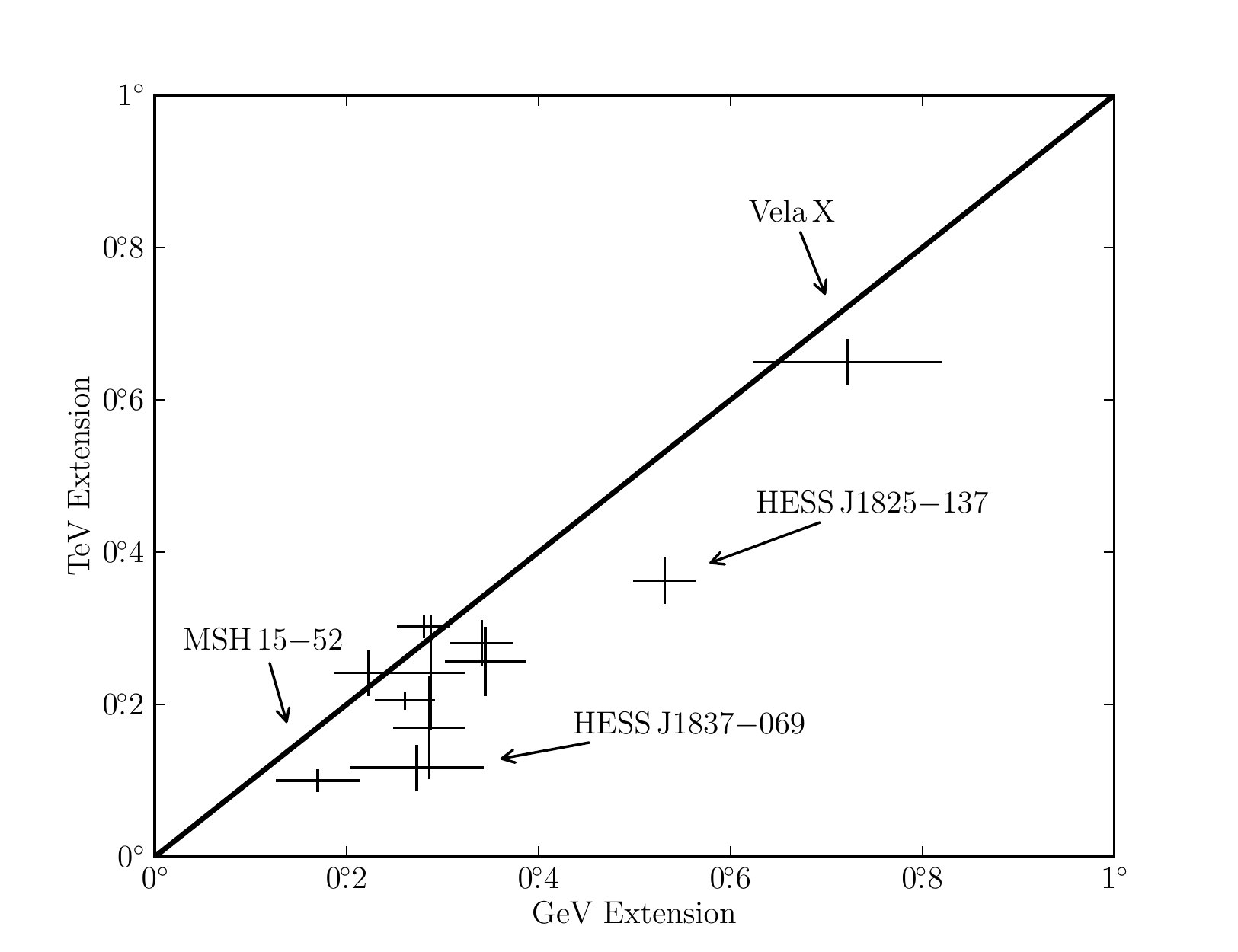}
  \caption{A comparison of the sizes of extended sources
  detected at both \gev and \tev energies.  The \tev sizes
  of W30, 2FGL\,J1837.3$-$0700c, 2FGL\,J1632.4$-$4753c,
  2FGL\,J1615.0$-$5051, and 2FGL\,J1615.2$-$5138 are from
  \cite{aharonian_2006a_h.e.s.s.-survey}.  The \tev sizes of MSH\,15$-$52,
  HESS\,J1825$-$137, Vela X, Vela Jr., RX\,J1713.7$-$3946 and W28 are from
  \cite{aharonian_2005a_discovery-extended,aharonian_2006a_energy-dependent,aharonian_2006a_first-detection,aharonian_2007a_h.e.s.s.-observations,aharonian_2007a_primary-particle,aharonian_2008a_discovery-energy}.
  The \tev size of IC~443 is from
  \cite{acciari_2009a_observation-extended} and W51C is from
  \cite{krause_2011a_probing-proton}.  The \tev sizes of MSH\,15$-$52,
  HESS\,J1614$-$518, HESS\,J1632$-$478, and HESS\,J1837$-$069 have only
  been reported with an elliptical 2D Gaussian fit and so the plotted
  sizes are the geometric mean of the semi-major and semi-minor axis.  The
  LAT extension of Vela X is from \cite{abdo_2010c_fermi-large}.  The \tev
  sources were fit assuming a 2D Gaussian surface brightness profile so
  the plotted \gev and \tev extensions were first converted to \rsixeight
  (see \subsecref{compare_source_size}).  Because of their large sizes,
  the shape of RX\,J1713.7$-$3946 and Vela Jr.  were not directly fit
  at \tev energies and so are not included in this comparison. On
  the other hand, dedicated publications by the LAT collaboration
  on these sources showed that their morphologies are consistent
  \citep{abdo_2011a_observations-young,tanaka_2011a_gamma-ray-observations}.
  The LAT extension errors are the statistical and systematic errors
  added in quadrature.}
  \figlabel{gev_vs_tev_plot}
\end{figure}

For the LAT extended sources also seen at \tev energies,
\figref{gev_vs_tev_plot} shows that there is a good correlation between
the sizes of the sources at \gev and \tev energies. Even so, the sizes
of PWNe are expected to vary across the \gev and \tev energy range and
the size of HESS\,J1825$-$137 is significantly larger at \gev than \tev
energies \citep{grondin_2011a_detection-pulsar}.  It is interesting to
compare the sizes of other PWN candidates at \gev and \tev energies, but
definitively measuring a difference in size would require a more in-depth
analysis of the LAT data using the same elliptical Gaussian spatial model.

\begin{figure}[htbp]
  \includegraphics{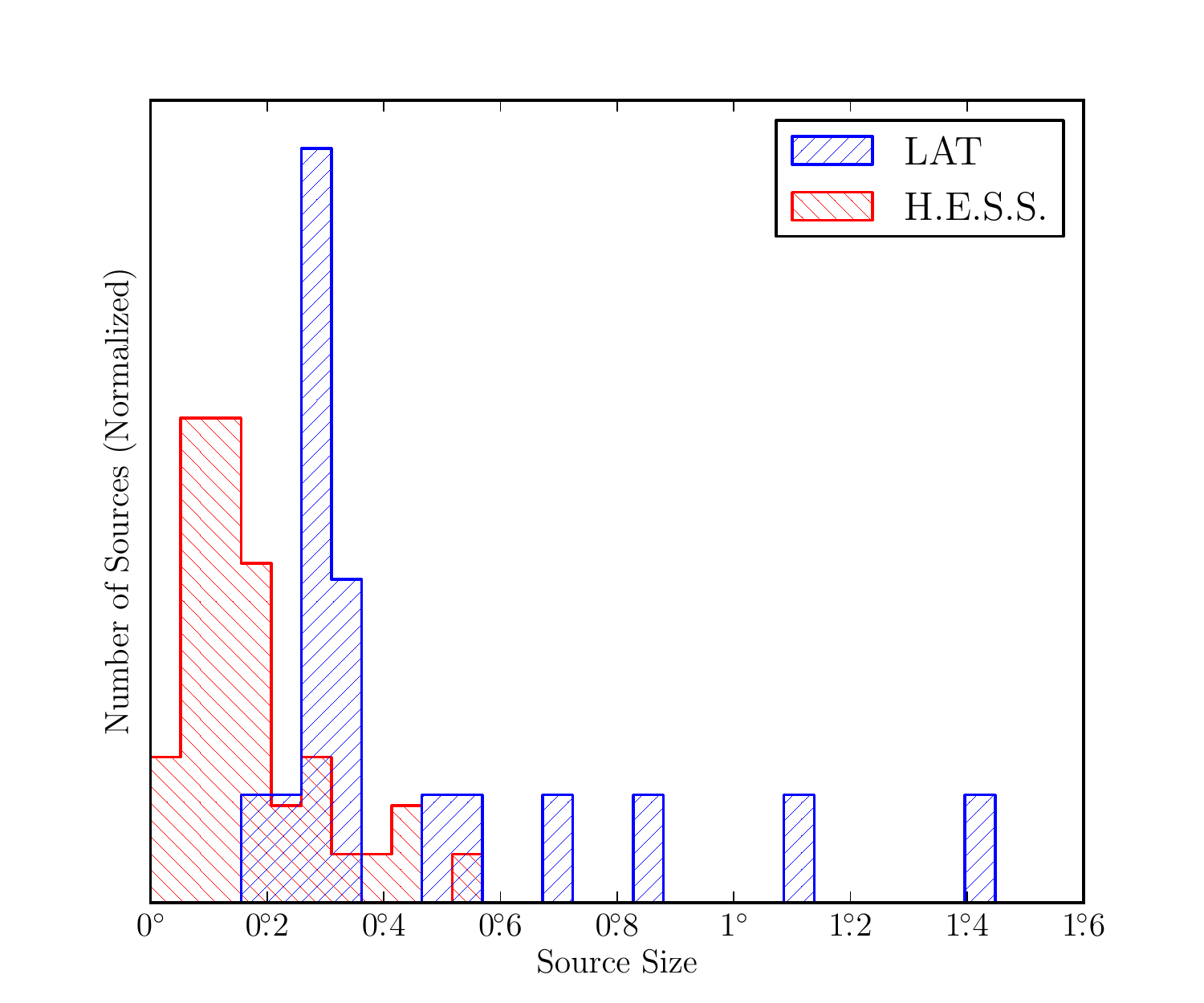}
  \caption{The distributions of the sizes of 18 extended LAT sources at
  \gev energies (colored blue in the electronic version) and the sizes of
  the 40 extended \ac{HESS} sources at \tev energies (colored red).  The
  \ac{HESS} sources were fit with a 2D Gaussian surface brightness profile
  so the LAT and \ac{HESS} sizes were first converted to \rsixeight.
  The \gev size of Vela X is taken from \cite{abdo_2010c_fermi-large}.
  Because of their large sizes, the shape of RX\,J1713.7$-$3946 and Vela
  Jr. were not directly fit at \tev energies and are not included in this
  comparison.  Centaurus A is not included because of its large size.}
  \figlabel{gev_vs_tev_histogram}
  \end{figure}

\figref{gev_vs_tev_histogram} compares the sizes of the 21 extended LAT
sources to the 42 extended \ac{HESS} sources.\footnote{The \tev extension
of the 42 extended \ac{HESS} sources comes from the \ac{HESS} Source
Catalog \url{http://www.mpi-hd.mpg.de/hfm/HESS/pages/home/sources/}.}
Because of the large field of view and all-sky coverage, the LAT can
more easily measure larger sources.  On the other hand, the better
angular resolution of \acp{IACT} allows them to measure a population
of extended sources below the resolution limit of the LAT (currently
about $\sim0\fdg2$).  \fermi has a 5 year nominal mission lifetime with
a goal of 10 years of operation.  As \figref{time_sensitivity} shows,
the low background of the LAT at high energies allows its sensitivity
to these smaller sources to improve by a factor greater than the square
root of the relative exposures.  With increasing exposure, the LAT will
likely begin to detect and resolve some of these smaller \tev sources.

\begin{figure}[htbp]
  \includegraphics{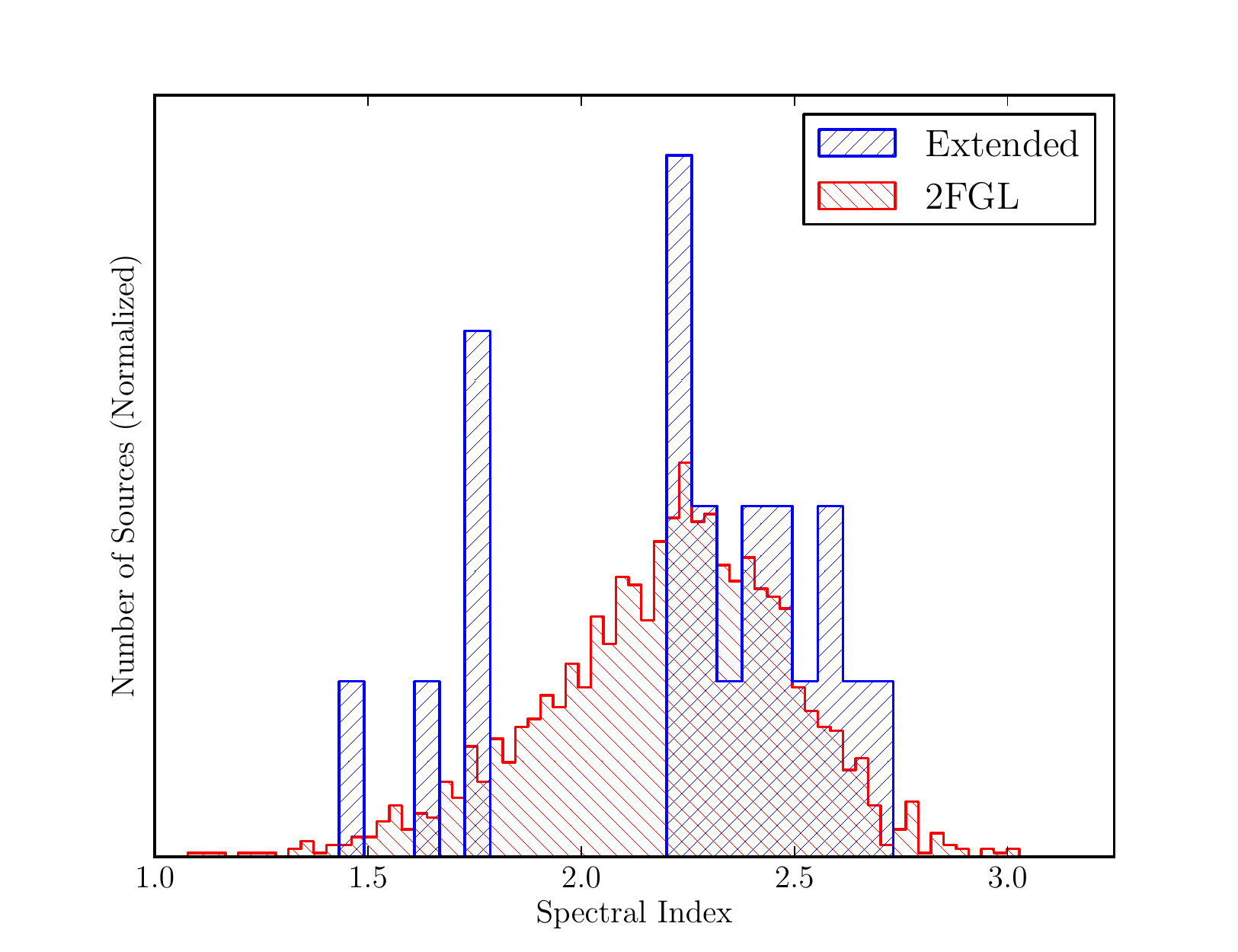}
  \caption{The distribution of spectral indices of the 1873 2FGL sources
  (colored red in the electronic version) and the 21 spatially extended
  sources (colored blue).  The index of Centaurus A is taken from
  \cite{nolan_2012_fermi-large} and the index of Vela X is taken from
  \cite{abdo_2010c_fermi-large}.}
  \figlabel{compare_index_2FGL}
\end{figure}

\figref{compare_index_2FGL} compares the spectral indices of LAT
detected extended sources and of all sources in the 2FGL catalog. This,
and \tabref{known_extended_sources} and \tabref{new_ext_srcs_table},
show that the LAT observes a population of hard extended sources at
energies above 10 \gev.  \figref{hess_seds} shows that the spectra of
four of these sources (2FGL\,J1615.0$-$5051, 2FGL\,J1615.2$-$5138,
2FGL\,J1632.4$-$4753c, and 2FGL\,J1837.3$-$0700c) at \gev energies
connects to the spectra of their \ac{HESS} counterparts at
\tev energies. This is also true of Vela Jr., HESS\,J1825$-$137
\citep{grondin_2011a_detection-pulsar}, and RX\,J1713.7$-$3946
\citep{abdo_2011a_observations-young}.  It is likely that the \gev and
\tev emission from these sources originates from the same population of
high-energy particles.

Many of the \tev-detected extended sources now seen at \gev energies
are currently unidentified and further multiwavelength follow-up
observations will be necessary to understand these particle accelerators.
Extending the spectra of these \tev sources towards lower energies with
LAT observations may help to determine the origin and nature of the
high-energy emission.

\chapter{Search for \Acsptitle{PWN} Associated with Gamma-loud Pulsars}
\chaplabel{offpeak}

\paperref{
  This chapter is based on section seven ``Unpulsed Magnetospheric Emission'' from  
  the paper ``The Second \fermi Large Area Telescope Catalog of Gamma-ray Pulsars''
  \citep{abdo_2013a_second-fermi}.}

Some pulsars have magnetospheric emission over their full rotation phase
with similar spectral characteristics to the emission seen through
their peaks.  This emission appears in the observed light curves as
a low-level, unpulsed component above the estimated background level
(i.e., not attributable to diffuse emission or nearby point sources)
and can be a powerful discriminator for the emission models.

On the other hand a PWN around the pulsar, or a photon excess due
to imprecise knowledge of diffuse emission around the pulsar, would
not be modulated at the rotational period and could be confused with
a constant magnetospheric signal.  Including the discovery of the GeV
PWN 3C 58 associated with PSR J0205+6449 described in this section, the
LAT sees 17 sources potentially associated with PWNe at GeV energies
\citep{acero_2013a_constraints-galactic}.  Some are highlighted in
\secref{off_peak_individual_source_discussion}.  This off-peak emission
should be properly modeled when searching for pulsar emission at all
rotation phases.

We can discriminate between these two possible signals through
spectral and spatial analysis.  If the emission is magnetospheric,
it is more likely to appear as a non-variable point source with
an exponentially cutoff spectrum with a well-known range of cutoff
energies.  On the other hand, PWNe and diffuse excesses have spectra
with a power-law shape and either a hard index continuing up to tens
of GeV in the PWN case or present only at lower energies with a very
soft index in the diffuse case.  In addition, PWNe are often spatially
resolvable at GeV energies \citep[e.g., Vela-X has been spatially
resolved with the LAT and \textit{AGILE} and HESS J1825$-$137 with the
LAT;][respectively]{abdo_2010c_fermi-large,pellizzoni_2010a_detection-gamma-ray,grondin_2011a_detection-pulsar}
so an extended source would argue against a magnetospheric origin of the
emission.  However, given the finite angular resolution of the LAT not all
PWNe will appear spatially extended at GeV energies.  The Crab Nebula,
for instance, cannot be resolved by the LAT but can be distinguished
from the gamma-bright Crab pulsar, in the off-peak interval, by its hard
spectrum above $\sim$1 GeV \citep{abdo_2010a_fermi-large}.  In addition,
GeV emission from the Crab Nebula was discovered to be time-variable
\citep[e.g.,][]{abdo_2011a_gamma-ray-flares} providing another possible
way to discern the nature of any observed off-peak signal.

Therefore, to identify pulsars with magnetospheric emission across
the entire rotation, we define and search the off-peak intervals
of the pulsars in this catalog for significant emission, except PSR
J2215+5135 for which the rotation ephemeris covers a short time interval
and the profile is noisy.  We then evaluate the spectral and spatial
characteristics of any off-peak emission to determine if it is likely
magnetospheric, related to the pulsar wind, or physically unrelated to
the pulsar (e.g., unmodeled diffuse emission).

\section{Off-peak Phase Selection}
\seclabel{peak_definition}

We first developed a systematic, model-independent, and
computationally-efficient method to define the off-peak interval of a
pulsar light curve.

We begin by deconstructing the light curve into simple Bayesian Blocks
using the algorithm described in \citet{jackson_2005a_algorithm-optimal}
and \citet{scargle_2013a_studies-astronomical}.  We could not apply the
Bayesian Block algorithm to the weighted-counts light curves because
they do not follow Poisson statistics, required by the algorithm.
We therefore use an unweighted-counts light curve in which the angular
radius and minimum energy selection have been varied to maximize the
H-test statistic.  To produce Bayesian Blocks on a periodic light curve,
we extend the data over three rotations, by copying and shifting the
observed phases to cover the phase range from $-$1 to 2.  We do, however,
define the final blocks to be between phases 0 and 1.  To avoid potential
contamination from the trailing or leading edges of the peaks, we reduce
the extent of the block by 10\% on either side, referenced to the center
of the block.

There is one free parameter in the Bayesian Block algorithm called ncp$\rm
_{prior}$ which modifies the probability that the algorithm will divide
a block into smaller intervals.  We found that, in most cases, setting
ncp$\rm _{prior}=8$ protects against the Bayesian Block decomposition
containing unphysically small blocks.  For a few marginally-detected
pulsars, the algorithm failed to find more than one block and we had to
decrease ncp$\rm _{prior}$ until the algorithm found a variable light
curve. Finally, for a few pulsars the Bayesian-block decomposition of the
light curves failed to model weak peaks found by the light-curve fitting
method presented in \citep{abdo_2013a_second-fermi} or extended too far
into the other peaks. For these pulsars, we conservatively shrink the
off-peak region.

For some pulsars, the observed light curve has two well-separated peaks
with no significant bridge emission, which leads to two well-defined
off-peak intervals.  We account for this possibility by finding the
second-lowest Bayesian block and accepting it as a second off-peak
interval if the emission is consistent with that in the lowest block
(at the 99\% confidence level) and if the extent of the second block is
at least half that of the first block.

\figref{off_peak_select} shows the energy-and-radius optimized light
curves, the Bayesian block decompositions, and the off-peak intervals for
six pulsars.  \citep{abdo_2013a_second-fermi} overlay off-peak intervals
over the weighted light curves of several pulsars.  The off-peak intervals
for all pulsars are given in \citep{abdo_2013a_second-fermi}.

\begin{figure}
  \includegraphics{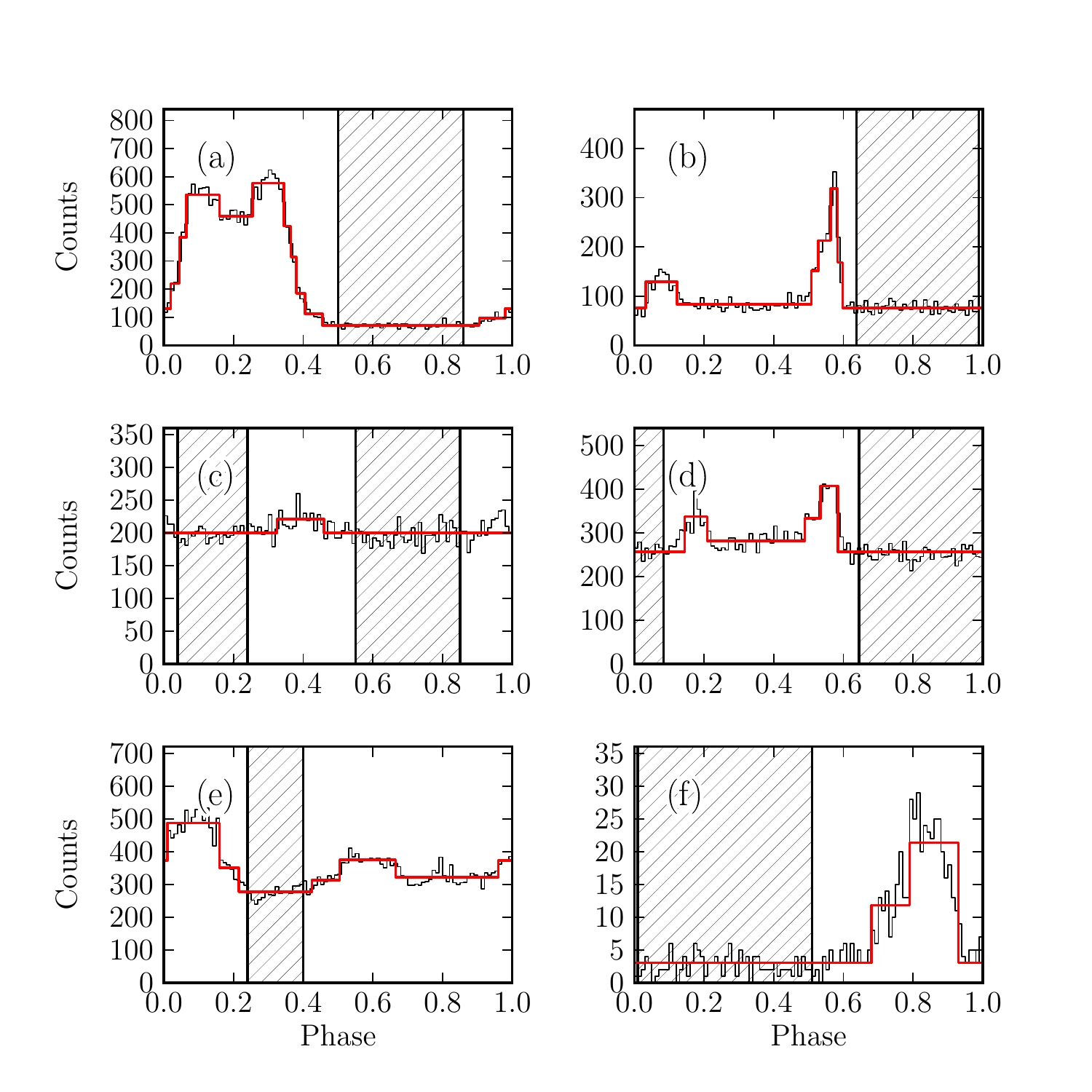}
  \caption{The energy-and-radius optimized light curve, Bayesian
  block decomposition of the light curve, and off-peak interval for
  (a) PSR J0007+7303, (b) PSR J0205+6449, (c) PSR J1410$-$6132, (d) PSR
  J1747$-$2958, (e) PSR J2021+4026, and (f) PSR J2124$-$3358.  The black
  histograms represent the light curves, the gray lines (colored red in
  the electronic version) represent the Bayesian block decompositions of
  the pulsar light curves, and the hatched areas represent the off-peak
  intervals selected by this method.}
  \figlabel{off_peak_select}
\end{figure}

\section{Off-peak Analysis Method}
\seclabel{off_peak_analysis}

Characterizing both the spatial and spectral characteristics of any
off-peak emission helps discern its origin.  We employ a somewhat
different analysis procedure here than for the phase-averaged analysis
described in \citep{abdo_2013a_second-fermi}.  To evaluate the spatial
characteristics of any off-peak emission we use the likelihood fitting
package \pointlike \citep[detailed in][]{lande_2012_search-spatially}, and
to fit the spectrum we use \gtlike in binned mode via {\it pyLikelihood}
as was done for the phase-averaged analysis.

For each pulsar we start from the same temporal and spatial event
selections described in \citep{abdo_2013a_second-fermi} but we
increase the maximum energy to 400 GeV (the highest event energy for
any ROI under this selection is $\sim$316 GeV).  For the \pointlike
analysis we further select a 10$\degree$ radius ROI and for \gtlike
a $14\degree\times14\degree$ square ROI, both centered on the pulsar
position.  Finally, we only consider photons with pulse phases within
the corresponding off-peak interval.

We search for off-peak emission assuming a point source and (except for
the Crab Nebula and Vela-X, described below) a power-law spectral model.
We fit the position of this putative off-peak source using \pointlike
as described by \citet{nolan_2012_fermi-large} and then use the best-fit
position in a spectral analysis with \gtlike.  From the spectral analysis
we require $TS\geq25$ (just over $4\sigma$) to claim a detection.
If $TS<25$, we compute upper limits on the flux in the energy range
from 100 \mev to 316 \gev assuming a power law with photon index fixed
to 2.0 and a PLEC1 model with $\Gamma=1.7$ and $E_{\rm cut}=3$ GeV.

The spectrum of the Crab Nebula (associated with PSR J0534+2200)
is uniquely challenging because the GeV spectrum contains
both a falling synchrotron and a rising inverse Compton
component \citep{abdo_2010a_fermi-large}.  For this particular
source we used the best-fit two-component spectral model from
\cite{buehler_2012a_gamma-ray-activity} and fit only the overall
normalization of the source.  In addition, for Vela-X (associated
with PSR J0835-4510) we took the best-fit spectral model from
\cite{grondin_2013a_vela-x-pulsar} and fit only the overall normalization
of this source. This spectrum has a smoothly broken power law spectral
model and was fit assuming Vela-X to have an elliptical disk spatial
model.

If the off-peak source is significant, we test whether
the spectrum shows evidence for a cutoff, as described in
\citet{ackermann_2011a_fermi-lat-search}, assuming the source is at the
pulsar position.  We say that the off-peak emission shows evidence for
a cutoff if $TS_{\rm cut}\geq9$, corresponding to a $3\sigma$ detection.

For a significant off-peak point source, we use \pointlike to test if the
emission is significantly extended.  We assume a radially-symmetric
Gaussian source and fit the position and extension parameter
($\sigma$) as described in \citet{lande_2012_search-spatially}.
The best-fit extended source parameters are then given to \gtlike,
which is used to fit the spectral parameters and the significance of the
extension over a point source, $TS_{\rm ext}$, evaluated as described in
\citet{lande_2012_search-spatially}.  That paper established that $TS_{\rm
ext}\geq16$ means highly probable source extension.  In the present work
we aim only to flag possible extension, and use  $TS_{\rm ext}\geq9$.

To test for variability, even without significant emission over the
3-year time range, we divide the dataset into 36 intervals and fit the
point-source flux independently in each interval, computing $TS_{\rm
var}$ as in 2FGL.  For sources with potential magnetospheric off-peak
emission and for regions with no detection, we performed the test at
the pulsar's position.  Otherwise, we test at the best-fit position.
The off-peak emission is said to show evidence for variability if
$TS_{\rm var}\geq91.7$, corresponding to a $4\sigma$ significance.  %
As noted in \citep{abdo_2013a_second-fermi}, our timing solutions for
PSRs J0205+6449 and J1838$-$0537 are not coherent across all three years.
For these two pulsars, we excluded the time ranges without ephemerides and
only tested for variability during months that were completely covered.
For J1838$-$0537 only one month is lost, whereas for J0205+6449 the 7\%
data loss is spread across three separate months.  As a result, $TS_{\rm
var}$ for these pulsars is a conservative estimate of variability
significance.

The procedure described above, especially the extension analysis, is
particularly sensitive to sources not included in 2FGL that are near the
pulsar of interest, for two reasons.  First, we are using an additional
year of data and second, when ``turning off'' a bright pulsar nearby,
faint sources become more important to the global fit.  Therefore, in
many situations we had to run the analysis several times, iteratively
improving the model by including new sources, until we removed all $TS>25$
residuals. The final \gtlike-formatted XML source model for each off-peak
region is included in the auxiliary material.

There are still, however, pulsars for which we were unable to obtain an
unbiased fit of the off-peak emission, most likely due to inaccuracies in
the model of the Galactic diffuse emission and incorrectly modeled
nearby sources.  The most common symptom of a biased fit is an
unphysically large extension.  In these cases, the extended source
attempts to account for multiple point sources or incorrectly-modeled
diffuse emission, not just the putative off-peak emission.  Systematics
associated with modeling extended sources are discussed more thoroughly
in \citet{lande_2012_search-spatially}.  For the purposes of this catalog,
we have flagged the pulsars where off-peak analysis suffered from these
issues and do not attempt a complete understanding of the emission.

Observations of magnetospheric off-peak emission can be used to constrain
pulsar geometry. Therefore, it is important to know if off-peak emission
that is otherwise pulsar-like might instead be incorrectly-modeled
Galactic diffuse emission.  We therefore performed a limited study of the
systematics associated with our model of the Galactic diffuse emission.

For sources which otherwise would be classified as magnetospheric,
we tested the significance of the emission assuming eight
different Galactic diffuse emission models as described in
\citet{de-palma_2013a_method-exploring}.  These models were constructed
using a different model building strategy, vary parameters of the Galactic
diffuse emission model, and have additional degrees of freedom in the fit.

We define $\tsaltdiff$ as the minimum test statistic of any of
the eight alternate diffuse models. This test statistic is computed
assuming the emission to be pointlike at the best-fit position and is
therefore comparable to \tspoint.  For sources which would otherwise be
considered magnetospheric, we flag the emission as potentially spurious
if $\tsaltdiff<25$.  We caution that although this test can help to
flag problematic regions, these models do not probe the entirety of the
uncertainty associated with our model of the Galactic diffuse emission.
Therefore, some diffuse emission could still be incorrectly classified
as magnetospheric.

\section{Off-peak Results}
\seclabel{off_peak_results}

The off-peak intervals of 54 LAT-detected pulsars have been evaluated
by \citet{ackermann_2011a_fermi-lat-search} using 16 months of sky
survey observations.  This led to the discovery of PWN-like emission
in the off-peak interval of PSR J1023$-$5746, coincident with HESS
J1023$-$575, and identification of 5 pulsars that appear to have near
100\% duty cycles.  Our results, summarized in \tabref{off_peak_table},
extend the analysis to 116 pulsars over 3 years of data.  Sample off-peak
spectra are shown in \figref{off_peak_seds}.  Using the procedures
outlined in \secref{peak_definition} and \secref{off_peak_analysis}, we
have identified 34 pulsars that have significant emission ($TS\geq25$)
in their off-peak intervals.  We classify the likely nature of the
emission as follows.

If the emission has $TS_{\rm cut}\geq9$, we consider the emission to be
either magnetospheric (`M') or possibly magnetospheric (`M*').  As was
discussed in \secref{off_peak_analysis}, we flag the emission as `M*'
if the source is formally spatially extended ($TS_{\rm ext}>16$) or if
the source is not robust against varying the diffuse emission models
($\tsaltdiff<25$).  On consideration of the angular extent of the PSF
of the LAT and inaccuracies in the Galactic diffuse emission model,
we caution against considering the `M*' sources to be definitively
classified.  If the source is significantly cutoff, not significantly
extended, and is significant when varying the alternative diffuse models,
we classify the emission as `M'.

On the other hand, if the emission has $TS_{\rm ext}\geq16$,
and does not suffer from confusion as discussed at the end of
\secref{off_peak_analysis}, and/or has a hard photon index, we say it is
likely to originate in the pulsar wind and identify these sources as type
`W'.  The remaining sources with off-peak emission not satisfying any of
the previous criteria are identified as type `U' to indicate that the
nature of the emission is unidentified and we do not speculate about
its origin.

We identify 9 type `M' sources, significantly expanding the number of
pulsars that perhaps have detectable magnetospheric emission across
all rotational phases.  One caution is that many of these `M' pulsars,
especially the young objects, are in regions of particularly bright
diffuse gamma-ray emission, where small fractional uncertainties in
the level of diffuse emission can account for much of the apparent
unpulsed emission.  However, if established as true magnetospheric
components, these will be important test cases for pulsar emission
models. In addition, we identify ten `M*'-type regions.  For type `M'
and `M*' sources, we present the best-fit spectral parameters using a
point source at the pulsar's position with a PLEC1 spectral model in
\tabref{off_peak_table}.  For all other sources (except the Crab Nebula
described in \secref{off_peak_analysis}), we present the spectral results
using a power-law spectral model and the best-fit spatial representation.

Additionally, we identified four off-peak emissions consistent with
a PWN hypothesis, one of them being a new detection at GeV energies
(PSR J0205+6449).  Only one of these four, the previously identified
Vela$-$X PWN \citep{abdo_2010c_fermi-large}, is spatially extended for
the LAT.  Similarly, we detect six type `U' regions. Three of these are
formally spatially extended but because of the spatial systematics we
assume point-like emission for the spectral analysis.

We mention that for a few sources, the spectral analysis
performed here is in disagreement with the analysis presented in
\citet{ackermann_2011a_fermi-lat-search}. For soft and faint sources
(including J1044$-$5737 and J1809$-$2332), the spectral discrepancy is
mainly caused by our use of a newer Galactic diffuse model. At lower
energies, small changes in the diffuse model can have a significant
impact on the analysis of a region.  For bright magnetospheric sources
(including J0633+1746 and J2021+4026), the spectral discrepancy is mainly
due to using different phase ranges (see \secref{peak_definition}).

\figref{off_peak_luminosity_vs_edot} shows that only a small fraction of
the spindown power goes into the gamma-ray emission from LAT-detected
PWNe.  Similarly, \cite{abdo_2013a_second-fermi} includes a plot
of $\sqrt{\energydot}/d^2$ vs $\PulsarAge$ which shows that the LAT
only detects PWNe from the youngest pulsars with the highest spindown
power.  GeV emission from the Crab Nebula is highly time variable
(\secref{off_peak_analysis}).  Indeed, we find $TS_{\rm var}=373$ for
the Crab Nebula; however no other source demonstrated flux variability
(all have $16 <TS_{\rm var} <65)$.  Other GeV PWNe may be variable,
but the combination of lower fluxes and less-extreme variations limits
our ability identify them as such.

The off-peak results for several interesting sources are presented in
\secref{off_peak_individual_source_discussion}.  The complete off-peak
search results will be included in the auxiliary information accompanying
\cite{abdo_2013a_second-fermi}.  For regions where we find $TS<25$,
the auxiliary information contains upper limits computed for both a
power-law spectral model and a PLEC1 model with $E_{\rm cut}=3$ GeV and
$\Gamma=1.7$.  The auxiliary information also contains $TS_{\rm var}$
for each off-peak interval.

\begin{deluxetable}{ll*{8}c}
  \tablecolumns{8}
  \tablewidth{0pt}
  \tabletypesize{\scriptsize}
  \tablecaption{Off-Peak Spatial and Spectral Results
  \tablabel{off_peak_table}
  }
  \tablehead{\colhead{PSR} & \colhead{Type} & \colhead{$\tspoint$} & \colhead{$\tsext$} & \colhead{$\tscutoff$} & \colhead{$\tsaltdiff$} & \colhead{Energy Flux} & \colhead{$\Gamma$} & \colhead{$\Ecutoff$}\\ \colhead{ } & \colhead{ } & \colhead{ } & \colhead{ } & \colhead{ } & \colhead{ } & \colhead{($10^{-11}\,\efluxunits$)} & \colhead{ } & \colhead{(GeV)}}
\startdata
\multicolumn{8}{c}{Young Pulsars} \\[3pt]
\hline
J0007+7303 & U & 71.4 & 10.8 & 0.0 &  & $1.98 \pm 0.26$ & $2.61 \pm 0.14$ &  \\
J0205+6449 & W & 33.7 & 0.5 & 0.0 &  & $1.75 \pm 0.68$ & $1.61 \pm 0.21$ &  \\
J0534+2200 & W & 5247. & 0.0 & 0.3 &  & $67.2 \pm 1.6$ & \tablenotemark{a} &  \\
J0631+1036 & U & 33.1 & 0.0 & 5.4 &  & $1.70 \pm 0.33$ & $2.38 \pm 0.14$ &  \\
J0633+1746 & M & 3666. & 2.3 & 239. & 3369. & $41.4 \pm 1.1$ & $1.37 \pm 0.09$ & $0.93 \pm 0.10$ \\
J0734$-$1559 & M* & 28.3 & 12.4 & 30.8 & 0.0 & $1.61 \pm 0.24$ & $0.01 \pm 0.08$ & $0.17 \pm 0.03$ \\
J0835$-$4510 & W & 473. & 283. & 22.8 &  & $30.3 \pm 1.2$ & \tablenotemark{b} &  \\
J0908$-$4913 & M* & 65.1 & 41.4 & 60.4 & 3.1 & $3.04 \pm 1.07$ & $0.15 \pm 0.59$ & $0.30 \pm 0.01$ \\
J1023$-$5746 & M* & 59.7 & 30.0 & 10.9 & 72.5 & $5.35 \pm 1.17$ & $0.57 \pm 0.80$ & $0.49 \pm 0.21$ \\
J1044$-$5737 & M* & 42.0 & 98.1 & 22.4 & 25.6 & $3.12 \pm 0.75$ & $0.80 \pm 0.93$ & $0.40 \pm 0.18$ \\
J1105$-$6107 & M* & 33.3 & 37.5 & 21.7 & 39.4 & $3.81 \pm 0.77$ & $0.92 \pm 0.56$ & $0.48 \pm 0.22$ \\
J1112$-$6103 & U & 65.0 & 71.1 & 0.9 &  & $5.10 \pm 0.74$ & $2.17 \pm 0.09$ &  \\
J1119$-$6127 & U & 61.3 & 1.0 & 0.9 &  & $4.11 \pm 0.63$ & $2.22 \pm 0.09$ &  \\
J1124$-$5916 & M & 95.9 & 0.0 & 18.2 & 59.4 & $2.87 \pm 0.71$ & $1.31 \pm 0.91$ & $1.43 \pm 1.42$ \\
J1410$-$6132 & U & 27.5 & 71.2 & 0.4 &  & $4.29 \pm 1.05$ & $1.90 \pm 0.15$ &  \\
J1513$-$5908 & W & 102. & 3.5 & 0.0 &  & $4.95 \pm 0.83$ & $1.78 \pm 0.12$ &  \\
J1620$-$4927 & M* & 28.9 & 0.5 & 35.2 & 0.0 & $5.25 \pm 0.96$ & $0.35 \pm 0.94$ & $0.57 \pm 0.29$ \\
J1746$-$3239 & M* & 53.3 & 34.3 & 34.2 & 0.0 & $3.65 \pm 0.59$ & $0.94 \pm 0.31$ & $0.60 \pm 0.10$ \\
J1747$-$2958 & M & 45.5 & 5.4 & 49.8 & 50.4 & $8.41 \pm 2.84$ & $0.02 \pm 0.32$ & $0.28 \pm 0.01$ \\
J1809$-$2332 & M* & 32.5 & 13.6 & 21.9 & 0.0 & $4.10 \pm 0.80$ & $0.24 \pm 0.83$ & $0.31 \pm 0.11$ \\
J1813$-$1246 & M & 62.8 & 0.0 & 9.0 & 49.7 & $6.31 \pm 1.40$ & $1.60 \pm 0.73$ & $0.99 \pm 0.95$ \\
J1836+5925 & M & 10407. & 0.0 & 365. & 10401. & $36.9 \pm 0.7$ & $1.47 \pm 0.03$ & $1.98 \pm 0.09$ \\
J1838$-$0537 & M* & 51.3 & 32.9 & 21.9 & 41.9 & $8.35 \pm 1.31$ & $1.39 \pm 0.54$ & $2.55 \pm 2.48$ \\
J2021+4026 & M & 1717. & 8.7 & 244. & 1978. & $64.0 \pm 1.4$ & $1.64 \pm 0.02$ & $1.82 \pm 0.04$ \\
J2032+4127 & U & 53.6 & 76.1 & 1.5 &  & $4.36 \pm 0.77$ & $2.07 \pm 0.12$ &  \\
J2055+2539 & M & 123. & 0.0 & 30.0 & 101. & $1.63 \pm 0.19$ & $1.05 \pm 0.28$ & $0.64 \pm 0.12$ \\
\hline\\[-5pt]
\multicolumn{8}{c}{Millisecond Pulsars} \\[3pt]
\hline
J0034$-$0534 & U & 41.0 & 0.0 & 6.0 &  & $0.82 \pm 0.16$ & $2.40 \pm 0.19$ &  \\
J0102+4839 & U & 49.7 & 0.0 & 7.4 &  & $1.29 \pm 0.20$ & $2.51 \pm 0.14$ &  \\
J0218+4232 & U & 50.1 & 0.0 & 6.8 &  & $2.13 \pm 0.33$ & $2.72 \pm 0.26$ &  \\
J0340+4130 & M* & 26.9 & 0.1 & 16.3 & 11.9 & $0.53 \pm 0.11$ & $0.02 \pm 0.22$ & $0.94 \pm 0.28$ \\
J1658$-$5324 & U & 42.3 & 0.0 & 1.9 &  & $1.69 \pm 0.29$ & $2.52 \pm 0.76$ &  \\
J2043+1711 & U & 52.5 & 0.0 & 8.8 &  & $1.46 \pm 0.27$ & $2.29 \pm 0.14$ &  \\
J2124$-$3358 & M & 129. & 0.0 & 19.8 & 118. & $1.08 \pm 0.15$ & $0.70 \pm 0.51$ & $1.21 \pm 0.49$ \\
J2302+4442 & M & 114. & 0.0 & 9.8 & 105. & $1.45 \pm 0.20$ & $1.54 \pm 0.40$ & $1.61 \pm 0.82$ \\
\enddata

  \tablenotetext{a}{The spectral shape of the Crab Nebula
  was taken from \citet{buehler_2012a_gamma-ray-activity}.}
  \tablenotetext{b}{The spectral shape of Vela$-$X was taken from
  \cite{grondin_2013a_vela-x-pulsar}.}

  \tablecomments{Off-peak regions with a significant detection of
  emission.  The source classification is `M' for likely magnetospheric,
  `M*' for possibly magnetospheric but with a problematic spatial analysis
  or in a region with possibly poorly-modeled Galactic diffuse emission,
  `W' for likely pulsar wind, and `U' for unidentified.  The table
  includes the significance of the source ($TS$), of the source extension
  ($TS_{\rm ext}$), of a spectral cutoff ($TS_{\rm cut}$), and with the
  alternative diffuse models ($\tsaltdiff$).  The best-fit energy flux and
  photon index are computed in the energy range from 100 \mev to 316 \gev.
  For sources with large $TS_{\rm cut}$, the exponential cutoff energies
  are presented.  The quoted errors are statistical only. A few sources
  are discussed in \secref{off_peak_individual_source_discussion}.}

\end{deluxetable}

\begin{figure}
  \includegraphics{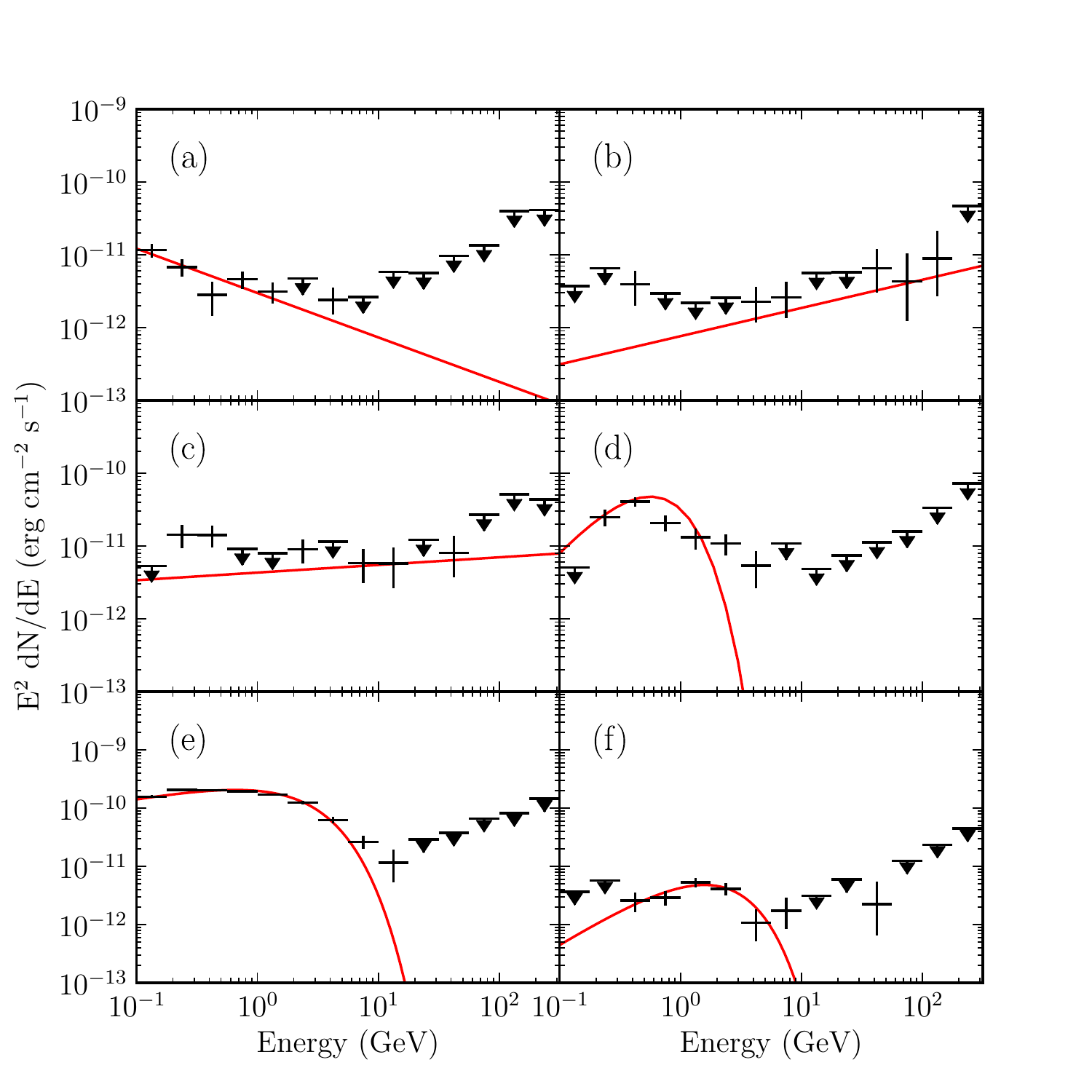}
  \caption{Spectral energy distributions for the off-peak phase intervals
  around (a) PSR J0007+7303 (b) PSR J0205+6449, (c) PSR J1410$-$6132,
  (d) PSR J1747$-$2958, (e) PSR J2021+4026, and (f) PSR J2124$-$3358.
  We plot a detection in those energy bands in which the source is found
  with $TS\geq4$ (a $2\sigma$ detection) and report a Bayesian 95\%
  confidence-level upper limit otherwise.  The best-fit spectral model,
  using the full energy range, is also shown for comparison.}
  \figlabel{off_peak_seds}
\end{figure}

\begin{figure}
  \includegraphics{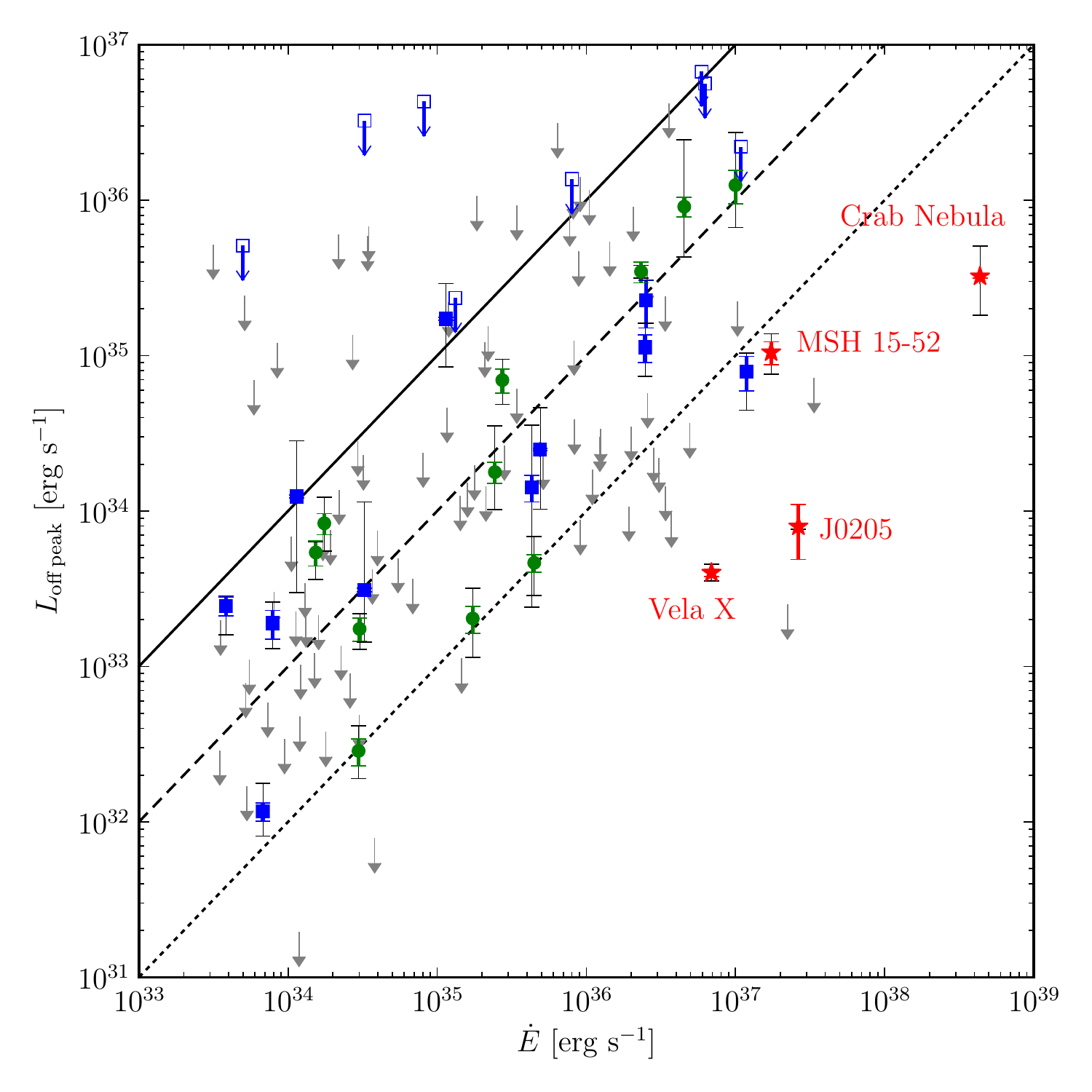}
  \caption{The off-peak luminosity compared to the observed pulsar
  spindown power.  The luminosity is computed and plotted with the
  same convention as luminosities in \cite{abdo_2013a_second-fermi}.
  A luminosity upper limit is plotted when there is no significant
  off-peak emission or when there is only a distance upper limit.  The
  star-shaped markers (colored red in the online version) represent type
  `W' sources, the square-shaped markers (colored blue) represent type `M'
  and `M*' sources, circular markers (colored green) represent type `U'
  sources, and the gray arrows represent non-detections.  The filled blue
  square-shaped markers represent `M' and `M*' sources with a detected
  luminosity and the unfilled markers represent luminosity upper limits
  where there is only a distance upper limit.  The solid, dashed, and
  dotted diagonals show 100\%, 10\%, and 1\% efficiency (respectively).}
  \figlabel{off_peak_luminosity_vs_edot}
\end{figure}

\section{Off-Peak Individual Source Discussion}
\seclabel{off_peak_individual_source_discussion}

Here we discuss several interesting sources found in the off-peak
analysis.

The off-peak emission from PSR J0007+7303 in the SNR CTA1 was previously
studied by \cite{abdo_2012a_j00077303-supernova}.  They found a soft and
not-significantly cut off source in the off-peak region that is marginally
extended.  We find a similar spectrum and extension significance ($TS_{\rm
ext}=10.8$), and therefore classify this source as type `U'.

The new type `W' source is associated with PSR J0205+6449
\citep{abdo_2009a_discovery-pulsations}.  The off-peak spectrum for this
source is shown in panel b of \figref{off_peak_seds}.  The emission
is best fit as a point source at $(l,b)=(130\fdg73,3\fdg11)$ with a
95\% confidence-level radius of $0\fdg03$.  The source has a hard
spectrum (power law with $\Gamma=1.61\pm0.21$) and is therefore
consistent with a PWN hypothesis.  This nebula has been observed
at infrared \citep{slane_2008a_infrared-detection} and X-ray
\citep{slane_2004a_constraints-structure} energies. This suggests
that we could be observing the inverse Compton emission from the same
electrons powering synchrotron emission at lower energies.  The PWN
hypothesis is supported by the associated pulsar's very high $\dot
E=2.6\times10^{36}$ erg s$^{-1}$ and relatively young characteristic
age, $\tau_c = 5400$ yr. This is consistent with the properties of
other pulsars with LAT-detected PWN, and we favor a PWN interpretation.
We note that the discrepancy between our spectrum and the upper limit
quoted in \citet{ackermann_2011a_fermi-lat-search} is mainly caused by
our expanded energy range and because the flux upper limit was computed
assuming a different spectral index.

However, we note that PSR J0205+6449 is associated to the
SNR 3C58 (G130.7+3.1).  Given the 2 kpc distance estimate from
\cite{abdo_2013a_second-fermi} and the density of thermal material
estimated by \cite{slane_2004a_constraints-structure}, we can estimate
the energetics required for the LAT emission to originate in the SNR.
Following the prescription in \cite{drury_1994a_gamma-ray-visibility},
we assume the LAT emission to be hadronic and estimate a cosmic-ray
efficiency for the SNR of $\sim10$\%, which is energetically allowed.
We therefore cannot rule out the SNR hypothesis.

No TeV detection of this source has been reported, but given the hard
photon index at GeV energies this is a good candidate for observations
by an atmospheric Cherenkov telescope. Improved spectral and spatial
observations at TeV energies might help to uniquely classify the emission.

We obtain a flux for Vela-X which is $\sim10\%$ larger than the flux
obtained in \cite{grondin_2013a_vela-x-pulsar}. This discrepancy is
most-likey due to assuming a different spatial model for the emission
(radially-symmetric Gaussian compared to elliptical Gaussian).

PSR J1023$-$5746 is associated with the TeV PWN HESS J1023$-$575
\citep{aharonian_2007a_detection-extended}.  LAT emission from this
PWN was first reported in \citet{ackermann_2011a_fermi-lat-search}.
Because of the dominant low-energy magnetospheric emission, we
classify this as type `M' and not as a PWN.  A phase-averaged
analysis of this source for energies above 10 GeV is reported in
\citet{acero_2013a_constraints-galactic}.

PSR J1119$-$6127 \citep{parent_2011a_observations-energetic}
is associated with the TeV source HESS J1119$-$614\footnote{The
discovery of HESS J1119$-$614 was presented at the ``Supernova
Remnants and Pulsar Wind Nebulae in the Chandra Era'' in 2009. See
\url{http://cxc.harvard.edu/cdo/snr09/pres/DjannatiAtai\_Arache\_
v2.pdf}.}. Our off-peak analysis classifies this source as `U' because
its spectrum is soft and not significantly cut off. However, the SED
appears to represent a cutoff spectrum at low energy and a hard rising
spectrum at high energy.  \citet{acero_2013a_constraints-galactic}
significantly detect this PWN using the analysis procedure as described
for J1023$-$575.  We are likely detecting a composite of magnetospheric
emission at low energy and pulsar-wind emission at high energy.

PSR J1357$-$6429 \citep{lemoine-goumard_2011a_discovery-gamma-}
has an associated PWN HESS J1356$-$645 detected at TeV energies
\citep{h.e.s.s.collaboration_2011a_discovery-source}.  Our analysis
of the off-peak regions surrounding PSR J1357$-$6429 shows a source
positionally and spectrally consistent with HESS J1356$-$645,
but with significance just below detection threshold ($TS=21.0$).
\citet{acero_2013a_constraints-galactic} present significant emission
from this source.

The off-peak region of PSR J1410$-$6132
\citep{obrien_2008a_j1410-6132:-young} shows a relatively hard spectral
index of $1.90\pm0.15$, and the spectrum is not significantly cut off.
There is no associated TeV PWN and enough low-energy GeV emission is
present to caution against a clear PWN interpretation.  We classify this
source as `U', but further observations could reveal interesting emission.

PSR J2021+4026 is spatially coincident with the LAT-detected and
spatially extended Gamma Cygni SNR \citep{lande_2012_search-spatially}.
The off-peak emission from this pulsar is consistent with an
exponentially-cutoff spectrum and is therefore classified as type `M'.
The source's marginal extension ($TS_{\rm ext}=8.7$) is likely due to
some contamination from the SNR.

\chapter{Search for \Acsptitle{PWN} associated with \tev Pulsars}
\chaplabel{tevcat}

\paperref{This chapter is based the first part of the paper
  ``Constraints on the Galactic Population of TeV Pulsar Wind Nebulae using Fermi Large Area Telescope Observations''
  \citep{acero_2013a_constraints-galactic}.}

Spatial analysis of \fermi data is important in identifying $\gamma$-ray
emitting \acp{PWN}.  In \chapref{extended_analysis}, we developed a new
method to study spatially-extended sources.  In \chapref{extended_search},
we searched for extended \ac{2FGL} sources.  In the process, we analyzed
the $\gamma$-ray emitting \acp{PWN} \hessj{1825} and \mshfifteenfiftytwo
which had previously been detected.  In addition, we discovered three
additional spatially-extended \fermi sources coincident with \acp{PWN}
candidates (\hessj{1616}, \hessj{1632}, \hessj{1837}).

In \chapref{offpeak}, we then searched for $\gamma$-ray emitting \acp{PWN}
by looking in the off-peak emission of \ac{LAT}-detected pulsars. In
that analysis, we detected four $\gamma$-ray emitting \acp{PWN} (\velax,
the Crab Nebula, \mshfifteenfiftytwo, and \threecfiftyeight).

In this chapter, we continue the search for $\gamma$-ray emitting
\acp{PWN} by searching for \acp{PWN} which had previously been
detected at \ac{VHE} energies by \acp{IACT}. We note that the work
presented here is a very condensed version of the results presented in
\citep{acero_2013a_constraints-galactic}.  We refer to that publication
for a more detailed discussion of the analysis.

\section{List of \ac{VHE} \Acstitle{PWN} Candidates}
\seclabel{tevcat_list_vhe_pwn_candidates}

We took all sources detected at \ac{VHE} energies and potentially
associated with \acp{PWN} and performed a search at \gev energies for
$\gamma$-ray emission. As was seen in previous chapters, many \acp{PWN}
have been detected both at \gev and \ac{VHE} energies, we suspect that
we might find new $\gamma$-ray emitting \ac{PWN} by searching \ac{LAT}
data in the regions of \ac{VHE} \acp{PWN}.

In addition, there are several \acp{PWN} which have been detected at
\ac{VHE} energies which do not have associated $\gamma$-ray pulsars
(such as \hessj{1825} and \hessj{1837}).  We therefore suspected that
this search could find new $\gamma$-ray emitting \ac{PWN} which were
not previously discovered either in the off-peak search discussed in
\chapref{offpeak} or in another dedicated analyses.

We used \tevcat\footnote{\tevcat can be found at
\url{http://tevcat.uchicago.edu}.} to define our list of \ac{VHE}
sources.  \tevcat is a catalog of sources detected at \ac{VHE} energies by
\acp{IACT}. We selected all sources from this catalog where the emission
was classified as being due to a \ac{PWN}. In addition, we included
all \ac{UNID} sources within 5\degree of the galactic plane since they
could be due to a \ac{PWN}.  Finally, we included \hessj{1023} because,
although this source is classified as a massive star cluster in the
\tevcat, \cite{de-naurois_2013a_galactic-h.e.s.s.} suggested that the
emission could originate in a \ac{PWN}.  \tabref{TeV_sources} presents
a list of all sources included in our analysis

\begin{deluxetable}{llrrlll}
\tabletypesize{\tiny}
\tablecaption{List of analyzed \ac{VHE} sources
\label{tab:TeV_sources}}
\tablewidth{0pt}
\tablecolumns{7}

\tablehead{\colhead{Name} & \colhead{Class} & \colhead{$l$} & \colhead{$b$} & \colhead{Pulsar} & \colhead{\acs{2PC}} & \colhead{Reference}\\
\colhead{} & \colhead{} & \colhead{(deg.)} & \colhead{(deg.)} & \colhead{} & \colhead{} & \colhead{}}
\startdata
  VER~J0006$+$727 &  PWN & 119.58 &   10.20 &   PSR~J0007+7303 & Y &    \cite{mcarthur_2011a_observation-veritas} \\
 MGRO~J0631$+$105 &  PWN & 201.30 &    0.51 &   PSR~J0631+1036 & Y &    \cite{abdo_2009a_milagro-observations} \\
  MGRO~J0632$+$17 &  PWN & 195.34 &    3.78 &   PSR~J0633+1746 & Y &   \cite{abdo_2009a_milagro-observations}  \\
 HESS~J1018$-$589 & UNID & 284.23 & $-1.72$ & PSR~J1016$-$5857 & Y &    \cite{h.e.s.s.collaboration_2012a_discovery-emission}  \\
 HESS~J1023$-$575 &  MSC & 284.22 & $-0.40$ & PSR~J1023$-$5746 & Y &     \cite{h.e.s.s.collaboration_2011a_revisiting-westerlund} \\
 HESS~J1026$-$582 &  PWN & 284.80 & $-0.52$ & PSR~J1028$-$5819 & Y &    \cite{h.e.s.s.collaboration_2011a_revisiting-westerlund}  \\
 HESS~J1119$-$614 &  PWN & 292.10 & $-0.49$ & PSR~J1119$-$6127 & Y & Presentation\tablenotemark{a} \\
 HESS~J1303$-$631 &  PWN & 304.24 & $-0.36$ & PSR~J1301$-$6305 & N &     \cite{aharonian_2005a_serendipitous-discovery} \\
 HESS~J1356$-$645 &  PWN & 309.81 & $-2.49$ & PSR~J1357$-$6429 & Y &     \cite{h.e.s.s.collaboration_2011a_discovery-source} \\
 HESS~J1418$-$609 &  PWN & 313.25 &    0.15 & PSR~J1418$-$6058 & Y &     \cite{aharonian_2006a_discovery-wings} \\
 HESS~J1420$-$607 &  PWN & 313.56 &    0.27 & PSR~J1420$-$6048 & Y &     \cite{aharonian_2006a_discovery-wings} \\
 HESS~J1427$-$608 & UNID & 314.41 & $-0.14$ &          \nodata & N &     \cite{aharonian_2008a_very-high-energy-gamma-ray} \\
 HESS~J1458$-$608 &  PWN & 317.75 & $-1.70$ & PSR~J1459$-$6053 & Y &    \cite{de-los-reyes_2012a_newly-discovered} \\
 HESS~J1503$-$582 & UNID & 319.62 &    0.29 &          \nodata & N &    \cite{renaud_2008a_nature-j1503-582} \\
 HESS~J1507$-$622 & UNID & 317.95 & $-3.49$ &          \nodata & N &     \cite{h.e.s.s.collaboration_2011a_discovery-follow-up} \\
 HESS~J1514$-$591 &  PWN & 320.33 & $-1.19$ & PSR~J1513$-$5908 & Y &     \cite{aharonian_2005a_discovery-extended} \\
 HESS~J1554$-$550 &  PWN & 327.16 & $-1.07$ &          \nodata & N &    \cite{acero_2012a_detection-emission} \\
 HESS~J1616$-$508 &  PWN & 332.39 & $-0.14$ & PSR~J1617$-$5055 & N &    \cite{aharonian_2006a_h.e.s.s.-survey} \\
 HESS~J1626$-$490 & UNID & 334.77 &    0.05 &          \nodata & N &     \cite{aharonian_2008a_very-high-energy-gamma-ray} \\
 HESS~J1632$-$478 &  PWN & 336.38 &    0.19 &          \nodata & N &    \cite{aharonian_2006a_h.e.s.s.-survey} \\
 HESS~J1634$-$472 & UNID & 337.11 &    0.22 &          \nodata & N &    \cite{aharonian_2006a_h.e.s.s.-survey} \\
 HESS~J1640$-$465 &  PWN & 338.32 & $-0.02$ &          \nodata & N &    \cite{aharonian_2006a_h.e.s.s.-survey} \\
 HESS~J1702$-$420 & UNID & 344.30 & $-0.18$ & PSR~J1702$-$4128 & Y &    \cite{aharonian_2006a_h.e.s.s.-survey} \\
 HESS~J1708$-$443 &  PWN & 343.06 & $-2.38$ & PSR~J1709$-$4429 & Y &     \cite{h.e.s.s.collaboration_2011a_detection-very-high-energy} \\
 HESS~J1718$-$385 &  PWN & 348.83 & $-0.49$ & PSR~J1718$-$3825 & Y &     \cite{aharonian_2007a_discovery-candidate} \\
 HESS~J1729$-$345 & UNID & 353.44 & $-0.13$ &          \nodata & N &     \cite{h.e.s.s.collaboration_2011a_shell-type-morphology:} \\
 HESS~J1804$-$216 & UNID &   8.40 & $-0.03$ & PSR~J1803$-$2149 & Y &   \cite{aharonian_2006a_h.e.s.s.-survey}  \\
 HESS~J1809$-$193 &  PWN &  11.18 & $-0.09$ & PSR~J1809$-$1917 & N &     \cite{aharonian_2007a_discovery-candidate} \\
 HESS~J1813$-$178 &  PWN &  12.81 & $-0.03$ & PSR~J1813$-$1749 & N &    \cite{aharonian_2006a_h.e.s.s.-survey} \\
 HESS~J1818$-$154 &  PWN &  15.41 &    0.17 & PSR~J1818$-$1541 & N &   \cite{hofverberg_2011a_discovery-gamma-ray}  \\
 HESS~J1825$-$137 &  PWN &  17.71 & $-0.70$ & PSR~J1826$-$1334 & N &     \cite{aharonian_2006a_energy-dependent} \\
 HESS~J1831$-$098 &  PWN &  21.85 & $-0.11$ & PSR~J1831$-$0952 & N &    \cite{sheidaei_2011a_discovery-very-high-energy} \\
 HESS~J1833$-$105 &  PWN &  21.51 & $-0.88$ & PSR~J1833$-$1034 & Y &    \cite{djannati-atai_2008a_companions-lonely} \\
 HESS~J1834$-$087 & UNID &  23.24 & $-0.31$ &          \nodata & N &    \cite{aharonian_2006a_h.e.s.s.-survey} \\
 HESS~J1837$-$069 & UNID &  25.18 & $-0.12$ & PSR~J1836$-$0655 & N &    \cite{aharonian_2006a_h.e.s.s.-survey} \\
 HESS~J1841$-$055 & UNID &  26.80 & $-0.20$ & PSR~J1838$-$0537 & Y &     \cite{aharonian_2008a_very-high-energy-gamma-ray} \\
 HESS~J1843$-$033 & UNID &  29.30 &    0.51 &          \nodata & N &    \cite{hoppe_2008a_h.e.s.s.-survey} \\
 MGRO~J1844$-$035 & UNID &  28.91 & $-0.02$ &          \nodata & N &    \cite{abdo_2009a_milagro-observations} \\
 HESS~J1846$-$029 &  PWN &  29.70 & $-0.24$ & PSR~J1846$-$0258 & N &    \cite{djannati-atai_2008a_companions-lonely} \\
 HESS~J1848$-$018 & UNID &  31.00 & $-0.16$ &          \nodata & N &    \cite{chaves_2008a_j1848-018:-discovery} \\
 HESS~J1849$-$000 &  PWN &  32.64 &    0.53 &  PSR~J1849$-$001 & N &    \cite{terrier_2008a_discovery-pulsar} \\
 HESS~J1857$+$026 & UNID &  35.96 & $-0.06$ &   PSR~J1856+0245 & N &     \cite{aharonian_2008a_very-high-energy-gamma-ray} \\
 HESS~J1858$+$020 & UNID &  35.58 & $-0.58$ &          \nodata & N &     \cite{aharonian_2008a_very-high-energy-gamma-ray} \\
 MGRO~J1900$+$039 & UNID &  37.42 & $-0.11$ &          \nodata & N &    \cite{abdo_2009a_milagro-observations} \\
  MGRO~J1908$+$06 & UNID &  40.39 & $-0.79$ &   PSR~J1907+0602 & Y &     \cite{aharonian_2009a_detection-energy} \\
 HESS~J1912$+$101 &  PWN &  44.39 & $-0.07$ &   PSR~J1913+1011 & N &     \cite{aharonian_2008a_discovery-very-high-energy} \\
  VER~J1930$+$188 &  PWN &  54.10 &    0.26 &   PSR~J1930+1852 & N &   \cite{acciari_2010a_discovery-energy}  \\
MGRO~J1958$+$2848 &  PWN &  65.85 & $-0.23$ &   PSR~J1958+2846 & Y &    \cite{abdo_2009a_milagro-observations} \\
  VER~J1959$+$208 &  PSR &  59.20 & $-4.70$ &   PSR~J1959+2048 & Y &    \cite{hall_2003a_search-emissions} \\
  VER~J2016$+$372 & UNID &  74.94 &    1.15 &          \nodata & N &    \cite{aliu_2011a_observations-region} \\
  MGRO~J2019$+$37 &  PWN &  75.00 &    0.39 &   PSR~J2021+3651 & Y &    \cite{abdo_2007a_gamma-ray-sources} \\
 MGRO~J2031$+$41A & UNID &  79.53 &    0.64 &          \nodata & N &    \cite{abdo_2007a_gamma-ray-sources} \\
 MGRO~J2031$+$41B & UNID &  80.25 &    1.07 &   PSR~J2032+4127 & Y &    \cite{bartoli_2012a_observation-gamma} \\
  MGRO~J2228$+$61 &  PWN & 106.57 &    2.91 &   PSR~J2229+6114 & Y &    \cite{abdo_2009a_milagro-observations} \\
\enddata

\tablenotetext{a}{This source was presented at the "Supernova Remnants and Pulsar Wind
Nebulae in the Chandra Era", 2009. See \url{http://cxc.harvard.edu/cdo/snr09/pres/DjannatiAtai\_Arache\_v2.pdf}.}

\tablecomments{The \ac{VHE} sources that we search for at \gev
energies.  This classifications comes from \tevcat: \acs{PWN}
for \acl{PWN}, \ac{UNID} for \acl{UNID}, and \acs{MSC} for \acl{MSC}.
We include \hessj{1023} because it is potentially \aac{PWN}
\citep{de-naurois_2013a_galactic-h.e.s.s.}.  For sources with an
associated pulsar, column 4 includes the pulsar's name.  Column 5
describes if the pulsar has been detected by the LAT and included in
\ac{2PC} (See \chapref{offpeak}).}
\end{deluxetable}
\clearpage

\section{Analysis Method} 
\seclabel{tevcat_analysis_method}

Our spectral and spatial analysis method was very similar to the analysis
in \chapref{offpeak}. We used the same hybrid \pointlike/\gtlike approach
for fitting \ac{LAT} data and modeled the regions using the same standard
background models.

The largest difference was that this analysis was performed only for
$\energy>10\unitspace\gev$.  As can be seen in \chapref{offpeak},
for energies much lower than $10\unitspace\gev$ source analysis becomes
strongly biased by systematic errors associated with incorrectly modeling
the Galactic-diffuse emission.  On the other hand, the $\gamma$-ray
emission from \ac{PWN} is expected to be the rising component of an
\ac{IC} peak which falls at \ac{VHE} energies. Therefore, the emission
observed by the \ac{LAT} is expected to be hard and most significant
at higher energies. Therefore, we expect that starting the analysis  at
$10\unitspace\gev$ will significantly reduce systematics associated with
this analysis while preserving most of the space for discovery.

Because the analysis was performed only in this high energy range where
the \ac{PSF} of the \ac{LAT} is much improved, we used a smaller region
of interest (a radius of $5\degree$ in \pointlike and a square of size
$7\degree\times7\degree$ in \gtlike).  Another differences is that we
used an event class with less background contamination (Pass 7 Clean
instead of Pass 7 Source) and modeled nearby background sources using
\ac{1FHL} \citep{ackermann_2013a_first-fermi-lat}.

For our analysis, we assume the \gev emission from our source to have
a power-law spectral model and that the \gev spatial model was the
same as the published \ac{VHE} spatial model.  We define \tstev as
the likelihood-ratio test for the significance of the source assuming
this source model and claim a detection when $\tstev > 16$.  Since our
significance test has two degrees of freedom, the flux and spectral
index, following this corresponds to a $3.6\sigma$ detection threshold
(see \subsecref{monte_carlo_validation}). When a source is significantly
detected, we quote the best-fit spectral parameters.  Otherwise, we
derive an upper limit on the flux of any potential emission.  We note
that \cite{acero_2013a_constraints-galactic} performed a more detailed
morphological analysis which studied the overlap between the \gev and
\ac{VHE} emission for these sources. For brevity, we omit the details
and simply use the results.

Many of these \acp{PWN} candidates are in regions with \ac{LAT}-detected
pulsars.  For these sources, \cite{acero_2013a_constraints-galactic}
included the spectral and spatial results both with and without the
\ac{LAT}-detected pulsar in the background model. For simplicity,
we include only the analysis with the pulsar included in the
background model. We caution that this method could be biased in either
oversubtracting or undersubtracting the pulsar depending upon systematics
associated with the \ac{2FGL} fits of the pulsars.

There are three major sources of systematic uncertainties effecting
the spectrum of these sources. The first is due to uncertainty in our
modeling of the Galactic diffuse emission, which we estimate following the
method of \secref{systematic_errors_on_extension}.  The second is due to
uncertainty in the effective area, which We estimated using the method
described in \cite{ackermann_2012a_fermi-large}.  The final systematic
is due to our uncertainty in the true morphology of of the source. We
use as our systematic error the difference in spectrum when the source
is fit assuming the published \ac{VHE} spatial model and spatial model
fit from \ac{LAT} data.

\section{Sources Detected} 
\seclabel{tevcat_sources_detected}

\thispagestyle{empty}
\tabletypesize{\scriptsize}
\begin{deluxetable}{llccc}
\tablewidth{0pt}
\tablecolumns{5}
\tablecaption{Spatial and spectral results for detected \ac{VHE} sources
\tablabel{tevcat_spatial_spectral}}

\tablehead{ \colhead{Name} & \colhead{ID} & \colhead{\tstev} & \colhead{F$_{10\,\text{GeV}}^{316\,\text{GeV}}$} & \colhead{$\Gamma$} \\ 
\colhead{} & \colhead{} & \colhead{} & \colhead{($10^{-10}\fluxunits$)} & \colhead{}}
\startdata
HESS~J1018$-$589 &    O &  25 &   $1.5 \pm 0.5 \pm 0.7$ & $2.31 \pm 0.50 \pm 0.49$ \\
HESS~J1023$-$575 & PWNc &  52 &   $4.6 \pm 0.9 \pm 1.2$ & $1.99 \pm 0.24 \pm 0.32$ \\
HESS~J1119$-$614 & PWNc &  16 &   $2.0 \pm 0.6 \pm 0.8$ & $1.83 \pm 0.41 \pm 0.36$ \\
HESS~J1303$-$631 & PWNc &  37 &   $3.6 \pm 0.9 \pm 2.1$ & $1.53 \pm 0.23 \pm 0.37$ \\
HESS~J1356$-$645 &  PWN &  24 &   $1.1 \pm 0.4 \pm 0.5$ & $0.94 \pm 0.40 \pm 0.40$ \\
HESS~J1420$-$607 & PWNc &  36 &   $3.4 \pm 0.9 \pm 1.1$ & $1.81 \pm 0.29 \pm 0.31$ \\
HESS~J1507$-$622 &    O &  21 &   $1.5 \pm 0.5 \pm 0.5$ & $2.33 \pm 0.48 \pm 0.48$ \\
HESS~J1514$-$591 &  PWN & 156 &   $6.2 \pm 0.9 \pm 1.3$ & $1.72 \pm 0.16 \pm 0.17$ \\
HESS~J1616$-$508 & PWNc &  75 &   $9.3 \pm 1.4 \pm 2.3$ & $2.18 \pm 0.19 \pm 0.20$ \\
HESS~J1632$-$478 & PWNc & 137 &  $11.8 \pm 1.5 \pm 5.3$ & $1.82 \pm 0.14 \pm 0.19$ \\
HESS~J1634$-$472 &    O &  33 &   $5.6 \pm 1.3 \pm 2.5$ & $1.96 \pm 0.25 \pm 0.29$ \\
HESS~J1640$-$465 & PWNc &  47 &   $5.0 \pm 1.0 \pm 1.7$ & $1.95 \pm 0.23 \pm 0.20$ \\
HESS~J1708$-$443 &  PSR &  33 &   $5.5 \pm 1.3 \pm 3.5$ & $2.13 \pm 0.31 \pm 0.33$ \\
HESS~J1804$-$216 &    O & 124 &  $13.4 \pm 1.6 \pm 3.1$ & $2.04 \pm 0.16 \pm 0.24$ \\
HESS~J1825$-$137 &  PWN &  56 &   $5.6 \pm 1.2 \pm 9.0$ & $1.32 \pm 0.20 \pm 0.39$ \\
HESS~J1834$-$087 &    O &  27 &   $5.5 \pm 1.2 \pm 2.5$ & $2.24 \pm 0.34 \pm 0.42$ \\
HESS~J1837$-$069 & PWNc &  73 &   $7.5 \pm 1.3 \pm 4.2$ & $1.47 \pm 0.18 \pm 0.30$ \\
HESS~J1841$-$055 & PWNc &  64 &  $10.9 \pm 0.8 \pm 4.1$ & $1.60 \pm 0.27 \pm 0.33$ \\
HESS~J1848$-$018 & PWNc &  19 &   $7.4 \pm 1.9 \pm 2.7$ & $2.46 \pm 0.50 \pm 0.51$ \\
  HESS~J1857+026 & PWNc &  53 &   $4.2 \pm 0.3 \pm 1.3$ & $1.01 \pm 0.24 \pm 0.25$ \\
   VER~J2016+372 &    O &  31 &   $1.8 \pm 0.5 \pm 0.8$ & $2.45 \pm 0.44 \pm 0.49$ \\
\enddata

\tablecomments{The \ac{VHE} \acp{PWN} candidates significantly detected by
the \ac{LAT}.  Column 2 is our classfication of the \ac{LAT} emission.
Column 3 is \tstev, Column 4 is the observed flux, and Column 5 is
the spectral index. The two errors on the flux and spectral index are
statistical and systematic.}
\end{deluxetable}
\normalsize
\noindent
\clearpage

We detected 22 sources at \gev energies.  For significantly-detected
sources, we present the spatial and spectral results for these sources
in \tabref{tevcat_spatial_spectral}.  Flux upper limits for non-detected
sources as well as spectral points in three independent energy bins are
can be found in \cite{acero_2013a_constraints-galactic}.

We attempt to classify the \gev emission into four categories:
\PWNClass-type for sources where the \gev emission is clearly identified
as \aac{PWN}, \PWNcClass for sources where the \gev emission could potentially
be due to a \ac{PWN}, \PSRClass-type for sources where the emission is
most likely due to pulsed emission inside the pulsar's magnetosphere,
and \OtherClass-type (for other) when the true nature of emission is
uncertain.

We categorize a source as \PWNClass-type or \PWNcClass-type when the
emission has a hard spectrum which connects spectrally to the \ac{VHE}
spectrum and when there is some multiwavelength evidence that the
\gev and \ac{VHE} emission should be due to a \ac{PWN}.  We label
a source as \PWNClass-type when the \ac{VHE} emission suggests
more strongly that the emission is due to a \ac{PWN}.  We include
in \tabref{tevcat_spatial_spectral} the source classifications
for each source.  We will discuss the \PWNClass-type source
in \subsecref{tevcat_pwn_pwnc_type_sources}.  We label a source
as \PSRClass-type if the emission is soft, point-like, and strongly
effected by our inclusion of its associated \ac{2FGL} pulsar in the
background model.  We label a source as \OtherClass-type otherwise.

\subsection{\PWNClass-type and \PWNcClass-type Sources}
\subseclabel{tevcat_pwn_pwnc_type_sources}

In total, we detect fourteen sources which we classify as \PWNClass-type
or \PWNcClass-type. Five of these \ac{PWN} and \ac{PWN} candidates are
first reported in this analysis.

Of these fourteen sources, three are classified as \PWNClass-type.  They
are \hessj{1356}, \mshfifteenfiftytwo (\hessj{1514}), and \hessj{1825}.
\hessj{1356} and \mshfifteenfiftytwo are classified as \PWNClass-type
because of the correlation between the X-ray and \ac{VHE} emission
\citep{h.e.s.s.collaboration_2011a_discovery-source,aharonian_2005a_discovery-extended}.
\hessj{1825} is classified as \PWNClass-type because of
the energy-dependent morphology observed at \ac{VHE} energies
\citep{aharonian_2006a_energy-dependent}.  We note that \hessj{1356}
is first presented as a $\gamma$-ray emitting \ac{PWN} in this work.
Once we add the Crab Nebula and \velax (not analyzed in this work) to
this list, the total number of clearly-identified \acp{PWN} detected at
\gev energies is five.

In addition, we detect eleven \PWNcClass-type sources.  Four of these
sources are first reported in this work: \hessj{1119}, \hessj{1303},
\hessj{1420}, and \hessj{1841}.  These sources are all powered by
pulsars energetic enough to power the observed emission (\psrj{1119},
\psrj{1301}, \psrj{1420}, \psrj{1838}), and they all have a hard
spectrum which connects to the spectra observed at \ac{VHE} energies.  The
multiwavelength interpretation of the new \ac{PWN} and \ac{PWN} candidates
is discussed more thoroughly in \cite{acero_2013a_constraints-galactic}.

The remaining seven \PWNcClass-type sources have been previously
published: \hessj{1023} \citep{ackermann_2011a_fermi-lat-search},
\hessj{1640} \citep{slane_2010a_fermi-detection}, \hessj{1616}
\citep{lande_2012_search-spatially}, \hessj{1632}
\citep{lande_2012_search-spatially}, \hessj{1837}
\citep{lande_2012_search-spatially}, \hessj{1848}
\citep{tam_2010a_search-counterparts}, and \hessj{1857}
\citep{rousseau_2012a_fermi-lat-constraints}.  We classify \hessj{1848}
as \PWNcClass even though the \gev emission has a soft spectrum based
on the analysis from \cite{lemoine-goumard_2011a_fermi-lat-detection}.

We mention that three of these \PWNcClass-type sources have
\ac{LAT}-detected pulsars (\psrj{1119}, \psrj{1420}, and \psrj{1838}) and
therefore were also studied in \chapref{offpeak}.  In \chapref{offpeak},
\psrj{1119} has $\ts=61.3$ and is classified as a ``U''-type source
because the spectrum is relatively soft (spectral index $\sim2.2$).
The off-peak spectrum of this source shows both a low-energy component and
high-energy component, so most likely the off-peak emission is composted
of the pulsar at low energy and the \ac{PWN} at high energy.

For \psrj{1420}, the off-peak emission (at the position of the pulsar)
has $\tspoint=8.1$ which is significantly less then the emission observed
in the high-energy analysis (\tstev=36).  It is possible that for this
source, \tstev is overestimated due to undersubtracting the emission
of \psrj{1420}.

Finally, \psrj{1838} is significantly detected in the off-peak, but
as a soft and significantly-cutoff spectrum. This emission is also
spatially-extended and the best-fit extension incorporates both the
emission at the position of \psrj{1838} and also residual towards the
center of \hessj{1841}. Most likely, the off-peak emission of \psrj{1838}
includes both a magnetospheric component at the position of the pulsar
and \aac{PWN} component from \hessj{1841}.

\subsection{\OtherClass-type Sources}

We detected six \OtherClass-type sources.  Two of these sources
(\hessj{1634} and \hessj{1804}) have a hard spectrum which connects
spectrally to the \ac{VHE} emission but are not classified as \ac{PWN}
based upon multiwavelength considerations.  \hessj{1634} is not
a \ac{PWN} candidate because there are no pulsar counterparts able to
power it.  \hessj{1804} was suggested to be \snrg{8.7}
\citep[W30,][]{ajello_2012a_fermi-large}.

The remaining four \OtherClass-type sources have a soft spectrum
which does not connect with the \ac{VHE} emission: \hessj{1018},
\hessj{1507}, \hessj{1834}, and \verj{2016}.  \hessj{1018}
is in the region of the $\gamma$-ray binary \onefglj{1018.6}
\citep{the-fermi-lat-collaboration_2012a_periodic-emission} and also
\snrg{284.3}.  \gev emission from the region of \hessj{1507} is studied
in \cite{domainko_2012a_exploring-nature}.  \hessj{1834} and \verj{2016}
both lack pulsars energetic enough to power the observed emission.

\subsection{\PSRClass-type Sources}

In \citep{acero_2013a_constraints-galactic}, the
$\energy>10\unitspace\gev$ search for \ac{PWN} was performed both
with and without associated pulsars included in the background model.
When we did not include the \ac{LAT}-detected pulsars included in the
background model, we detected nine sources which were consistent with
magnetospheric emission.  After modeling the associated pulsars in the
background, only \hessj{1708} remained significant.  Even so, the source
was strongly influenced by the inclusion of the pulsar in the background
model, so we suspect that the emission primarily magnetospheric and that
our pulsar emission model underpredicts the true magnetospheric emission.

\chapter{Population Study of \Acstitle{LAT}-detected \Acsptitle{PWN}}
\chaplabel{population_study}

\paperref{This chapter is based the second part of the paper
  ``Constraints on the Galactic Population of TeV Pulsar Wind Nebulae using Fermi Large Area Telescope Observations''
  \citep{acero_2013a_constraints-galactic}.}

In \chapref{extended_search}, we search for new spatially-extended \fermi
sources and found that spatial extension was an important characteristic
for detecting new \acp{PWN}. In the process, we discovered three new
$\gamma$-ray emitting \acp{PWN} candidates (\hessj{1616}, \hessj{1632},
\hessj{1837}).  In \chapref{offpeak}, we then searched in the off-peak
phase interval of \ac{LAT}-detected pulsars for new \acp{PWN} and
discovered \threecfiftyeight.  Finally, in \chapref{tevcat} we searched
in the regions surrounding \acp{PWN} candidates detected at \tev energies
for \gev-emitting \acp{PWN} and detected four new \acp{PWN} candidates
(\hessj{1119}, \hessj{1303}, \hessj{1420}, and \hessj{1841}) and 1 new
PWN (\hessj{1356})

\begin{deluxetable}{lcclccc}

\tabletypesize{\tiny}
\tablecolumns{7}
\tablewidth{0pt}

\tablecaption{The muliwavelenth properties of the \ac{VHE} source and their associated \ac{LAT}-detected pulsars.
\tablabel{pwn_multiwavelenth_properties}}

\tablehead{\colhead{Source} & \colhead{\FluxPWNTeV} & \colhead{\FluxPWNKeV} & \colhead{PSR} & \colhead{$\dot{E}$} & \colhead{$\tau$} & \colhead{Distance}\\ \colhead{ } & \colhead{($10^{-12}\erg\unitspace\cm^{-2}\second^{-1}$)} & \colhead{($10^{-12}\erg\unitspace\cm^{-2}\second^{-1}$)} & \colhead{ } & \colhead{($\erg\unitspace\second^{-1}$)} & \colhead{(kyr)} & \colhead{(kpc)}}
\startdata
VER\,J0006+727 & \nodata & \nodata & PSR\,J0007+7303 & 4.5e+35 & 13.9 & $1.4 \pm 0.3$ \\
3C\,58 & $<18$ & 5.5 & PSR\,J0205+6449 & 2.6e+37 & 5.5 & 1.95 \\
Crab & $80 \pm 16$ & $21000 \pm 4200$ & PSR J0534+2200 & 4.6e+38 & 1.2 & $2.0 \pm 0.5$ \\
MGRO\,J0631+105 & \nodata & \nodata & PSR\,J0631+1036 & 1.7e+35 & 43.6 & $1.00 \pm 0.20$ \\
MGRO\,J0632+17 & \nodata & \nodata & PSR\,J0633+1746 & 3.2e+34 & 342 & $0.2_{-0.1}^{+0.2}$ \\
Vela$-$X & $79 \pm 21$ & $54 \pm 11$ & PSR\,J0835$-$4510 & 6.9e+36 & 11.3 & $0.29 \pm 0.02$ \\
HESS\,J1018$-$589 & $0.9 \pm 0.4$ & \nodata & PSR\,J1016$-$5857 & 2.6e+36 & 21 & 3 \\
HESS\,J1023$-$575 & $4.8 \pm 1.7$ & \nodata & PSR\,J1023$-$5746 & 1.1e+37 & 4.6 & 2.8 \\
HESS\,J1026$-$582 & $5.9 \pm 4.4$ & \nodata & PSR\,J1028$-$5819 & 8.4e+35 & 90 & $2.3 \pm 0.3$ \\
HESS\,J1119$-$614 & $2.3 \pm 1.2$ & \nodata & PSR\,J1119$-$6127 & 2.3e+36 & 1.6 & $8.4 \pm 0.4$ \\
HESS\,J1303$-$631 & $27 \pm 1$ & $0.16 \pm 0.03$ & PSR\,J1301$-$6305 & 1.7e+36 & 11 & $6.7_{-1.2}^{+1.1}$ \\
HESS\,J1356$-$645 & $6.7 \pm 3.7$ & $0.06 \pm 0.01$ & PSR\,J1357$-$6429 & 3.1e+36 & 7.3 & $2.5_{-0.4}^{+0.5}$ \\
HESS\,J1418$-$609 & $3.4 \pm 1.8$ & $3.1 \pm 0.1$ & PSR\,J1418$-$6058 & 4.9e+36 & 1 & $1.6 \pm 0.7$ \\
HESS\,J1420$-$607 & $15 \pm 3$ & $1.3 \pm 0.3$ & PSR\,J1420$-$6048 & 1.0e+37 & 13 & $5.6 \pm 0.9$ \\
HESS\,J1458$-$608 & $3.9 \pm 2.4$ & \nodata & PSR\,J1459$-$6053 & 9.1e+35 & 64.7 & 4 \\
HESS\,J1514$-$591 & $20 \pm 4$ & $29 \pm 6$ & PSR\,J1513$-$5906 & 1.7e+37 & 1.56 & $4.2 \pm 0.6$ \\
HESS\,J1554$-$550 & $1.6 \pm 0.5$ & $3.1 \pm 1.0$ & \nodata & \nodata & 18 & $7.8 \pm 1.3$ \\
HESS\,J1616$-$508 & $21 \pm 5$ & $4.2 \pm 0.8$ & PSR\,J1617$-$5055 & 1.6e+37 & 8.13 & $6.8 \pm 0.7$ \\
HESS\,J1632$-$478 & $15 \pm 5$ & $0.43 \pm 0.08$ & \nodata & 3.0e+36 & 20 & 3 \\
HESS\,J1640$-$465 & $5.5 \pm 1.2$ & $0.46 \pm 0.09$ & \nodata & 4.0e+36 & \nodata & \nodata \\
HESS\,J1646$-$458B & $5.0 \pm 2.0$ & \nodata & PSR\,J1648$-$4611 & 2.1e+35 & 110 & $5.0 \pm 0.7$ \\
HESS\,J1702$-$420 & $9.0 \pm 3.0$ & $0.01 \pm 0.00$ & PSR\,J1702$-$4128 & 3.4e+35 & 55 & $4.8 \pm 0.6$ \\
HESS\,J1708$-$443 & $23 \pm 7$ & \nodata & PSR\,J1709$-$4429 & 3.4e+36 & 17.5 & $2.3 \pm 0.3$ \\
HESS\,J1718$-$385 & $4.3 \pm 1.6$ & $0.14 \pm 0.03$ & PSR\,J1718$-$3825 & 1.3e+36 & 89.5 & $3.6 \pm 0.4$ \\
HESS\,J1804$-$216 & $12 \pm 2$ & $0.07 \pm 0.01$ & PSR\,J1803$-$2137 & 2.2e+36 & 16 & $3.8_{-0.5}^{+0.4}$ \\
HESS\,J1809$-$193 & $19 \pm 6$ & $0.23 \pm 0.05$ & PSR\,J1809$-$1917 & 1.8e+36 & 51.3 & $3.5 \pm 0.4$ \\
HESS\,J1813$-$178 & $5.0 \pm 0.6$ & \nodata & PSR\,J1813$-$1749 & 6.8e+37 & 5.4 & 4.7 \\
HESS\,J1818$-$154 & $1.3 \pm 0.9$ & \nodata & PSR\,J1818$-$1541 & 2.3e+33 & 9 & $7.8_{-1.4}^{+1.6}$ \\
HESS\,J1825$-$137 & $61 \pm 14$ & $0.44 \pm 0.09$ & PSR\,J1826$-$1334 & 2.8e+36 & 21 & $3.9 \pm 0.4$ \\
HESS\,J1831$-$098 & $5.1 \pm 0.6$ & \nodata & PSR\,J1831$-$0952 & 1.1e+36 & 128 & $4.0 \pm 0.4$ \\
HESS\,J1833$-$105 & $2.4 \pm 1.2$ & $40 \pm 0$ & PSR\,J1833$-$1034 & 3.4e+37 & 4.85 & $4.7 \pm 0.4$ \\
HESS\,J1837$-$069 & $23 \pm 9$ & $0.64 \pm 0.24$ & PSR\,J1836$-$0655 & 5.5e+36 & 2.23 & $6.6 \pm 0.9$ \\
HESS\,J1841$-$055 & $23 \pm 3$ & \nodata & PSR\,J1838$-$0537 & 5.9e+36 & 4.97 & 1.3 \\
HESS\,J1846$-$029 & $9.0 \pm 1.5$ & $29 \pm 1$ & PSR\,J1846$-$0258 & 8.1e+36 & 0.73 & 5.1 \\
HESS\,J1848$-$018 & $4.3 \pm 1.0$ & \nodata & \nodata & \nodata & \nodata & 6 \\
HESS\,J1849$-$000 & $2.1 \pm 0.4$ & $0.90 \pm 0.20$ & PSR\,J1849$-$001 & 9.8e+36 & 42.9 & 7 \\
HESS\,J1857+026 & $18 \pm 3$ & \nodata & PSR\,J1856+0245 & 4.6e+36 & 20.6 & $9.0 \pm 1.2$ \\
MGRO\,J1908+06 & $12 \pm 5$ & \nodata & PSR\,J1907+0602 & 2.8e+36 & 19.5 & $3.2 \pm 0.3$ \\
HESS\,J1912+101 & $7.3 \pm 3.7$ & \nodata & PSR\,J1913+1011 & 2.9e+36 & 169 & $4.8_{-0.7}^{+0.5}$ \\
VER\,J1930+188 & $2.3 \pm 1.3$ & $5.2 \pm 0.1$ & PSR\,J1930+1852 & 1.2e+37 & 2.89 & $9_{-2}^{+7}$ \\
VER\,J1959+208 & \nodata & \nodata & PSR\,J1959+2048 & 1.6e+35 & \nodata & $2.5 \pm 1.0$ \\
MGRO\,J2019+37 & \nodata & \nodata & PSR\,J2021+3651 & 3.4e+36 & 17.2 & $10_{-4}^{+2}$ \\
MGRO\,J2228+61 & \nodata & $0.88 \pm 0.02$ & PSR\,J2229+6114 & 2.2e+37 & 10.5 & $0.80 \pm 0.20$ \\
\enddata

\tablecomments{The multiwavelenth properties of \ac{LAT}-detected \ac{PWN}
candidates.  This table includes the
X-ray flux in the $2\unitspace\tev$ to $30\unitspace\tev$ energy range
(\FluxPWNKeV) and the \ac{VHE} flux in the $1\unitspace\tev$
to $30\unitspace\tev$ range (\FluxPWNTeV).  In addition, this table
includes the names of the associated pulsars and their spin-down energy,
age, and distance. For several sources, no associated pulsar has been
detected, but properties from an assumed pulsar can be estimated.
The references for all sources in this table for except 3C\,58 can be
found in \cite{acero_2013a_constraints-galactic}.  For 3C\,58, we took the
X-ray flux from \cite{torii_2000a_observations-crab-like}, the \ac{VHE}
flux upper limit from \cite{konopelko_2008a_observations-pulsar}, and the
pulsar properties from \citep{abdo_2013a_second-fermi}.  We note that
there is no error reported on the X-ray flux measurement.}

\end{deluxetable}

In this chapter, we take the population of $\gamma$-ray
emitting \acp{PWN} and \acp{PWN} candidates and study how their
multiwavelength properties evolve with properties of their host pulsars.
In \tabref{pwn_multiwavelenth_properties}, we compile the multiwavelength
properties of the \ac{VHE} sources studied in \chapref{tevcat}. In
particular, we include the spectra of the \acp{PWN} observed at X-ray and \ac{VHE}
energies, and the observed spin-down powers, ages, and distances of the associated
pulsars.

\begin{figure}[htbp]
  \centering
  \includegraphics{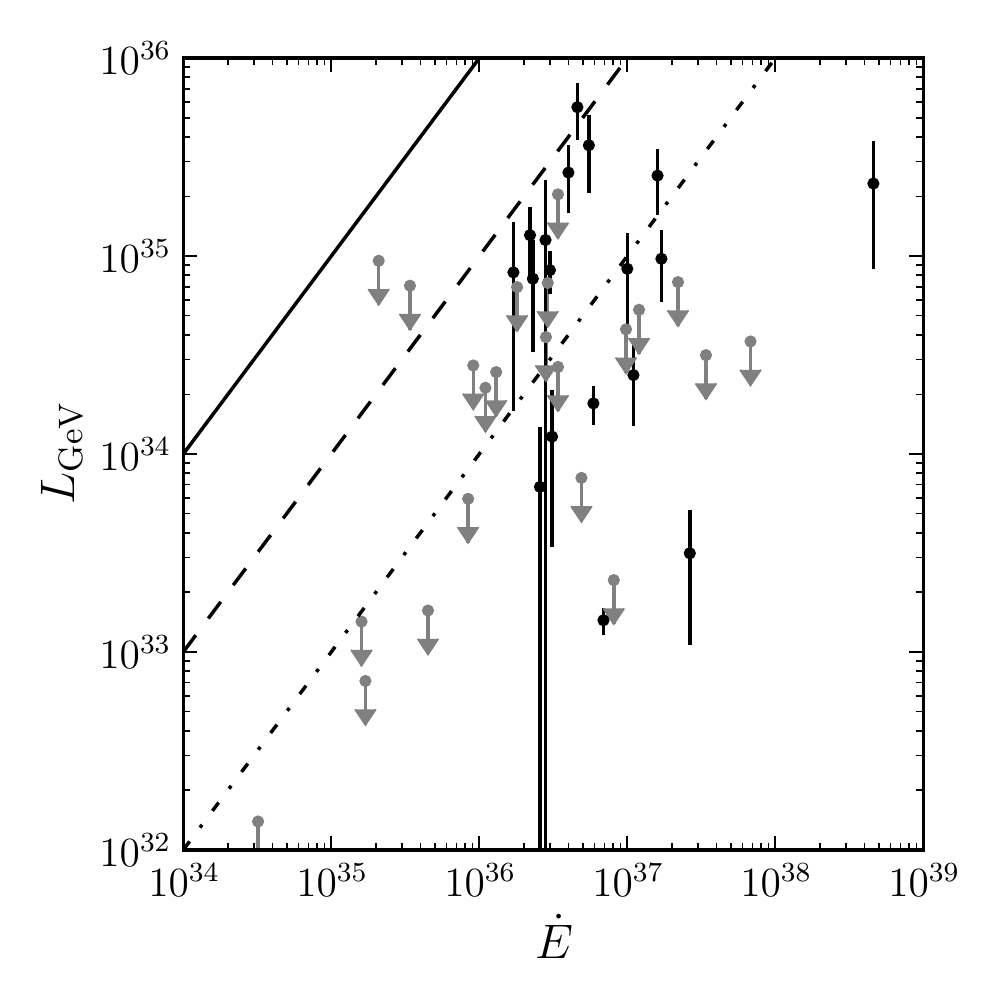}
  \caption{The observed $\gamma$-ray luminosities plotted against the
  observed spin-down luminosities for the \acp{PWN} candidates from
  \tabref{pwn_multiwavelenth_properties}.}
  \figlabel{pwn_luminosity_vs_edot}
\end{figure}

In \figref{pwn_luminosity_vs_edot}, we compare the observed luminosity at
\gev energies to the spin-down power of the observed pulsar.  This plot
shows that all \ac{LAT}-detected \acp{PWN} emit a fraction $\lesssim
10\%$ of their spin-down energy into the $\gamma$-ray emission from
the \ac{PWN}.

\begin{figure}[htbp]
  \centering
  \includegraphics{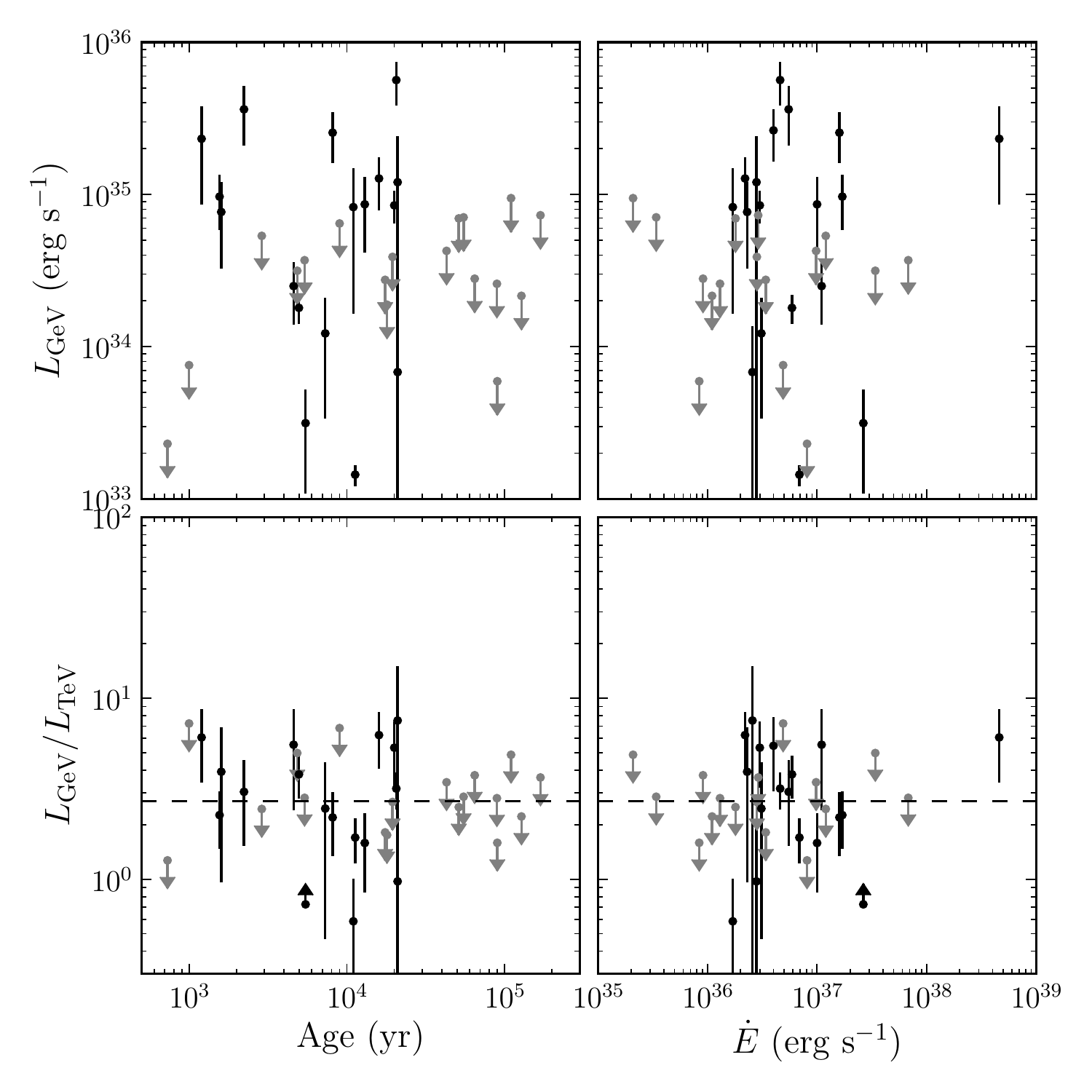}
  \caption{The observed $\gamma$-ray luminosities and the
  \gev to \tev luminosity ratios as a function of the ages
  and spin-down energies for the \ac{PWN} candidates 
  from \tabref{pwn_multiwavelenth_properties}.  The dotted line
  corresponds to the average luminosity ratio (\MeanLuminosityRatio).
  Because \hessj{1708} is classified as being a \PSRClass-type
  source in \chapref{tevcat}, we consider the observed $\gamma$-ray
  luminosity of it to be an upper limit on the \ac{PWN} emission.
  \figlabel{pwn_age_edot_vs_l_gev.pdf}.}
\end{figure}

Next, in \figref{pwn_age_edot_vs_l_gev.pdf} we compare compare the \gev
luminosities and \gev to \tev luminosity ratios as a function of ages
and spin-down energies of the host pulsars.  These plots shows that
there is no correlation between the \gev luminosities or \gev to \tev
luminosity ratios and the ages and spin-down energies of the associate
pulsars.  In addition, we overlay the mean \gev to \tev luminosity ratio
($\MeanLuminosityRatio=2.7_{-1.4}^{+2.7}$).

\begin{figure}[htbp]
  \centering
  \includegraphics{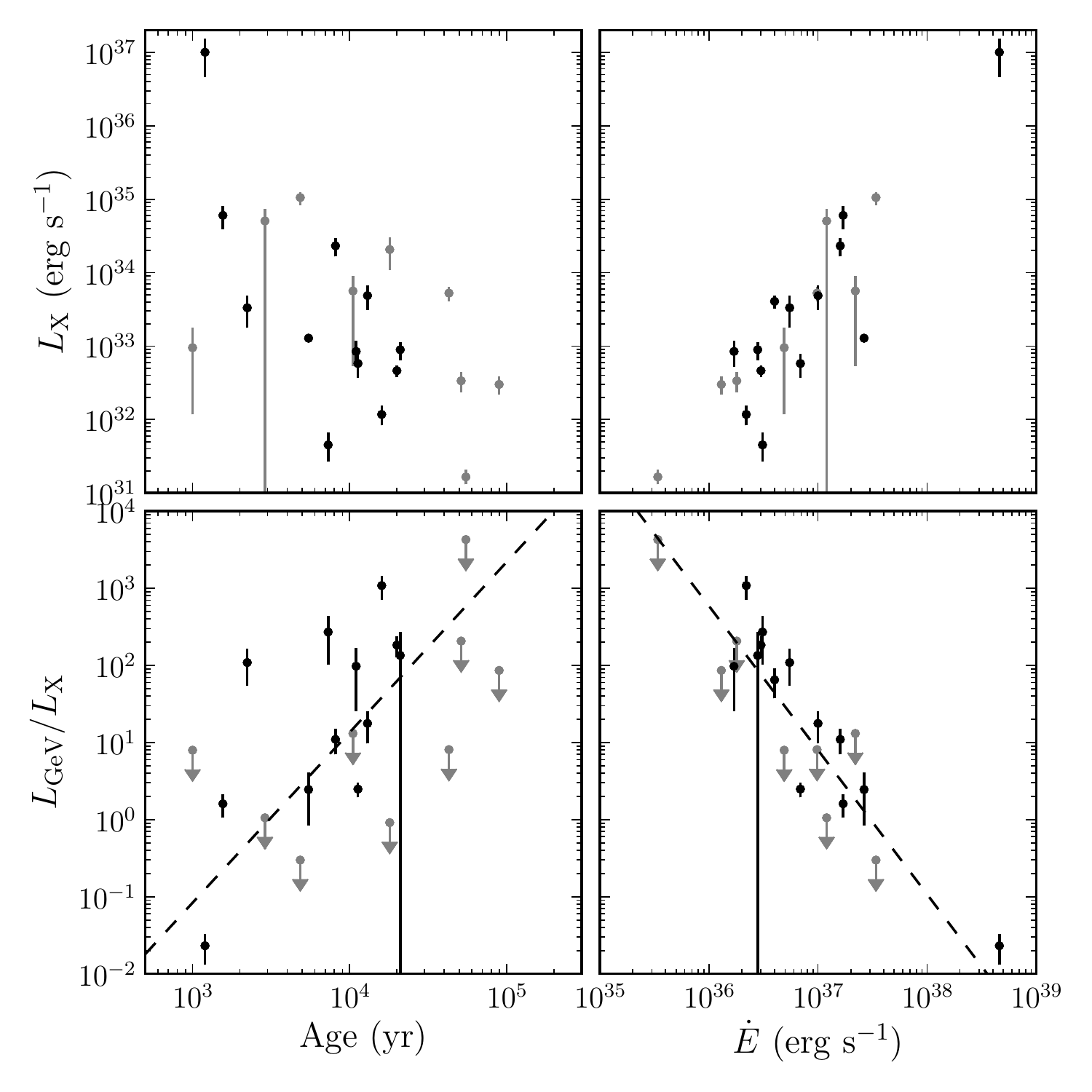}
  \caption{The observed X-ray flux and \gev to \tev luminosity ratios
  as a function of the age and spin-down energies for the \ac{PWN}
  candidates from \tabref{pwn_multiwavelenth_properties}.
  The dotted line corresponds to the scaling relationships from
  \cite{mattana_2009_evolution-gamma-} for the \tev to X-ray
  luminosity ratio scaled by the average \gev to \tev luminosity
  (\MeanLuminosityRatio).  We caution that \threecfiftyeight does not
  have a X-ray luminosity error.}
  \figlabel{pwn_age_edot_vs_l_xray}
\end{figure}

Finally, in \figref{pwn_age_edot_vs_l_xray} we compare the distribution
of the X-ray luminosities and the \gev to X-ray luminosity ratios as a
function of the ages and spin-down energies of the pulsars.  This plot
shows that the X-ray luminosity decreases with pulsar age and increases
with spin-down energy. Similarly, the \gev to X-ray luminosity increases
with age and decreases with energy.  These correlations are consistent
with the simple model predicted in \cite{mattana_2009_evolution-gamma-}
(See \secref{pwn_emission}) and also with the observed \ac{VHE}
relationships from the same paper.

\chapter{Outlook}
\chaplabel{outlook}

Since the observation of the Crab Nebula
in \citeyear{weekes_1989a_observation-gamma}
\citep{weekes_1989a_observation-gamma}, we have learned much about the
high-energy \ac{IC} emission from \ac{PWN}. The current generation of
\ac{VHE} experiments (\ac{HESS}, Magic, and Veritas) have drastically
expanded the population of \ac{PWN} observed at $\gamma$-ray energies and
\acp{PWN} are the most populous class of \ac{VHE} sources in the Galaxy.
Now, using the \ac{LAT} on board \fermi, we have detected a large
fraction of these \ac{VHE} \ac{PWN} at \gev energies and one \ac{PWN}
not yet detected at \ac{VHE} energies.

The next great improvement in our knowledge of \ac{PWN} will most
likely come from next-generation \acp{IACT}. The proposed \ac{CTA}
\citep{actis_2011a_design-concepts} will have a much improved effective
area and angular resolution, allowing for the discovery of more \ac{VHE}
\ac{PWN} as well as improved imaging of \ac{PWN} candidates.

As was the case for \hessj{1825}, energy-dependent morphology at \ac{VHE}
energies can be used to unambiguous identify \ac{VHE} emission as being
caused by a pulsar \citep{aharonian_2006a_energy-dependent}.  Similarly,
\cite{van-etten_2011a_multi-zone-modeling} showed for \hessj{1825} that
detailed spatial and spectral observations combined with multi-zone
modeling can constrain the properties of the \ac{PWN}.  Detailed
energy-dependent imaging of a larger sample of \ac{PWN} by \ac{CTA}
will allow us a greater understanding of the physics of pulsar winds.

In addition, the Crab nebula has challenged our basic understand of the
physics of \acp{PWN}. It is possible that more detailed observations could
uncover additional variable \ac{PWN} and this could help to explain the
nature of this variable emission.

Finally, because of the high density of \ac{VHE} \ac{PWN} in the galactic
plane, it is important to identify \ac{VHE} sources as \ac{PWN} to assist
in the search for new source classes.  There is significant potential
for the discovery of new \ac{VHE} source classes, but only after the
numerous \ac{VHE} \ac{PWN} are classified.  If the past is any guide to
the future, there is much still to be learned.

\bibliographystyle{macros/apj}

\bibliography{lande_thesis_stanford}

\end{document}